\pdfoutput=1

\documentclass[11pt,twoside,a4paper,cmspaper,final,collab]{cms-tdr}
\def\svnVersion{246091}\def\cmsCernNo{2014-001}\def\cmsCernDate{\today}\def\cmsMessage{Published in the Journal of High Energy Physics as \href{http://dx.doi.org/10.1007/JHEP05(2014)104}{\doi{10.1007/JHEP05(2014)104.}}}

\begin{document}\cmsNoteHeader{HIG-13-004}

\hyphenation{had-ron-i-za-tion}
\hyphenation{cal-or-i-me-ter}
\hyphenation{de-vices}

\RCS$Revision: 236066 $
\RCS$HeadURL: svn+ssh://svn.cern.ch/reps/tdr2/papers/HIG-13-004/trunk/HIG-13-004.tex $
\RCS$Id: HIG-13-004.tex 236066 2014-04-10 11:23:47Z cbern $
\newlength\cmsFigWidth
\ifthenelse{\boolean{cms@external}}{\setlength\cmsFigWidth{0.85\columnwidth}}{\setlength\cmsFigWidth{0.4\textwidth}}
\ifthenelse{\boolean{cms@external}}{\providecommand{\cmsLeft}{top}}{\providecommand{\cmsLeft}{left}}
\ifthenelse{\boolean{cms@external}}{\providecommand{\cmsRight}{bottom}}{\providecommand{\cmsRight}{right}}
\newcommand{\xsecbr}{\sigma_\phi\cdot B_{\Pgt\Pgt}}
\newcommand{\xsecbrH}{\sigma_H\cdot B_{\Pgt\Pgt}}
\newcommand{\xsecbshort}{\sigma \times B}
\newcommand{\MT}{\ensuremath{m_\mathrm{T}}\xspace}
\newcommand{\vecMET}{\ensuremath{\vec{E}_\mathrm{T}^\mathrm{miss}}\xspace}
\newcommand{\vecpt}{\ensuremath{\vec{p}_\mathrm{T}}\xspace}
\newcommand{\mvis}{\ensuremath{m_\text{vis}}\xspace}
\newcommand{\mtt}{\ensuremath{m_{\Pgt\Pgt}}\xspace}
\newcommand{\LT}{\ensuremath{L_\mathrm{T}}\xspace}
\newcommand{\Hi}{\ensuremath{\mathrm{H}}}
\newcommand{\W}{\PW}
\newcommand{\WW}{\PW\PW}
\newcommand{\emt}{\ensuremath{\Pe\Pgm\Pgt_h}}
\newcommand{\mmt}{\ensuremath{\Pgm\Pgm\Pgt_h}}
\newcommand{\WH}{\ensuremath{\W\Hi}}
\newcommand{\ZH}{\ensuremath{\Z\Hi}}
\newcommand{\PZ}{\cPZ}
\newcommand{\hww}{\Hi\to\WW}
\newcommand{\htt}{\ensuremath{\Hi\to\tau\tau}\xspace}
\newcommand{\pth}{\ensuremath{\pt^{\tau\tau}}\xspace}
\newcommand{\Pgth}{\ensuremath{\Pgt_{\rm h}}\xspace}
\newcommand{\mjj}{\ensuremath{m_\mathrm{jj}}\xspace}
\newcommand{\detajj}{\ensuremath{\Delta \eta_\mathrm{jj}}\xspace}
\renewcommand{\Pm}{\ensuremath{\mu}\xspace}
\newcommand{\bestfitmu}{\ensuremath{\hat \mu = 0.78 \pm 0.27}\xspace}
\newcommand{\kappav}{\ensuremath{\kappa_\mathrm{V}}\xspace}
\newcommand{\kappaf}{\ensuremath{\kappa_\mathrm{f}}\xspace}

\newcommand{\emu}{\Pe\Pgm}
\newcommand{\ee}{\Pe\Pe}
\newcommand{\mumu}{\Pgm\Pgm}
\newcommand{\mutau}{\Pgm\Pgth}
\newcommand{\etau}{\Pe\Pgth}
\newcommand{\tautau}{\Pgth\Pgth}

\providecommand{\mH}{\ensuremath{m_{\PH}}\xspace}
\cmsNoteHeader{HIG-13-004} 
\title{Evidence for the 125\GeV Higgs boson decaying to a pair of $\tau$ leptons}



























\date{\today}

\abstract{
A search for a standard model Higgs boson decaying into a pair of $\tau$ leptons is performed using events recorded by the CMS experiment at the LHC in 2011 and 2012. The dataset corresponds to an integrated luminosity of 4.9\fbinv at a centre-of-mass energy of 7\TeV and 19.7\fbinv at 8\TeV.
Each $\tau$ lepton decays hadronically or leptonically to an electron or a muon,
leading to six different final states for the $\tau$-lepton pair,
all considered in this analysis.
An excess of events is observed over the expected background contributions,
with a local significance larger than 3 standard deviations for \mH values between 115 and 130\GeV.
The best fit of the observed $\PH \to \tau \tau$ signal cross section times branching fraction for $\mH= 125$\GeV is
$0.78 \pm 0.27$
times the standard model expectation.
These observations constitute evidence for the 125\GeV Higgs boson decaying to a pair of $\tau$ leptons.
}

\hypersetup{%
pdfauthor={CMS Collaboration},%
pdftitle={Evidence for the 125 GeV Higgs boson decaying to a pair of tau leptons},%
pdfsubject={CMS},%
pdfkeywords={CMS, physics, software, computing}}

\maketitle 

\section{Introduction}

Elucidating the mechanism of electroweak symmetry breaking,
through which the $\PW$\ and $\cPZ$\ bosons become massive,
is an important goal of the Large Hadron Collider (LHC) physics programme.
In the standard model (SM)~\cite{SM1,SM3}, electroweak symmetry breaking is achieved via the  Brout--Englert--Higgs
mechanism~\cite{Englert:1964et,Higgs:1964ia,Higgs:1964pj,Guralnik:1964eu,Higgs:1966ev,Kibble:1967sv},
which also predicts the existence of a scalar Higgs boson.
On July 4, 2012, the discovery of a new boson with a mass around 125\GeV
was announced at CERN by the ATLAS and CMS
Collaborations~\cite{ATLASobservation125,CMSobservation125,CMSobservation125Long}.
The excess was most significant in the $\cPZ\cPZ$, $\gamma \gamma$, and $\PW\PW$ decay modes.
The spin and CP properties of the new boson are compatible with those of the SM
Higgs boson~\cite{Chatrchyan:2012jja,Aad:2013xqa}.
In the SM, the masses of the fermions are generated via the Yukawa couplings between the
Higgs field and the fermionic fields.
The measurement of these couplings is essential for identifying this boson as the SM Higgs boson.
The $\tau \tau$ decay mode is the most promising
because of the large event rate expected in the SM compared to the other leptonic decay modes
and the smaller contribution from background events with respect to the $\cPqb\cPaqb$ decay mode.

Searches for SM Higgs bosons decaying to a $\tau$-lepton pair have been performed at the LEP, Tevatron, and LHC colliders.
The collaborations at LEP have searched for associated $\PZ\PH$ production and found no significant excess of events over the background expectation~\cite{Barate:2000ts,Abdallah:2003ip,Achard:2001pj,Abbiendi:2000ac}.
Dedicated searches in the $\tau\tau$ final state have been carried out at the Tevatron and at the LHC,
placing upper limits on the Higgs boson production cross section times branching fraction,
denoted as $\xsecbshort$,
at the 95\% confidence level (CL).
Using $\Pp\Pap$ collisions at $\sqrt{s} = 1.96$\TeV,
the CDF Collaboration excluded values larger than 16 times $(\xsecbshort)^{125\GeV}_\text{SM}$~\cite{Aaltonen:2012jh}, with $(\xsecbshort)^{125\GeV}_\text{SM}$ denoting the SM prediction for $\xsecbshort$
with $\mH = 125$\GeV,
and the D0 Collaboration excluded values larger than 14 times $(\xsecbshort)^{125\GeV}_\text{SM}$~\cite{Abazov:2012zj}.
With $\Pp\Pp$ collisions at $\sqrt{s}=7$\,\TeV, the ATLAS Collaboration
found an observed (expected) upper limit of $3.7\,(3.5)$ times $( \xsecbshort)^{125\GeV}_\text{SM}$~\cite{Aad:2012mea},
whereas the CMS Collaboration placed an observed (expected) upper limit of $4.2\,(3.1)$   times $(\xsecbshort)^{\text{125\GeV}}_\text{SM}$~\cite{Chatrchyan:2012vp}.

This paper reports on the results of a search for a SM Higgs
boson using final states with a pair of $\tau$ leptons in proton-proton collisions at $\sqrt{s}=7$ and 8\TeV at the LHC.
We use the entire dataset collected in 2011 and 2012 by the CMS experiment
corresponding to an
integrated luminosity of 4.9\fbinv at a centre-of-mass energy of 7\TeV and 19.7\fbinv at 8\TeV.

The paper is organized as follows.
An overview of the analysis strategy is given in section~\ref{sec:analysis_overview},
while the CMS detector, the event reconstruction, and the Monte Carlo (MC) simulation are described in section~\ref{sec:reconstruction}.
The event selection is summarized in section~\ref{sec:event_selection},
followed by the description of the reconstruction of the $\tau$-lepton pair invariant mass in section~\ref{sec:htt_mass}
and the categorization of events in section~\ref{sec:event_categories}.
The background estimation is based on data control regions whenever possible and is explained in section~\ref{sec:background_estimation}.
Finally, systematic uncertainties are summarized in section~\ref{sec:systematics}
and the results are presented in section~\ref{sec:results}.

\section{Analysis overview}
\label{sec:analysis_overview}

Throughout this paper, the symbol $\Pgth$ denotes the reconstructed hadronic decay of a $\tau$ lepton.
The $\Pgth$ candidates are reconstructed in decay modes with one or three charged particles (see section~\ref{sec:reconstruction}).
The symbol $\ell$ refers to an electron or a muon, and the symbol $L$ to any kind of reconstructed charged lepton,
namely electron, muon, or $\Pgth$.

\begin{figure}[htbp]
\centering
\includegraphics[width=0.32\textwidth]{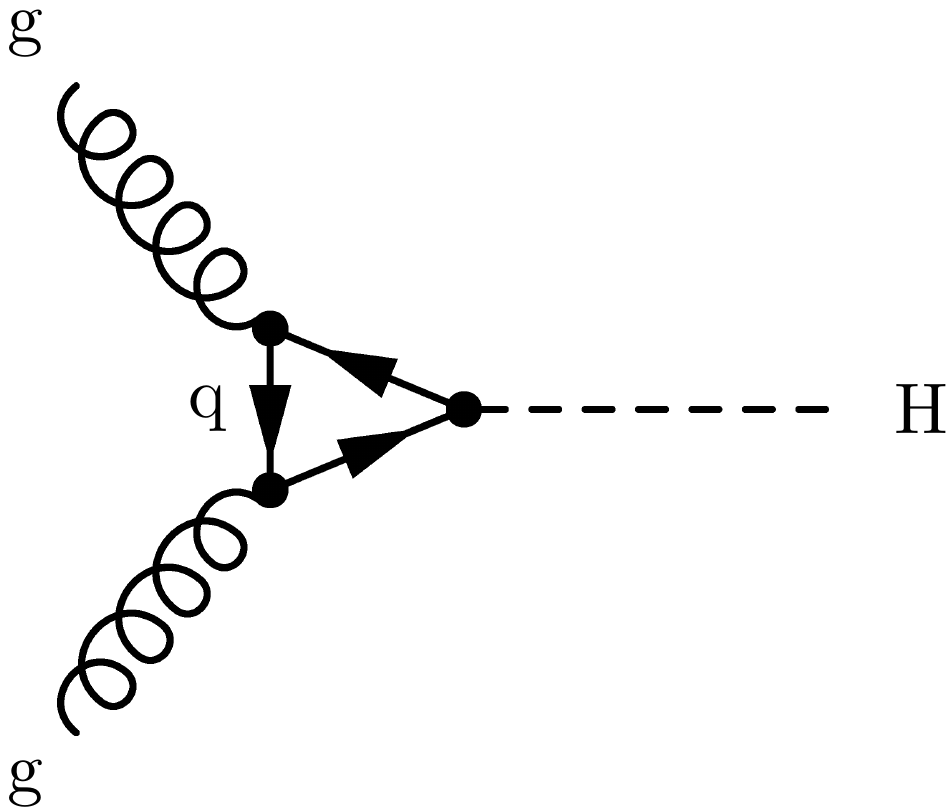}
\includegraphics[width=0.32\textwidth]{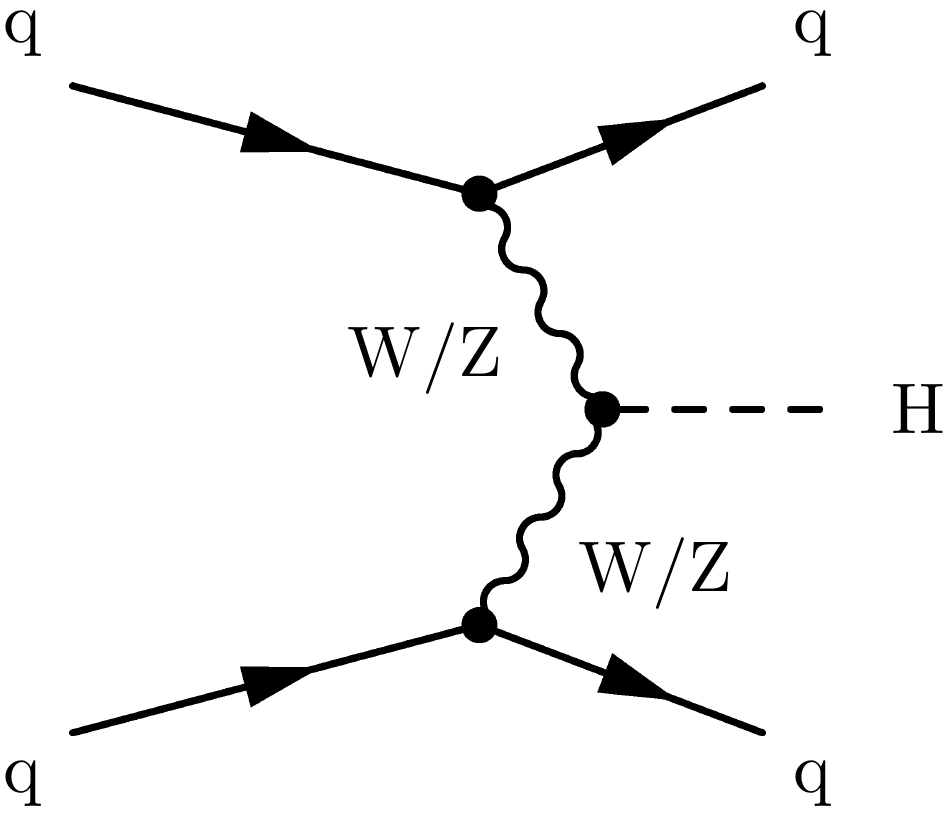}
\includegraphics[width=0.32\textwidth]{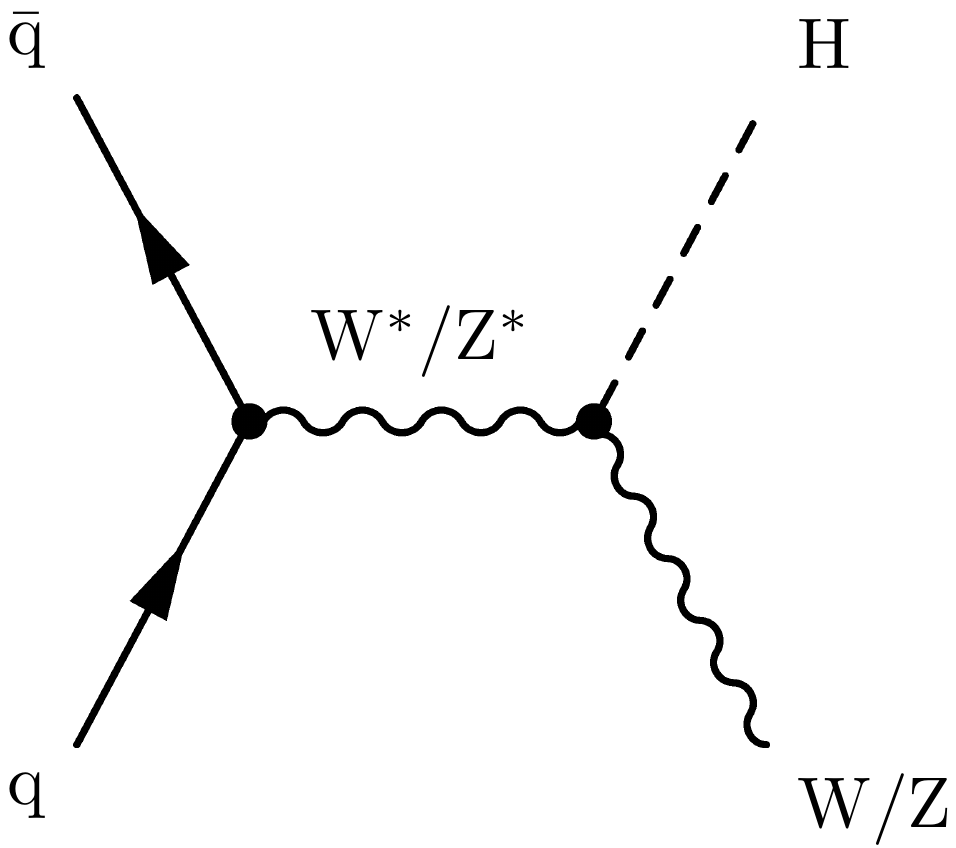}
\caption{
Leading-order Feynman diagrams for Higgs boson production through gluon-gluon fusion (left), vector boson fusion (middle), and the associated production with a $\PW$ or a $\PZ$ boson (right).
}
\label{fig:prodmech}
\end{figure}

The main Higgs boson production mechanisms, shown in figure~\ref{fig:prodmech}, lead to final states with a different number of charged leptons.
For Higgs boson production through gluon-gluon fusion and vector boson fusion (VBF),
final states with $\htt$ decays contain only two charged leptons,
defining the $LL'$ channels.
All six $\tau$-pair final states are studied: $LL' = \Pgm\Pgth$, $\Pe\Pgth$,
$\Pgth\Pgth$, $\Pe\Pgm$, $\Pgm\Pgm$, and $\Pe\Pe$.

Sensitivity to the associated production with a $\PW$ or a $\PZ$ boson is achieved by requiring one or two additional electrons or muons compatible with leptonic decays of the $\PW$ or $\PZ$ boson.
The four most sensitive final states are retained in the $\ell + L \Pgth$ channels aiming at the associated production with a $\PW$ boson, $\ell + L \Pgth = \Pgm + \Pgm\Pgth$, $\Pe + \Pgm\Pgth/\Pgm + \Pe\Pgth$, $\Pgm + \Pgth\Pgth$, and $\Pe + \Pgth\Pgth$.
In the $\ell \ell + LL'$ channels that target the associated production with a $\PZ$ boson decaying to $\ell \ell$, the $\tau$-pair final states $\Pgm\Pgth$, $\Pe\Pgth$,
$\Pe\Pgm$, and $\Pgth\Pgth$ are considered, leading to eight channels in total.
The $\Pe\Pe$ and $\Pgm\Pgm$ $\tau$-pair final states are excluded because the corresponding events are already used in the search for $\PH\to\PZ\PZ\to4\ell$~\cite{CMSobservation125Long}.

To maximize the sensitivity of the analysis in the $LL'$ channels, events are classified in categories according to the number of jets in the final state, excluding the jets corresponding to the $L$ and $L'$ leptons.
The events are further classified according to a number of kinematic quantities
that exhibit different distributions for signal and background events (see section~\ref{sec:event_categories}).
In particular, the contribution of the VBF production process is enhanced for events with two or more jets by requiring a large rapidity gap between the two jets with the highest transverse momentum.
For the remaining events with at least one jet, requiring a large $\pt$ of the reconstructed Higgs boson candidate increases the sensitivity to Higgs boson production through gluon fusion.
A complete listing of all lepton final states and
event categories
is given in appendix~\ref{sec:event_yields}.

With the exception of the $\ell + L \Pgth$, $\Pe\Pe$, and $\Pgm\Pgm$ channels,
the
signal is extracted from the distribution of the invariant mass of the $\tau$-lepton pair, $\mtt$,
calculated from the $L$ and $L'$ four-momenta and the missing transverse energy vector.
In the $\ell + L \Pgth$ channels, the signal extraction is instead based on the invariant mass, $\mvis$, of the visible $L\Pgth$ decay products
because the missing transverse energy does not entirely arise from the neutrinos produced in the decay of the two $\tau$ leptons.
In the $\Pe\Pe$ and $\Pgm\Pgm$ channels, a discriminating variable combining a number of kinematic quantities and other observables, including $\mtt$, is used.

The background composition depends on the channel and, in particular, on the number of electrons and muons in the final state.
The Drell--Yan production of a $\PZ$ boson decaying into a pair of $\tau$ leptons constitutes the main irreducible background in all $LL'$ channels.
Another source of background with the same leptonic final state
is the production of top-quark pairs ($\ttbar$), which is most important in the $\emu$ channel.
Reducible background contributions include QCD multijet production that is particularly relevant in the $\tautau$ channel and $\PW(\to \ell\nu)+\text{jets}$ production with a jet misidentified as a $\Pgth$ in the $\ell \Pgth$ channels.
In the $\ell+L\Pgth$ and $\ell \ell+LL'$ channels, diboson production is the largest irreducible background.

While the signal contribution is expected to be a pure sample of $\htt$ decays in many channels considered,
there is a significant contribution from $\hww$ decays in the $\ell + \ell' \Pgth$ and the $\ell \ell+LL'$ channels, and, most importantly, in the two-jet event samples of the $\emu$, $\ee$, and $\mumu$ channels.
The contribution from $\hww$ decays is treated as a background in the search for $\htt$ decays.
Given the discovery of a SM-like Higgs boson with the mass near $125$\GeV,
this contribution is taken from the expectation for a SM Higgs boson with $\mH = 125$\GeV.
On the other hand, the presence of a $\hww$ contribution provides additional sensitivity to the coupling of the Higgs boson to vector bosons.
Therefore, the $\hww$ contribution is treated as a signal process for the measurement of the fermionic and the bosonic couplings of the Higgs boson.

\section{The CMS experiment}
\label{sec:reconstruction}

The central feature of the CMS apparatus is a superconducting solenoid of 6\unit{m} internal diameter, providing a magnetic field of 3.8\unit{T}.
Within the volume of the superconducting solenoid are a silicon pixel and strip tracker, a lead tungstate crystal electromagnetic calorimeter, and a brass/scintillator hadron calorimeter.
The coverage of these calorimeters is complemented by extensive forward calorimetry.
Muons are detected in gas-ionization chambers embedded in the steel flux return yoke outside the solenoid.
The first level of the CMS trigger system (L1), composed of custom hardware processors, uses information from the calorimeters and muon detectors to select the most interesting events in a fixed time interval of less than 4\mus.
The high-level trigger (HLT) processor farm further decreases the event rate from around 100\unit{kHz} to around 300\unit{Hz}, before data storage.
A more detailed description of the CMS detector can be found in ref.~\cite{CMS-JINST}.

The CMS experiment uses a right-handed coordinate system, with the origin at the nominal interaction point, the $x$ axis pointing to the centre of the LHC, the $y$ axis pointing up (perpendicular to the LHC plane), and the $z$ axis along the anticlockwise-beam direction.
The polar angle $\theta$ is measured from the positive $z$ axis and the azimuthal angle $\phi$ is measured in the transverse $(x,y)$ plane.
The pseudorapidity is defined as $\eta \equiv - {\ln}[{\tan} (\theta / 2)]$.

The number of inelastic proton-proton collisions occurring per LHC bunch crossing was, on average, 9 in 2011 and 21 in 2012.
The tracking system is able to separate collision vertices as close as 0.5\mm along the beam direction~\cite{IEEE_DetAnnealing}.
For each vertex, the sum of the squared transverse momenta of all associated tracks is computed.
The vertex for which this quantity is the largest is assumed to correspond to the hard-scattering
process and is referred to as the primary vertex.
The additional proton-proton collisions happening in the same bunch crossing are termed pileup (PU).

A particle-flow (PF) algorithm~\cite{CMS-PAS-PFT-09-001, CMS-PAS-PFT-10-002, CMS-PAS-PFT-10-003} combines the information
 from the CMS subdetectors to identify and reconstruct the particles emerging from proton-proton collisions:
charged hadrons, neutral hadrons, photons, muons, and electrons.
These particles are then used to reconstruct the missing transverse energy vector $\vecMET$, the jets, the $\Pgth$ candidates,
and to quantify the lepton isolation.
Jets are reconstructed from all particles using the anti-$k_T$ jet clustering algorithm
implemented in \textsc{fastjet}~\cite{Cacciari:fastjet1, Cacciari:fastjet2}, with a
distance parameter of 0.5.  
The jet energy scale is calibrated through correction factors that depend on the \pt and $\eta$ of the jet~\cite{CMS-JME-10-011}.
Jets originating from the hadronization of b quarks are identified using the combined secondary vertex (CSV) algorithm~\cite{Chatrchyan:2012jua} which exploits observables related to the long lifetime of b hadrons.
The b-tagging efficiencies in simulation are corrected for differences between simulated and recorded events.
Jets originating from PU are identified and rejected based on both vertex information and jet shape information~\cite{CMS-PAS-JME-13-005}.
All particles reconstructed in the event are used to determine the \vecMET (and its magnitude, \MET) with a high, PU-independent, resolution~\cite{CMS-JME-12-002} using a multivariate regression technique based on a boosted decision tree (BDT)~\cite{Hocker:2007ht}.

\begin{figure}[htbp]
\centering
\includegraphics[width=0.8\textwidth]{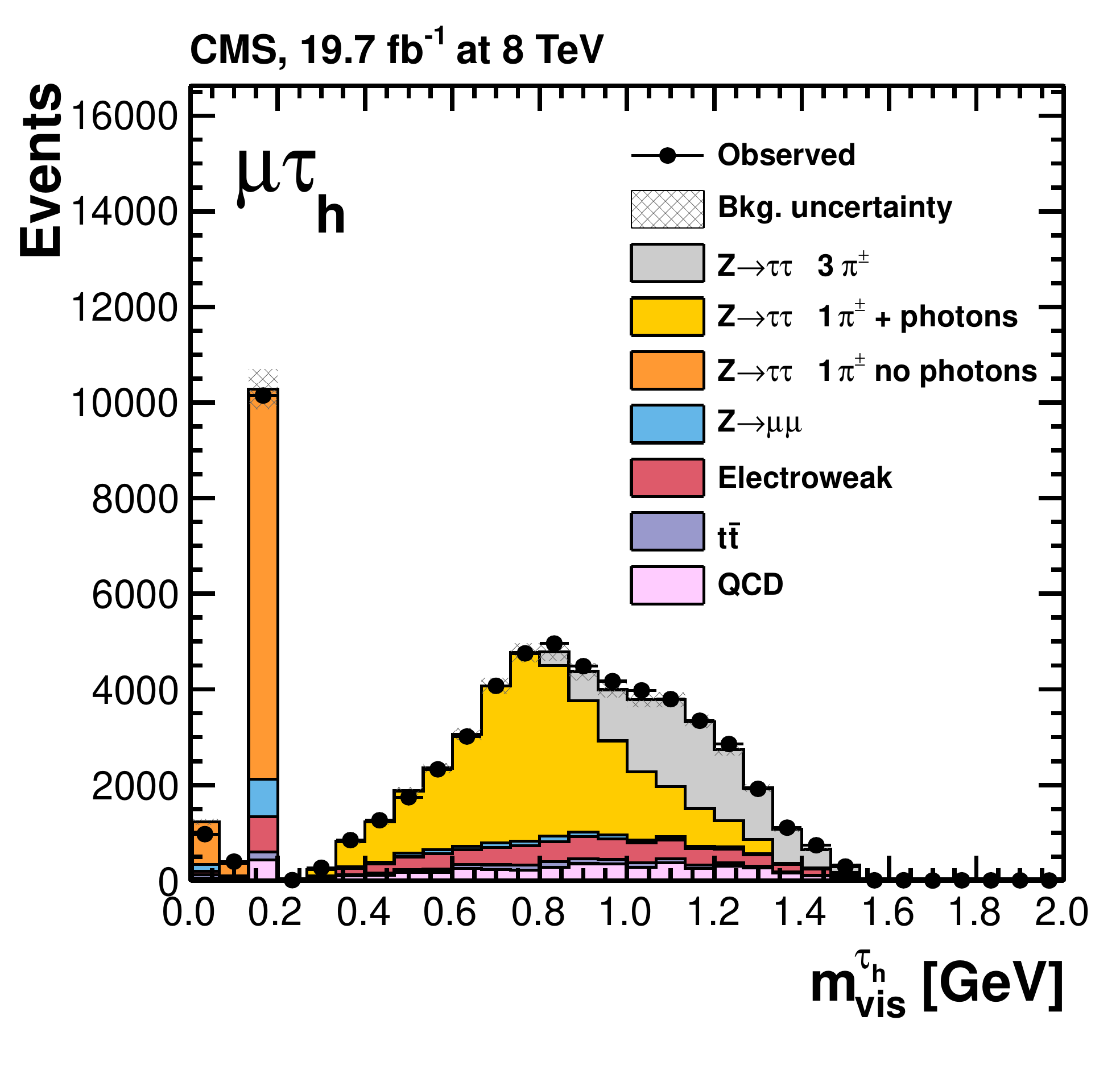}
\caption{
Observed and predicted distributions for the visible $\Pgth$ mass, $\mvis^{\Pgth}$,
in the $\mutau$ channel after the baseline selection described in section~\ref{sec:event_selection}.
The yields predicted for the $\PZ \to \tau \tau$, $\PZ \to \mu \mu$, electroweak, $\ttbar$, and QCD multijet background contributions correspond to the result of the final fit presented in Section~\ref{sec:results}.
The $\PZ \to \tau \tau$ contribution is then split according to the decay mode reconstructed by the hadron-plus-strips algorithm as shown in the legend.
The mass distribution of the \Pgth built from one charged hadron and photons peaks near the mass of the intermediate $\Pgr$ resonance; the mass distribution of the \Pgth built from three charged hadrons peaks around the mass of the intermediate $\Pai$ resonance.
The \Pgth built from one charged hadron and no photons are reconstructed with the $\pi^\pm$ mass, assigned to all charged hadrons by the PF algorithm, and constitute the main contribution to the third bin of this histogram.
The first two bins correspond to $\tau^\pm$ leptons decaying into $\Pe^\pm\nu\nu$ and $\Pgm^\pm\nu\nu$, respectively, and for which the electron or muon is misidentified as a $\Pgth$.
The electroweak background contribution is dominated by $\PW+\text{jets}$ production.
In most selected $\PW+ \text{jets}$, $\ttbar$,  and QCD multijet events,
a jet is misidentified as a $\Pgth$.
The ``bkg.\ uncertainty'' band represents the combined statistical and systematic uncertainty in the background yield in each bin.
The expected contribution from the SM Higgs signal is negligible.
}
\label{fig:taumass}
\end{figure}

Muons are identified with additional requirements on the quality of the track reconstruction and on the number of measurements in the tracker and the muon systems~\cite{Chatrchyan:2012xi}.
Electrons are identified with a multivariate discriminant combining several quantities describing the track quality, the shape of the energy deposits in the electromagnetic calorimeter, and the compatibility of the measurements from the tracker and the electromagnetic calorimeter~\cite{CMS-PAS-EGM-10-004}.
The \Pgth are reconstructed and identified using the ``hadron-plus-strips'' algorithm~\cite{CMS-PAS-TAU-11-001}
which uses charged hadrons and photons to reconstruct the main decay modes of the $\tau$ lepton: one charged hadron, one charged hadron + photons, and three charged hadrons.
Electrons and muons misidentified as $\Pgth$ are suppressed using dedicated criteria based on the consistency between the measurements in the tracker, the calorimeters, and the muon detectors.
Figure~\ref{fig:taumass} shows the resulting $\Pgth$ mass distribution reconstructed from the visible decay products, $\mvis^{\Pgth}$, in the $\mutau$ channel after the baseline selection described in section~\ref{sec:event_selection}, illustrating the different decay modes.

To reject non-prompt or misidentified leptons, the absolute lepton isolation is defined as

\begin{equation}
I^{L} \equiv \sum_\text{charged}  \PT + \max\left( 0, \sum_\text{neutral}  \PT
                                        +  \sum_{\gamma} {\PT} - \frac{1}{2} \sum_\text{charged, PU} \PT  \right ).
\label{eq:reconstruction_isolation}
\end{equation}

In this expression, $\sum_\text{charged}  \PT$ is the scalar sum of the transverse momenta of the charged hadrons, electrons, and muons originating from the primary vertex and located in a cone of size  $\Delta R = \sqrt{\smash[b]{(\Delta \eta)^2 + (\Delta \phi)^2}} = 0.4$ centred on the lepton direction.
The sums $\sum_\text{neutral}  \PT$ and $\sum_{\gamma} \PT$ represent the same quantity for neutral hadrons and photons, respectively.
In the case of $\Pgth$, the particles used in the reconstruction of the $\Pgth$ are excluded from the sums.
The contribution of pileup photons and neutral hadrons is estimated from the scalar sum of the transverse momenta of charged hadrons from pileup vertices in the cone, $\sum_\text{charged, PU} \PT$.
This sum is multiplied by a factor of $1/2$ which corresponds approximately to the ratio of neutral to charged hadron production in the hadronization process of inelastic proton-proton collisions, as estimated from simulation.
The relative lepton isolation is defined as $R^{L} \equiv I^L / \pt^L$, where $\pt^L$ is the lepton transverse momentum.

The signal event samples with a SM Higgs boson produced through gluon-gluon fusion or VBF are generated with \POWHEG 1.0~\cite{Nason:2004rx,Frixione:2007vw, Alioli:2010xd, Alioli:2010xa, Alioli:2008tz},
while \PYTHIA 6.4~\cite{pythia} is used for the production of a SM Higgs boson in association with a $\PW$ or $\PZ$ boson, or with a $\ttbar$ pair.
The \MADGRAPH 5.1~\cite{Alwall:2011uj} generator is used for $\PZ+ \text{jets}$, $\PW+ \text{jets}$, $\ttbar+ \text{jets}$, and diboson production, and \POWHEG for single-top-quark production.
The \POWHEG and \MADGRAPH generators are interfaced with \PYTHIA for parton shower and fragmentation.
The \PYTHIA parameters affecting the description of the underlying event are set to the Z2 tune for the 7\TeV samples and to the Z2$^\ast$ tune for the 8\TeV samples~\cite{Chatrchyan:2011id}.
All generators are interfaced with \textsc{tauola}~\cite{TAUOLA} for the simulation of the $\tau$-lepton decays.
The Higgs boson \pt spectrum from \POWHEG is reweighted to the spectrum obtained from a next-to-next-to-leading-order (NNLO) calculation using \textsc{hres}~\cite{deFlorian:2012mx}.
The reweighting increases by about 3\% the fraction of gluon-gluon fusion signal events with a Higgs boson mass of $125\GeV$ and $\pt > 100\GeV$.
The various production cross sections and branching fractions for SM processes and their corresponding uncertainties are taken from references~\cite{LHCHiggsCrossSectionWorkingGroup:2011ti,Dittmaier:2012vm,Heinemeyer:2013tqa,Djouadi:1991tka,Dawson:1990zj,Spira:1995rr,Harlander:2002wh,Anastasiou:2002yz,Ravindran:2003um,Catani:2003zt,Aglietti:2004nj,Degrassi:2004mx,Actis:2008ug,Anastasiou:2008tj,deFlorian:2009hc,Baglio:2010ae,Ciccolini:2007jr,Ciccolini:2007ec,Arnold:2008rz,Brein:2003wg,Ciccolini:2003jy,hdecay2,Denner:2011mq,Alekhin:2011sk,Botje:2011sn,Lai:2010vv,Ball:2011mu,Djouadi:1991tka,Aglietti:2004nj,Degrassi:2004mx,Baglio:2010ae}.

The presence of pileup interactions is incorporated by simulating additional proton--proton collisions with \PYTHIA.
All generated events are processed through a detailed simulation of the CMS detector based on
\GEANTfour~\cite{Agostinelli:2002hh} and are reconstructed with the same algorithms as for data.
Simulated and recorded $\PZ+ \text{jets}$ events are compared to extract event weighting factors and energy correction factors for the various physics objects. These are then applied to all simulated events in order to minimize the remaining discrepancies with data.
In particular, (i)~a recoil correction is applied to the response and resolution of the components of the \vecMET~\cite{CMS-JME-12-002},
(ii)~energy correction factors are applied to the leptons,
and (iii)~simulated events are weighted by the ratio between the observed and expected lepton selection efficiencies,
which can differ by a few percent.

\section{Baseline event selection}
\label{sec:event_selection}

Events are selected and classified in the various channels according to the number of selected electrons, muons, and $\Pgth$ candidates.
The resulting event samples are independent.
Using simulated event samples, the trigger and offline selection criteria have been optimized for each channel to maximize the sensitivity to a SM Higgs boson signal.
These criteria are summarized in table~\ref{tab:inclusive_selection}
for the $LL'$ and $\ell + L \Pgth$ channels,
and in table~\ref{tab:inclusive_selection_ZH} for the $\ell \ell + LL'$ channels.

\begin{table}[htbp]
\centering
\topcaption{ Lepton selection
for the $LL'$ and $\ell + L \Pgth$ channels.
The HLT requirement is defined by a combination of trigger objects with \pt over a given threshold.
The \pt and $I^{\Pgth}$ thresholds are given in \GeVns.
The indices 1 and 2 denote, respectively, the leptons with the highest and next-to-highest \pt.
The definitions of the lepton isolation, $R$ and $I$, are given in the text.
For a number of channels,
the isolation requirements depend on the lepton flavour, \pt, and $\eta$.
Similarly, a range of \pt thresholds is given when the HLT requirements change with the data-taking period.
\label{tab:inclusive_selection}
}
\begin{tabular}{l|l|lll}
\hline
  Channel           &         HLT requirement              &    \multicolumn{3}{c}{Lepton selection}                 \\
\hline
  $\mu\Pgth$       &         $\mu (12\text{--}18)\,\&\,\Pgth (10\text{--}20)$     &     $\pt^\mu>17\text{--}20$  &  $\abs{\eta^\mu}<2.1$   &   $R^{\mu}<0.1$       \\
                          &       &     $\pt^{\Pgth}>30$ &  $\abs{\eta^{\Pgth}}<2.4$ &   $I^{\Pgth}<1.5$   \\
\hline
  $\Pe\Pgth$        &         $\Pe (15\text{--}22)\,\&\,\Pgth (15\text{--}20)$     &     $\pt^\Pe>20\text{--}24$  & $\abs{\eta^\Pe}<2.1$  &   $R^{\Pe}<0.1$  \\
                          &               &     $\pt^{\Pgth}>30$ & $\abs{\eta^{\Pgth}}<2.4$                        &   $I^{\Pgth}<1.5$\\
\hline
  $\Pgth\Pgth$    &         $\Pgth (35)\,\&\,\Pgth (35)$              &     $\pt^{\Pgth}>45$ & $\abs{\eta^{\Pgth}}<2.1$  &    $I^{\Pgth}<1$               \\
(2012 only)                          &         $\Pgth (30)\,\&\,\Pgth (30)\,\&\,$jet$(30)$     &                           \\
\hline
  $\Pe\mu$        &         $\Pe (17)\,\&\,\mu (8)$    &     $\pt^{\ell_1}>20$ & $\abs{\eta^\mu}<2.1$             & $R^{\ell}<0.1\text{--}0.15$   \\
                          &         $\Pe (8)\,\&\,\mu (17)$                                &     $\pt^{\ell_2}>10$ & $\abs{\eta^\Pe}<2.3$  &    \\
\hline
  $\mu\mu$          &       $\mu (17)\,\&\,\mu (8)$     &     $\pt^{\mu_1}>20$ & $\abs{\eta^{\mu_1}}<2.1$  &  $R^{\mu}<0.1$  \\
                          &                              &     $\pt^{\mu_2}>10$ & $\abs{\eta^{\mu_2}}<2.4$  &    \\
\hline
  $\Pe\Pe$             &         $\Pe (17)\,\&\,\Pe (8)$    &     $\pt^{\Pe_1}>20$ & $\abs{\eta^\Pe}<2.3$ &  $R^{\Pe}<0.1\text{--}0.15$  \\
                          &                                          &     $\pt^{\Pe_2}>10$ &                &    \\
\hline
\hline
  $\mu + \mu \Pgth$      &         $\mu (17)\,\&\,\mu (8)$       &       $\pt^{\mu_1}>20$       &  $\abs{\eta^\mu}<2.4$   &       $R^{\mu}<0.1\text{--}0.2$             \\
                                  &                                  &       $\pt^{\mu_2}>10$       &                                   &                                                             \\
                                  &                                  &       $\pt^{\Pgth}>20$  &    $\abs{\eta^{\Pgth}}<2.3$  &      $I^{\Pgth}<2$             \\

\hline
  $\Pe + \mu \Pgth$/    &         $\Pe (17)\,\&\,\mu (8)$         &       $\pt^{\ell_1}>20$       &  $\abs{\eta^\Pe}<2.5$         &       $R^{\ell}<0.1\text{--}0.2$         \\
  $\mu + \Pe \Pgth$    &         $\Pe (8)\,\&\,\mu (17)$         &       $\pt^{\ell_2}>10$       &  $\abs{\eta^\mu}<2.4$       &                                                         \\
                                    &                                  &       $\pt^{\Pgth}>20$    &  $\abs{\eta^{\Pgth}}<2.3$  &      $I^{\Pgth}<2$                  \\
\hline
  $\mu + \Pgth\Pgth$    &         $\mu (24)$              &       $\pt^\mu>24$  &  $\abs{\eta^\mu}<2.1$   &       $R^{\mu}<0.1$             \\
                                  &                                  &       $\pt^{\Pgt_{h,1}}>25$  &    $\abs{\eta^{\Pgth}}<2.3$  &     $I^{\Pgth}<2\text{--}3$               \\
                                  &                                  &       $\pt^{\Pgt_{h,2}}>20$  &                                   &                                                              \\
\hline
  $\Pe + \Pgth\Pgth$ &         $\Pe (20)\,\&\,\Pgth (20)$       &       $\pt^\Pe>24$  &  $\abs{\eta^\Pe}<2.1$   &       $R^{\Pe}<0.1\text{--}0.15$             \\
                                  &         $\Pe (22)\,\&\,\Pgth (20)$       &       $\pt^{\Pgt_{h,1}}>25$  &    $\abs{\eta^{\Pgth}}<2.3$  &     $I^{\Pgth}<2$               \\
                                  &                                  &       $\pt^{\Pgt_{h,2}}>20$  &                                   &                                                              \\
\hline
\end{tabular}
\end{table}

\begin{table}[htbp]
\centering
\topcaption{ Lepton selection for the $\ell\ell + LL'$ channels.
The HLT requirement is defined by a combination of trigger objects over a \pt threshold indicated in \GeV.
The \pt and $I^{\Pgth}$ thresholds are given in \GeVns.
The indices 1 and 2 denote, respectively, the leptons with the highest and next-to-highest \pt.
\label{tab:inclusive_selection_ZH}
}\begin{tabular}{l|l|llll}
\hline
  Resonance                &         HLT requirement              &    \multicolumn{3}{c}{Lepton selection}                     \\
\hline
  $\PZ \to \mu\mu$      &         $\mu (17)\,\&\,\mu (8)$     &     $\pt^{\mu_1}>20$  &  $\abs{\eta^\mu}<2.4$   &   $R^{\mu}<0.3$      \\
                                   &                               &     $\pt^{\mu_2}>10$ &                                    &                                           \\
\hline
  $\PZ \to \Pe\Pe$          &         $\Pe (17)\,\&\,\Pe (8)$       &     $\pt^{\Pe_1}>20$  &  $\abs{\eta^\Pe}<2.5$   &   $R^{\Pe}<0.3$       \\
                                   &                               &     $\pt^{\Pe_2}>10$ &                                    &                                           \\
\hline
\hline
  $\PH \to \mu\Pgth$     &        &     $\pt^\mu>10$  &  $\abs{\eta^\mu}<2.4$      &   $R^{\mu}<0.3$       \\
                                       &        &     $\pt^{\Pgth}>15$ &   $\abs{\eta^{\Pgth}}<2.3$   &   $I^{\Pgth}<2$      \\
\hline
  $\PH \to \Pe\Pgth$      &        &     $\pt^\Pe>10$  &  $\abs{\eta^\Pe}<2.5$      &   $R^{\Pe}<0.2$       \\
                                       &        &     $\pt^{\Pgth}>15$ &   $\abs{\eta^{\Pgth}}<2.3$   &   $I^{\Pgth}<2$      \\
\hline
  $\PH \to \Pgth\Pgth$      &        &     $\pt^{\Pgth}>15$  &  $\abs{\eta^{\Pgth}}<2.3$      &   $I^{\Pgth}<1$      \\
                                          &        &                                  &                                          &                                                    \\
\hline
  $\PH \to \Pe\mu$      &        &     $\pt^\ell>10$  &  $\abs{\eta^\Pe}<2.5$      &   $R^{\ell}<0.3$      \\
                                  &        &                             &  $\abs{\eta^\mu}<2.4$     &                                                    \\
\hline
\end{tabular}
\end{table}

The HLT requires a combination of electron, muon, and $\Pgth$ trigger objects~\cite{CMS-PAS-EGM-10-004,Chatrchyan:2012xi,CMS-EWK-TAU}.
A specific version of the PF algorithm is used in the HLT to quantify the isolation of $\Pgth$ trigger objects as done in the offline reconstruction.
Channels with two $\ell$ are based on a di-$\ell$ trigger.
Channels with a single $\ell$ are based on a $\ell\Pgth$ trigger
except for the $\mu + \Pgth \Pgth$ channel which uses a single-muon trigger.
The fully hadronic $\Pgth \Pgth$ channel relies on di-$\Pgth$ and di-$\Pgth+{\rm jet}$ triggers,
implemented for the 8\TeV data taking period.
For these triggers, the reconstruction of the two $\Pgth$ trigger objects is seeded by objects from the L1 trigger system.
These objects can either be two calorimeter jets with $\pt > 64\GeV$ and $\abs{\eta}<3.0$ or two narrow calorimeter jets with $\pt>44\GeV$ and $\abs{\eta}<2.17$.
The offline isolation requirements range within the values given in table~\ref{tab:inclusive_selection} depending on the lepton flavour, \pt, and $\eta$.
For the channels considered in this analysis, the efficiency to reconstruct and select offline a $\tau$ decaying hadronically ranges from 60 to 70\%, with a jet misidentification probability around 1\%.
For $\Pgth$ candidates selected offline, the efficiency of the HLT selection plateaus at 90\%.
For the $\mutau$ channel, the muon \pt threshold was raised in 2012 to 20\GeV to cope with the increased instantaneous luminosity.
For the same reason, the electron \pt threshold was raised to 24\GeV in the $\etau$ channel.
For the $\ell \ell + LL'$ channels, the selection proceeds by first identifying a $\PZ$ boson candidate ($\PZ \to \ell \ell$) with mass between 60 and 120\GeV from opposite-charge electron or muon pairs, and then a Higgs boson candidate ($\PH \to LL'$) from the remaining leptons.
With some variations among the channels,
all leptons meet the minimum requirement that the distance of closest approach to the primary vertex satisfies $d_z<0.2\unit{cm}$ along the beam direction,
and $d_{xy}<0.045\unit{cm}$ in the transverse plane.
The two leptons assigned to the Higgs boson decay are required to be of opposite charge.

In the $\ell \Pgth$ channels, the large $\PW+ \text{jets}$ background is reduced by requiring
\begin{equation}
\MT \equiv \sqrt{\smash[b]{2 \pt^\ell \MET (1-\cos(\Delta\phi))}} < 30\GeV,
\end{equation}
where $\pt^\ell$ is the $\ell$ transverse momentum
and $\Delta\phi$ is the difference in azimuthal angle between the $\ell$ direction and the \vecMET.
In the $\emu$ channel, the \ttbar background is reduced using a BDT discriminant that makes use of
kinematic variables related to the $\emu$ system and the \vecMET,
the distance of closest approach between the leptons and the primary vertex,
and the value of the CSV b-tagging discriminator for the leading jet with $\pt > 20\GeV$, if any.

In the $\ell + \Pgth \Pgth$ channels, the background from QCD multijet, $\PW+ \text{jets}$, and $\PZ+ \text{jets}$ production is suppressed using a BDT discriminant based on the \MET and on kinematic variables related to the $\Pgth \Pgth$ system.
With $\Pgt_\text{h,1}$ and $\Pgt_\text{h,2}$ denoting the $\Pgth$ with highest and second-highest $\pt$, respectively, these variables are $\pt^{\Pgt_\text{h,1}}$, $\pt^{\Pgt_\text{h,2}}$, $\Delta R(\Pgt_\text{h,1}, \Pgt_\text{h,2} )$, and $\pt^{\Pgt_\text{h,1},\Pgt_\text{h,2}}/(\pt^{\Pgt_\text{h,1}} +\pt^{\Pgt_\text{h,2}})$.
For the chosen threshold on the BDT score, the signal efficiency is $\sim$60\% whereas the efficiency for the reducible background components is $\sim$13\%.

In the $\ell + \ell' \Pgth$ channels, the large background from $\PZ$ and \ttbar production is strongly reduced by requiring the $\ell$ and $\ell'$ leptons to have the same charge.
For the 7\TeV dataset, the requirement $\LT \equiv \pt^\ell + \pt^{\ell'} + \pt^{\Pgth} > 80\GeV$ is imposed to further suppress the reducible background components.
For the 8\TeV dataset, the $\LT$ variable is instead used to divide the data into two event categories,
one with high $\LT$ ($\geq$130\GeV) and one with low $\LT$ ($<$130\GeV).
The $\PZ+ \text{jets}$ background in the $\ell\ell + LL'$ channels is reduced by selecting events with high
$L_\mathrm{T}^{LL'} \equiv \pt^L + \pt^{L'}$.
The requirements are
$L_\mathrm{T}^{\mu\Pgth} > 45\GeV$ for $\ell\ell + \mu\Pgth$,
$L_\mathrm{T}^{\Pe\Pgth} > 30\GeV$ for $\ell\ell + \Pe\Pgth$,
$L_\mathrm{T}^{\Pgth\Pgth} > 70\GeV$ for $\ell\ell + \Pgth\Pgth$,
and $L_\mathrm{T}^{\emu} > 25\GeV$ for $\ell\ell + \emu$.

\section[The tau-pair invariant-mass reconstruction]{The $\tau$-pair invariant-mass reconstruction}
\label{sec:htt_mass}

The visible mass, $\mvis$, of the $LL'$ system could be used to separate the $\PH\to \tau \tau$ signal events
from the $\PZ \to \tau \tau$ events, which constitute an important irreducible background.
However, the neutrinos from the $\tau$-lepton decay can take away a large amount of energy,
thereby limiting the separation power of the $\mvis$ variable.
In $\PZ \to \tau \tau$ events and in $\PH \to \tau \tau$ events where the Higgs boson is produced through gluon-gluon fusion, VBF, or in association with a $\PZ$ boson, the $\tau$-lepton decay is the only source of neutrinos.
The \textsc{svfit} algorithm described below combines the \MET with the $L$ and $L'$ momenta to calculate a more precise estimator of the mass of the parent boson, the $\tau$-pair invariant mass \mtt.

Six parameters are needed to specify a hadronic $\tau$-lepton decay:
the polar and azimuthal angles of the visible decay product system in the $\tau$-lepton rest frame, the three boost parameters from the $\tau$-lepton rest frame to the laboratory frame,
and the invariant mass $\mvis$ of the visible decay products.
In the case of a leptonic $\tau$ decay two neutrinos are produced
and the invariant mass of the two-neutrino system is the seventh parameter.
The unknown parameters are constrained by four observables that are the components of the four-momentum
of the system formed by the visible decay products of the $\tau$ lepton, measured in the laboratory frame.
For each hadronic (leptonic) $\tau$-lepton decay, 2 (3) parameters are thus left unconstrained.
These parameters are chosen to be:
\begin{itemize}
\item $x$, the fraction of the $\tau$-lepton energy in the laboratory frame carried by the visible decay products;
\item $\phi$, the azimuthal angle of the $\tau$-lepton direction in the laboratory frame;
\item $m_{\nu\nu}$, the invariant mass of the two-neutrino system in leptonic $\Pgt$ decays; for hadronic $\tau$-lepton decays, we take $m_{\nu\nu} \equiv 0$ in the fit described below.
\end{itemize}
The two components $E_{x}^\text{miss}$ and $E_{y}^\text{miss}$ of the $\vecMET$ provide two further constraints,
albeit each with an experimental resolution of 10--15 \GeV~\cite{PFMEtSignAlgo,CMS-JME-12-002}.

The fact that the reconstruction of the $\tau$-pair decay kinematics is underconstrained by the measured observables
is addressed by a maximum likelihood fit method.
The mass, $\mtt$, is reconstructed by combining the measured observables $E_{x}^\text{miss}$ and $E_{y}^\text{miss}$ with a likelihood model
that includes terms for the $\tau$-lepton decay kinematics and the \MET resolution.
The likelihood function $f(\vec{z}, \vec{y}, \vec{a_1}, \vec{a_2})$
of the parameters $\vec{z} = (E_{x}^\text{miss}, E_{y}^\text{miss})$ in an event is constructed,
given that the unknown parameters specifying the kinematics of the two $\tau$-lepton decays have values
$\vec{a_1} = (x_{1}, \phi_{1}, m_{\nu\nu,1})$ and $\vec{a_2} = (x_{2}, \phi_{2}, m_{\nu\nu,2})$,
and that the four-momenta of the visible decay products have the measured values $\vec{y} = (p^\text{vis}_{1}, p^\text{vis}_{2})$.
This likelihood model is used to compute the probability

\begin{equation}
P(\mtt^{i}) = \int \delta \left( \mtt^{i} - \mtt(\vec{y}, \vec{a_1}, \vec{a_2}) \right) f(\vec{z}, \vec{y}, \vec{a_1}, \vec{a_2})\,\rd\vec{a_1}\,\rd\vec{a_2},
\label{eq:mtt}
\end{equation}

as a function of the mass hypothesis $\mtt^{i}$.
The best estimate, $\hat{m}_{\tau\tau}$, for $\mtt$ is taken to be the value of $\mtt^{i}$ that maximizes $P(\mtt^{i})$.

The likelihood $f(\vec{z}, \vec{y}, \vec{a_1}, \vec{a_2})$ is the product of three likelihood functions:
the first two functions model the decay parameters $\vec{a_1}$ and $\vec{a_2}$ of the two $\tau$ leptons,
and the last one quantifies the compatibility of a $\tau$-pair decay hypothesis with the measured $\vecMET$.
The likelihood functions modelling the $\tau$-lepton decay kinematics are different for leptonic and hadronic $\tau$-lepton decays.
Matrix elements for unpolarized $\tau$-lepton decays from ref.~\cite{TauPol} are used to model the differential distributions in the leptonic decays,

\begin{equation}
\mathcal{L}_{\tau,\text{l}} = \frac{\rd\Gamma}{\rd{}x\,\rd{}m_{\nu\nu}\,\rd\phi} \propto \frac{m_{\nu\nu}}{4m_{\tau}^2} [(m_{\tau}^2 +2m_{\nu\nu}^2 )(m_{\tau}^2 - m_{\nu\nu}^2)],
\label{eq:likelihoodLepTauDecay}
\end{equation}

within the physically allowed region $0 \leq x \leq 1 \mbox{ and } 0 \leq m_{\nu\nu} \leq m_{\tau}\sqrt{1-x}$.
For hadronic $\tau$-lepton decays, a model based on the two-body phase space~\cite{PDG} is used,
treating all the visible decay products of the $\tau$ lepton as a single system,

\begin{equation}
\mathcal{L}_{\tau,\text{h}} = \frac{\rd\Gamma}{\rd{}x\,\rd\phi} \propto \frac{1}{1- m^2_\text{vis}/m^2_{\tau}},
\label{eq:likelihoodHadTauDecay}
\end{equation}

within the physically allowed region $m_\text{vis}^{2}/m_{\tau}^{2} \leq x \leq 1$.
It has been verified that the two-body phase space model is adequate for representing hadronic $\tau$-lepton decays
by comparing distributions generated by a parameterized MC simulation based on the two-body phase space model
with results from the detailed simulation implemented in \textsc{tauola}.
The likelihood functions for hadronic (leptonic) $\tau$-lepton decays do not depend on the parameters $x$, $\phi$, and $m_{\nu\nu}$ ($x$ and $\phi$).
The dependence on $x$ enters via the integration boundaries.
The dependence on $\phi$ comes from the likelihood function $\mathcal{L}_{\nu}$,
which quantifies the compatibility of a $\tau$-lepton decay hypothesis
with the reconstructed \vecMET in an event,
assuming the neutrinos from the $\tau$-lepton decays to be the only source of missing transverse energy.
This likelihood function is defined as

\begin{equation}
\mathcal{L}_{\nu} (E_{x}^\text{miss}, E_{y}^\text{miss}) = \frac{1}{2 \pi \sqrt{\abs{V}}}
 \exp \left[ -\frac{1}{2}
 \left( \begin{array}{c} E_{x}^\text{miss} - \sum p_{x}^{\nu} \\ E_{y}^\text{miss} - \sum p_{y}^{\nu} \end{array} \right)^{T}
 V^{-1}
 \left( \begin{array}{c} E_{x}^\text{miss} - \sum p_{x}^{\nu} \\ E_{y}^\text{miss} - \sum p_{y}^{\nu} \end{array} \right)
\right].
\end{equation}

In this expression, the expected \vecMET resolution is represented by the covariance matrix $V$, estimated on an event-by-event basis using a
\vecMET significance algorithm~\cite{PFMEtSignAlgo};
$\abs{V}$ is the determinant of this matrix.

The relative $\mtt$ resolution achieved by the \textsc{svfit} algorithm is estimated from simulation and found to be
about $10\%$ in the $\Pgth\Pgth$ decay channel, $15\%$ in the $\ell \Pgth$ channels,
and $20\%$ in the $\ell\ell'$ channels.
The resolution varies at the level of a few percent between the different event categories defined in section~\ref{sec:event_categories} because in some categories events with a boosted (\ie high-\pt) Higgs boson candidate and thus a better \vecMET resolution are selected.
The $\mtt$ resolution for each channel and each category is listed in table~\ref{tab:event_yields} of appendix~\ref{sec:event_yields}.
Figure~\ref{fig:htt_svfitperf} shows the normalized distributions of $m_\text{vis}$ and $\mtt$ in the $\Pgm\Pgth$ channel after the baseline selection for simulated $\PZ \to \tau \tau$ events and simulated SM Higgs boson events with $\mH=125\GeV$.
The \textsc{svfit} mass reconstruction allows for a better separation between signal and background than $m_\text{vis}$ alone,
yielding an improvement in the final expected significance of $\sim$40\%.

\begin{figure}[htbp]
\centering
\includegraphics[width=0.45\textwidth]{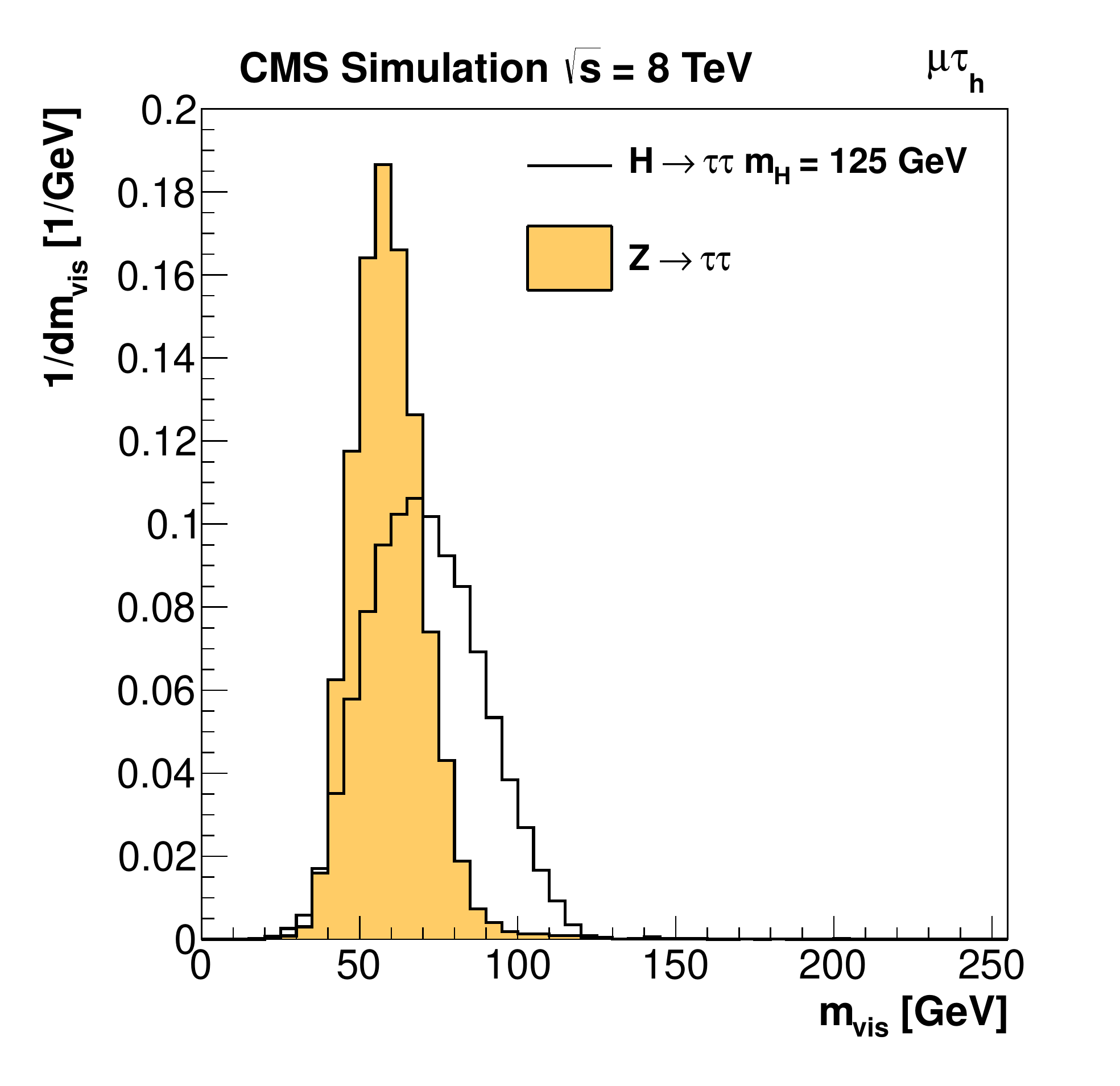}
\includegraphics[width=0.45\textwidth]{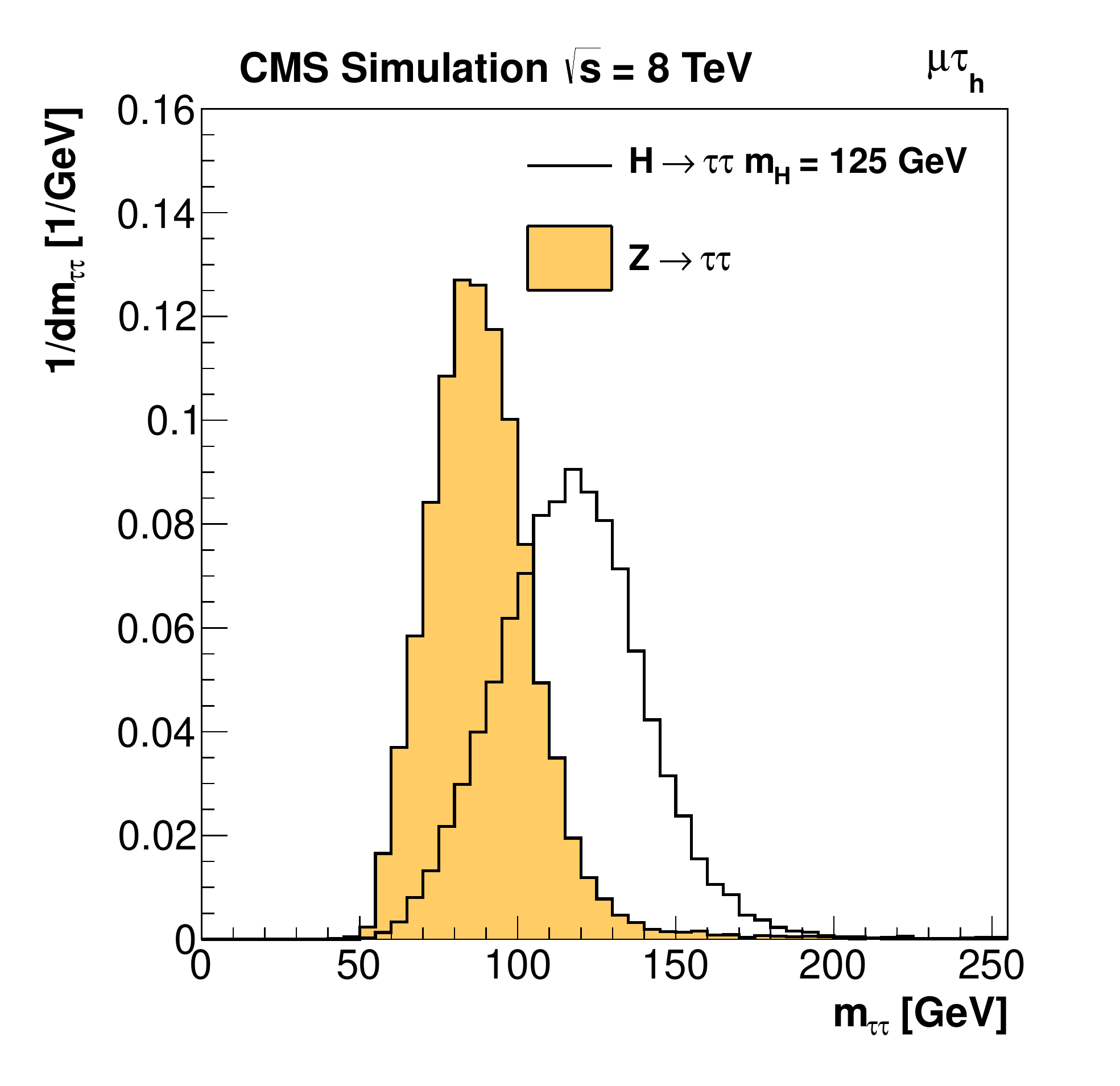} \\
\caption{Normalized distributions obtained in the $\mutau$ channel after the baseline selection for
(left) the invariant mass, $\mvis$, of the visible decay products of the two $\tau$ leptons, and (right) the \textsc{svfit} mass, $\mtt$.
The distribution obtained for a simulated sample of $\PZ\to \tau \tau$ events (shaded histogram) is compared to the one obtained for a signal sample with a SM Higgs boson of mass $\mH=125\GeV$ (open histogram).}
\label{fig:htt_svfitperf}
\end{figure}

In the case of Higgs boson production in association with a $\PW$~boson, the neutrino from the $\PW$-boson decay is an additional source of \MET.
Therefore, in the $\ell + L \Pgth$ channels, the signal is extracted from the distribution of the visible mass, $\mvis$, of the $L \Pgth$ system.
In the $\ell + \ell'\Pgth$ channels, the visible mass is calculated from the $\Pgth$ and the  electron or muon with smaller $\pt$.

\section{Event categories}
\label{sec:event_categories}

\begin{figure}[tbhp]
\centering
\includegraphics[width=\textwidth]{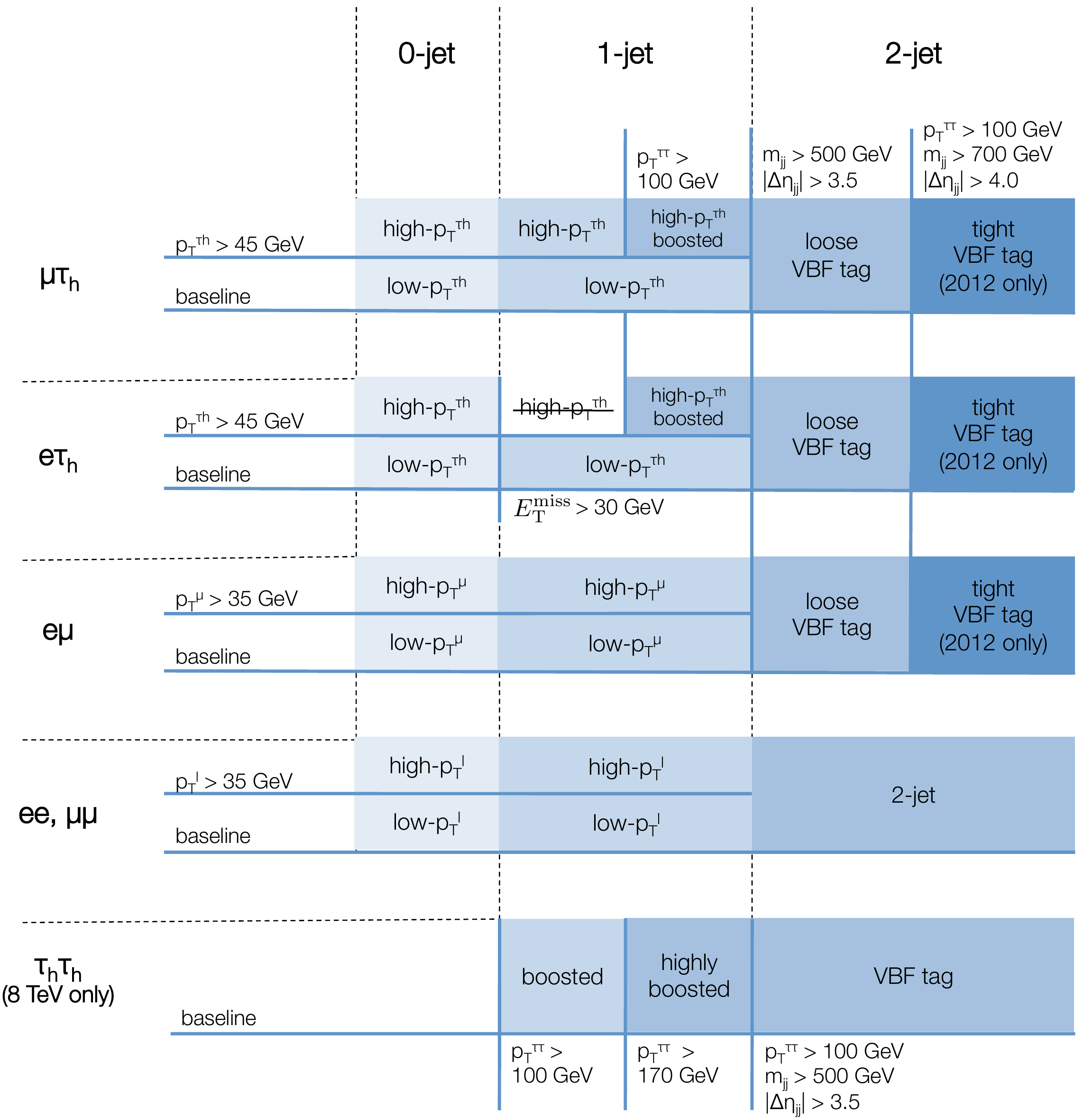}
\caption{Event categories for the $LL'$ channels.
The $\pth$ variable is the transverse momentum of the Higgs boson candidate.
In the definition of the VBF-tagged categories, $\abs{\detajj}$ is the difference in pseudorapidity between the two highest-\pt jets, and $\mjj$ their invariant mass.
In the $\mumu$ and $\ee$ channels, events with two or more jets are not required to fulfil any additional VBF tagging criteria.
For the analysis of the 7\TeV $\etau$ and $\mutau$ data,
the loose and tight VBF-tagged categories are merged into a single VBF-tagged category.
In the $\etau$ channel, the \MET is required to be larger than $30\GeV$ in the 1-jet category.
Therefore, the high-$\pt^{\Pgth}$ category is not used and is accordingly crossed out.
The term ``baseline'' refers to the baseline selection described in section~\ref{sec:event_selection}.
}
\label{fig:LL_categories}
\end{figure}

The event sample is split into mutually exclusive categories,
defined to maximize the sensitivity of the analysis to the presence of a SM Higgs boson with a mass, \mH, between 110 and 145\GeV.
The categories for the $LL'$ channels
are schematically represented in figure~\ref{fig:LL_categories} and described below.

In each channel, events are first classified according to the number of reconstructed jets with transverse momentum and pseudorapidity $\pt^\mathrm{j}>30\GeV$ and $\abs{\eta^\mathrm{j}}<4.7$,
and a separation in $(\eta, \phi)$ space between the jet and all selected leptons of $\Delta R^{\mathrm{j}L}>0.5$.
In all categories, events containing at least one b-tagged jet with $\pt^\mathrm{j}>20\GeV$ are rejected to reduce the $\ttbar$ background.

In the $\mutau$, $\etau$, $\tautau$ and $\emu$ channels, events with at least two jets are further required to pass a set of criteria targeting signal events where the Higgs boson is produced via VBF,
\ie in association with two jets separated by a large pseudorapidity gap.
This ``VBF tag'' strongly suppresses the background, in particular the irreducible $\PZ \to \tau \tau$ background.
This background is suppressed for two main reasons: first, because the requirement of two high-\pt jets is effective in rejecting the gluon-initiated jets from initial state radiation in the Drell-Yan production process; second, because such jets are typically produced in the central region of the detector.
The VBF-tagged category consists of events for which the two highest-\pt jets have a large invariant mass, $\mjj$, and a large separation in pseudorapidity, $\abs{\detajj}$.
A central-jet veto is applied by not allowing any additional jet in the pseudorapidity region delimited by the two highest-\pt jets.
For the analysis of the 8\TeV data, the VBF-tagged category is further split into tight and loose sub-categories.
In the $\ee$ and $\mumu$ channels, events with two jets or more are not required to fulfil any additional selection criteria,
except for the central-jet veto.
Instead, a multivariate discriminant involving $\mjj$ and $\abs{\detajj}$ is used to extract the signal, as described in section~\ref{sec:results}.

Events failing the VBF tag requirements, or the 2-jet category selection in the case of the $\ee$ and $\mumu$ channels, are collected in the 1-jet category if they contain at least one jet, and in the 0-jet category otherwise.
The latter has low sensitivity to the presence of a SM Higgs boson and is mainly used to constrain the $\PZ \to \tau \tau$ background for the more sensitive categories.
The $\tautau$ channel does not feature a 0-jet category because of the large background from QCD multijet events.

The 1-jet and 2-jet categories are further split according to the transverse momentum of the Higgs boson candidate, defined as
\begin{equation}
\pth = \abs{ \vec \pt^L+ \vec \pt^{L'} + \vecMET},
\end{equation}
where $\vec \pt^L$ and $\vec \pt^{L'}$ denote the transverse momenta of the two leptons.
The \pth variable is used to select sub-categories in which the Higgs boson candidate is boosted in the transverse plane.
The $\mtt$ resolution is improved for such events and a better separation between the $\PH \to \Pgt \Pgt$ signal and the $\PZ \to \Pgt \Pgt$ background is achieved.
This selection also has the advantage of reducing the QCD multijet background which is especially large in the $\tautau$ channel.

\begin{figure}[htbp]
\centering
\includegraphics[width=0.49\textwidth]{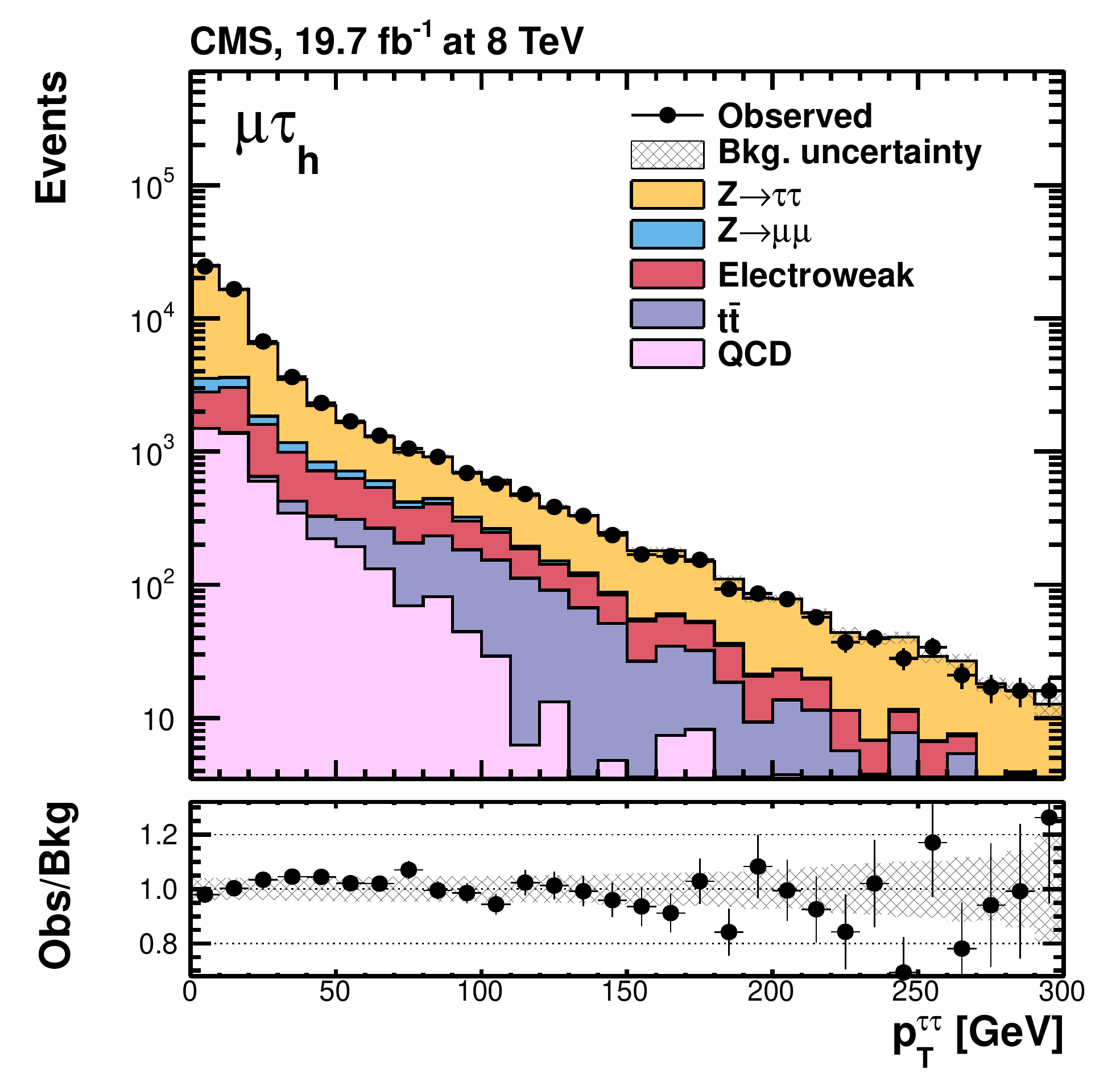}
\includegraphics[width=0.49\textwidth]{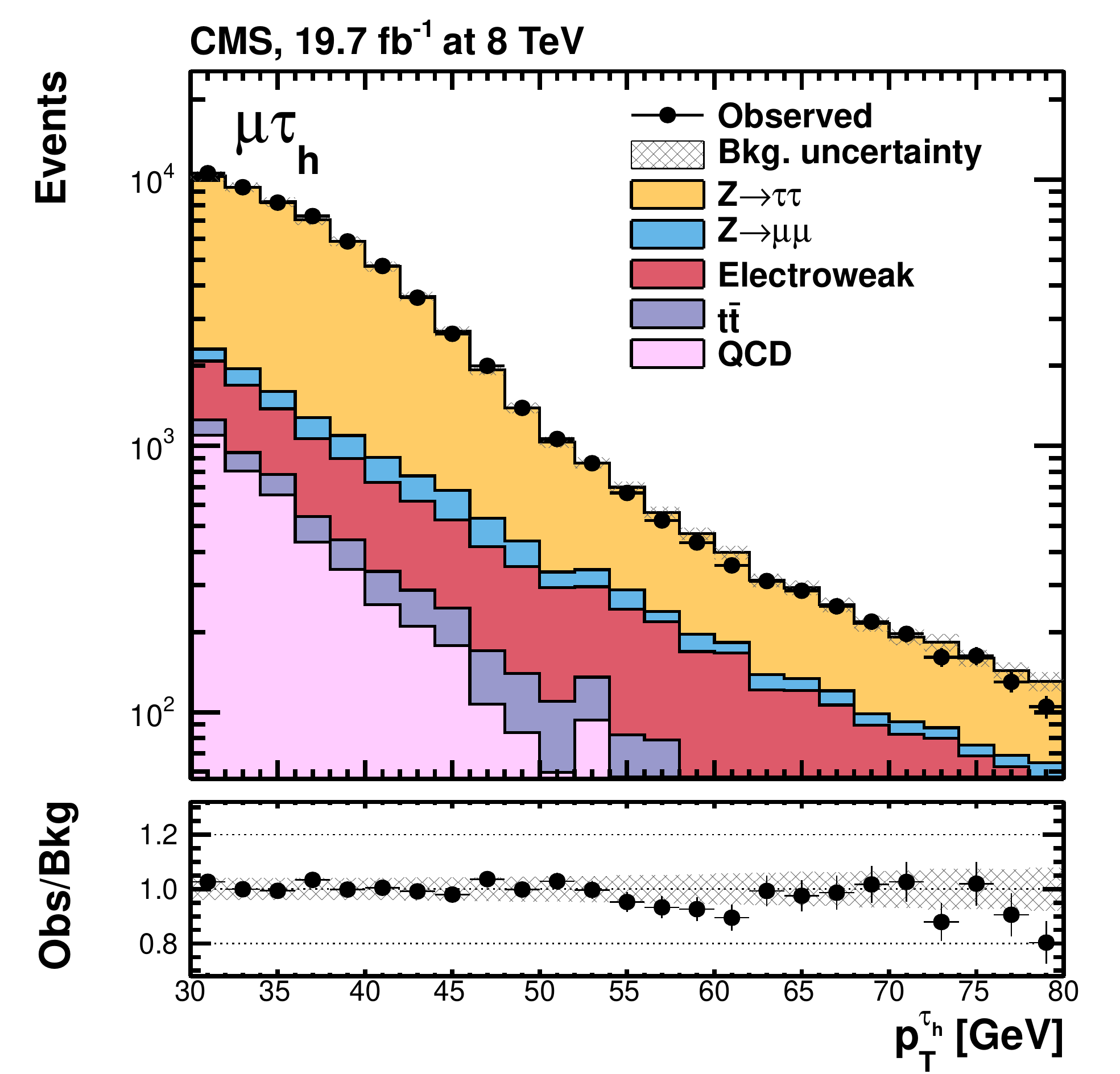}
\caption{
Observed and predicted distributions in the $\mutau$ channel
after the baseline selection, for (left) the transverse momentum of the Higgs boson candidates
and (right) the transverse momentum of the $\Pgth$. The yields predicted for the various background contributions correspond to the result of the final fit presented in Section~\ref{sec:results}.
The electroweak background contribution includes events from $\PW+ \text{jets}$, diboson, and single-top-quark production. The ``bkg.\ uncertainty'' band represents the combined statistical and systematic uncertainty in the background yield in each bin. In each plot, the bottom inset shows the ratio of the observed and predicted numbers of events. The expected contribution from the SM Higgs signal is negligible.
}
\label{fig:pth}
\end{figure}

The 0-jet and 1-jet categories are further divided into low and high $\pt^L$ categories, where
(i)~$L=\Pgth$ in the $\ell \Pgth$ channels,
(ii)~$L=\mu$ in the $\emu$ channel,
and (iii)~$L$ is the highest-\pt lepton in the $\ee$ and $\mumu$ channels.
For $\mH > m_Z$, higher-\pt leptons are produced in the $\PH \to \tau \tau$ process than in the $\PZ \to \tau \tau$ process.
Selecting high-\pt leptons also reduces the contribution of background events in which a jet is misidentified as a lepton.
Figure~\ref{fig:pth} demonstrates a good modelling of the \pth and $\pt^{\Pgth}$ distributions for the $\mutau$ channel,
after the baseline selection.

In the 1-jet category of the $\etau$ channel, the background from $\PZ \to \Pe \Pe$ events in which an electron is misidentified as a $\Pgth$ is reduced by requiring $\MET>30\GeV$.
This extra requirement makes it difficult to predict the $\mtt$ distribution for the $\PZ \to \tau \tau$ background events in the 1-jet high-$\pt^{\Pgth}$ category.
This category, which has relatively low sensitivity, is therefore ignored in the $\etau$ channel.

For the 8\TeV dataset, events in the $\ell + \ell' \Pgth$ channels are categorized according to $\LT$ with a threshold at $130\GeV$.
The $\ell + \Pgth \Pgth$ and $\ell \ell + L L'$ samples are not split into categories.

\section{Background estimation}
\label{sec:background_estimation}

The estimation of the shape and yield of the major backgrounds in each channel is based on the observed data.
The experimental systematic uncertainties affecting the background shapes and yields are thus directly related to the background estimation techniques and are also discussed in this section.

In the $\mutau$, $\etau$, $\tautau$, and $\emu$ channels,
the largest source of background is the Drell--Yan production of $\PZ\to\Pgt\Pgt$.
This contribution is greatly reduced by the 1-jet and VBF tag selection criteria as the jet-multiplicity distribution in Drell--Yan production falls off steeply.
It is modelled using ``embedded'' event samples recorded in each data-taking period under a loose $\PZ \to \Pgm \Pgm$ selection. In each event, the PF muons are replaced by the PF particles reconstructed from the visible decay products of the $\tau$ leptons in simulated $\PZ \to \tau \tau$ events, before reconstructing the $\vecMET$, the jets, the $\Pgth$ candidates, and the lepton isolation. The Drell-Yan event yield is rescaled to the observed yield using the inclusive sample of $\PZ\to\Pgm\Pgm$ events; thus, for this dominant background, the systematic uncertainties in the jet energy scale, the missing transverse energy, and the luminosity measurement are negligible. Additional uncertainties arise due to the extrapolation to the different event categories. These include uncertainties in the event reconstruction and acceptance of the ``embedded'' event samples that are estimated in simulated events as well as statistical uncertainties due to the limited number of events in these samples. In the $\Pe\Pgth$ and $\Pgm\Pgth$ channels, the largest remaining systematic uncertainty affecting the $\PZ\to\Pgt\Pgt$ background yield is due to the $\Pgth$ selection efficiency. This uncertainty, which includes the uncertainty in the efficiency to trigger on a $\Pgth$, is estimated to be 8\% in an independent study based on a tag-and-probe method~\cite{CMS-EWK-WZ} and, in addition, a $\mutau$ event sample recorded with single-muon triggers.

The Drell--Yan production of $\PZ\to \ell\ell$ is the largest background in the $\ell\ell$ channels.
The $\PZ\to \ell\ell$ event yield is normalized to the data in each category after subtracting all backgrounds. In the $\Pe\Pgth$ channel, $\PZ\to \ell\ell$ production is also an important source of background because of the 2--3\% probability for electrons to be misidentified as a $\Pgth$~\cite{CMS-PAS-TAU-11-001}
and the fact that the reconstructed $m_{\tau\tau}$ distribution peaks in the Higgs boson mass search range.
Because of the lower $\Pgm\to\Pgth$ misidentification rate, the $\PZ\to \ell\ell$ contribution in the $\Pgm\Pgth$ channel is small. The contribution of this background is estimated from simulation in both channels, after rescaling the simulated Drell--Yan yield to the one derived from $\PZ\to\Pgm\Pgm$ data.
The dominant systematic uncertainty in the $\PZ\to \ell\ell$ background yields arises from the $\ell \to \Pgth$ misidentification rate. This uncertainty is estimated using the tag-and-probe method with $\PZ\to \ell\ell$ event samples, and is found to be 20\% for electrons and 30\% for muons.
The small contribution from $\PZ\to \ell\ell$ events in the $\Pgm\Pgth$, $\Pe\Pgth$, and $\Pgth\Pgth$ channels where a lepton is lost and a jet is misidentified as a $\Pgth$ candidate is also estimated from simulation.
Depending on the event category, the uncertainties range from 20\% to 80\%, including uncertainties in the $\text{jet} \to \Pgth$ misidentification rate and statistical uncertainties due to the limited number of simulated events.

\begin{figure}[htb]
\centering
\includegraphics[width=0.7\textwidth]{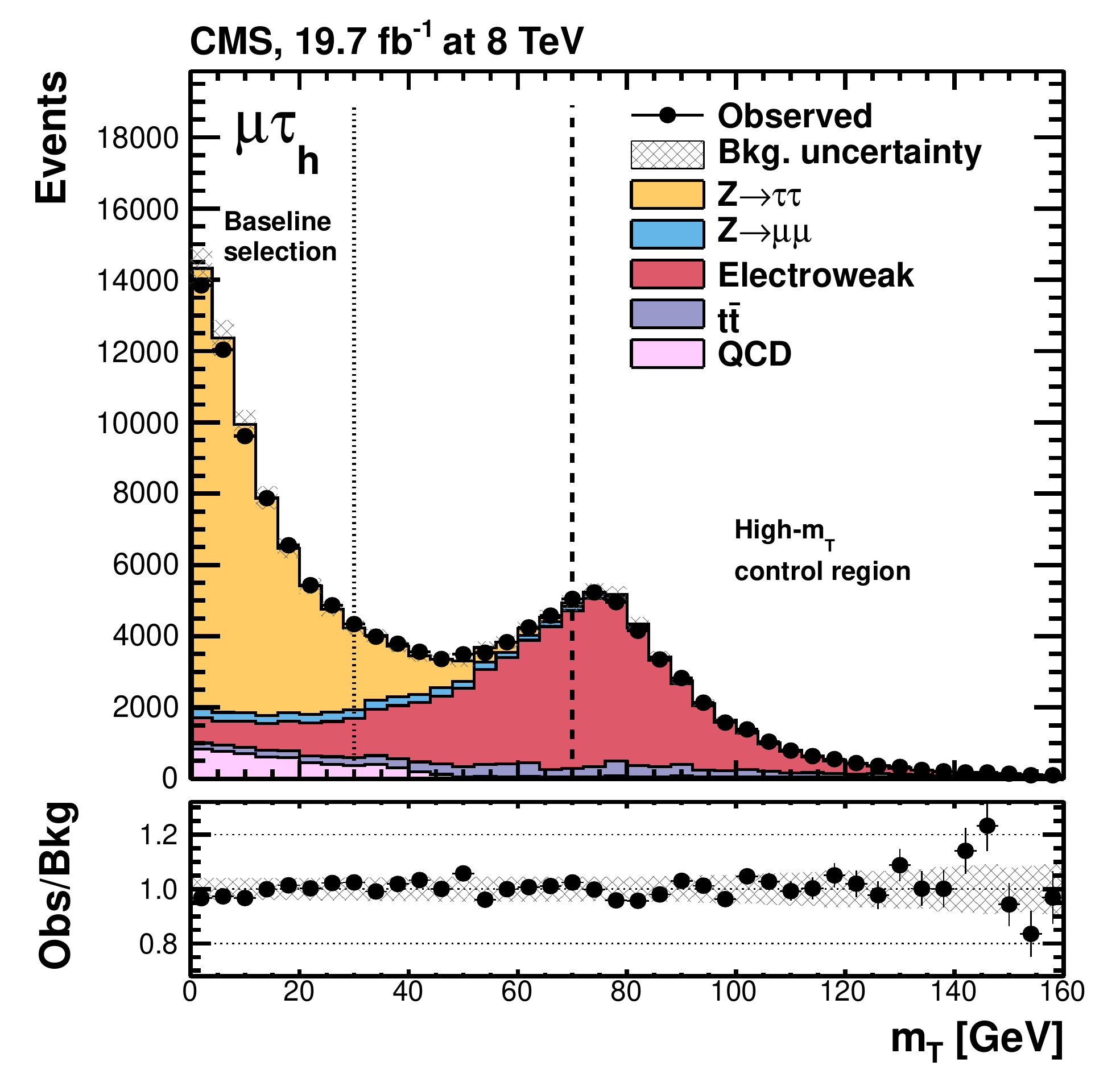}
\caption{Observed and predicted \MT distribution in the 8\TeV $\mutau$ analysis
after the baseline selection but before applying the $\MT<30\GeV$ requirement, illustrated as a dotted vertical line. The dashed line delimits the high-\MT control region that is used to normalize the yield of the $\PW+\text{jets}$ contribution in the analysis as described in the text. The yields predicted for the various background contributions correspond to the result of the final fit presented in Section~\ref{sec:results}.
The electroweak background contribution includes events from $\PW+\text{jets}$, diboson, and single-top-quark production. The ``bkg.\ uncertainty'' band represents the combined statistical and systematic uncertainty in the background yield in each bin. The bottom inset shows the ratio of the observed and predicted numbers of events.
The expected contribution from a SM Higgs signal is negligible.}
\label{fig:htt_control_mutau}
\end{figure}

The background from $\PW+\text{jets}$ production contributes significantly to the $\Pe\Pgth$ and $\Pgm\Pgth$ channels when the $\PW$ boson decays leptonically and a jet is misidentified as a $\Pgth$.
The background shape for these channels is modelled using the simulation.
Figure~\ref{fig:htt_control_mutau} shows the observed and predicted $\MT$ distribution obtained in the 8\TeV $\Pgm\Pgth$ channel after the baseline selection but without the $\MT<30\GeV$ requirement. In each category, the $\PW+\text{jets}$ background yield in a high-$\MT$
control region is normalized to the observed yield.
The extrapolation factor to the low-$\MT$ signal region is obtained from the simulation
and has an estimated systematic uncertainty of 10\% to 25\%, depending on the event category.
The uncertainty is estimated by comparing the $\MT$ distribution in simulated and recorded $\PZ(\to\Pgm\Pgm)+\text{jets}$ events in which a reconstructed muon is removed from the event to emulate $\PW+\text{jets}$ events. In the high-$\MT$ region of figure~\ref{fig:htt_control_mutau} the observed and predicted yields match by construction,
and the agreement in shape indicates good modelling of the \vecMET in the simulation.
In the VBF-tagged categories, where the number of simulated $\PW+\text{jets}$ events is small,
smooth \mtt templates are obtained by loosening the VBF selection criteria.
The $\mtt$ bias introduced in doing so was found to be negligible in a much larger sample of events obtained by relaxing the $\MT$ selection. For the $\Pgth\Pgth$ channel, a $\Pgm\Pgth$ control sample is used to define the same categories as in the $\Pgth\Pgth$ channel. In each of these categories,  the $\PW+\text{jets}$ background is normalized to the yield observed in the high-$\MT$ control region through a factor that is then used to scale the $\PW+\text{jets}$ background in the $\Pgth\Pgth$ sample, with a 30\% systematic uncertainty.

\begin{figure}[htb]
\centering
\includegraphics[width=0.7\textwidth]{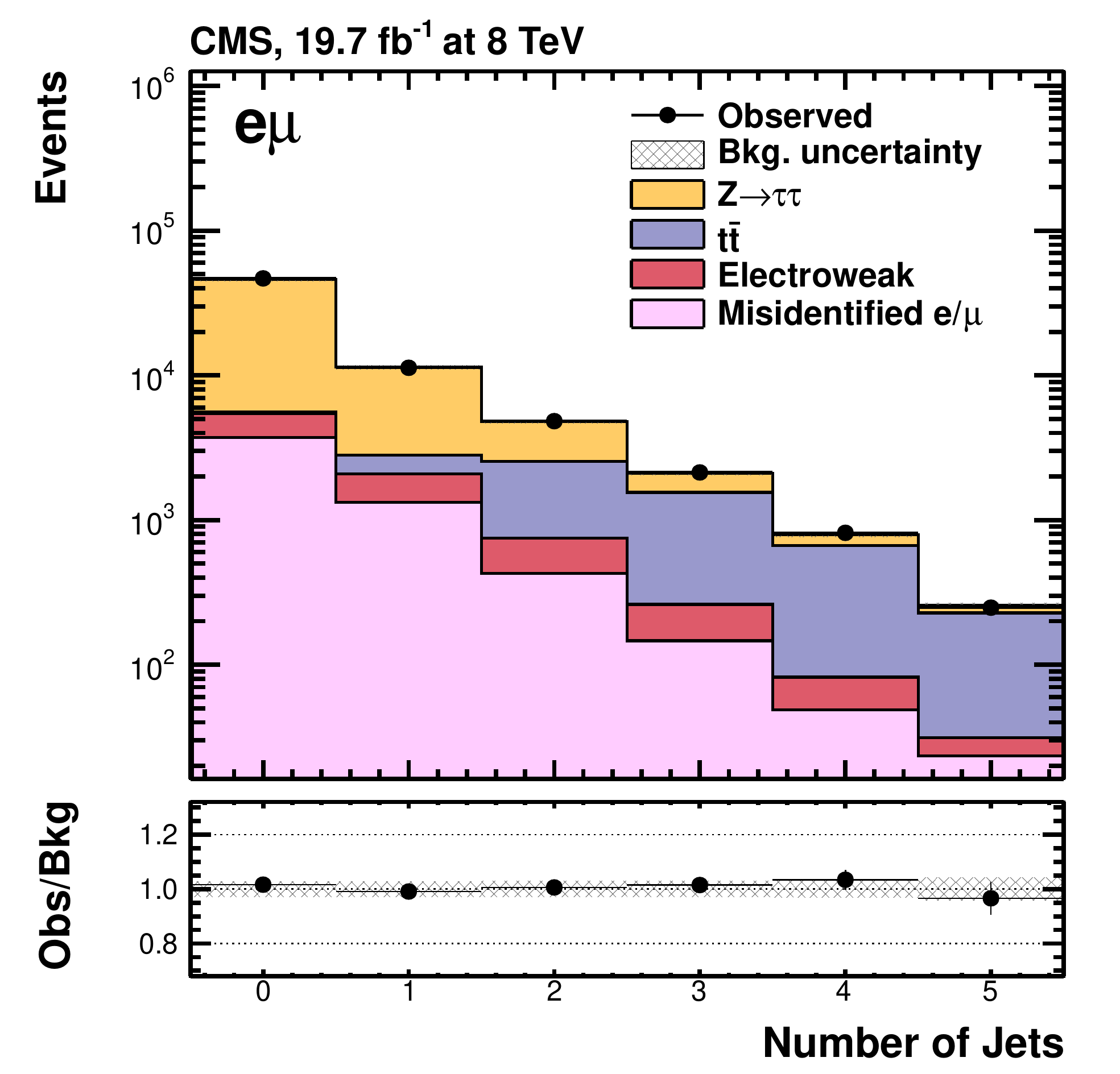}
\caption{Observed and predicted distribution for the number of jets in the 8\TeV $\emu$ analysis
after the baseline selection described in section~\ref{sec:event_selection}. The yields predicted for the various background contributions correspond to the result of the final fit presented in Section~\ref{sec:results}. The electroweak background contribution includes events from diboson and single-top-quark production. The ``bkg.\ uncertainty'' band represents the combined statistical and systematic uncertainty in the background yield in each bin. The bottom inset shows the ratio of the observed and predicted numbers of events. The expected contribution from a SM Higgs signal is negligible.}
\label{fig:htt_control_emu}
\end{figure}

The \ttbar production process is one of the main backgrounds in the $\Pe\Pgm$ channel.
Its shape for all $LL'$ channels is predicted by the simulation, and the yield is adjusted to the one observed using
a \ttbar-enriched control sample, extracted by requiring b-tagged jets in the final state.
The systematic uncertainty in the yield includes, among others, the systematic uncertainty in the b-tagging efficiency,
which ranges from 1.5\% to 7.4\% depending on the b-tagged jet \pt~\cite{Chatrchyan:2012jua}.
Furthermore, it is affected by systematic uncertainties in the jet energy scale, the \MET scale, and the background yields in the control sample.
Figure~\ref{fig:htt_control_emu} shows a good agreement between the observed and predicted distributions for the number of jets after the baseline selection in the 8\TeV $\emu$ analysis, in particular for events with three or more jets, for which the \ttbar process dominates.

QCD multijet events, in which one jet is misidentified as a $\Pgth$ and another as an $\ell$, constitute another important source
of background in the $\ell \Pgth$ channels.
In the 0-jet and 1-jet low-$\pt^{\Pgth}$ categories that have a high event yield, the QCD multijet background yield is obtained using a control sample where the
$\ell$ and the $\Pgth$ are required to have the same charge.
In this control sample, the QCD multijet  yield is obtained by subtracting from the data the contribution of the Drell--Yan, \ttbar,
and $\PW+ \text{jets}$ processes, estimated as explained above.
The expected contribution of the QCD multijet  background in the opposite-charge signal sample is then derived by rescaling the yield obtained in the same-charge control sample by a factor of 1.06,
which is measured using a pure QCD multijet sample obtained by inverting the $\ell$ isolation requirement
and relaxing the $\Pgth$ isolation requirement.
The 10\% systematic uncertainty in this factor accounts for a small dependence on $\pt^{\Pgth}$ and the statistical uncertainty in the measurement, and dominates the uncertainty in the yield of this background contribution.
In the VBF-tagged and 1-jet high-$\pt^{\Pgth}$ boosted categories,
the number of events in the same-charge control sample is too small to use this procedure.
Instead, the QCD multijet  yield is obtained by multiplying the QCD multijet yield estimated after the baseline selection
by the category selection efficiency.
This efficiency is measured using a sample dominated by QCD multijet production in which the $\ell$ and the $\Pgth$ are not isolated.
The yield is affected by a 20\% systematic uncertainty.
In all categories, the \mtt template is obtained from a same-charge control region in which the $\ell$ isolation requirement is inverted.
In addition, the VBF tagging and the $\Pgth$ isolation criteria are relaxed in the VBF-tagged and 1-jet high-$\pt^{\Pgth}$ boosted categories, respectively, to obtain a smooth template shape.

In the $\Pgth\Pgth$ channel, the large QCD multijet background is estimated from a control region with a relaxed $\Pgth$ isolation requirement, disjoint from the signal region.
In this region, the QCD multijet background shape and yield are obtained by subtracting from the observed data the contribution of the
Drell--Yan, \ttbar, and $\PW+ \text{jets}$ processes, estimated as explained above.
The QCD multijet background yield in the signal region is obtained by multiplying the yield in the control region by an extrapolation factor,
obtained using identical signal region and control region definitions applied to a sample of same-charge $\Pgth\Pgth$ events.
Depending on the event category,
the systematic uncertainty in this yield is estimated to range from 35\% to 50\%.
The uncertainty includes contributions from the limited number of events in the control region and uncertainties in the expected yields of the subtracted background components.
The QCD multijet background shape in the signal region is taken to be the same as in the control region with a relaxed $\Pgth$ isolation requirement,
an assumption that is verified by comparing the QCD multijet shapes obtained in the signal region and in the control region of the same-charge sample.

The small background due to $\PW+ \text{jets}$ and QCD multijet production in the $\Pe\Pgm$
channel corresponds to events in which one or two jets are misidentified as leptons, and is denoted as the ``misidentified-$\ell$'' background.
A misidentified-$\ell$ control region is defined by requiring the $\ell$ to pass relaxed selection criteria,
and to fail the nominal selection criteria.
The expected contribution from processes with a dilepton final state is subtracted.
The number of events $N_\ell$ in the signal region in which a jet is misidentified as an $\ell$
is estimated as the yield in the misidentified-$\ell$ control region multiplied by the ratio between the yields measured in the signal and control regions using a multijet sample.
The procedure is applied separately for electrons and muons, leading to the estimation of $N_\Pe$ and $N_\Pm$.
The number of events for which two jets are misidentified as an electron and a muon, $N_{\Pe\Pgm}$,
is estimated from a control region in which both the electron and the muon pass the relaxed selection criteria and fail the nominal selection criteria.
The background yield in the signal region is then estimated as $N_\Pe + N_{\Pgm} - N_{\Pe\Pgm}$.
In the 0-jet and 1-jet low-$\pt^{\Pgth}$ categories, the $\mtt$ template is taken from a same-charge control region with inverted electron isolation requirement.
In the 1-jet high-$\pt^{\Pgth}$ and VBF-tagged categories, the number of events in this control region is too small and the template is instead taken from the opposite-charge misidentified-electron control region.

The small contributions from diboson and single-top-quark production in the $LL'$ channels are taken from simulation. For $\mH = 125$\GeV, the contribution from $\hww$ decays amounts to up to 45\% of the expected SM Higgs boson signal in the VBF-tagged categories in the $\emu$ channel, and to up to 60\% of the complete expected SM Higgs boson signal in the 2-jet categories of the $\ee$ and $\mumu$ channels. In all other $LL'$ channels, the contribution from $\hww$ decays is negligible.

In the $\ell + L \Pgth$ channels, the irreducible background is due to $\PW\PZ$ and $\PZ\PZ$ production,
while the reducible background comes from QCD multijet, $\PW+ \text{jets}$, $\PZ+ \text{jets}$, $\PW+\PGg$, $\PZ+\PGg$, and \ttbar production.
In the $\ell\ell + LL'$ channels, $\PZ\PZ$ production and $\ttbar$ production in association with a $\PZ$ boson constitute the sources of irreducible background;
the reducible background comes from $\PW\PZ+ \text{jets}$, $\PZ+ \text{jets}$, and \ttbar production.
Events from reducible background sources contain at least one jet misidentified as a lepton.
For each channel, the reducible background contribution is estimated from sidebands in which one or more lepton candidates satisfy relaxed selection criteria but do not satisfy the nominal selection criteria.
The number of reducible background events for which all leptons satisfy the nominal selection criteria is obtained by weighting the sideband events according to the probability for a lepton passing the relaxed selection criteria to also pass the nominal selection criteria.
These misidentification probabilities are measured in independent control samples of QCD multijet, $\PW(\to \ell \nu) + \text{jets}$, and $\PZ(\to \ell\ell) + \text{jets}$ events with one lepton passing the relaxed selection criteria in addition to the well-identified leptons corresponding to the decay of the $\PW$ or $\PZ$ boson, if any.
The misidentification probabilities are parameterized as a function of the lepton \pt, the number of jets in the event, and, in case a jet is found close to the lepton, the jet \pt.
To obtain a smooth template shape, the isolation criteria for the leptons associated to the Higgs boson decay are relaxed in the $\ell\ell + LL'$ channels.
The systematic uncertainties in the event yield of the reducible background components range from 15 to 30\%.
They include contributions from the limited number of events in the sideband, uncertainties in the estimation of the misidentification probabilities, and uncertainties in the background composition in the sidebands.

\section{Systematic uncertainties}
\label{sec:systematics}

The values of a number of imprecisely known quantities can affect the rates and shapes of the $m_{\tau\tau}$ distributions for the signal and background processes.
These systematic uncertainties can be grouped into theory related uncertainties, which are predominantly relevant for the expected signal yields, and into uncertainties from experimental sources, which can further be subdivided into uncertainties related to the reconstruction of physics objects and uncertainties in the background estimation.
The uncertainties related to the reconstruction of physics objects apply to processes estimated with simulated samples, most importantly the signal processes.
As outlined in the previous section, the distributions for several background processes are estimated from data, and the corresponding systematic uncertainties are therefore mostly uncorrelated with the ones affecting the signal distributions.

The main experimental uncertainties in the decay channels involving a $\Pgth$ are related to the reconstruction of this object.
The $\Pgth$ energy scale is obtained from a template fit to the $\Pgth$ mass distribution, such as the one shown in figure~\ref{fig:taumass}.
In this fit, the shape of the mass distribution is morphed as a function of the $\Pgth$ energy scale parameter.
The uncertainty of $\pm$3\% in the energy scale of each $\Pgth$ affects both the shape and the rate of the relevant signal and background distributions in each category.
The $\Pgth$ identification and trigger efficiencies for genuine $\tau$ leptons sum up to an overall rate uncertainty of 6 to~10\%
per $\Pgth$, depending on the decay channel, due to the different trigger and $\ell$ rejection criteria and additional uncertainties for higher $\pt^{\Pgth}$.
For $\PZ \to \ell\ell$ events where jets, muons, or electrons are misidentified as $\Pgth$, the estimation of the $\Pgth$ identification efficiency leads to rate uncertainties of 20 to~80\%, including the statistical uncertainty due to the limited number of simulated events.

In the decay channels with muons or electrons, the uncertainties in the muon and electron identification, isolation, and trigger efficiencies lead to rate uncertainties of 2 to~4\% for muons and 2 to~6\% for electrons.
The uncertainty in the electron energy scale is relevant only in the $\Pe\mu$ and $\Pe\Pe$ channels, where it affects the normalization and shape of the simulated \mtt and final discriminant distributions.
The uncertainty in the muon energy scale is found to have a negligible effect for all channels.
The relative \MET scale uncertainty of 5\% affects the event yields
for all channels making use of the \MET in the event selection, in particular for the $\ell \Pgth$ channels due to the $\MT$ selection~\cite{CMS-JME-12-002}.
This translates into yield uncertainties of 1 to~12\%, depending on the channel and the event category.
The uncertainties are largest for event categories with a minimum \MET requirement and for background contributions with no physical source of \MET,
e.g. the $\PZ\to\Pe\Pe$ contribution in the high-$\pt^{\Pgth}$ boosted category in the $\Pe\Pgth$ channel.
The uncertainty in the jet energy scale varies with jet\ \pt and jet\ $\eta$~\cite{CMS-JME-10-011} and leads to rate uncertainties for the signal contributions of up to 20\% in the VBF-tagged categories.
For the most important background samples, the effect on the rate is, however, well below 5\%.
Because of the veto of events with b-tagged jets,
uncertainties in the tagging efficiency for b-quark jets and in the mistagging efficiency for c-quark, light-flavour, and gluon jets result in rate uncertainties of up to 8\% for the different signal and background components.
The uncertainty in the integrated luminosity amounts to 2.2\% for the 7\TeV analysis~\cite{CMS-PAS-SMP-12-008} and 2.6\% for the 8\TeV analysis~\cite{CMS-PAS-LUM-13-001},
yielding corresponding rate uncertainties for the affected signal and background samples.

The uncertainties related to the estimation of the different background processes are discussed in detail in the previous section, and only a summary is given here.
For the different Drell--Yan decay modes, the uncertainty in the inclusive $\PZ\to\Pgt\Pgt$ yield is 3\%, with additional extrapolation uncertainties in the different categories in the range of 2 to~14\%.
The uncertainties in the $\PW+ \text{jets}$ event yields estimated from data are in the range of 10--100\%.
The values are dominated by the statistical uncertainties involved in the extrapolation from high to low \MT and due to the limited number of data events in the high-\MT control region.
As a consequence, they are treated as uncorrelated with any other uncertainty.
The QCD multijet background estimation results in 6 to~35\% rate uncertainties for
the $LL'$ channels, except for the very pure dimuon final state and the VBF-tagged categories where uncertainties of up to 100\% are estimated.
Additional shape uncertainties are included in the $\Pe\Pgth$, $\mu\Pgth$, and $\Pe\mu$ channels to account for the uncertainty in the shape estimation from the control regions.

The rate and acceptance uncertainties for the signal processes related to the theoretical calculations are due to uncertainties in the parton distribution functions (PDF), variations of the renormalization and factorization scales,
and uncertainties in the modelling of the underlying event and parton showers.
The magnitude of the rate uncertainty depends on the production process and on the event category.
In the VBF-tagged categories, the theoretical uncertainties
concerning the $\Pq\Pq' \to \PH$ process are 4\% from the PDFs and 3\% from scale variations.
The rate and acceptance uncertainties in the $\Pg\Pg\to \PH$ process in the VBF-tagged categories are estimated by comparing the four different MC generators
\POWHEG, \MADGRAPH, \POWHEG interfaced with \textsc{Minlo}~\cite{Hamilton:2012np}, and a\MCATNLO~\cite{Frixione:2002ik}.
They amount to 30\% and thus become of similar absolute size as the uncertainties in the $\Pq \Pq' \to \PH$ process.

For the $\Pg\Pg\to \PH$ process, additional uncertainties are incorporated to account for missing higher-order corrections ranging from 10 to 41\% depending on the category and on the decay channel.
The combined systematic uncertainty in the background yield arising from diboson and single-top-quark production processes is estimated to be 15\%
for the $LL'$ channels
based on recent CMS measurements~\cite{Chatrchyan:2013oev,Chatrchyan:2012ep}.
In the $\ell + L\Pgth$ and $\ell \ell + LL'$ channels, the uncertainties in the event yields of $\PW\PZ$ production and $\PZ\PZ$ production arise from scale variations and uncertainties in the PDFs,
including the PDF uncertainties in the $\Pg\Pg \to \PZ\PZ$ event yields which are 44\%.
The resulting overall uncertainties range from 4 to 8\%.
The uncertainties in the small background from $\ttbar+\PZ$ production in the $\ell \ell + LL'$ channels amount to 50\%~\cite{Chatrchyan:2013qca}.

In addition, uncertainties due to the limited number of simulated events or due to the limited number of events in data control regions are taken into account.
These uncertainties are uncorrelated across bins in the individual templates. A summary of the considered systematic uncertainties is given in table~\ref{tab:uncertainties}.

\begin{table}[!ht]
\centering
\topcaption{Systematic uncertainties, affected samples, and change in acceptance resulting from a variation of the nuisance parameter equivalent to one standard deviation. Several systematic uncertainties are treated as (partially) correlated for different decay channels and/or categories.}
\begin{tabular}{lcc}
 Uncertainty                    & Affected processes            & Change in acceptance  \\
\hline
 Tau energy scale               & signal \& sim.\ backgrounds & 1--29\% \\
 Tau ID (\& trigger)               & signal \& sim.\ backgrounds & 6--19\% \\
 $\Pe$ misidentified as $\tau_{h}$         & $\PZ\to \Pe \Pe$            & 20--74\% \\
 $\mu$ misidentified as $\tau_{h}$         & $\PZ\to \mu \mu$            & 30\% \\
 Jet misidentified as $\tau_{h}$           & $\PZ+ \text{jets}$       & 20--80\% \\
 Electron ID \& trigger         & signal \& sim.\ backgrounds & 2--6\%\\
 Muon ID \& trigger             & signal \& sim.\ backgrounds & 2--4\%\\
 Electron energy scale          & signal \& sim.\ backgrounds & up to 13\% \\
 Jet energy scale               & signal \& sim.\ backgrounds & up to 20\% \\
 \MET scale                     & signal \& sim.\ backgrounds & 1--12\% \\
 $\varepsilon_\text{b-tag}$ b jets        & signal \& sim.\ backgrounds & up to 8\%\\
 $\varepsilon_\text{b-tag}$ light-flavoured jets    & signal \& sim.\ backgrounds & 1--3\%\\
\hline
 Norm. $\PZ$ production                  & $\PZ$ & 3\%\\
 $\PZ\to\tau\tau$ category               & $\PZ\to\Pgt\Pgt$ & 2--14\%\\
 Norm.\ $\PW+ \text{jets}$                        & $\PW+ \text{jets}$ & 10--100\% \\
 Norm.\ $\ttbar$                        & $\ttbar$ & 8--35\% \\
 Norm.\ diboson                           & diboson & 6--45\% \\
 Norm.\ QCD multijet                      & QCD multijet & 6--70\%\\
 Shape QCD multijet                      & QCD multijet & shape only \\
 Norm.\ reducible background         & Reducible bkg. & 15--30\% \\
 Shape reducible background         & Reducible bkg. & shape only \\
 Luminosity 7\TeV(8\TeV)                 & signal \& sim.\ backgrounds & 2.2\% (2.6\%) \\

\hline
 PDF ($\Pq\Pq$)                          & signal \& sim.\ backgrounds  & 4--5\% \\
 PDF ($\Pg\Pg$)                          & signal \& sim.\ backgrounds  & 10\% \\
 Norm.\ $\PZ\PZ/\PW\PZ$            & $\PZ\PZ/\PW\PZ$ & 4--8\% \\
 Norm.\ $\ttbar + \PZ$            & $\ttbar + \PZ$ & 50\% \\
 Scale variation                         & signal                       & 3--41\% \\
 Underlying event \& parton shower      & signal                       & 2--10\% \\
\hline
 Limited number of events                & all                 & shape only \\
\end{tabular}
\label{tab:uncertainties}
\end{table}

\section{Results}
\label{sec:results}

The search for an excess of SM Higgs boson events over the expected background involves a global maximum likelihood fit based on final discriminating variables which are either \mtt or \mvis in all channels except for ee and $\mumu$~\cite{LHC-HCG-Report,Chatrchyan:2012tx}.
In these two channels the final discriminating variable $D$ is built for a given event from the output of two boosted decision trees $B_1$ and $B_2$.
The two BDTs are based on kinematic variables related to the $\ell\ell$ system and the \vecMET,
on the distance of closest approach between the leptons, and,
in the 2-jet category, the $\mjj$ and $\abs{\Delta \eta_\text{jj}}$ variables.
The first BDT is trained to discriminate $\PZ \to \tau \tau$ from $\PZ \to \ell \ell$ events,
whereas the second BDT is trained to discriminate $\PH \to \tau \tau$ from $\PZ \to \tau \tau$ events.
Both BDTs are separately trained in the 2-jet category and in the combined 0-jet and 1-jet categories.
The final discriminant is defined as

\begin{equation}
D = \int_{-\infty}^{B_1} \int_{-\infty}^{B_2} f_\text{sig} (B_1', B_2')\,\rd{}B_1'\,\rd{}B_2'.
\end{equation}

In this expression, $f_\text{sig}$ is the two-dimensional joint probability density for the signal.
Therefore, $D$ represents the probability for a signal event to have a value lower than $B_1$ for the first BDT and $B_2$ for the second BDT.

\begin{figure}[htbp]
\centering
     \includegraphics[width=0.42\textwidth]{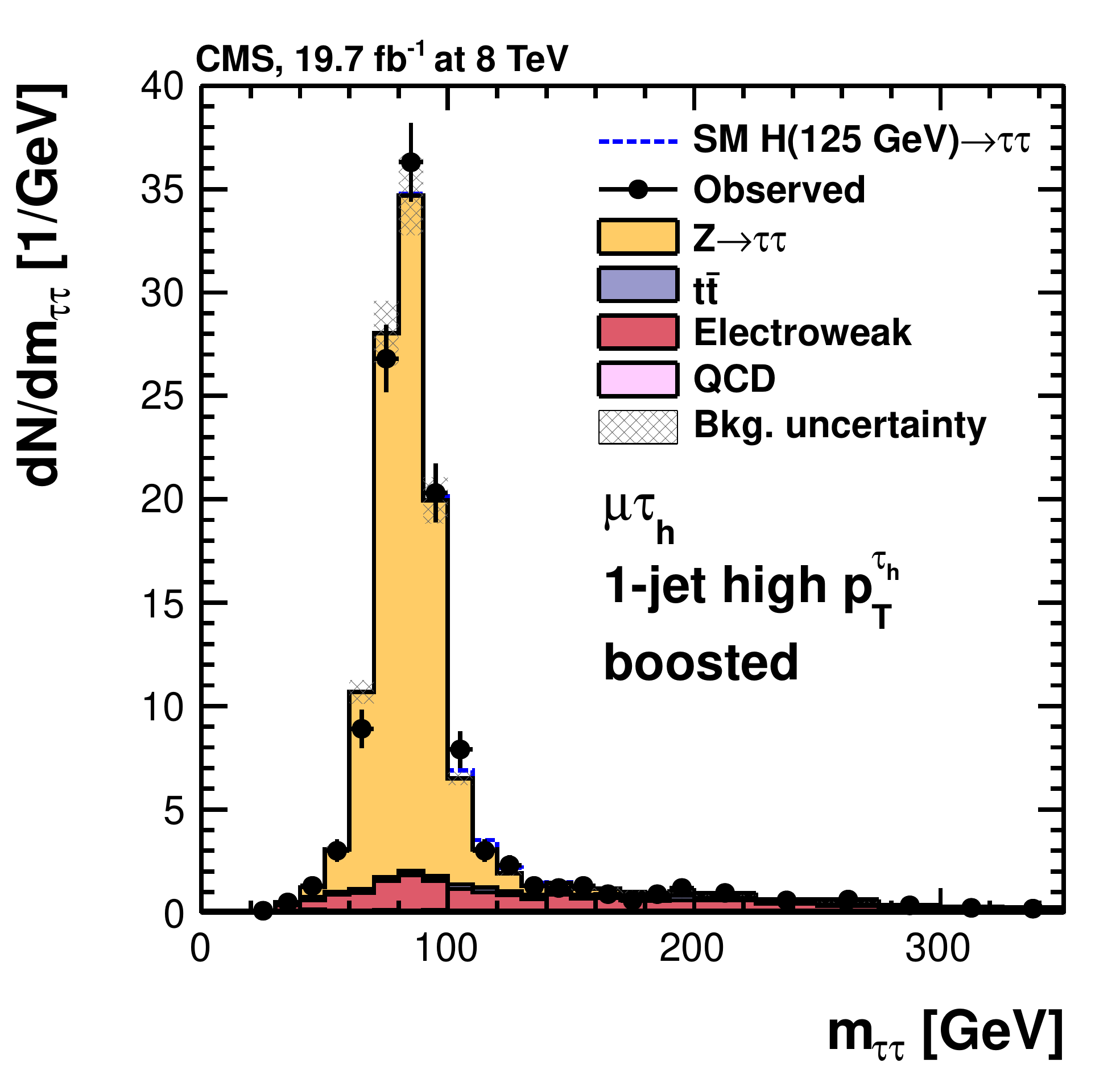}
     \includegraphics[width=0.42\textwidth]{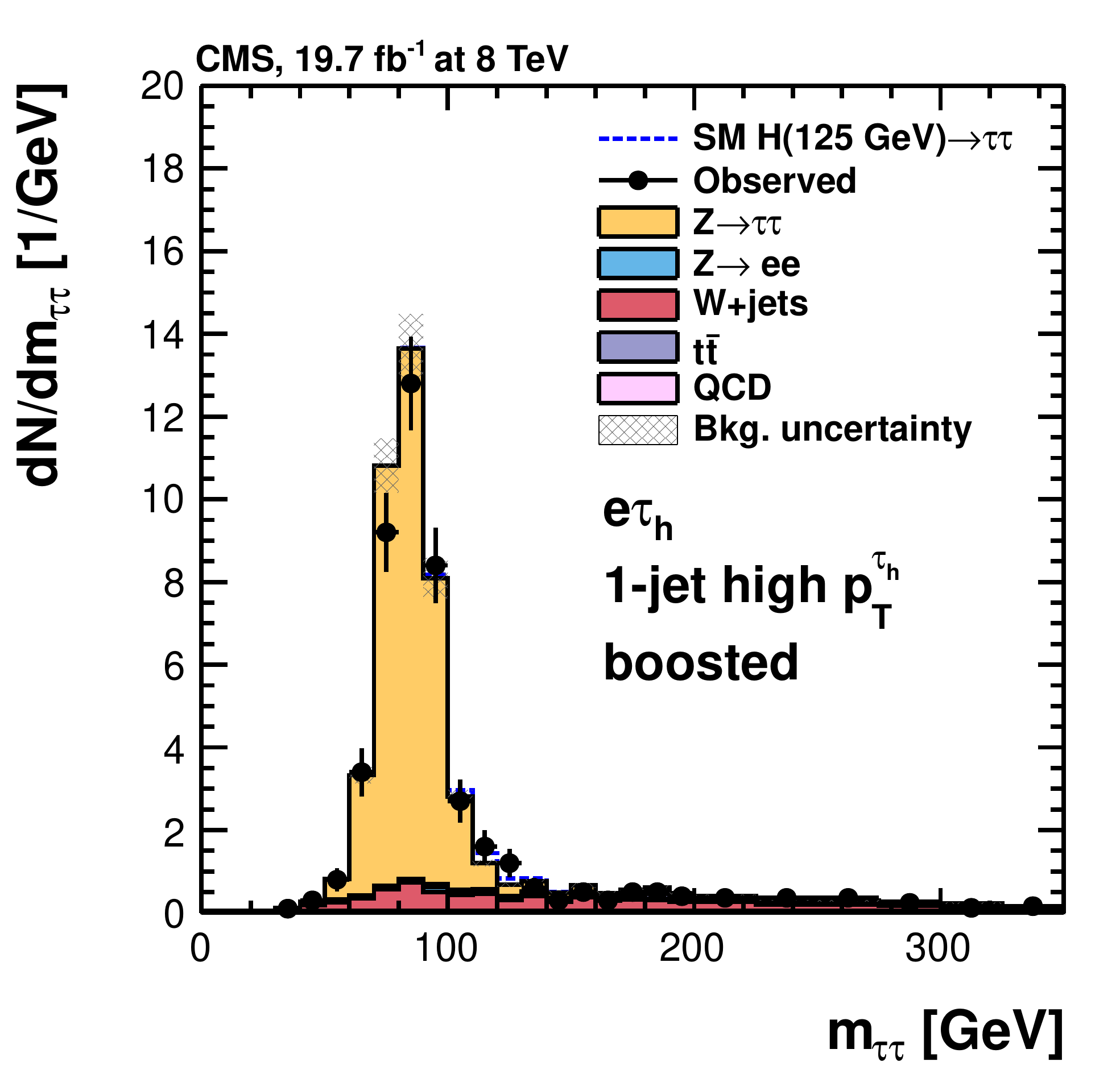} \\
     \includegraphics[width=0.42\textwidth]{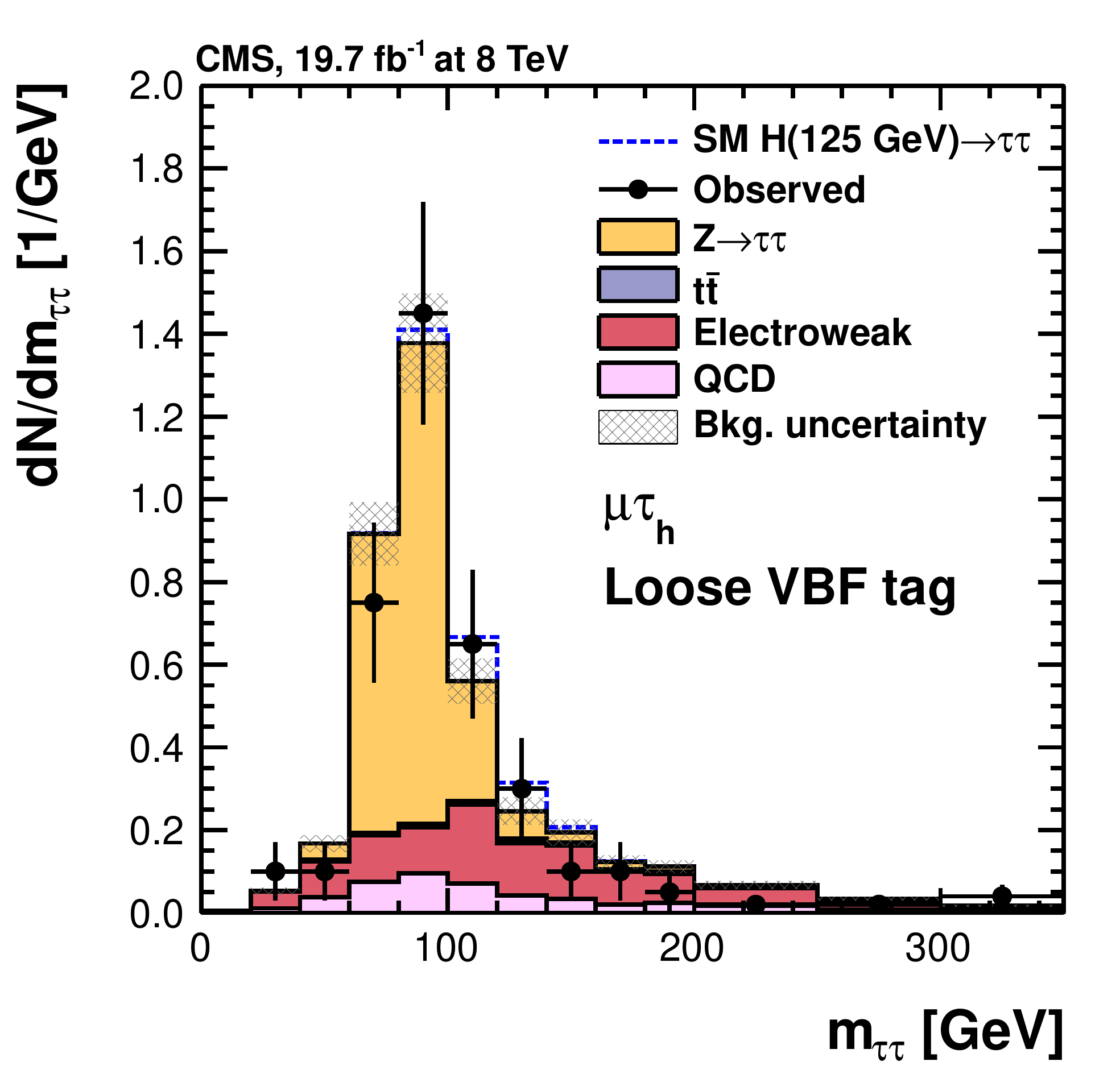}
     \includegraphics[width=0.42\textwidth]{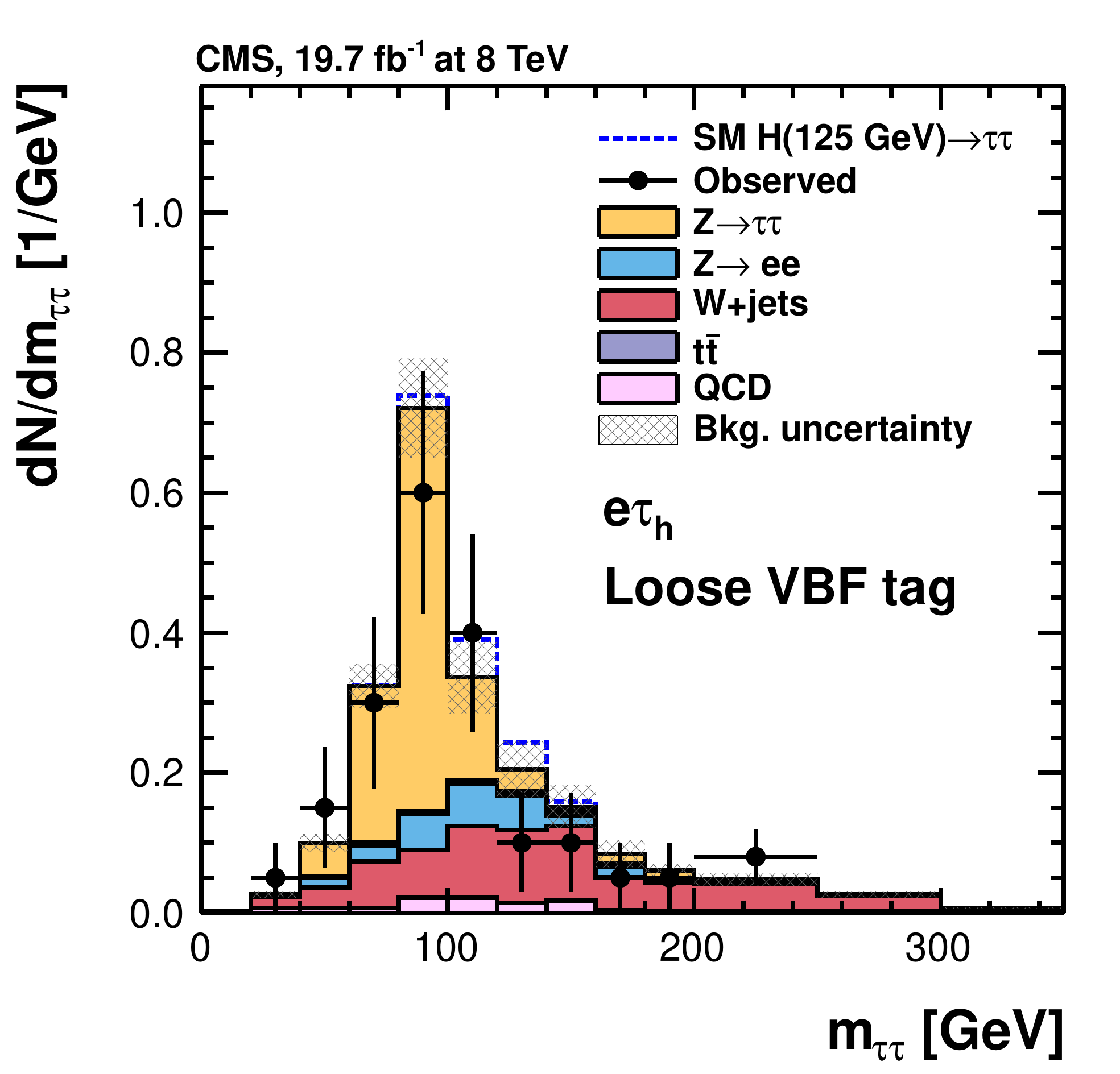} \\
     \includegraphics[width=0.42\textwidth]{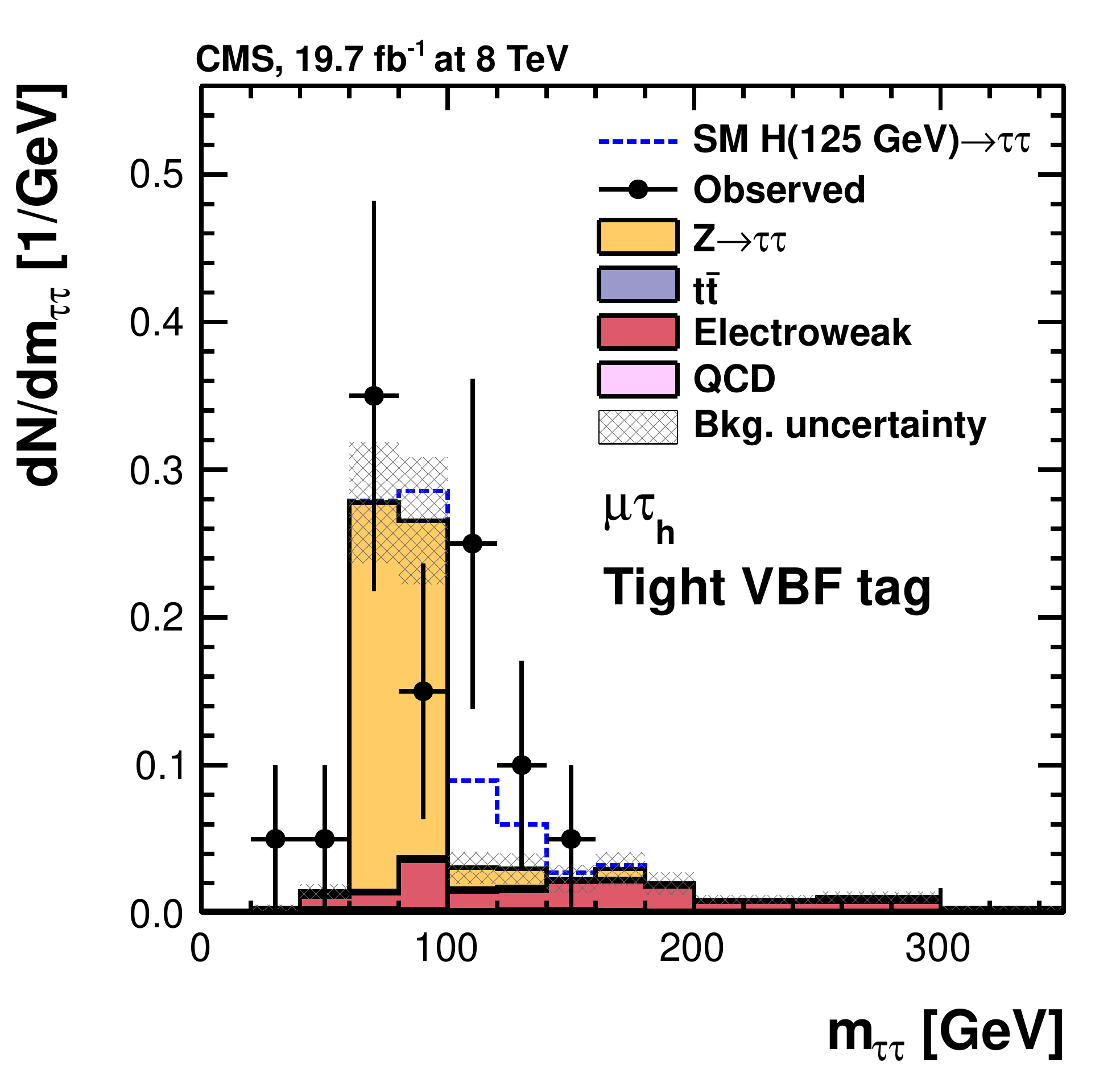}
     \includegraphics[width=0.42\textwidth]{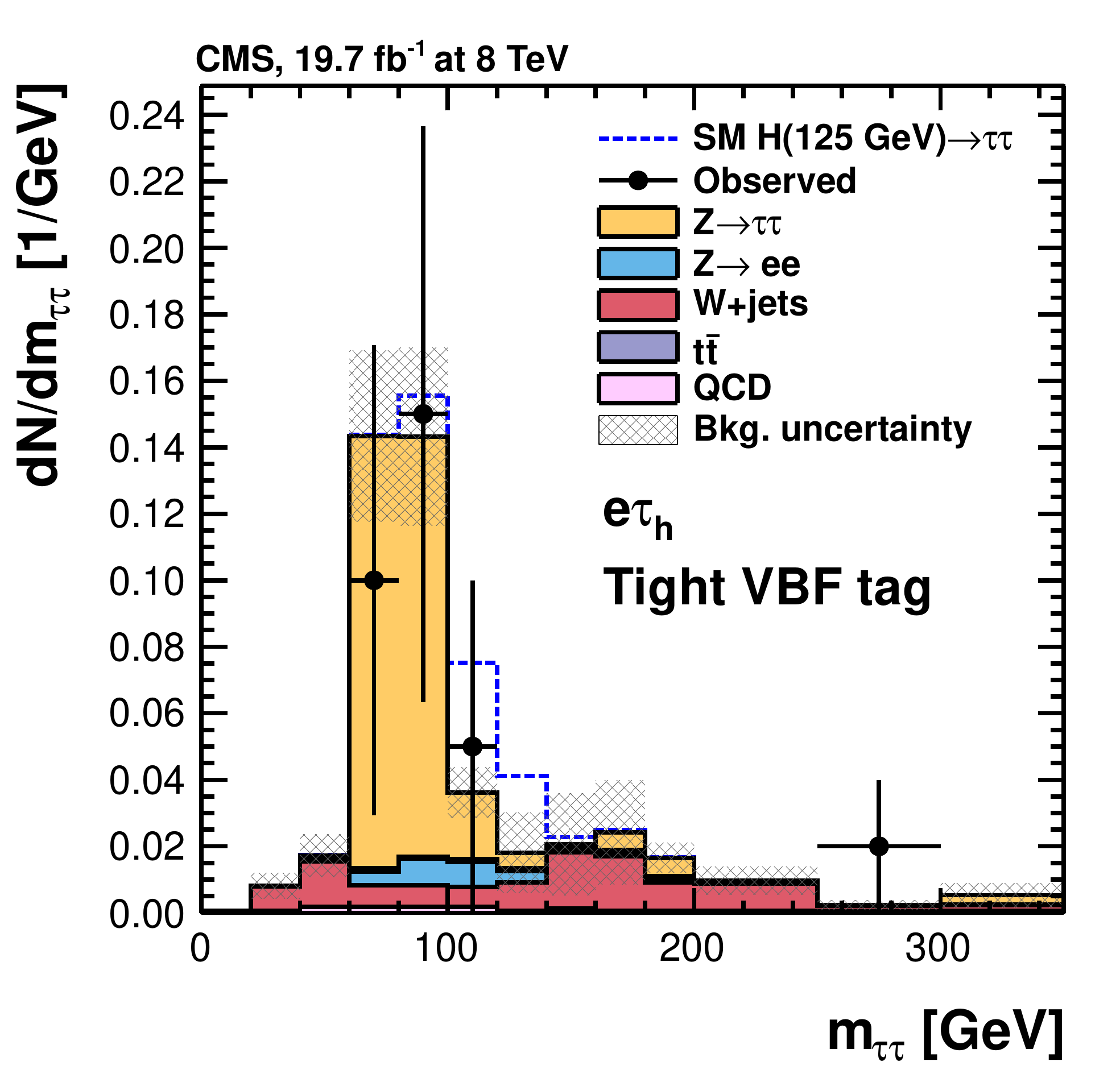} \\
     \caption{Observed and predicted $\mtt$ distributions in the 8\TeV $\Pgm\Pgth$ (left) and $e\Pgth$ (right) channels, and for the 1-jet high-$\pt^{\Pgth}$ boosted (top), loose VBF tag (middle), and tight VBF tag (bottom) categories. The normalization of the predicted background distributions corresponds to the result of the global fit. The signal distribution, on the other hand, is normalized to the SM prediction. The signal and background histograms are stacked.}
     \label{fig:svfit_lt}
\end{figure}

\begin{figure}[htbp]
\centering
     \includegraphics[width=0.42\textwidth]{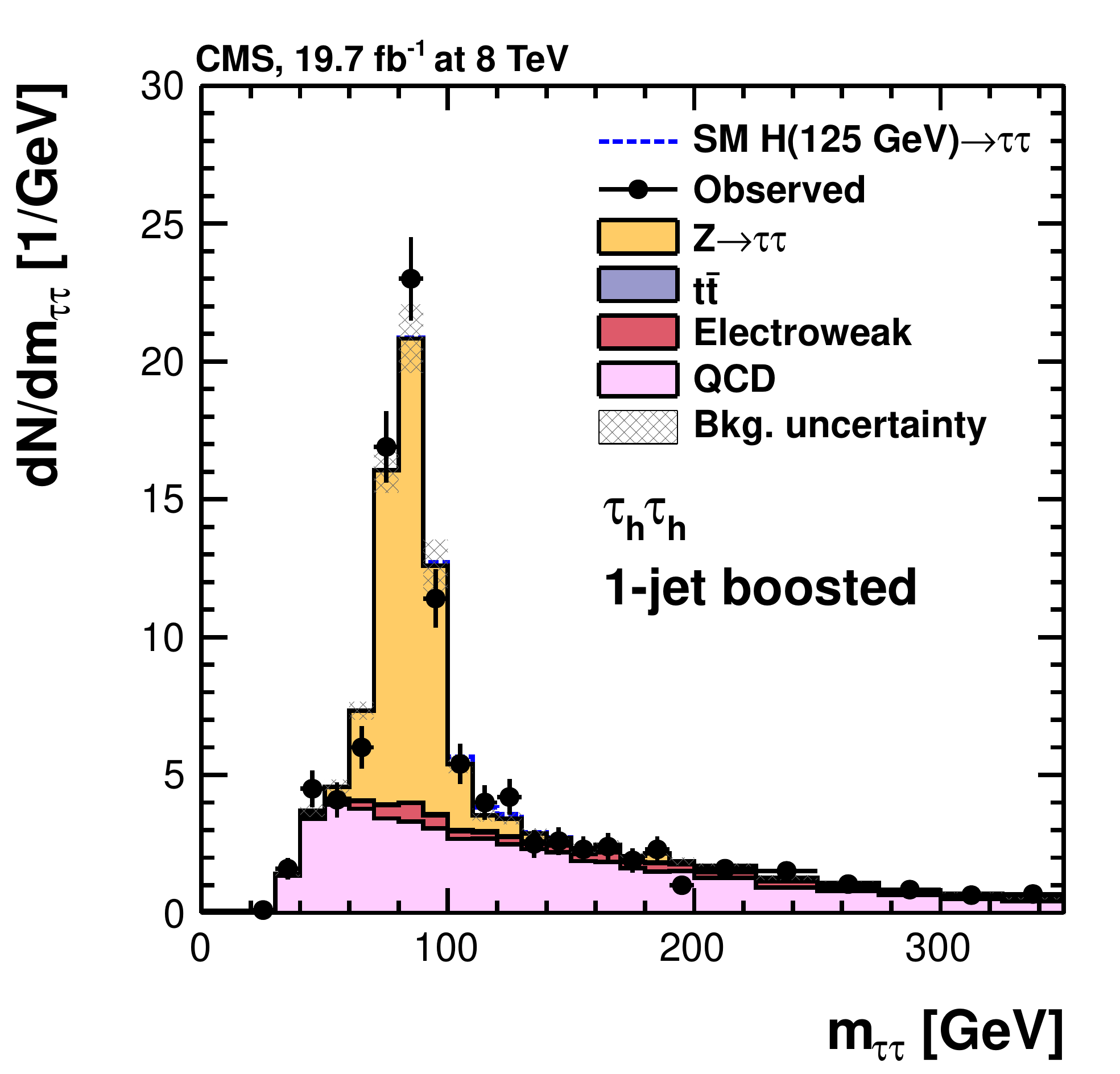}
     \includegraphics[width=0.42\textwidth]{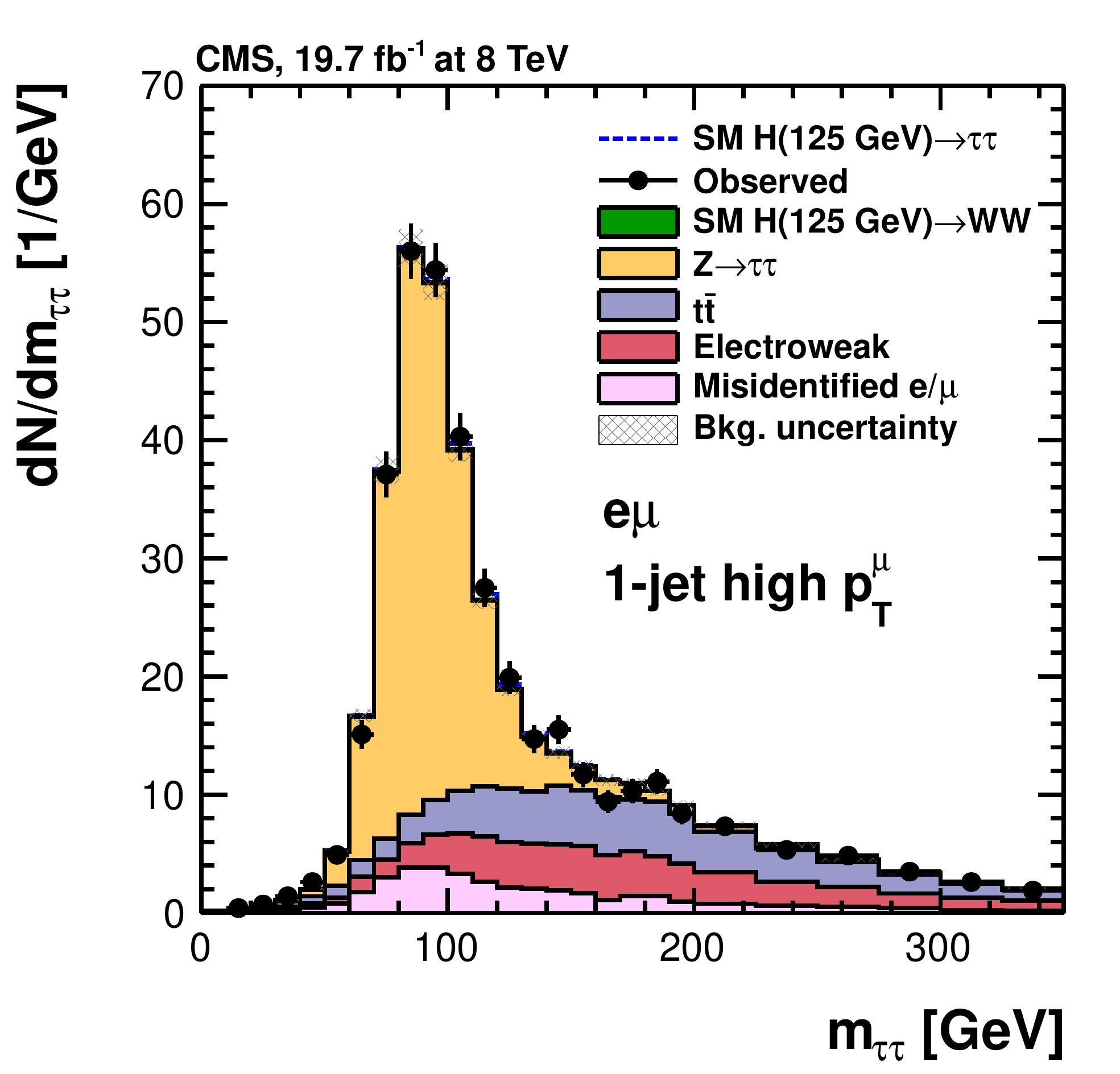} \\
     \includegraphics[width=0.42\textwidth]{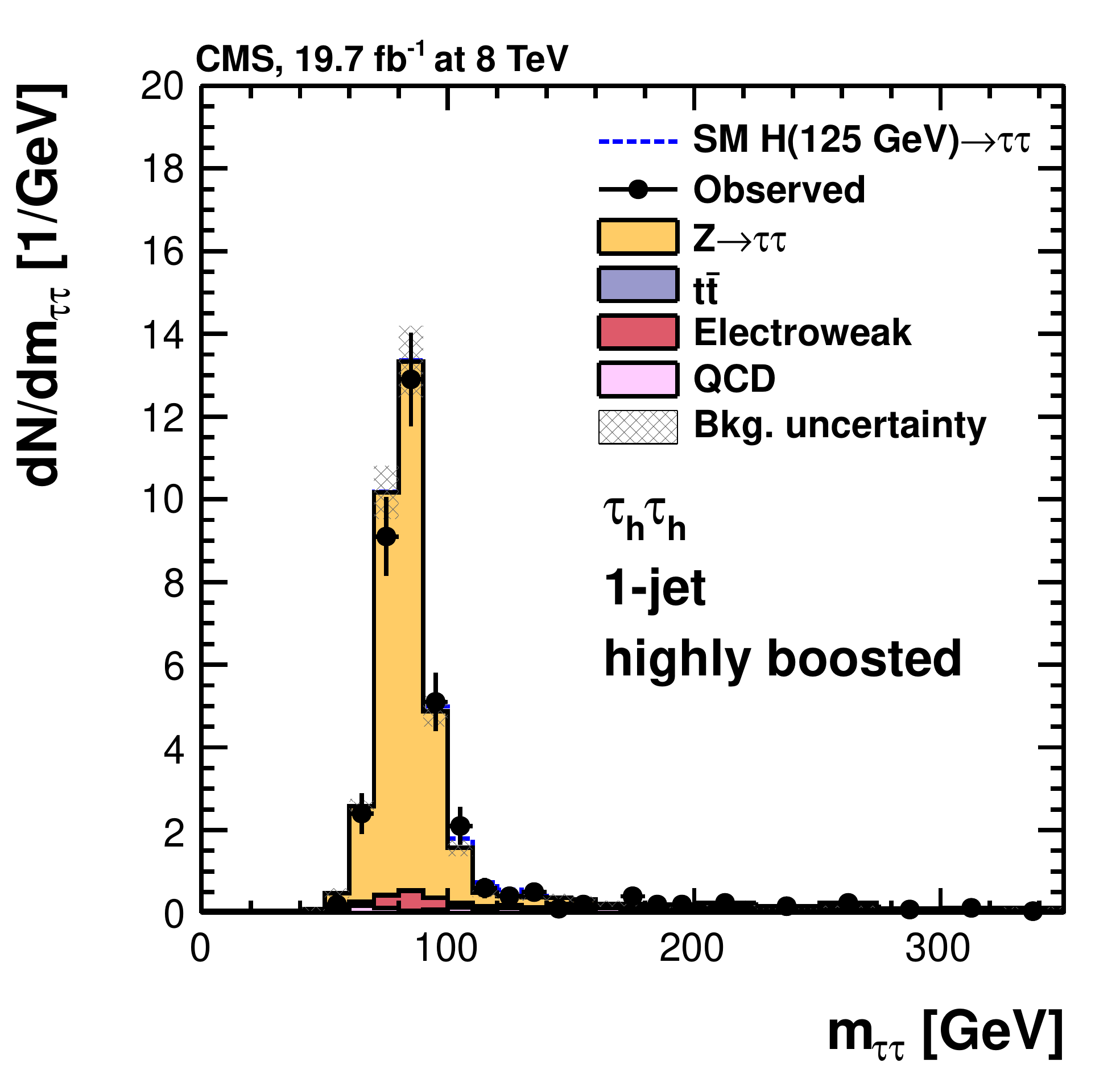}
     \includegraphics[width=0.42\textwidth]{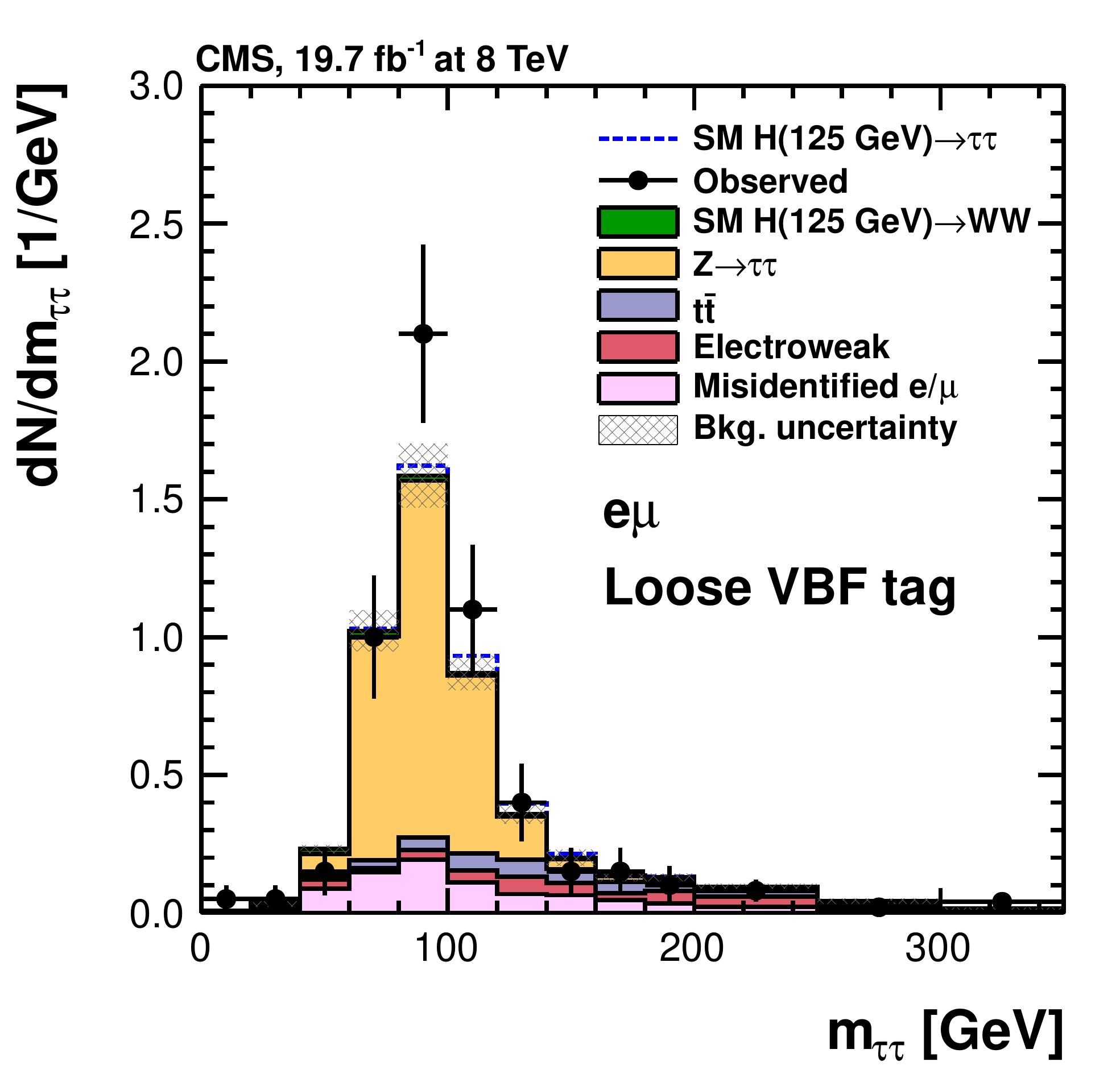} \\
     \includegraphics[width=0.42\textwidth]{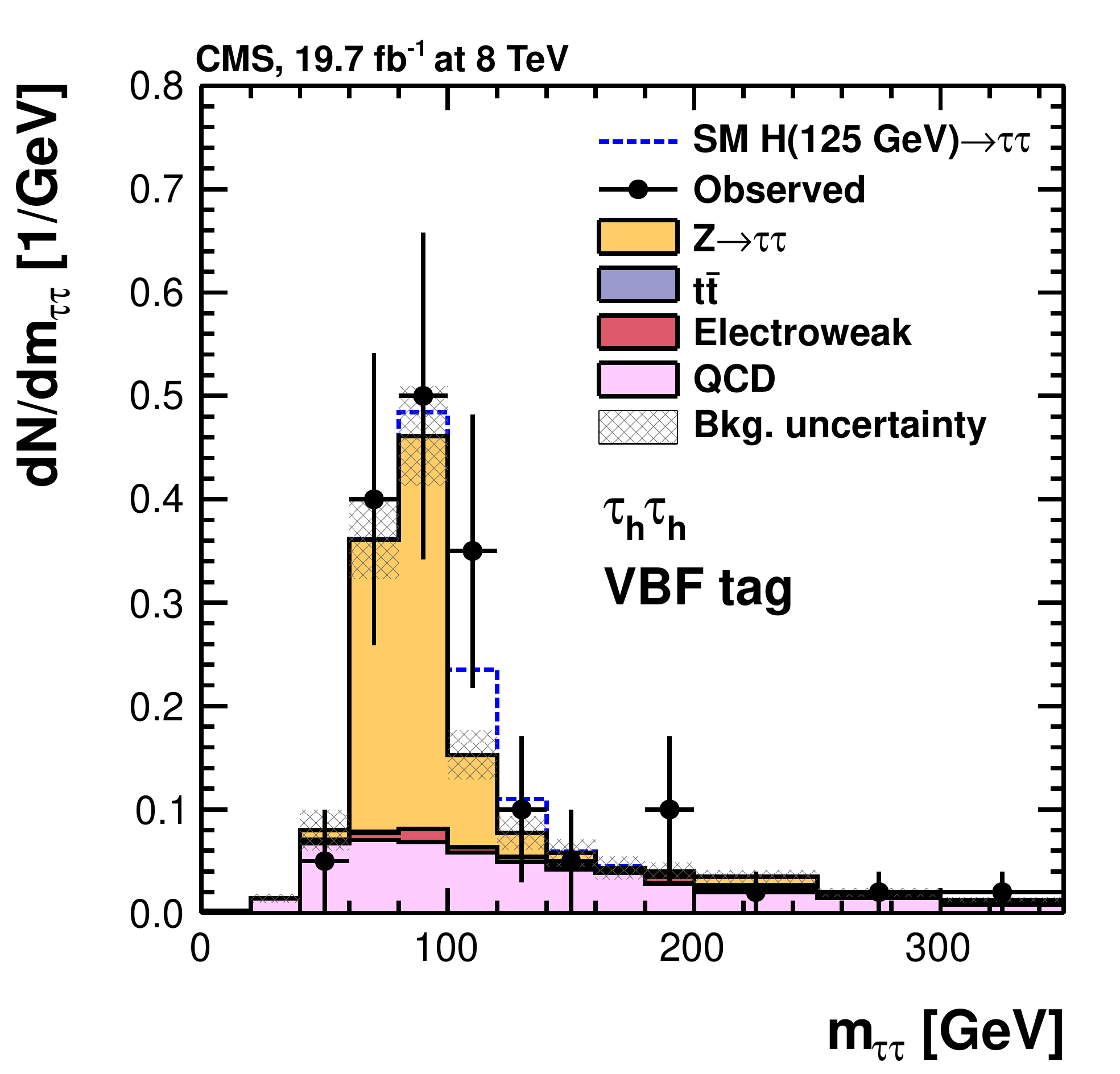}
     \includegraphics[width=0.42\textwidth]{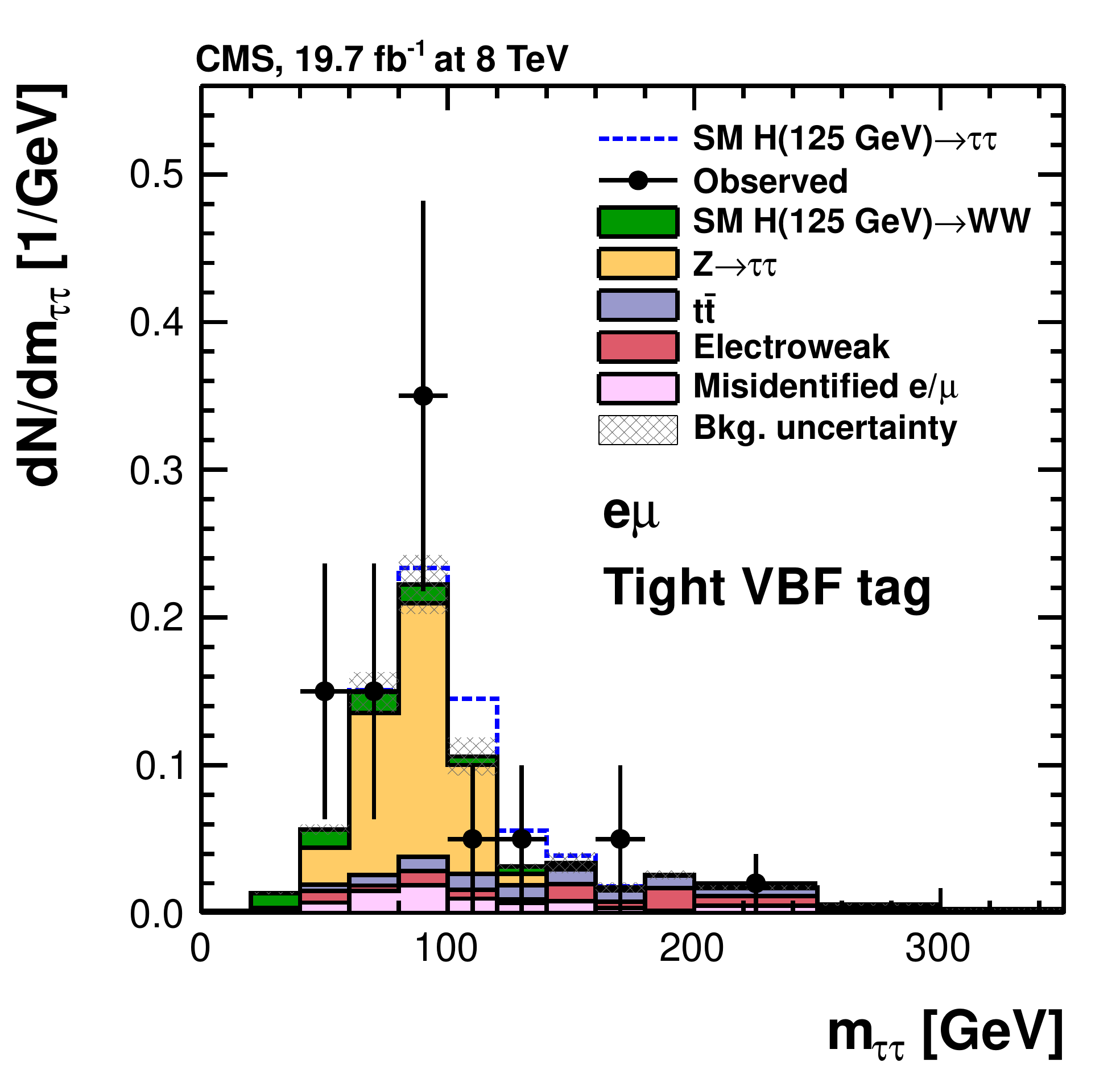} \\
     \caption{Observed and predicted $\mtt$ distributions
in the 8\TeV $\Pgth\Pgth$ (left) channel for the 1-jet boosted (top), 1-jet highly-boosted (middle), and VBF-tagged (bottom) categories,
and in the 8\TeV $\Pe\Pgm$ (right) channel for the 1-jet high-$\pt^\Pgm$ (top), loose VBF tag  (middle) and tight VBF tag (bottom) categories.
The normalization of the predicted background distributions corresponds to the result of the global fit. The signal distribution, on the other hand, is normalized to the SM prediction.
In the $\Pe\Pgm$ channel, the expected contribution from $\hww$ decays is shown separately. The signal and background histograms are stacked.
}
     \label{fig:svfit_emtt}
\end{figure}

Figures~\ref{fig:svfit_lt} and \ref{fig:svfit_emtt} show the $\mtt$ distributions observed for the 8\TeV dataset in the most sensitive categories
of the $\mutau$, $\etau$, $\tautau$, and $\emu$ channels
together with the background distributions resulting from the global fit described in detail below.
The discriminator distributions for the 8\TeV dataset in the $\ell\ell$ channels are shown in figure~\ref{fig:discri_ll}.
The complete set of distributions is presented in appendix~\ref{sec:app_postfit}.
The signal prediction for a Higgs boson with $\mH = 125\GeV$ is normalized to the SM expectation.
The corresponding event yields for all event categories are given in tables~\ref{tab:event_yields} and~\ref{tab:event_yields_eemumu} in appendix~\ref{sec:event_yields}.

\begin{figure}[phtb]
\centering
     \includegraphics[width=0.42\textwidth]{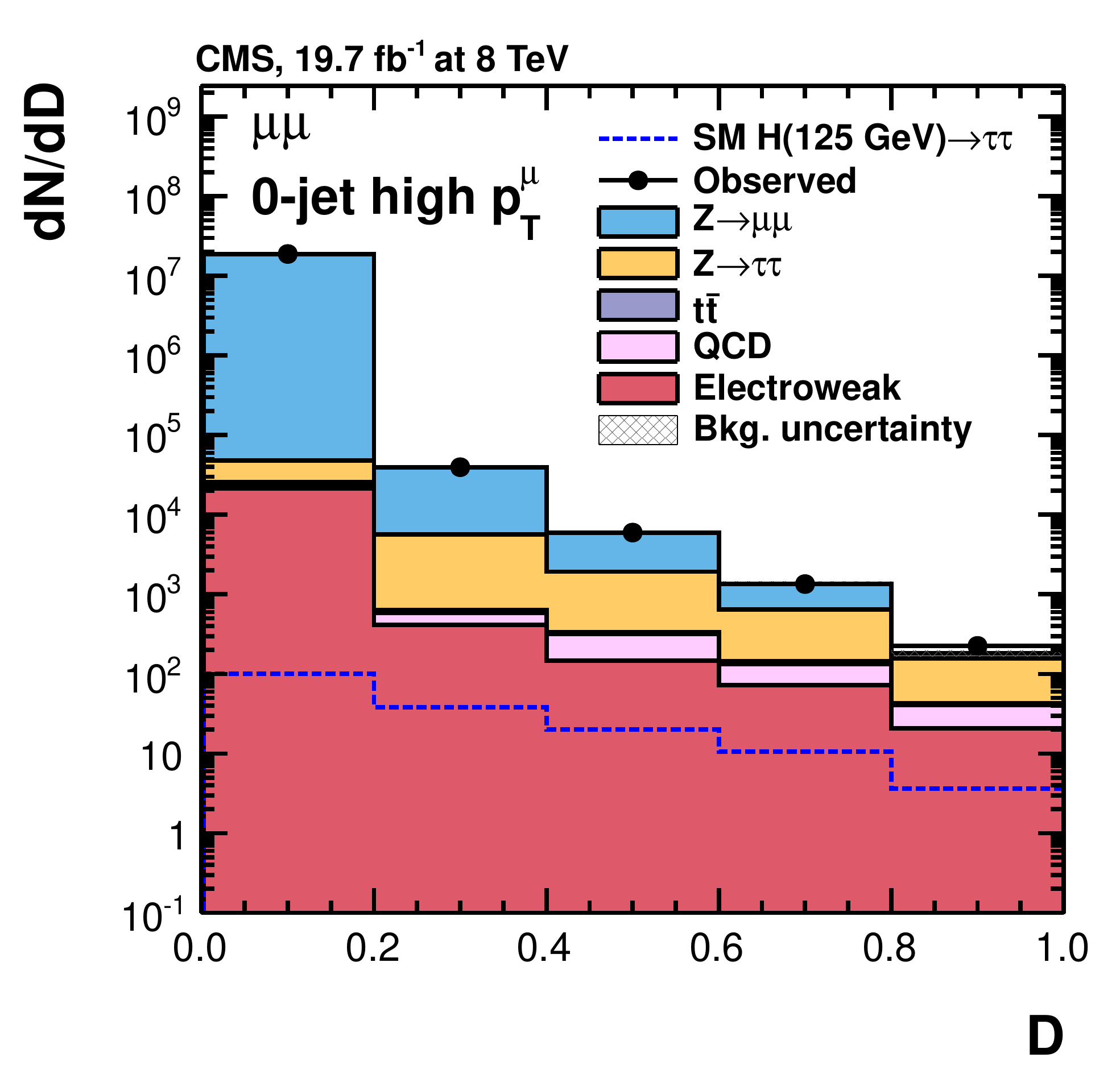}
     \includegraphics[width=0.42\textwidth]{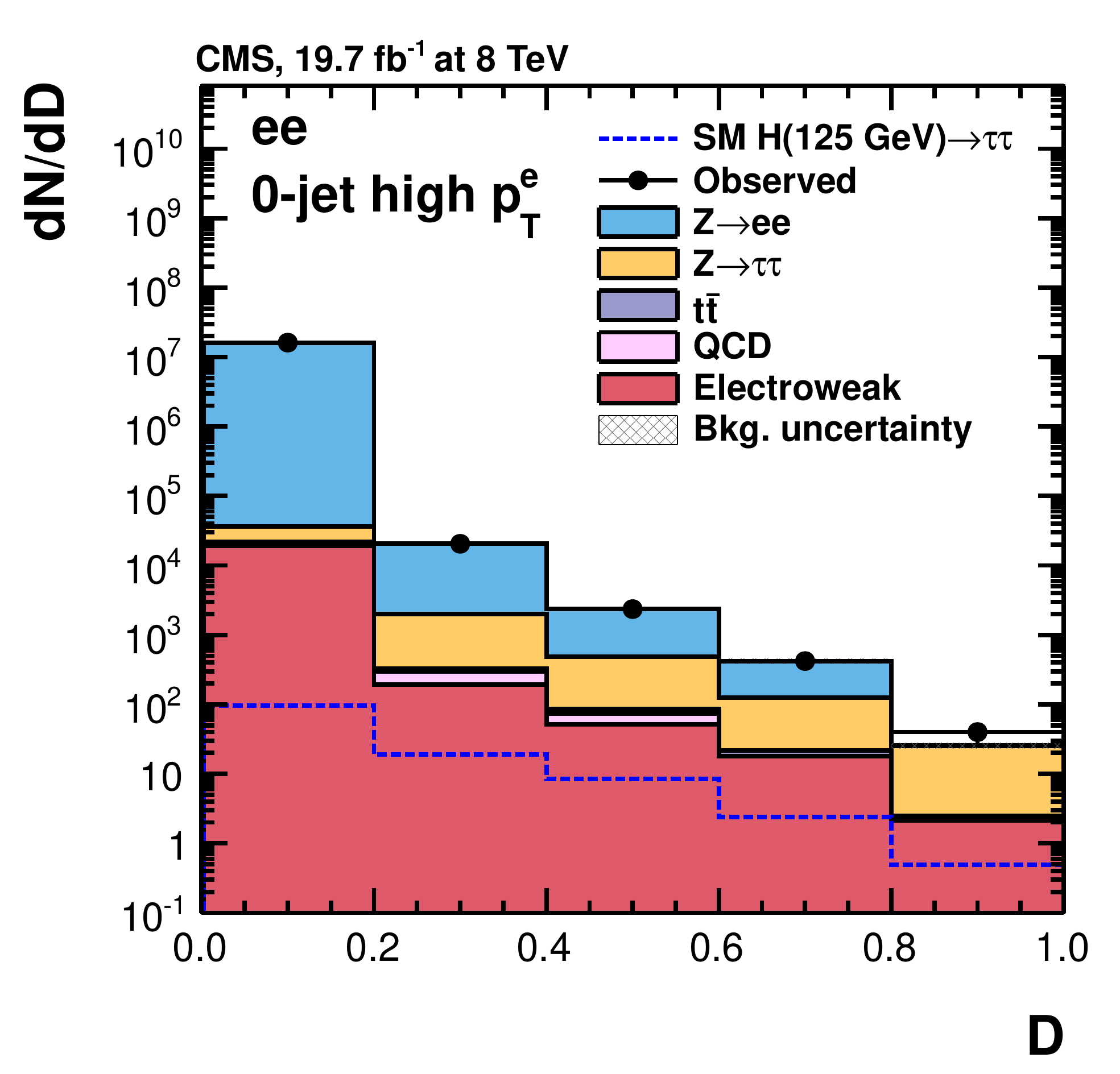} \\
     \includegraphics[width=0.42\textwidth]{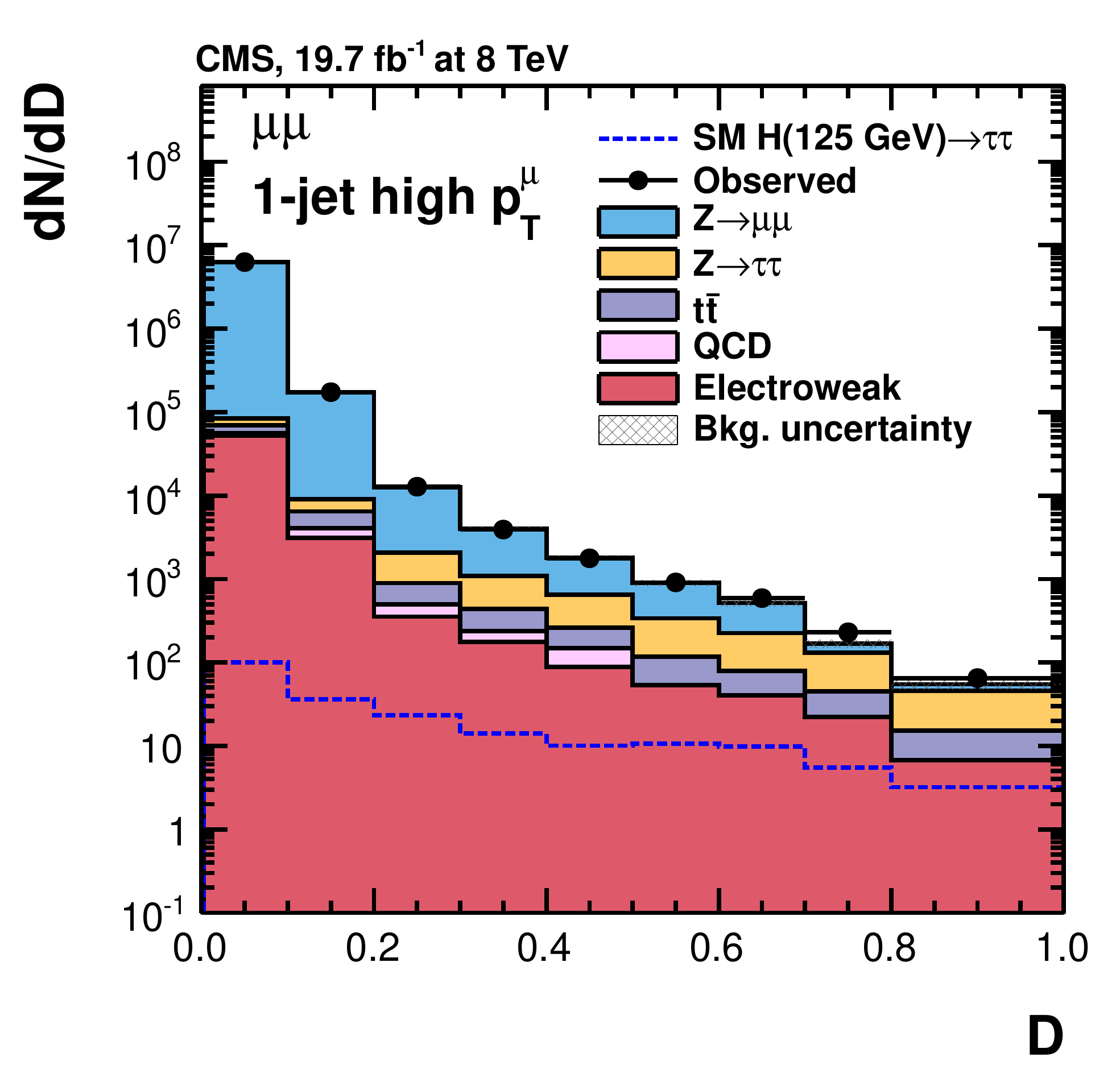}
     \includegraphics[width=0.42\textwidth]{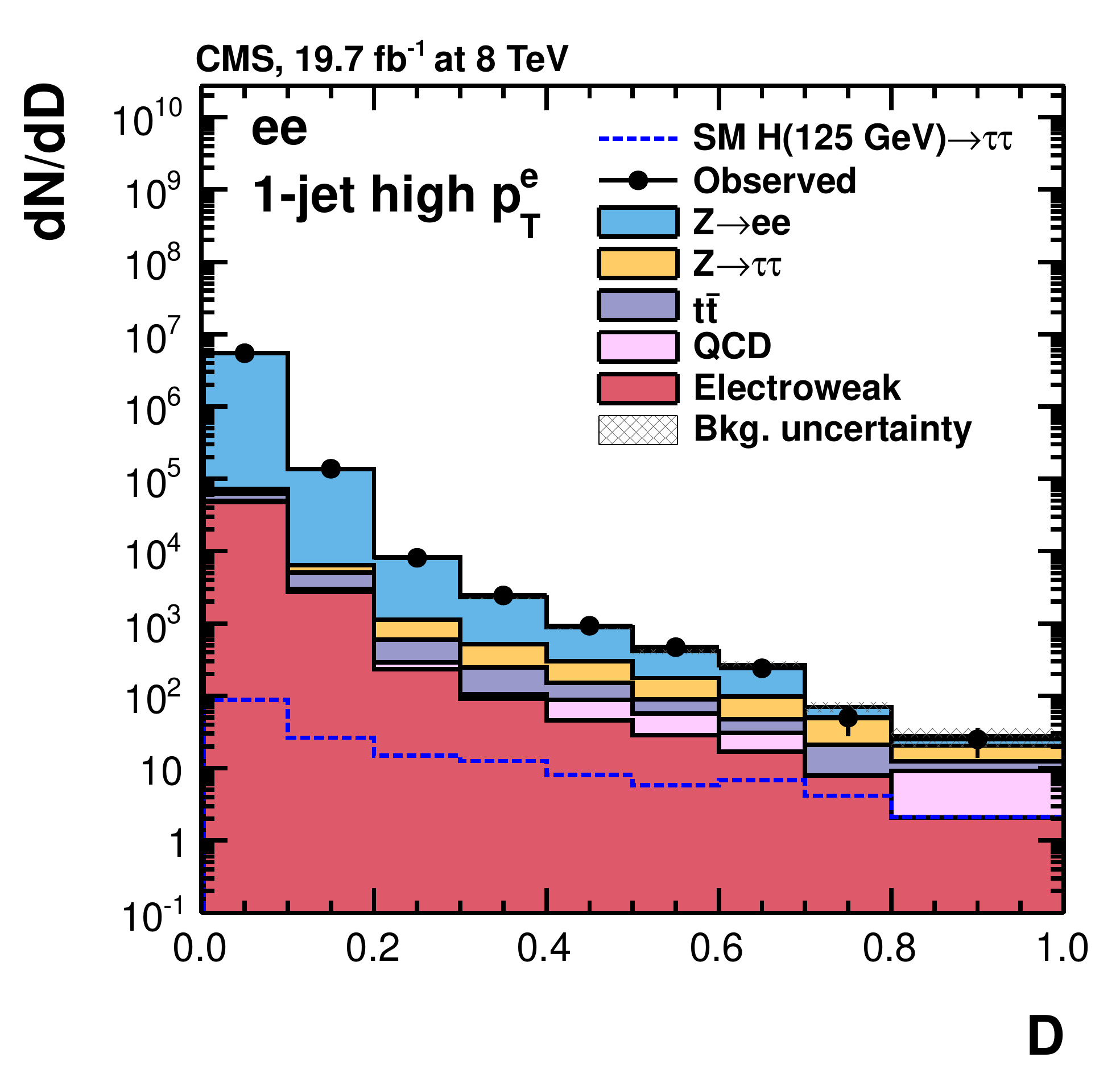} \\
     \includegraphics[width=0.42\textwidth]{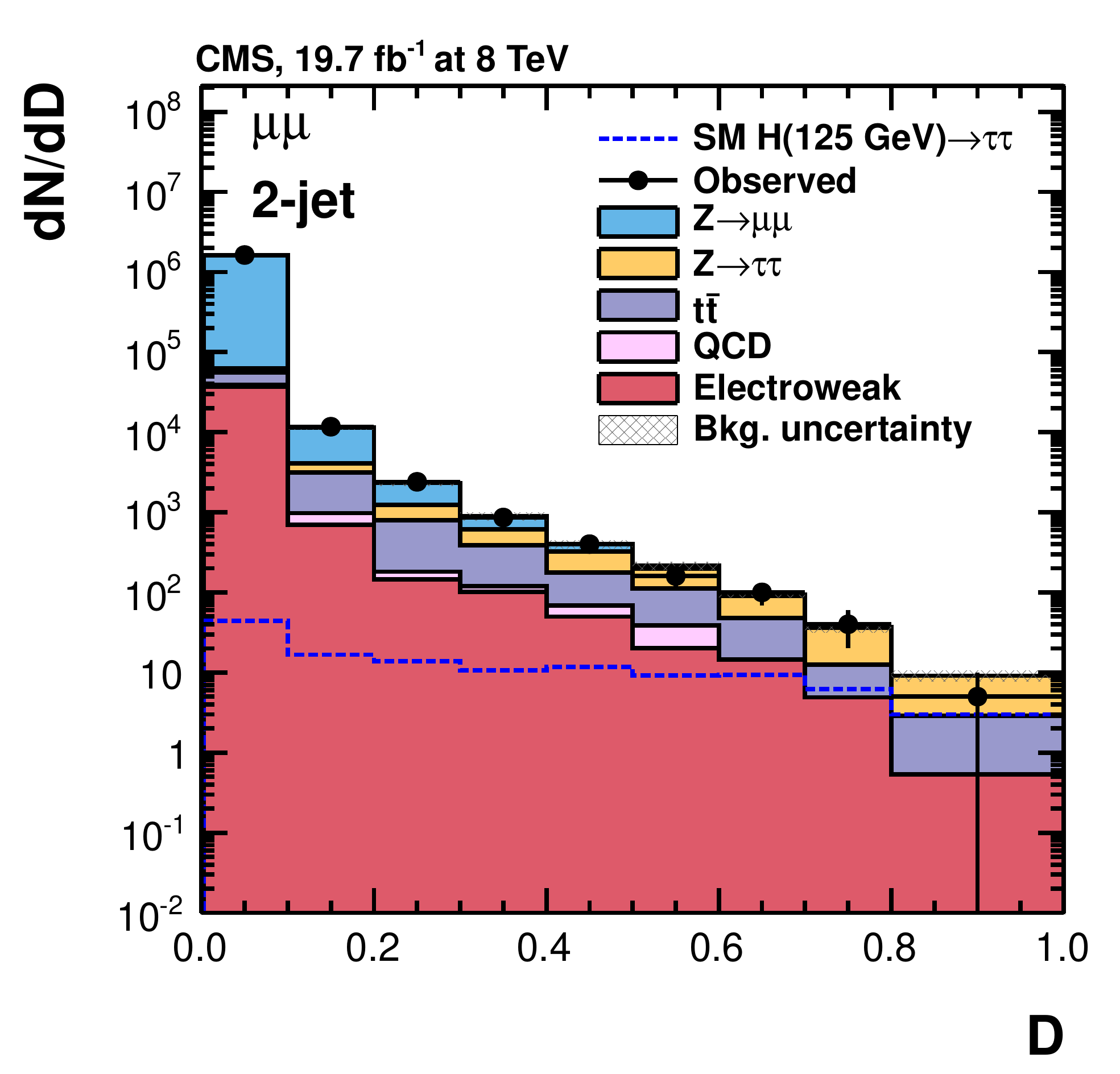}
     \includegraphics[width=0.42\textwidth]{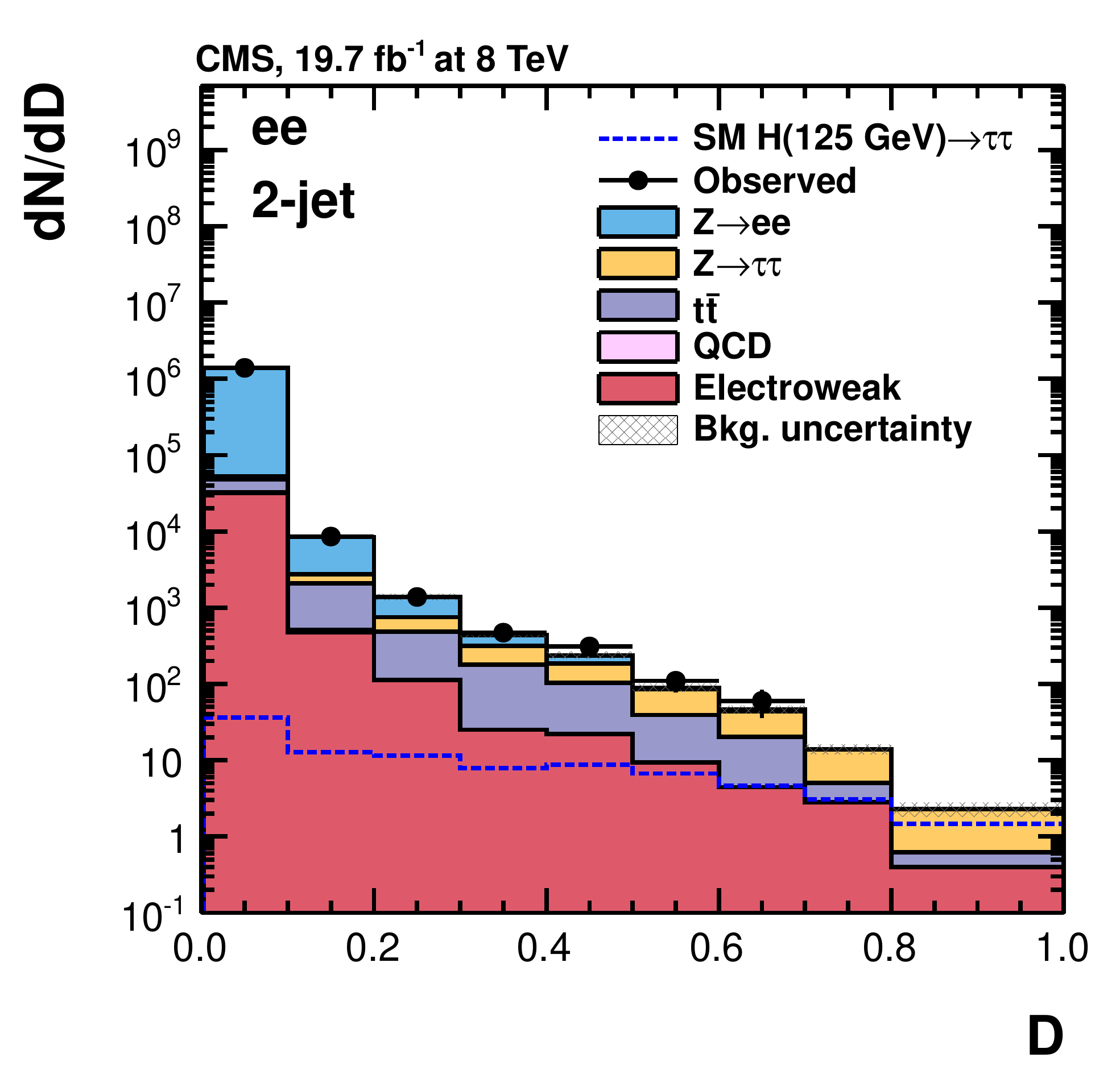} \\
     \caption{Observed and predicted distributions for the final discriminator $D$ in the 8\TeV $\Pgm\Pgm$ (left) and $\Pe\Pe$ (right) channels, and for the 0-jet high-$\pt^\ell$ (top), 1-jet high-$\pt^\ell$ (middle), and 2-jet (bottom) categories. The normalization of the predicted background distributions corresponds to the result of the global fit. The signal distribution, on the other hand, is normalized to the SM prediction. The open signal histogram is shown superimposed to the background histograms, which are stacked.}
     \label{fig:discri_ll}
\end{figure}

For the global fit, the distributions of the final discriminating variable obtained for each category and each channel at 7 and 8\TeV are combined in a binned likelihood,
involving the expected and observed numbers of events in each bin.
The expected number of signal events is the one predicted by the SM for the production of a Higgs boson of mass \mH decaying into a pair of $\tau$ leptons,
multiplied by a signal strength modifier $\mu$ treated as free parameter in the fit.

The systematic uncertainties are represented by nuisance parameters that are varied in the fit according to their probability density function.
A log-normal probability density function is assumed for the nuisance parameters affecting the event yields of the various background contributions.
Systematic uncertainties that affect the template shapes,
\eg the $\Pgth$ energy scale uncertainty,
are represented by nuisance parameters whose variation results in a continuous perturbation of the spectrum~\cite{Conway-PhyStat} and which are assumed to have a Gaussian probability density function.

Nuisance parameters affect the yields and template shapes across categories and channels when applicable.
For example, in the VBF-tagged categories of the $\Pm\Pgth$ channel,
the most important nuisance parameters related to background normalization are the ones affecting
the $\PZ\to \tau \tau$ yield ($\Pgth$ selection efficiency)
and the $\PW$ + jets yield (statistical uncertainty for the normalization to the yield in the high-\MT region, and extrapolation to the low-$\MT$ region).
The nuisance parameter affecting the $\PW$ + jets yield is constrained only by the events observed in the given category, in particular in the high-mass region.
The nuisance parameters related to the $\Pgth$ identification efficiency and to the \Pgth energy scale are, however,
mostly constrained by the 0-jet and 1-jet categories,
for which the number of events in the $\PZ\to \tau\tau$ peak is very large.
Overall, the statistical uncertainty in the observed event yields is the dominant source of uncertainty for all combined results.

\begin{figure}[htb]
  \begin{center}
      \includegraphics[width=0.8\textwidth]{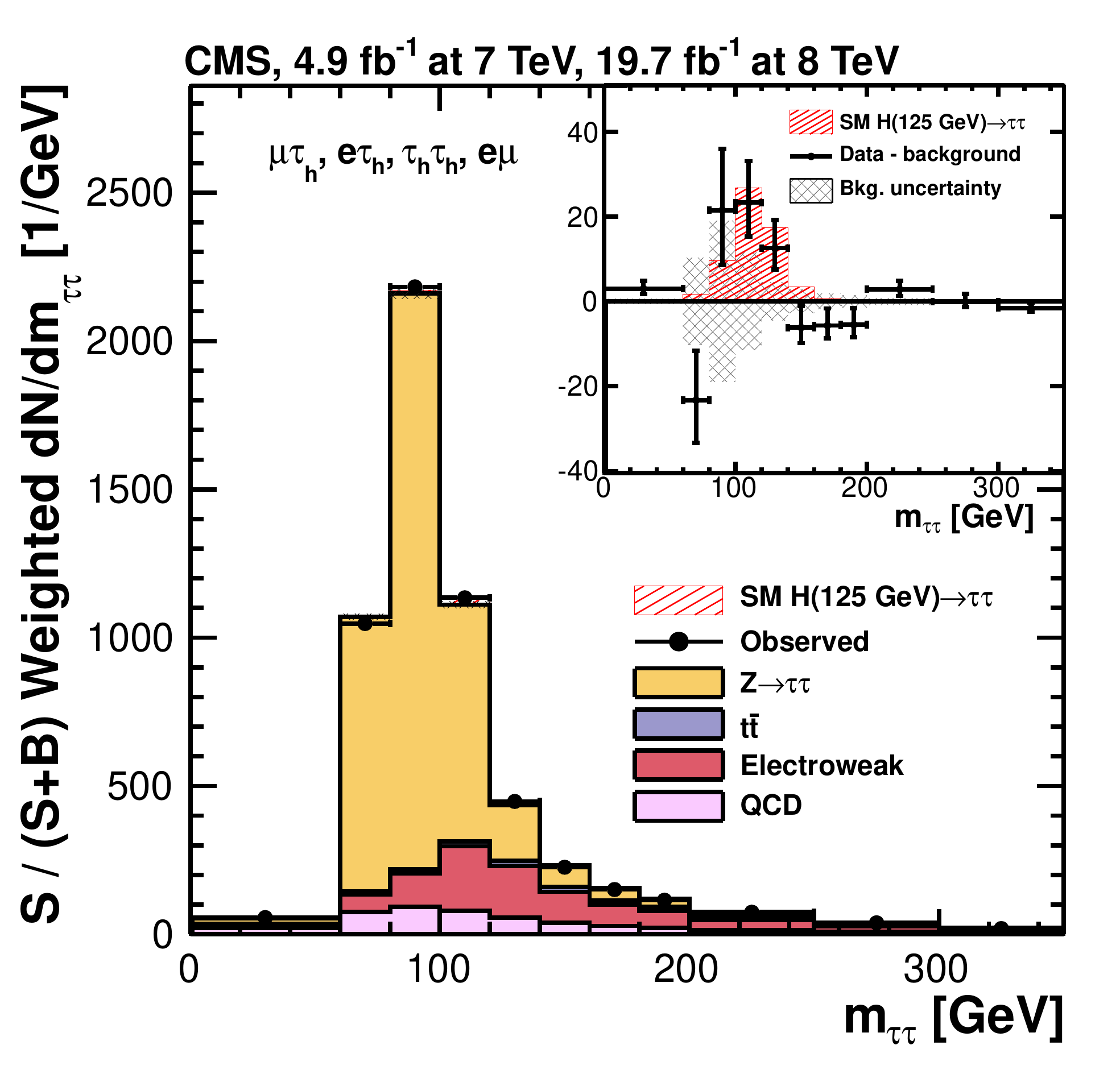}
   \caption{Combined observed and predicted $\mtt$ distributions for the $\Pgm\Pgth$, $\Pe\Pgth$, $\Pgth\Pgth$, and $\Pe\Pgm$ channels.
   The normalization of the predicted background distributions corresponds to the result of the global fit.
   The signal distribution, on the other hand, is normalized to the SM prediction ($\mu = 1$).
   The distributions obtained in each category of each channel are weighted by the ratio between the expected signal and signal-plus-background yields in the category, obtained in the central $\mtt$ interval containing 68\% of the signal events.
The inset shows the corresponding difference between the observed data and expected background distributions, together with the signal distribution for a SM Higgs boson at $\mH=125\GeV$.
The distribution from SM Higgs boson events in the $\WW$ decay channel does not
significantly contribute to this plot.
}
    \label{fig:weighted_distribution}
  \end{center}
\end{figure}

The excess of events observed in the most sensitive categories of figures~\ref{fig:svfit_lt} and \ref{fig:svfit_emtt} is highlighted in figure~\ref{fig:weighted_distribution}, which shows the observed and expected $\mtt$ distributions for all categories of the $\ell\Pgth$, $\Pe\Pgm$, and $\Pgth\Pgth$ channels combined.
The $\Pe\Pe$ and $\Pgm\Pgm$ channels are not included because the final discrimination is based on $D$ instead of \mtt in these channels.
The distributions are weighted in each category of each channel
by the $S/(S+B)$ ratio where $S$ is the expected signal yield for a SM Higgs boson with $\mH = 125\GeV$ ($\mu=1$)
and $B$ is the predicted background yield corresponding to the result of the global fit.
The ratio is obtained in the central $\mtt$ interval containing 68\% of the signal events.
The figure also shows the difference between the observed data and expected background distributions,
together with the expected distribution for a SM Higgs boson signal with $\mH=125\GeV$.

\begin{figure}[htb]
  \begin{center}
    \includegraphics[width=0.68\textwidth]{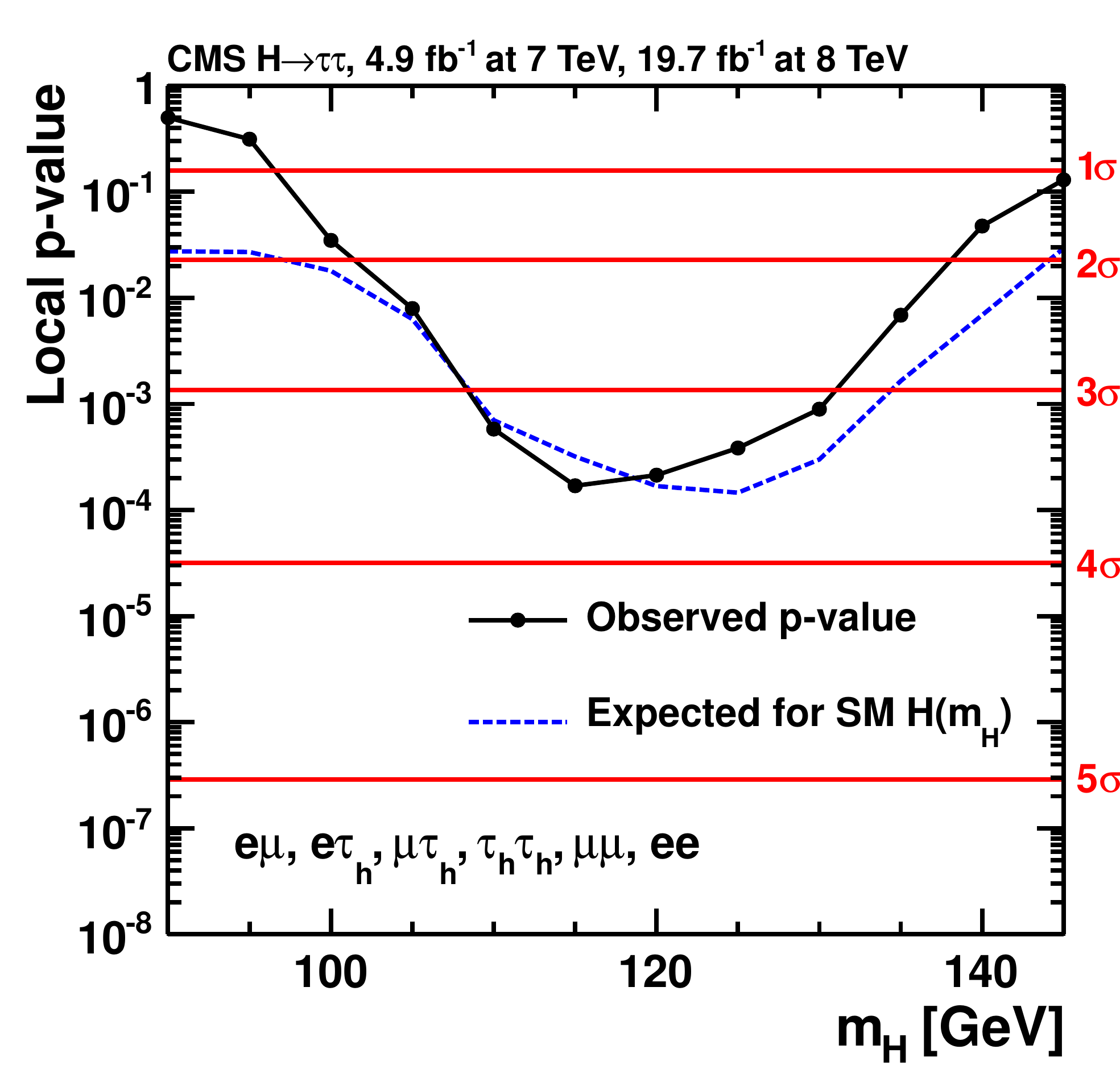}
   \caption{Local $p$-value and significance in number of standard deviations as a function of the SM Higgs boson mass hypothesis for the $LL'$ channels.
The observation (solid line) is compared to the expectation (dashed line) for a SM Higgs boson with mass \mH.
The background-only hypothesis includes the $\Pp\Pp\to\PH\text{(125\GeV)}\to\PW\PW$ process for every value of \mH.
   }
    \label{fig:p_value_nonvh}
  \end{center}
\end{figure}

The visible excess in the weighted $\mtt$ distribution is quantified by calculating the corresponding local $p$-values for the $LL'$ channels using a profile-likelihood ratio test statistics~\cite{LHC-HCG-Report,Chatrchyan:2012tx}.
Figure~\ref{fig:p_value_nonvh} shows the distribution of local $p$-values and significances as a function of the Higgs boson mass hypothesis.
The expected significance for a SM Higgs boson with $\mH = 125$\GeV is 3.6 standard deviations.
For \mH between 110 and 130\GeV, the observed significance is larger than three standard deviations,
and equals 3.4 standard deviations for $\mH = 125$\GeV.
The corresponding best-fit value for $\mu$ is $\hat \mu = 0.86 \pm 0.29$ at $\mH = 125\GeV$.

\begin{figure}[htbp]
     \includegraphics[width=0.47\textwidth]{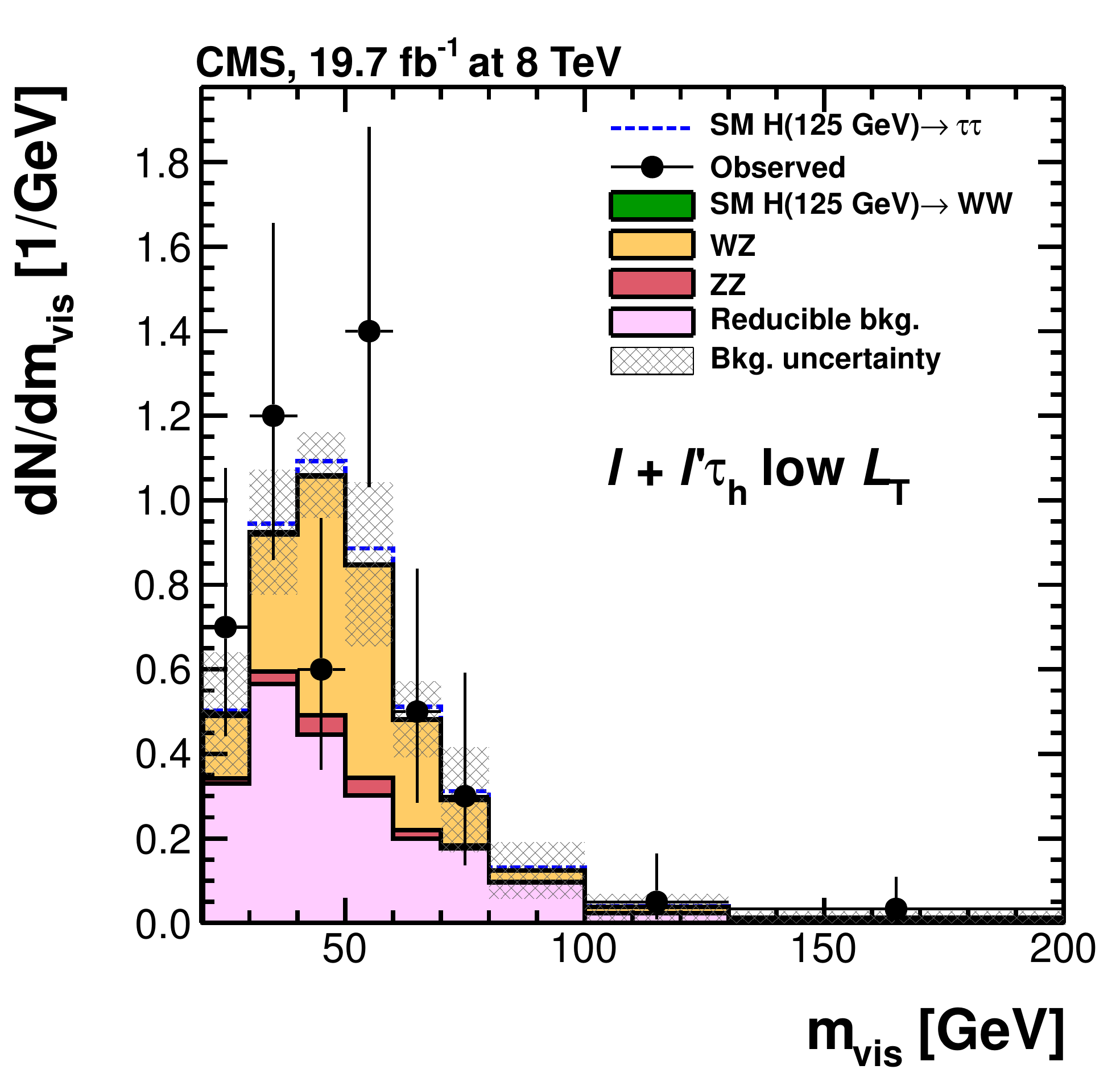}
     \includegraphics[width=0.47\textwidth]{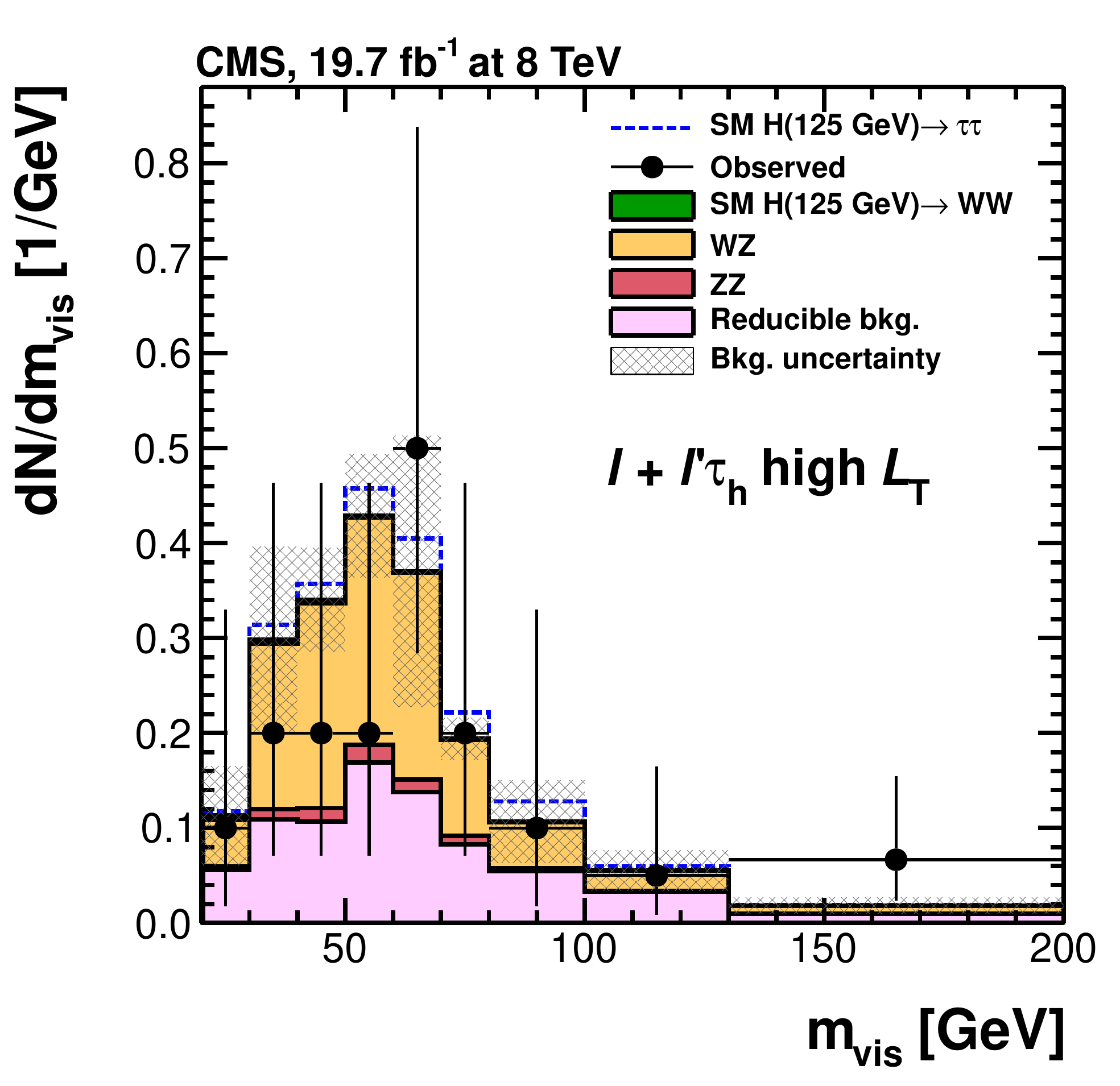} \\
     \includegraphics[width=0.47\textwidth]{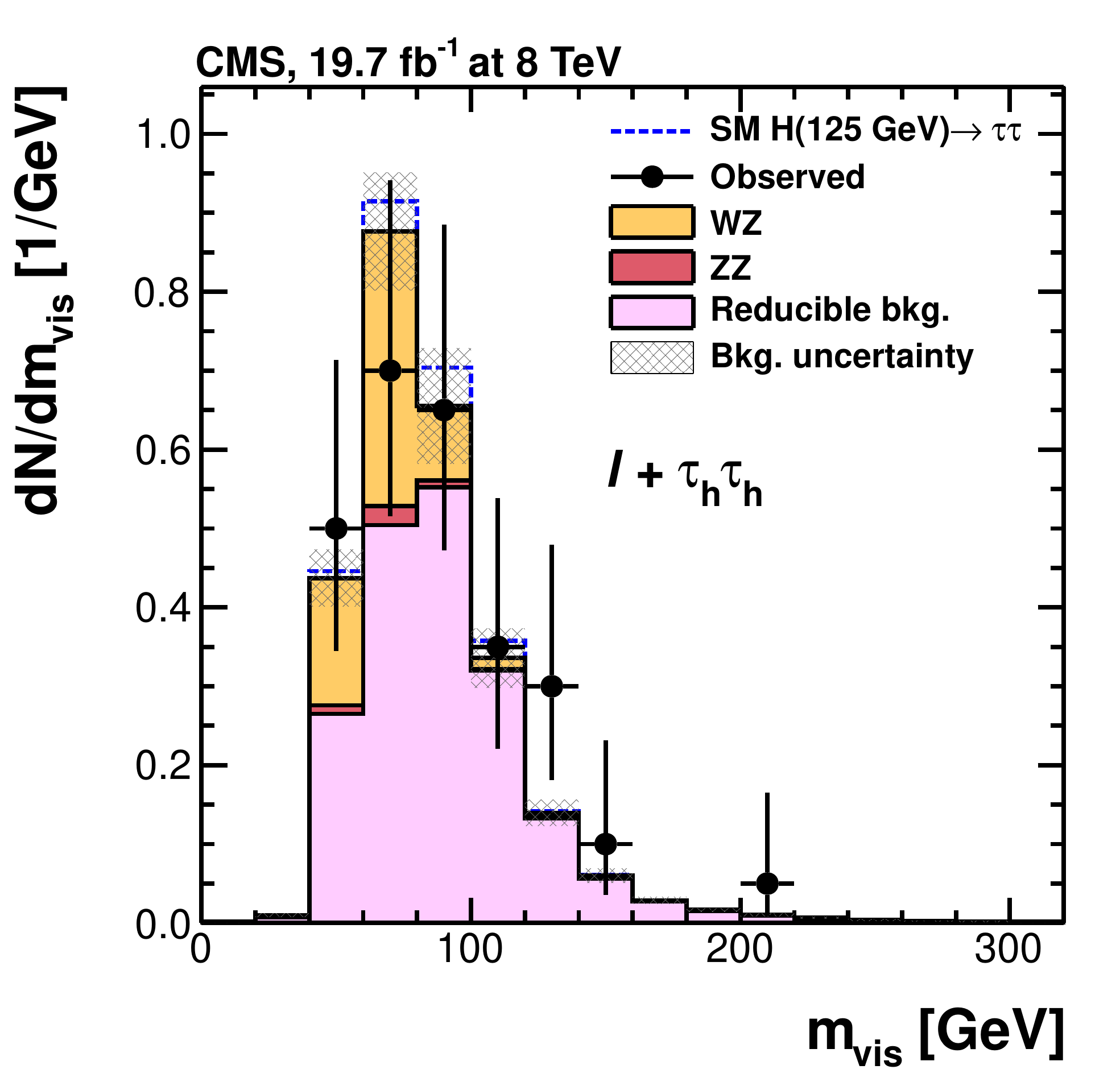}
     \includegraphics[width=0.47\textwidth]{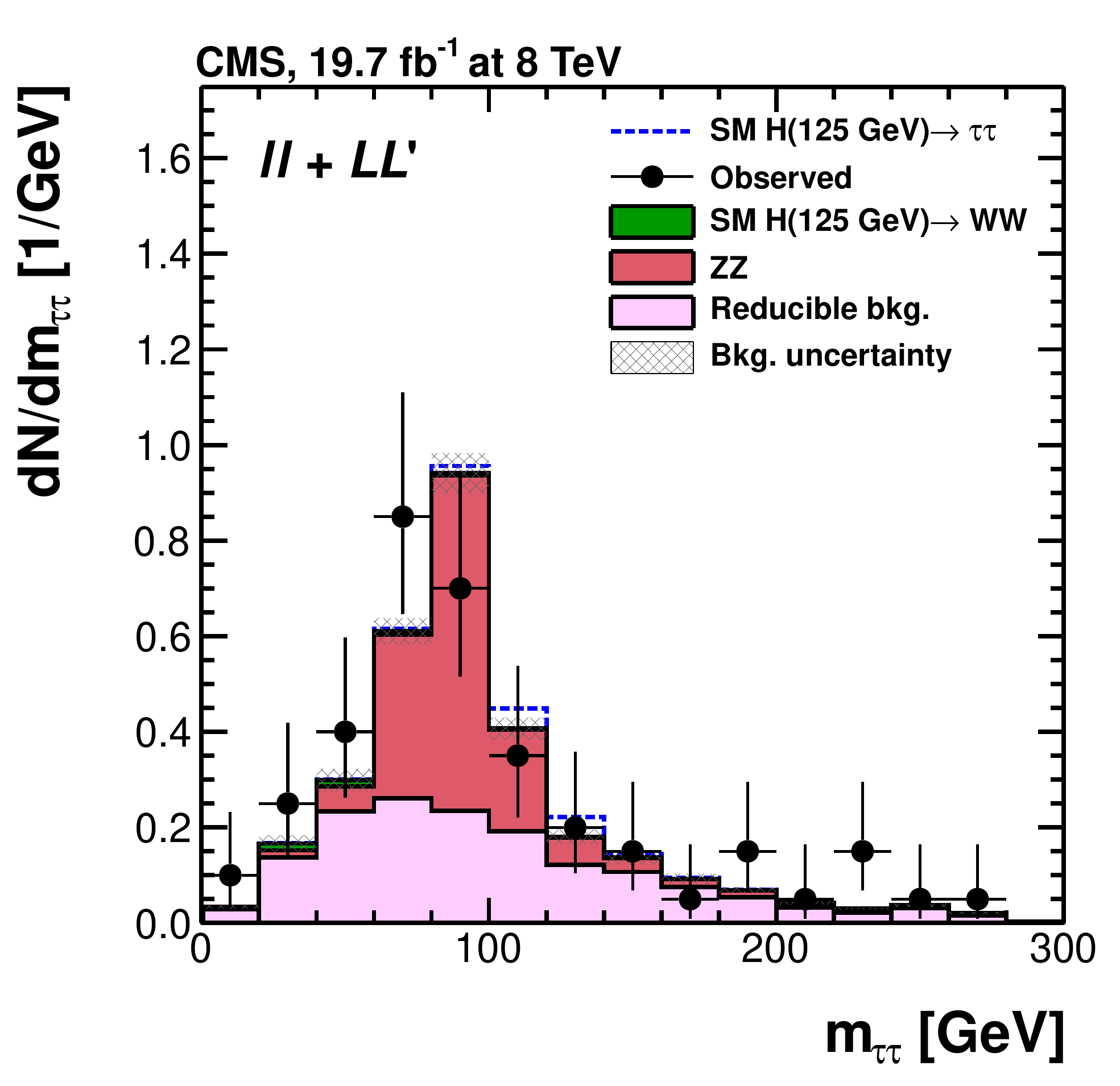}

     \caption{
Observed and predicted
$\mvis$ distributions in the $\ell+\ell'\Pgth$ channel in the low-$\LT$ (top left) and high-$\LT$ (top right) categories,
each for the 8\TeV dataset,
and in the $\ell+\Pgth\Pgth$ channel (bottom left);
observed and predicted $\mtt$ distributions in the $\ell\ell + LL'$ channel (bottom right).
The normalization of the predicted background distributions corresponds to the result of the global fit. The signal distribution, on the other hand, is normalized to the SM prediction ($\mu = 1$). The signal and background histograms are stacked.
}
     \label{fig:mass_vh}
\end{figure}

The $\mvis$ or $\mtt$ distributions obtained for the 8\TeV dataset in the $\ell + L\Pgth$ and $\ell\ell + LL'$ channels are shown in figure~\ref{fig:mass_vh}.
Because of the small number of expected events in each event category,
different event categories are combined.
The complete set of distributions is presented in appendix~\ref{sec:app_postfit},
and the event yields for the individual event categories are given in table~\ref{tab:event_yields_vh} in appendix~\ref{sec:event_yields}.

\begin{figure}[htbp]
  \begin{center}
     \includegraphics[width=0.49\textwidth]{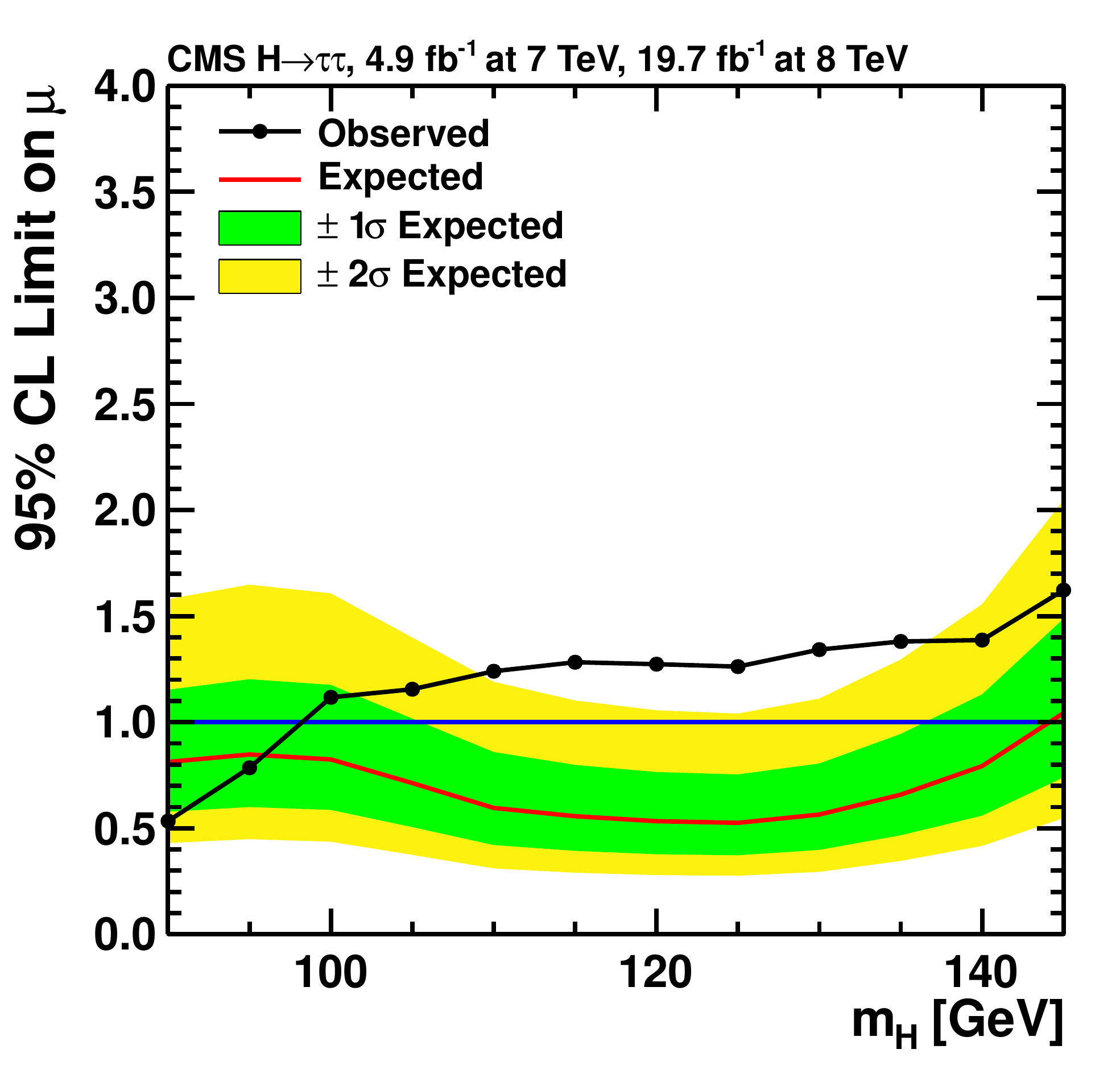}
     \includegraphics[width=0.49\textwidth]{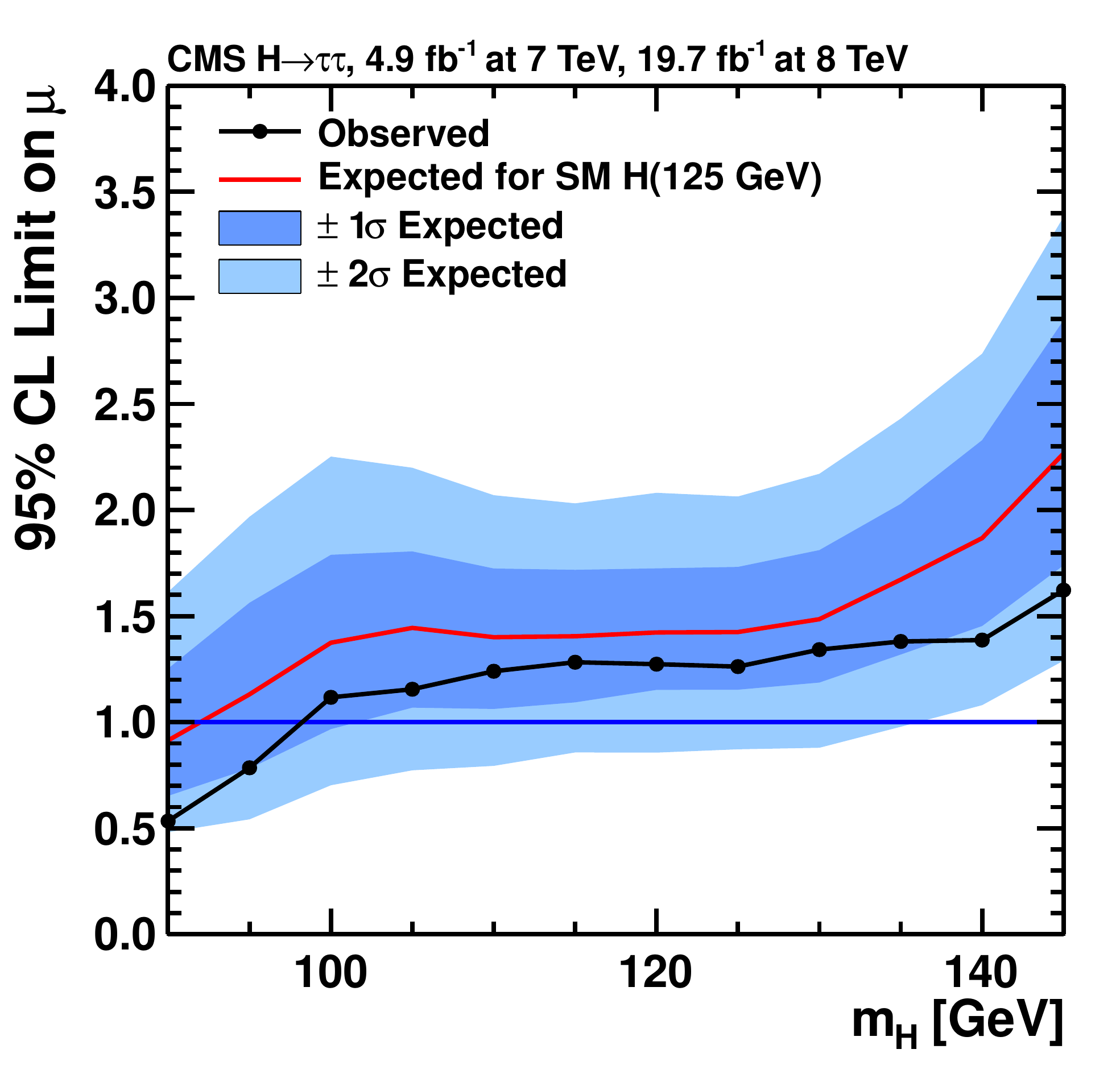}
   \caption{Combined observed 95\% CL upper limit on the signal strength parameter $\mu$,
together with the expected limit obtained in the background-only hypothesis (left), and the signal-plus-background hypothesis for a SM Higgs boson with $\mH=125\GeV$ (right).
The background-only hypothesis includes the $\Pp\Pp\rightarrow\PH\text{(125\GeV)}\to\WW$ process for every value of \mH.
The bands show the expected one- and two-standard-deviation probability intervals around the expected limit.
}
    \label{sm-limit}
  \end{center}
\end{figure}

\begin{figure}[htbp]
  \begin{center}
    \includegraphics[width=0.68\textwidth]{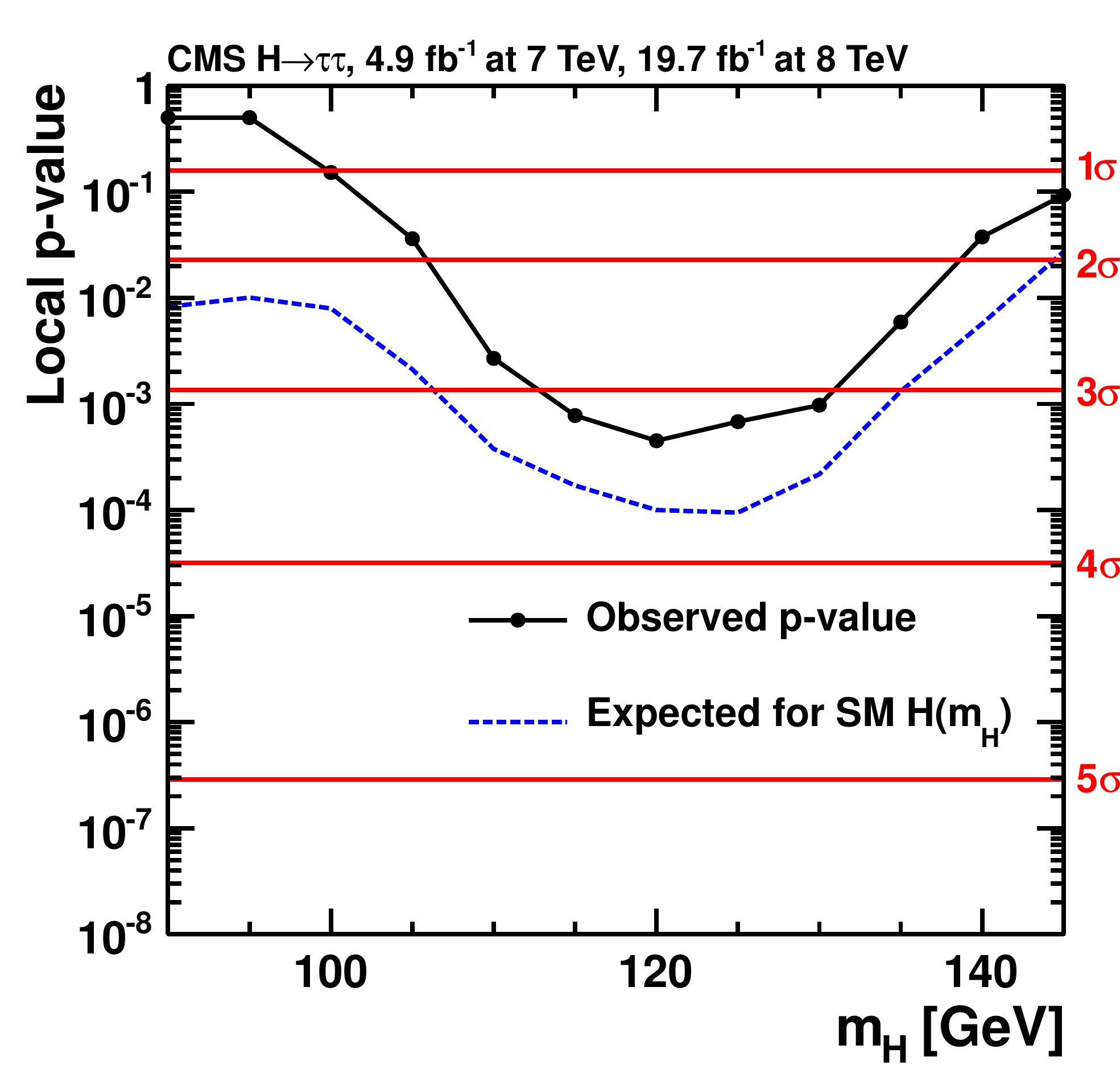}
   \caption{Local $p$-value and significance in number of standard deviations as a function of the SM Higgs boson mass hypothesis for the combination of all decay channels.
The observation (solid line) is compared to the expectation (dashed line) for a SM Higgs boson with mass \mH.
The background-only hypothesis includes the $\Pp\Pp\to\PH\text{(125\GeV)}\to\PW\PW$ process for every value of \mH.
   }
    \label{fig:p_value}
  \end{center}
\end{figure}

The following results include all decay channels considered.
Figure~\ref{sm-limit} left shows the observed 95\% CL upper limit obtained using the modified frequentist construction CL$_\text{s}$~\cite{Junk,Read2} together with the expected limit obtained for the background-only hypothesis for Higgs boson mass hypotheses ranging from 90 to 145\unit{\GeV}.
The background-only hypothesis includes the expected contribution from $\hww$ decays for $\mH = 125$\GeV.
The difference between evaluating this contribution at $\mH = 125$\GeV
or at the corresponding $\mH$ value for $\mH \neq 125$\GeV is less than 5\%.
An excess is visible in the observed limit with respect to the limit expected for the background-only hypothesis.
The observed limit is compatible with the expected limit obtained in the signal-plus-background hypothesis for a SM Higgs boson with $\mH=125\GeV$ (figure~\ref{sm-limit} right).
The excess is quantified in figure~\ref{fig:p_value} which shows the local $p$-value as a function of $\mH$.
For $\mH=125\GeV$, the expected $p$-value is smallest, corresponding to a significance of 3.7 standard deviations.
The expected $p$-value is slightly smaller when including the $\ell + L\Pgth$ and $\ell\ell + LL'$ channels.
The observed $p$-value is minimal for $\mH=120\GeV$ with a significance of 3.3 standard deviations.
The observed significance is larger than three standard deviations for \mH between 115 and 130\GeV,
and is equal to 3.2 standard deviations for $\mH = 125\GeV$.

\begin{figure} [htbp]
\centering
\includegraphics[width=0.47\textwidth]{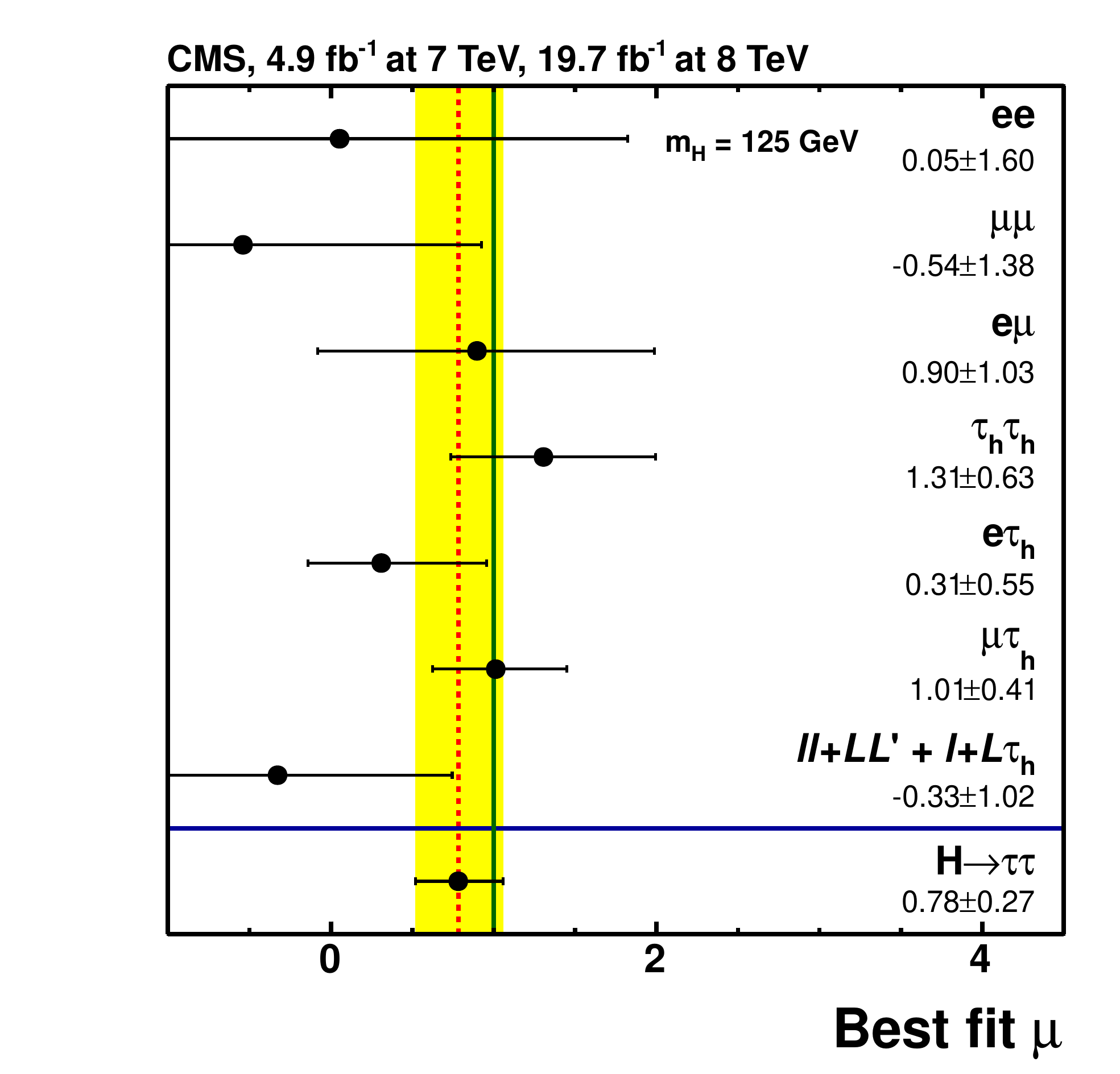}
\includegraphics[width=0.47\textwidth]{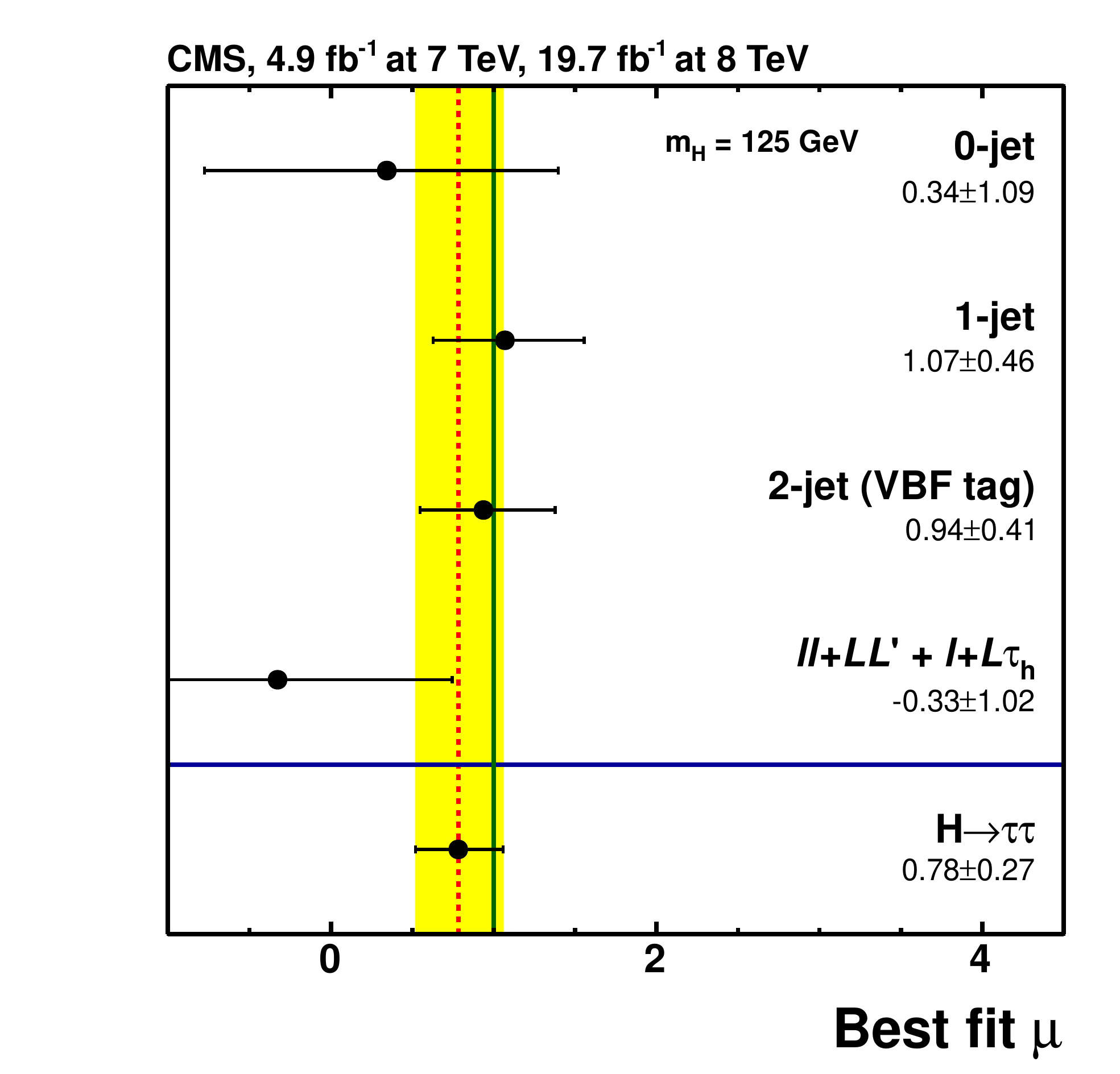}
\caption{
Best-fit signal strength values, for independent channels (left) and categories (right), for $\mH = 125\GeV$.
The combined value for the $\PH \to \tau \tau$ analysis in both plots corresponds to \bestfitmu, obtained in the global fit combining all categories of all channels.
The dashed line corresponds to the best-fit $\mu$ value.
The contribution from the $\Pp\Pp\to\PH\text{(125\GeV)}\to\PW\PW$ process is treated as background normalized to the SM expectation.
}
\label{fig:compatibility}
\end{figure}

The best-fit value for $\mu$, combining all channels, is \bestfitmu at $\mH = 125\GeV$. Figure~\ref{fig:compatibility} shows the results of the fits performed in each decay channel for all categories, and in each category for all decay channels.
These compatibility tests do not constitute measurements of any physical parameter per se, but rather
show the consistency of the various observations with the expectation for a SM Higgs boson with $\mH = 125$\GeV.
The uncertainties of the individual $\mu$ values in the 1-jet and 2-jet (VBF-tagged) categories are of similar size, showing that both contribute about equally to the sensitivity of the analysis.
The fraction of signal events from VBF production in the 1-jet categories and of signal events produced via gluon-gluon fusion in the 2-jet (VBF-tagged) categories are each of the order of 20 to 30\%;
hence, it is not possible to fully disentangle the two production modes.

\begin{figure}[htbp]
  \begin{center}
      \includegraphics[width=0.8\textwidth]{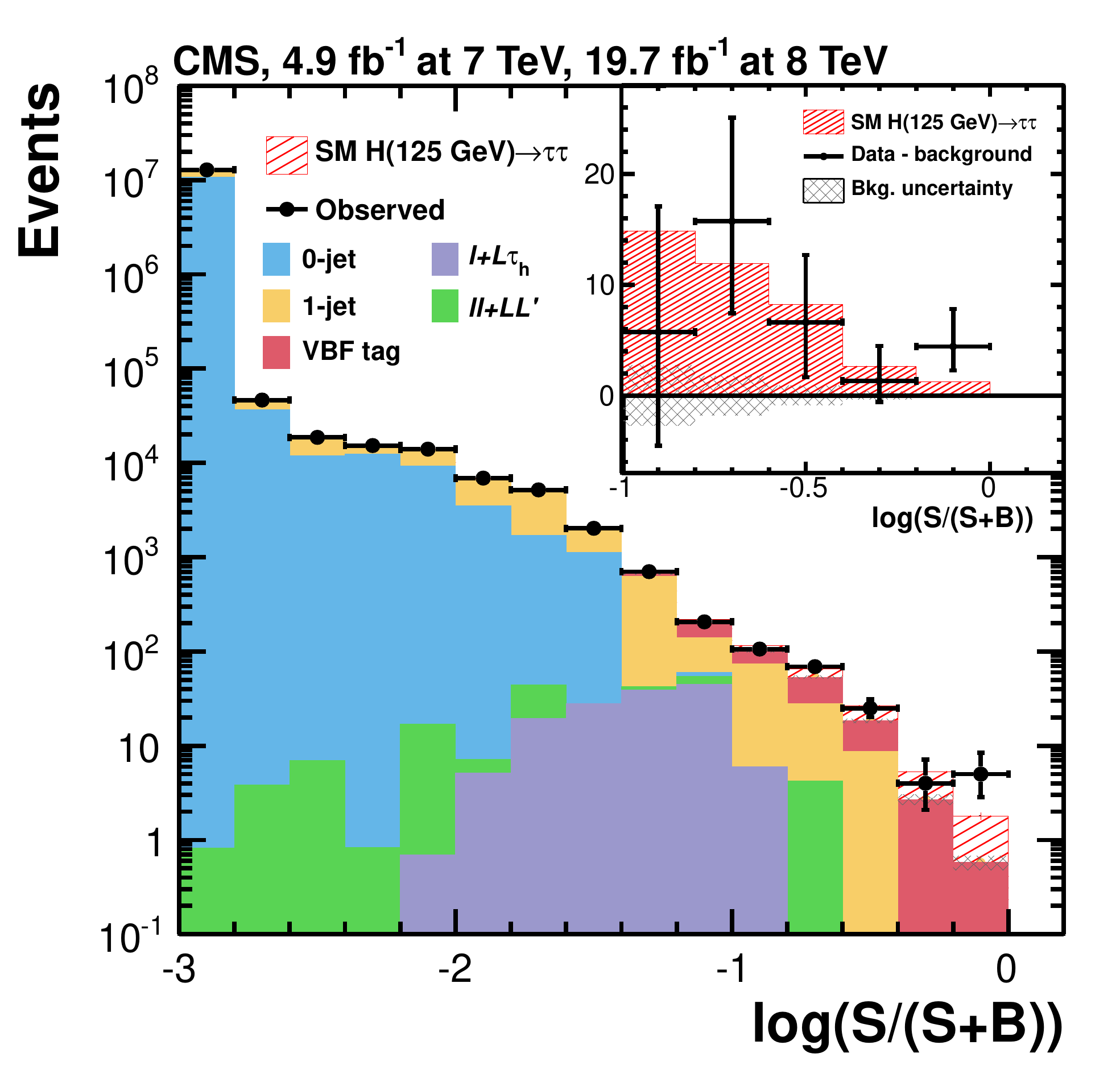}
   \caption{Combined observed and predicted distributions of the decimal logarithm $\log(S/(S+B))$ in each bin of the final $\mtt$, $\mvis$, or discriminator distributions obtained in all event categories and decay channels,
   with $S/(S+B)$ denoting the ratio of the predicted signal and signal-plus-background event yields in each bin.
   The normalization of the predicted background distributions corresponds to the result of the global fit.
   The signal distribution, on the other hand, is normalized to the SM prediction ($\mu = 1$).
   The inset shows the corresponding difference between the observed data and expected background distributions, together with the signal distribution for a SM Higgs boson at $\mH=125\GeV$.
   The distribution from SM Higgs boson events in the $\WW$ decay channel does not significantly contribute to this plot.
}
    \label{fig:soversplusb_distribution}
  \end{center}
\end{figure}

The combined distribution of the decimal logarithm $\log(S/(S+B))$ obtained in each bin of the final discriminating variables for all event categories and channels is shown in figure~\ref{fig:soversplusb_distribution}.
Here, $S$ denotes the expected signal yield for a SM Higgs boson with $\mH = 125\GeV$ ($\mu=1$) and $B$ denotes the expected background yield in a given bin.
The plot illustrates the contribution from the different event categories that are sensitive to the different Higgs boson production mechanisms.
In addition, it provides a visualization of the observed excess of data events over the background expectation in the region of high $S/(S+B)$.

\begin{figure}[htbp]
  \begin{center}
    \includegraphics[width=0.49\textwidth]{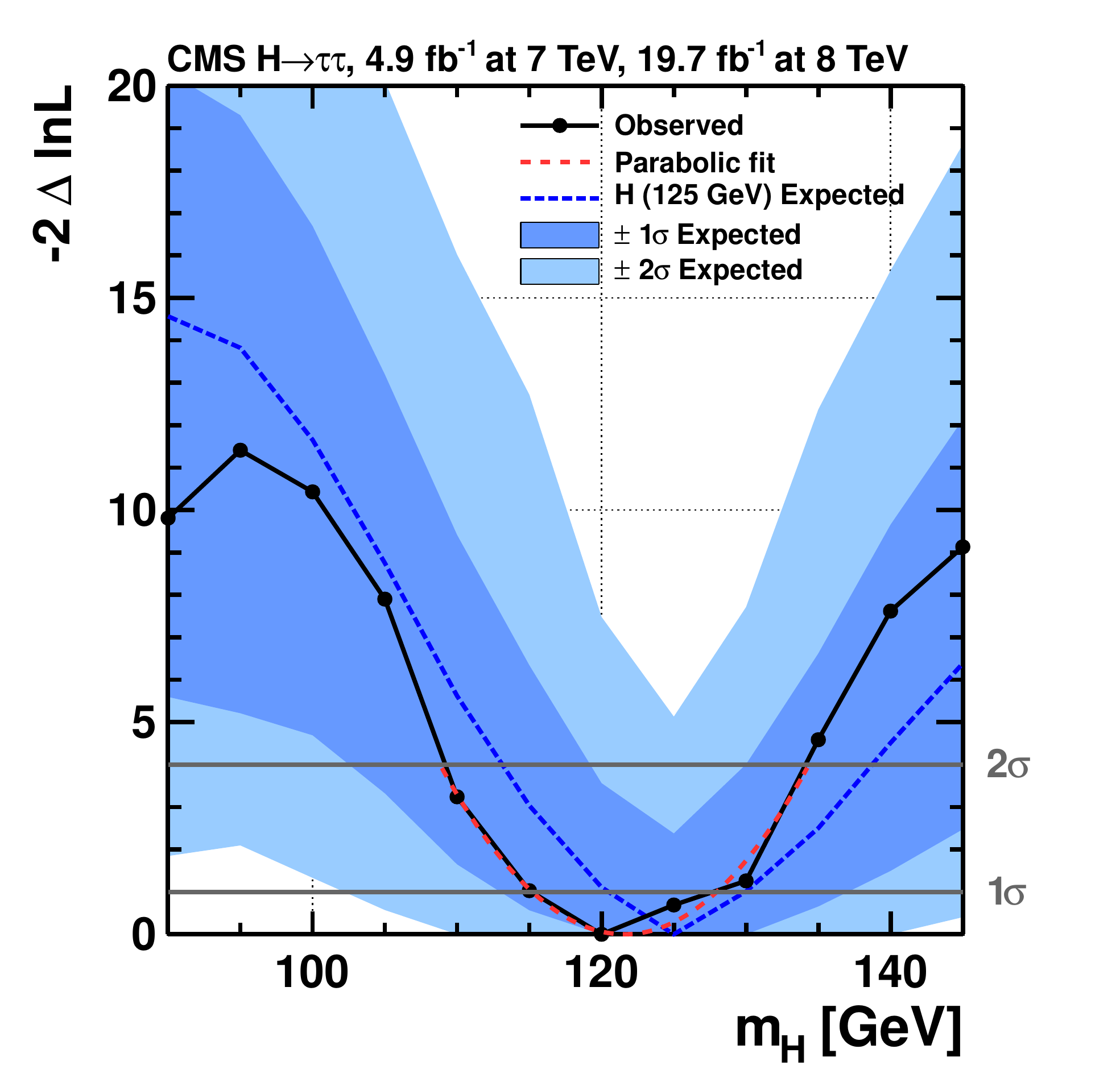}
    \includegraphics[width=0.49\textwidth]{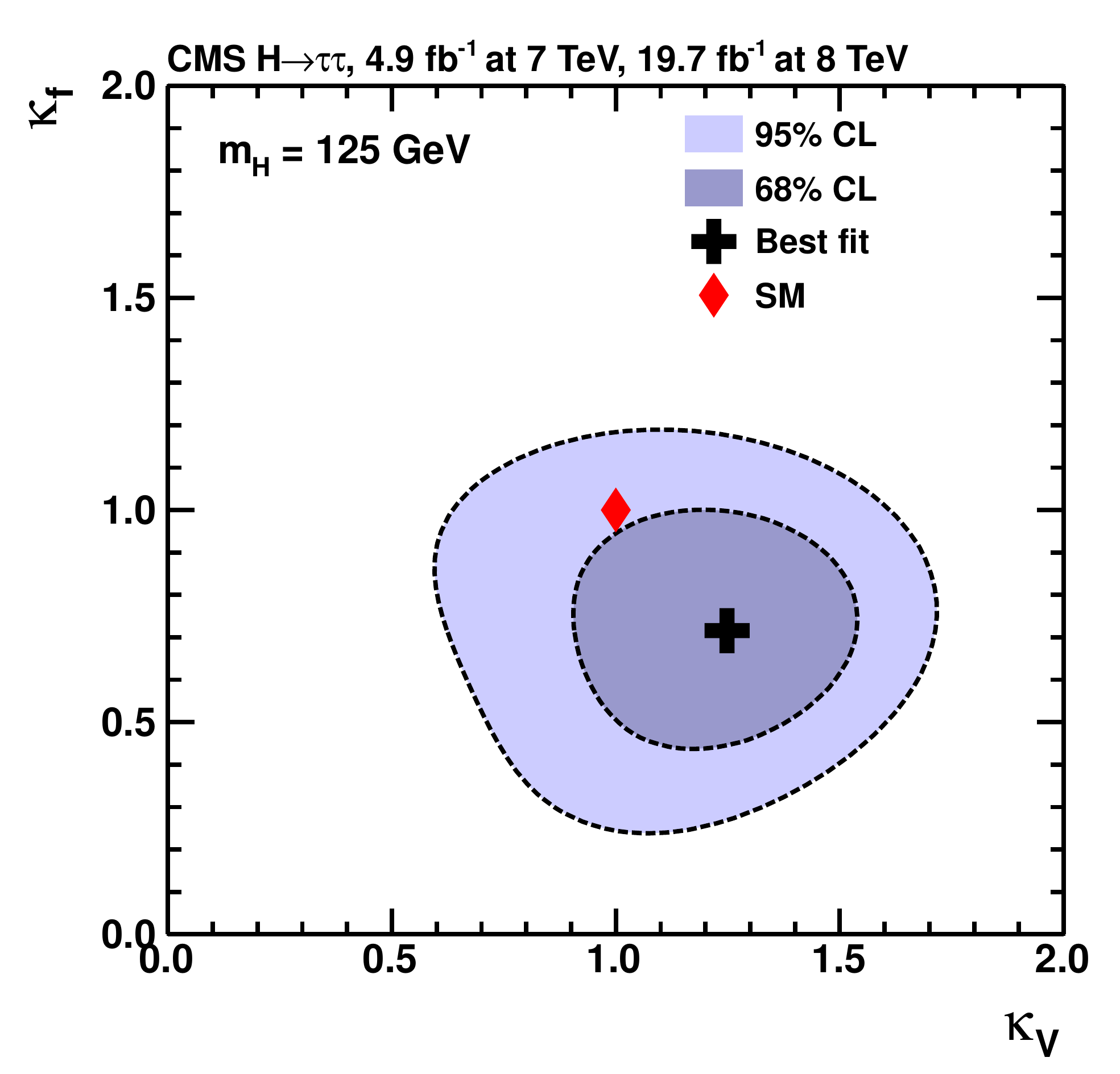}
   \caption{Scan of the negative log-likelihood difference, $-2\Delta\ln\mathcal{L}$,
as a function of $\mH$ (left) and as a function of $\kappav$ and $\kappaf$ (right).
For each point, all nuisance parameters are profiled.
For the likelihood scan as a function of $\mH$, the background-only hypothesis includes the $\Pp\Pp\to\PH\text{(125\GeV)}\to\PW\PW$ process for every value of \mH.
The observation (solid line) is compared to the expectation (dashed line) for a SM Higgs boson with mass $\mH = 125$\GeV.
For the likelihood scan as a function of $\kappav$ and $\kappaf$, the $\hww$ contribution is treated as a signal process.
}
    \label{fig:likelihood_scans}
  \end{center}
\end{figure}

Figure~\ref{fig:likelihood_scans} left presents a scan of the negative log-likelihood difference, $-2\Delta\ln\mathcal{L}$, as a function of \mH.
For each point in the parameter space, all nuisance parameters and the $\mu$ parameter are profiled.
The background-only hypothesis includes the contribution from the $\Pp\Pp\to\PH\text{(125\GeV)}\to\PW\PW$ process for every value of \mH.
The difference between evaluating this additional background at $\mH = 125$\GeV
or at the corresponding $\mH$ value for $\mH \neq 125$\GeV is small.
At the mass value corresponding to the minimum of the mass scan, the combined statistical and systematic uncertainties from the likelihood scan amount to 6\GeV.
Additional contributions to the overall uncertainty of the mass measurement arise due to uncertainties in the absolute energy scale and its variation with \pt of 1 to 2\% for $\Pgth$ candidates, electrons, muons, and the $\MET$, summing up to an uncertainty of 4\GeV.
Given the coarse $\mH$ granularity, a parabolic fit is performed to $-2\Delta\ln\mathcal{L}$ values below 4.
The combined measured mass of the Higgs boson is $\mH = 122 \pm 7$\GeV.

Figure~\ref{fig:likelihood_scans} right shows a likelihood scan in the two-dimensional $(\kappav, \kappaf)$ parameter space for $\mH = 125\GeV$.
The $\kappav$ and $\kappaf$ parameters quantify the ratio between the measured and the SM value for the coupling of the Higgs boson to vector bosons and fermions, respectively~\cite{Heinemeyer:2013tqa}.
To consistently measure deviations of the fermionic and the bosonic couplings of the Higgs boson, the $\hww$ contribution is considered as a signal process in this likelihood scan.
For the VBF production of a Higgs boson that decays to a $\WW$ pair, the bosonic coupling enters both in the production and in the decay, thus providing sensitivity to the bosonic coupling despite the small expected event rates.
All nuisance parameters are profiled for each point in the parameter space.
The observed likelihood contour is consistent with the SM expectation of $\kappav = \kappaf = 1$.

\section{Summary}

We report a search for the standard model Higgs boson decaying into a pair of $\tau$ leptons.
The search is based on the full proton-proton collision sample recorded by CMS in 2011 and 2012,
corresponding to an integrated luminosity of 4.9\fbinv at a centre-of-mass energy of 7\TeV and 19.7\fbinv at 8\TeV.
The analysis is performed in six channels corresponding to the final states $\Pgm\Pgt_h$, $\Pe\Pgt_h$,
$\Pgt_h\Pgt_h$, $\Pe\Pgm$, $\Pgm\Pgm$, and $\Pe\Pe$.
The gluon-gluon fusion and vector-boson fusion production of a Higgs boson are probed in the one-jet and two-jet final states, respectively, whereas the production of a Higgs boson in association with a $\PW$ or $\PZ$ boson decaying leptonically is targeted by requiring additional electrons or muons in the final state.
An excess of events over the background-only hypothesis is observed with a local significance in excess of 3 standard deviations for Higgs boson mass hypotheses between $\mH=115$ and 130\GeV,
and equal to 3.2 standard deviations at $\mH = 125\GeV$,
to be compared to an expected significance of 3.7 standard deviations.
The best fit of the observed $\PH \to \tau \tau$ signal cross section times branching fraction for $\mH = 125$\GeV is $0.78 \pm 0.27$ times the standard model expectation.
Assuming that this excess corresponds to a Higgs boson decaying to $\tau\tau$, its mass is measured to be $\mH = 122 \pm 7\GeV$.
These results constitute evidence for the coupling between the $\tau$ lepton and the 125\GeV Higgs boson discovered in 2012 by the ATLAS and CMS Collaborations.

\hyphenation{Bundes-ministerium Forschungs-gemeinschaft Forschungs-zentren}

\section*{Acknowledgements}

We congratulate our colleagues in the CERN accelerator departments for the excellent performance of the LHC and thank the technical and administrative staffs at CERN and at other CMS institutes for their contributions to the success of the CMS effort. In addition, we gratefully acknowledge the computing centres and personnel of the Worldwide LHC Computing Grid for delivering so effectively the computing infrastructure essential to our analyses. Finally, we acknowledge the enduring support for the construction and operation of the LHC and the CMS detector provided by the following funding agencies: the Austrian Federal Ministry of Science and Research and the Austrian Science Fund; the Belgian Fonds de la Recherche Scientifique, and Fonds voor Wetenschappelijk Onderzoek; the Brazilian Funding Agencies (CNPq, CAPES, FAPERJ, and FAPESP); the Bulgarian Ministry of Education and Science; CERN; the Chinese Academy of Sciences, Ministry of Science and Technology, and National Natural Science Foundation of China; the Colombian Funding Agency (COLCIENCIAS); the Croatian Ministry of Science, Education and Sport, and the Croatian Science Foundation; the Research Promotion Foundation, Cyprus; the Ministry of Education and Research, Recurrent financing contract SF0690030s09 and European Regional Development Fund, Estonia; the Academy of Finland, Finnish Ministry of Education and Culture, and Helsinki Institute of Physics; the Institut National de Physique Nucl\'eaire et de Physique des Particules~/~CNRS, and Commissariat \`a l'\'Energie Atomique et aux \'Energies Alternatives~/~CEA, France; the Bundesministerium f\"ur Bildung und Forschung, Deutsche Forschungsgemeinschaft, and Helmholtz-Gemeinschaft Deutscher Forschungszentren, Ger\-ma\-ny; the General Secretariat for Research and Technology, Greece; the National Scientific Research Foundation, and National Innovation Office, Hungary; the Department of Atomic Energy and the Department of Science and Technology, India; the Institute for Studies in Theoretical Physics and Mathematics, Iran; the Science Foundation, Ireland; the Istituto Nazionale di Fisica Nucleare, Italy; the Korean Ministry of Education, Science and Technology and the World Class University program of NRF, Republic of Korea; the Lithuanian Academy of Sciences; the Mexican Funding Agencies (CINVESTAV, CONACYT, SEP, and UASLP-FAI); the Ministry of Business, Innovation and Employment, New Zealand; the Pakistan Atomic Energy Commission; the Ministry of Science and Higher Education and the National Science Centre, Poland; the Funda\c{c}\~ao para a Ci\^encia e a Tecnologia, Portugal; JINR, Dubna; the Ministry of Education and Science of the Russian Federation, the Federal Agency of Atomic Energy of the Russian Federation, Russian Academy of Sciences, and the Russian Foundation for Basic Research; the Ministry of Education, Science and Technological Development of Serbia; the Secretar\'{\i}a de Estado de Investigaci\'on, Desarrollo e Innovaci\'on and Programa Consolider-Ingenio 2010, Spain; the Swiss Funding Agencies (ETH Board, ETH Zurich, PSI, SNF, UniZH, Canton Zurich, and SER); the National Science Council, Taipei; the Thailand Center of Excellence in Physics, the Institute for the Promotion of Teaching Science and Technology of Thailand, Special Task Force for Activating Research and the National Science and Technology Development Agency of Thailand; the Scientific and Technical Research Council of Turkey, and Turkish Atomic Energy Authority; the Science and Technology Facilities Council, UK; the US Department of Energy, and the US National Science Foundation.

Individuals have received support from the Marie-Curie programme and the European Research Council and EPLANET (European Union); the Leventis Foundation; the A. P. Sloan Foundation; the Alexander von Humboldt Foundation; the Belgian Federal Science Policy Office; the Fonds pour la Formation \`a la Recherche dans l'Industrie et dans l'Agriculture (FRIA-Belgium); the Agentschap voor Innovatie door Wetenschap en Technologie (IWT-Belgium); the Ministry of Education, Youth and Sports (MEYS) of Czech Republic; the Council of Science and Industrial Research, India; the Compagnia di San Paolo (Torino); the HOMING PLUS programme of Foundation for Polish Science, cofinanced by EU, Regional Development Fund; and the Thalis and Aristeia programmes cofinanced by EU-ESF and the Greek NSRF.

\bibliography{auto_generated}   
\clearpage
\appendix
\pagebreak
\section{Post-fit distributions}
\label{sec:app_postfit}

\begin{figure}[hb!tp]
\makebox[0.32\textwidth]{}\includegraphics[width=0.32\textwidth]{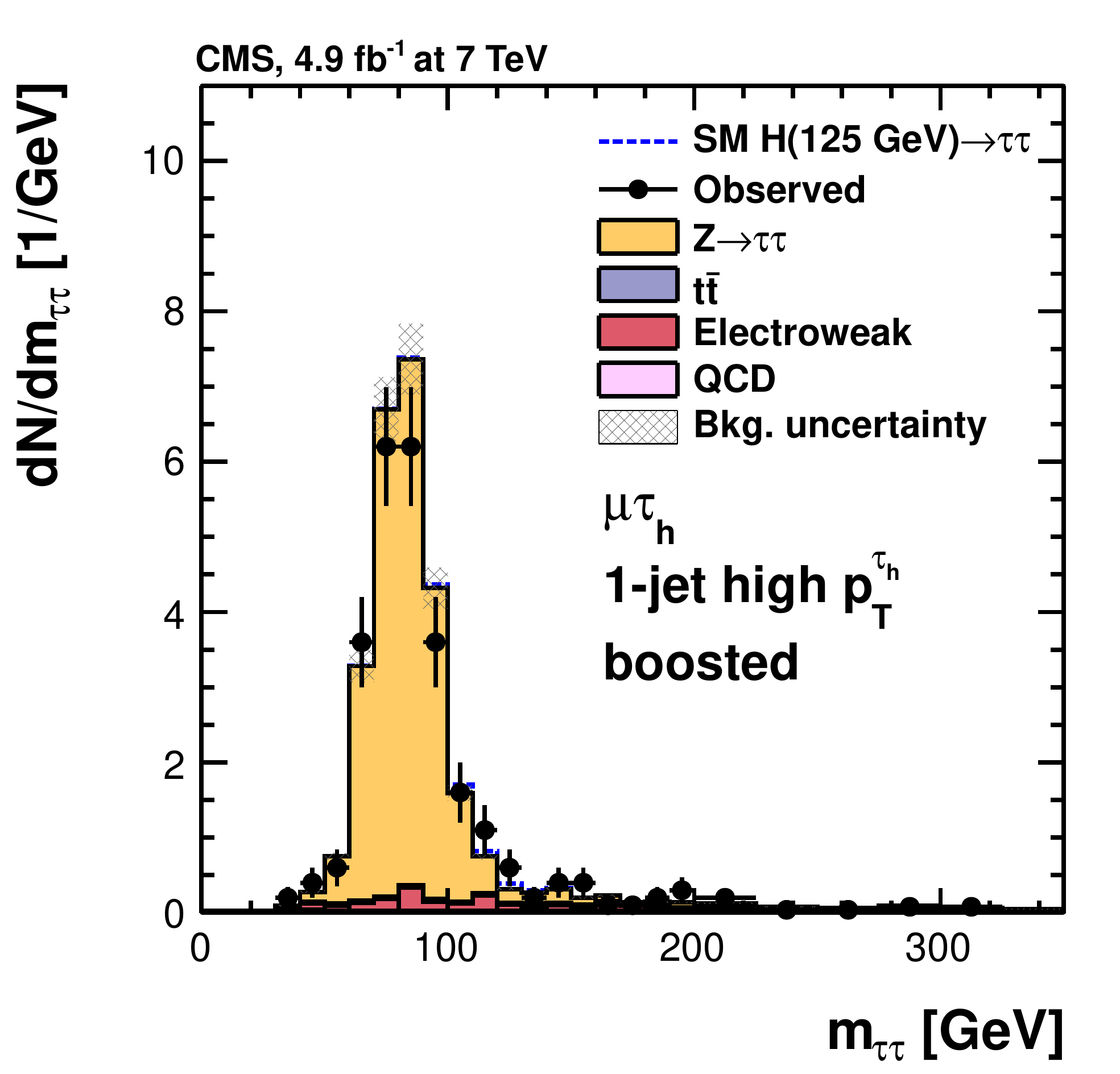}\makebox[0.32\textwidth]{}
\includegraphics[width=0.32\textwidth]{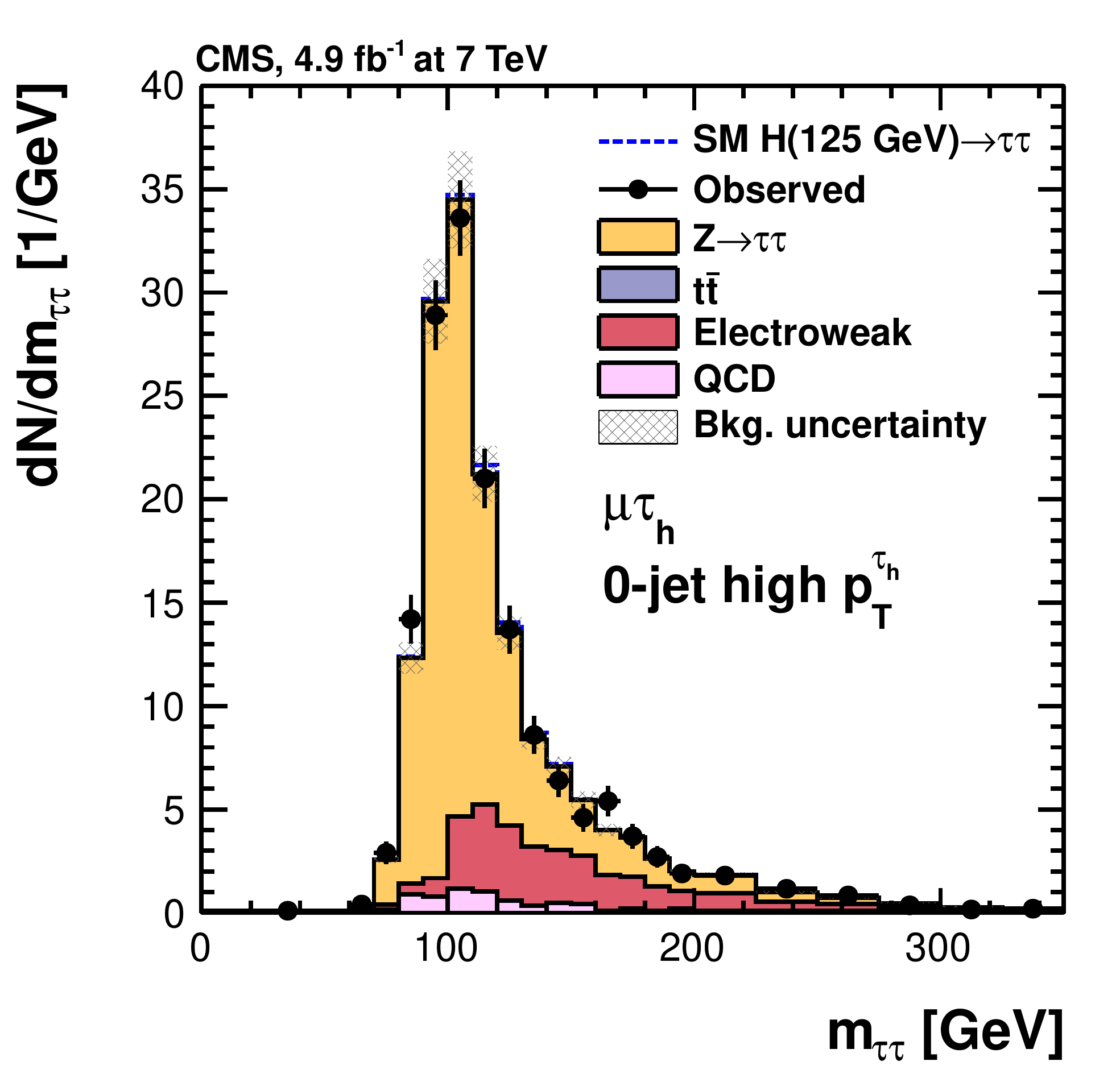}
\includegraphics[width=0.32\textwidth]{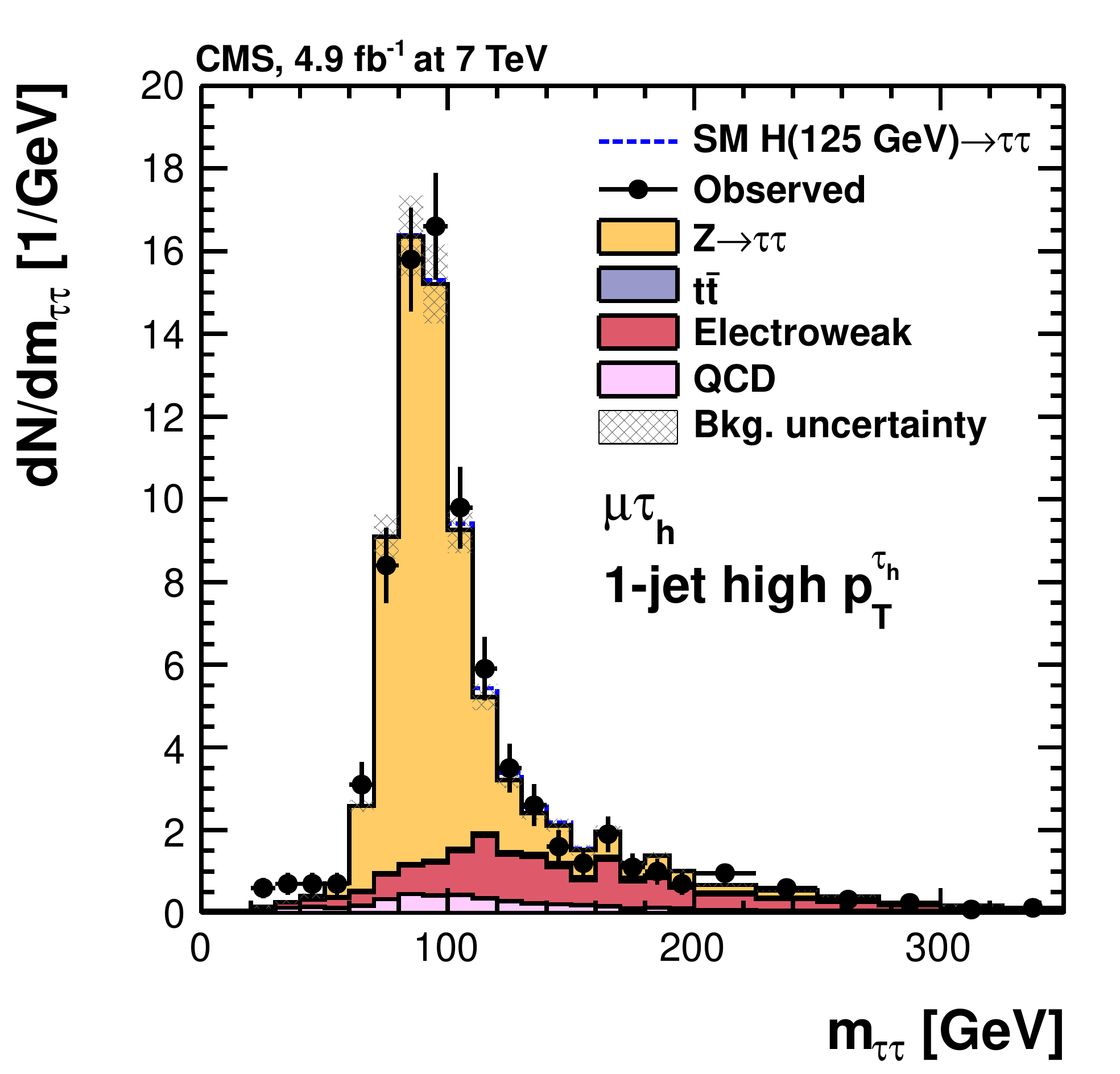}
\makebox[0.32\textwidth]{}
\includegraphics[width=0.32\textwidth]{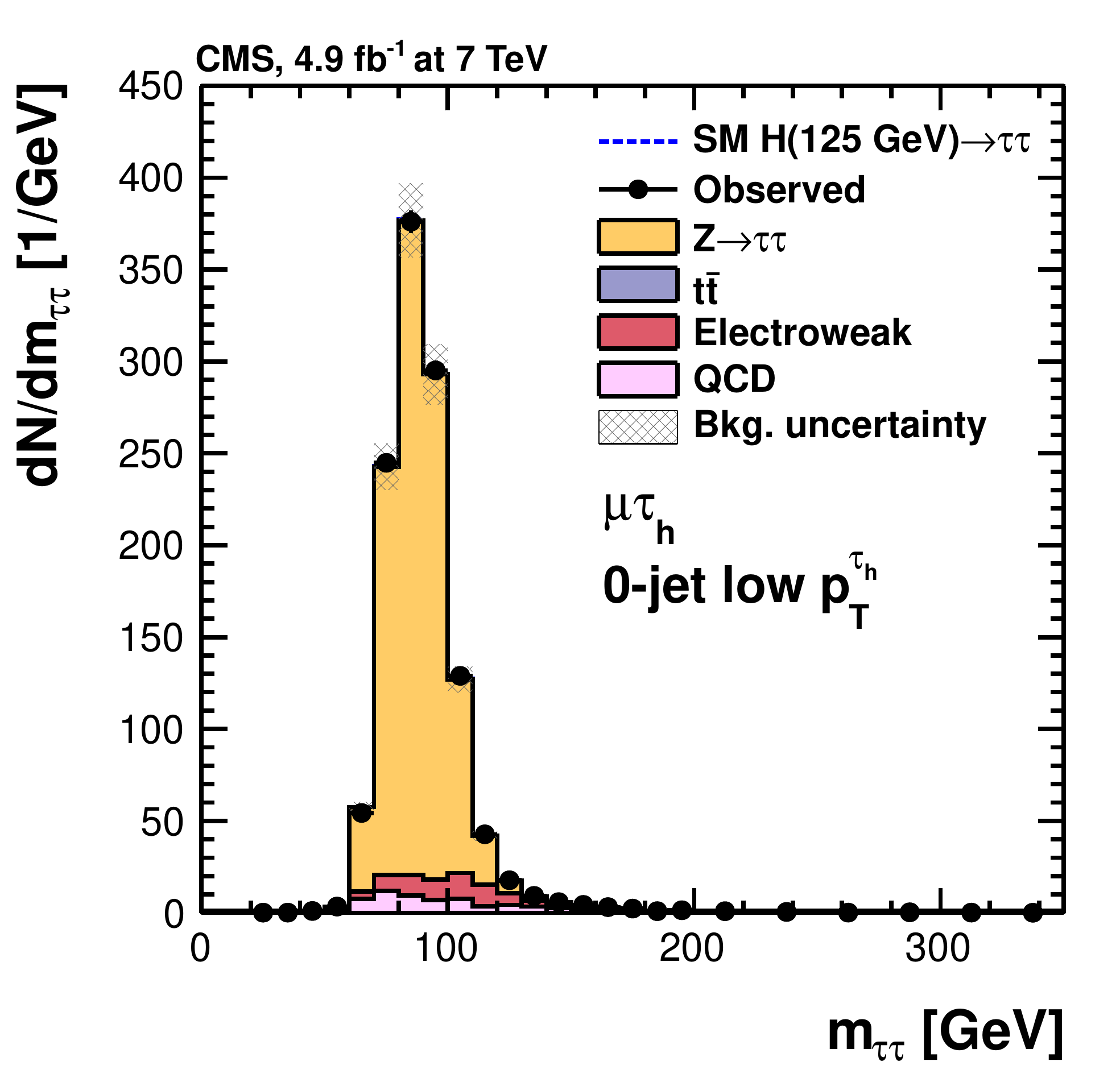}
\includegraphics[width=0.32\textwidth]{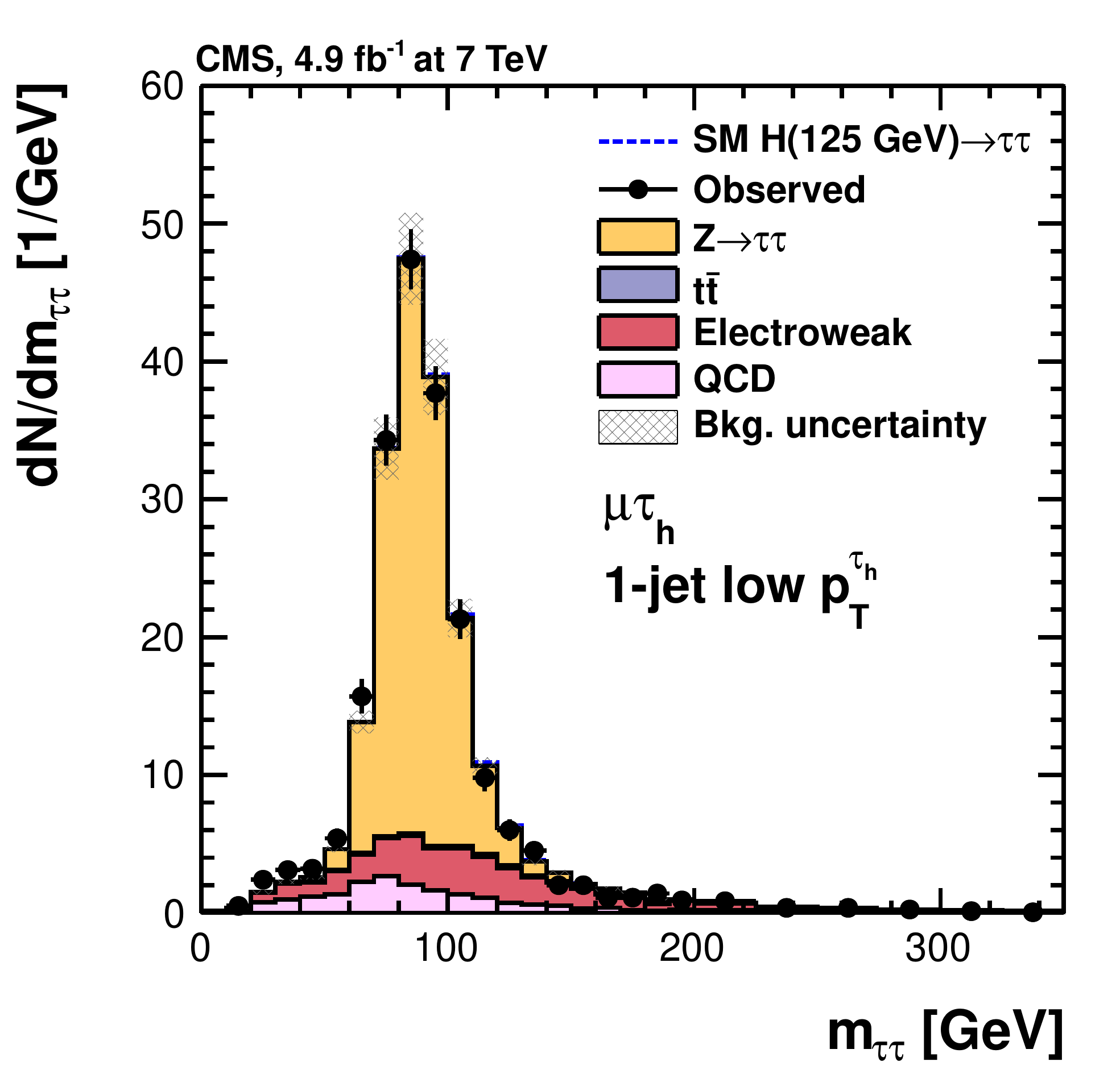}
\includegraphics[width=0.32\textwidth]{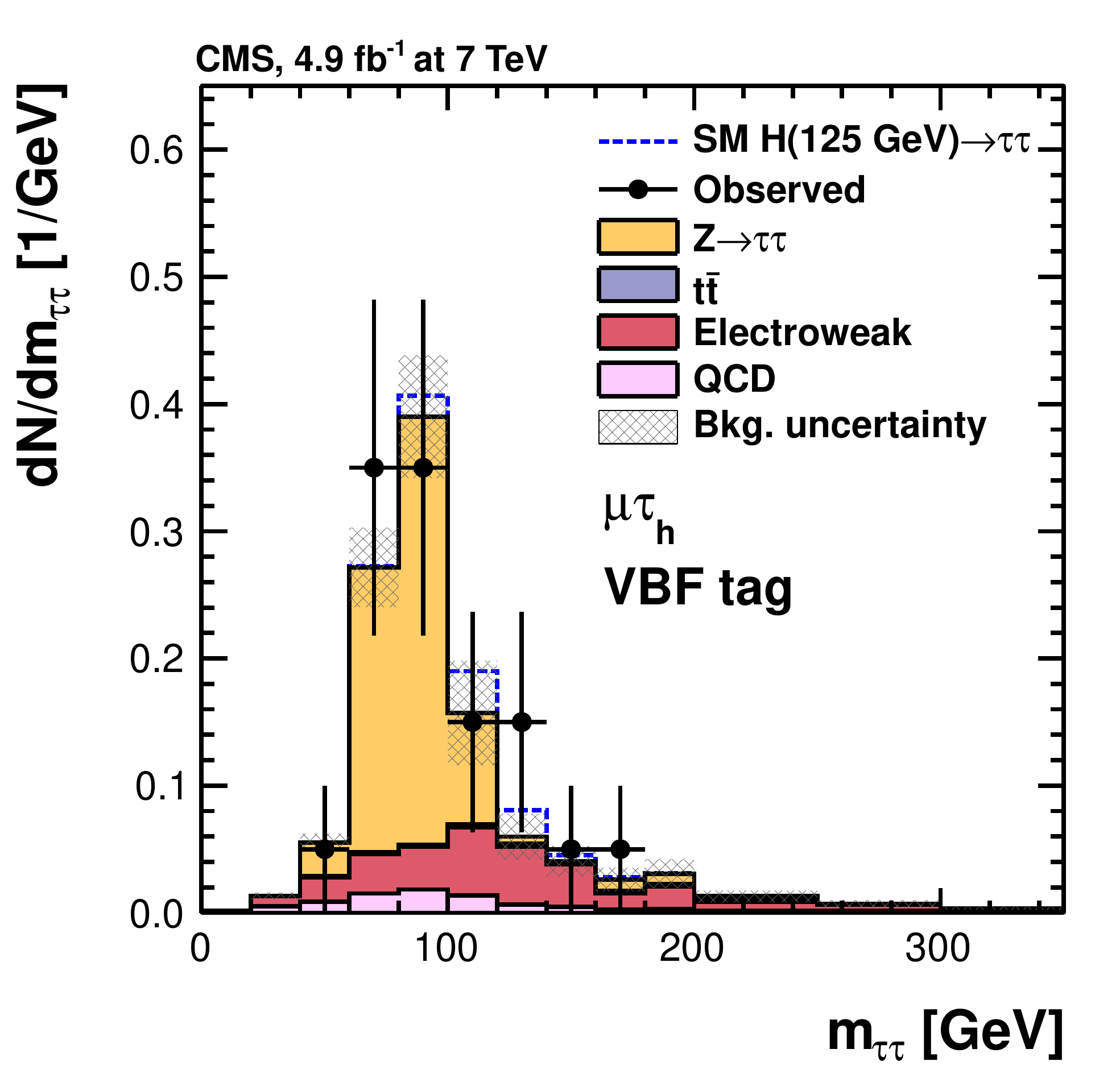}
\centering
\caption{Observed and predicted $\mtt$ distributions in the $\mutau$ channel, for all categories used in the 7\TeV data analysis. The normalization of the predicted background distributions corresponds to the result of the global fit. The signal distribution, on the other hand, is normalized to the SM prediction. The signal and background histograms are stacked.}
     \label{fig:app_mt_7}
\end{figure}

\begin{figure}[b!htp]
\makebox[0.32\textwidth]{}
\includegraphics[width=0.32\textwidth]{figures/plots_131206/mt/muTau_1jet_high_mediumhiggs_postfit_8TeV_LIN.pdf}
\makebox[0.32\textwidth]{}
\includegraphics[width=0.32\textwidth]{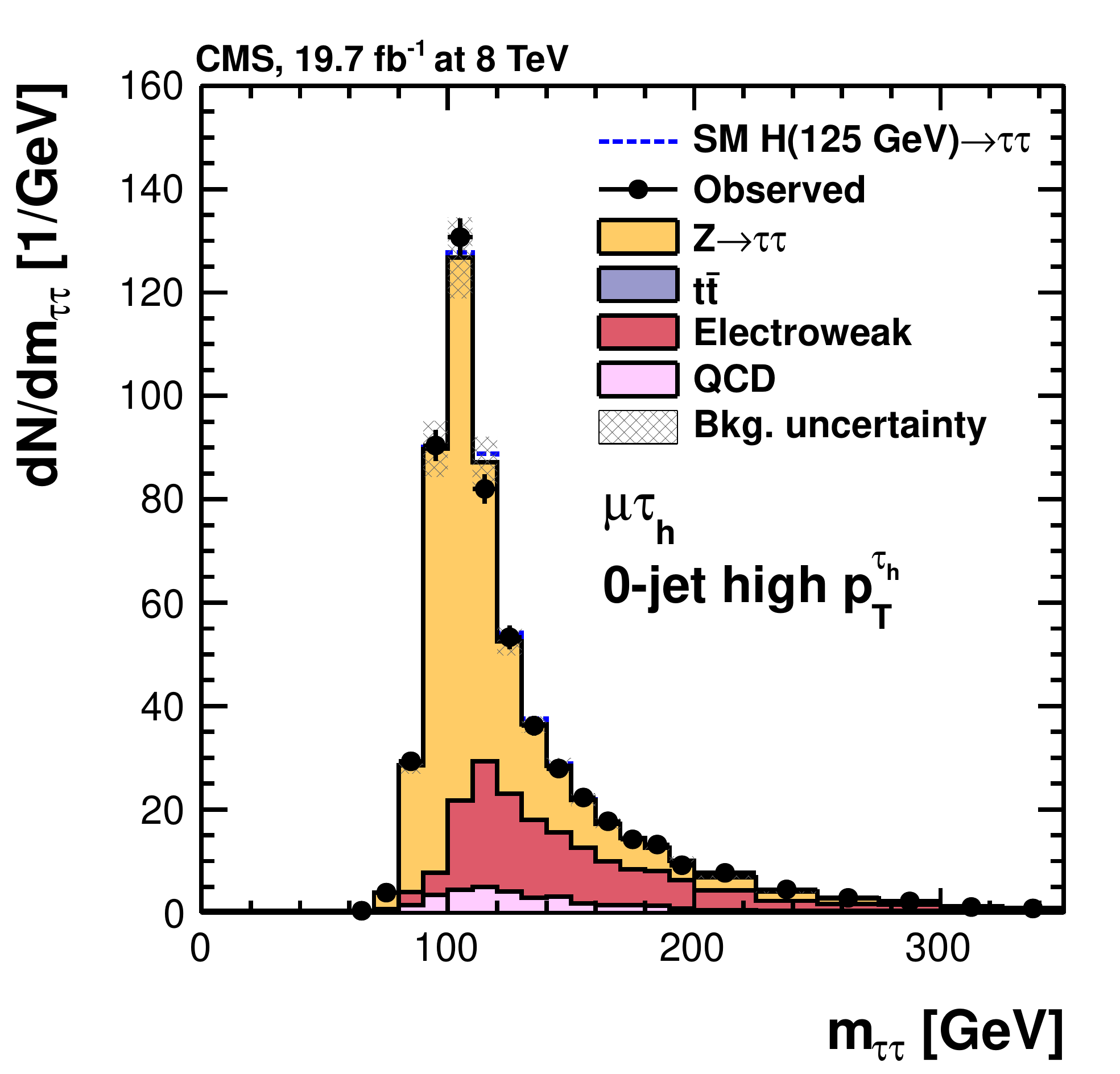}
\includegraphics[width=0.32\textwidth]{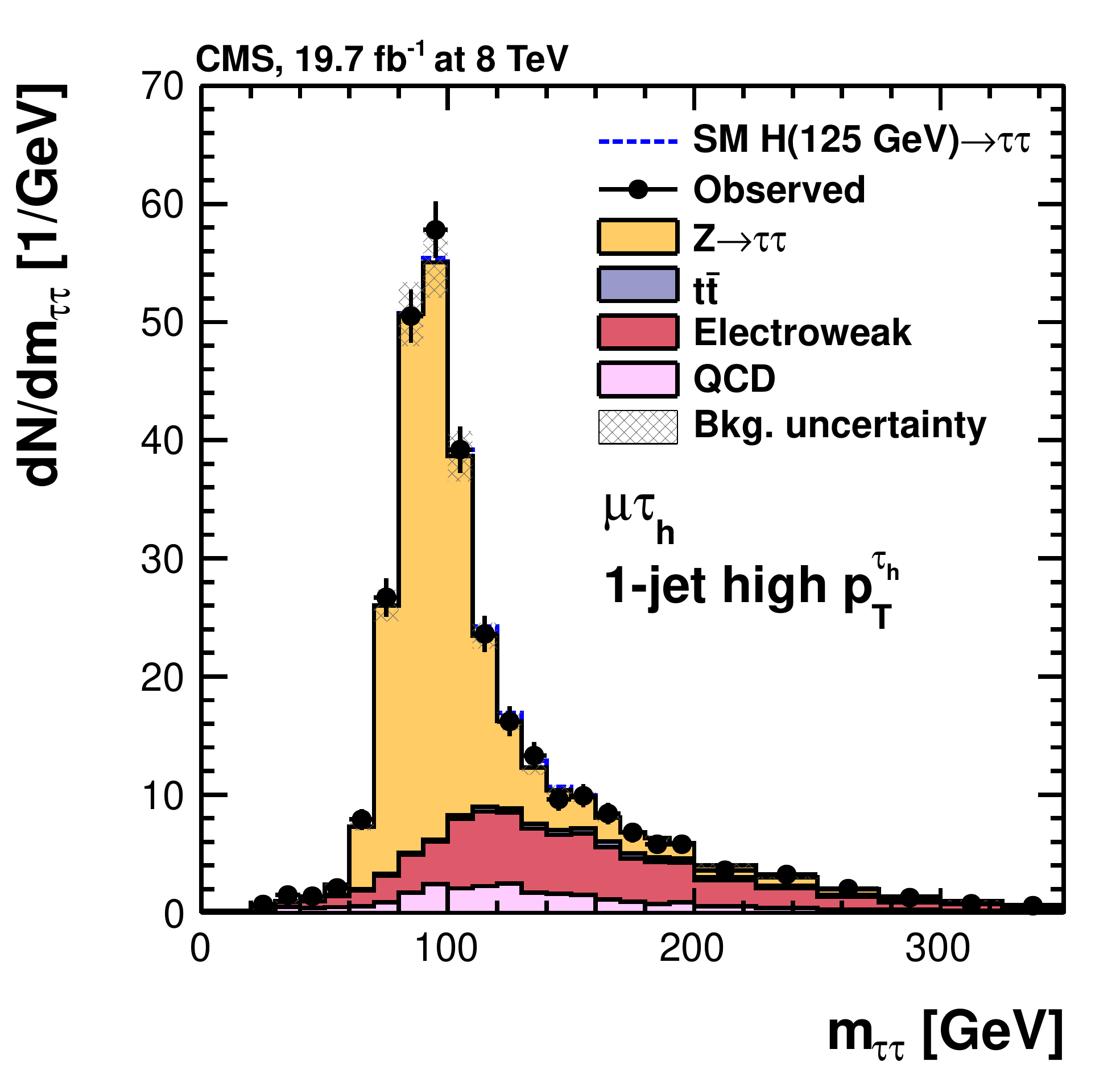}
\includegraphics[width=0.32\textwidth]{figures/plots_131206/mt/muTau_vbf_tight_postfit_8TeV_LIN.pdf}
\includegraphics[width=0.32\textwidth]{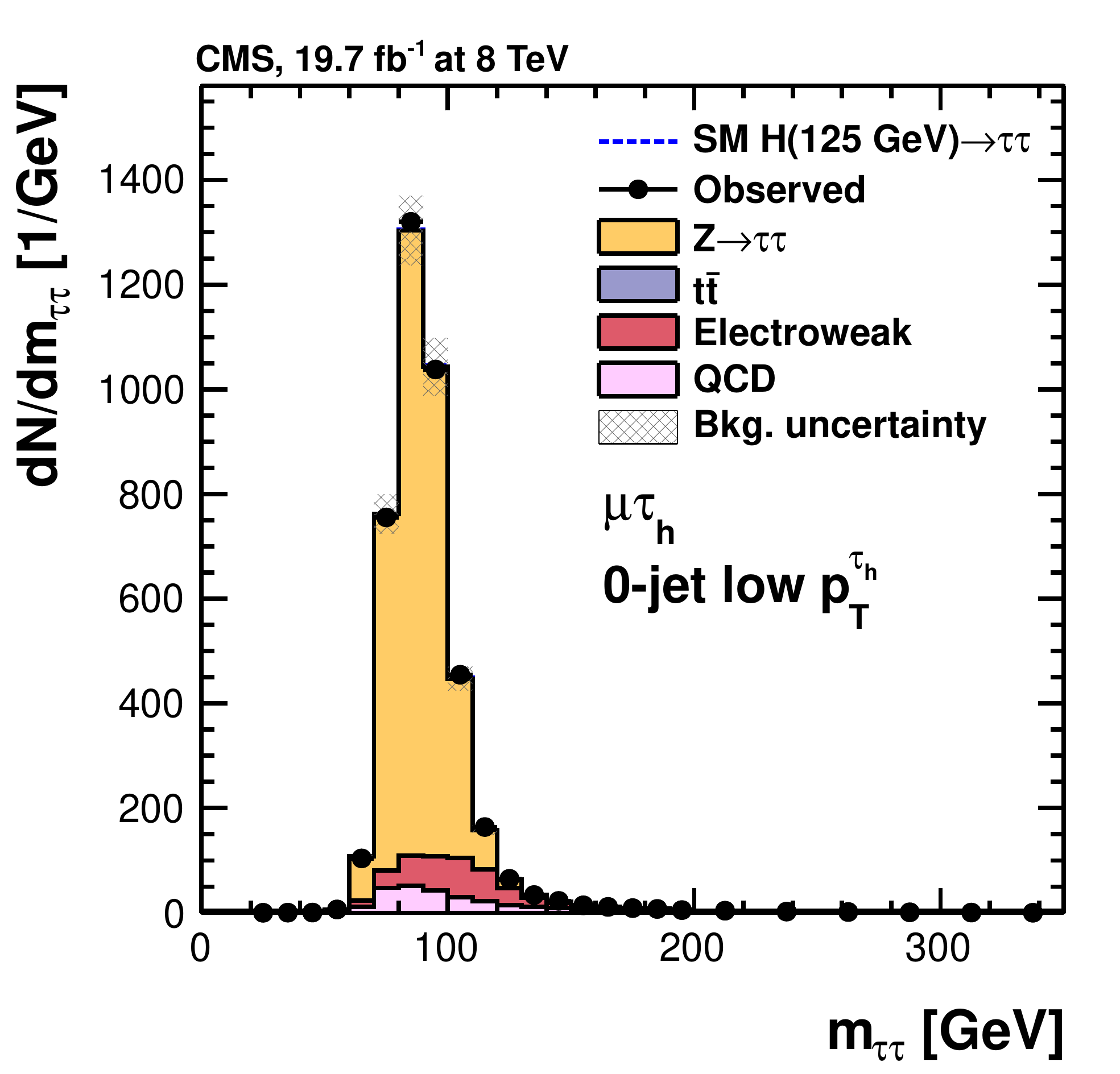}
\includegraphics[width=0.32\textwidth]{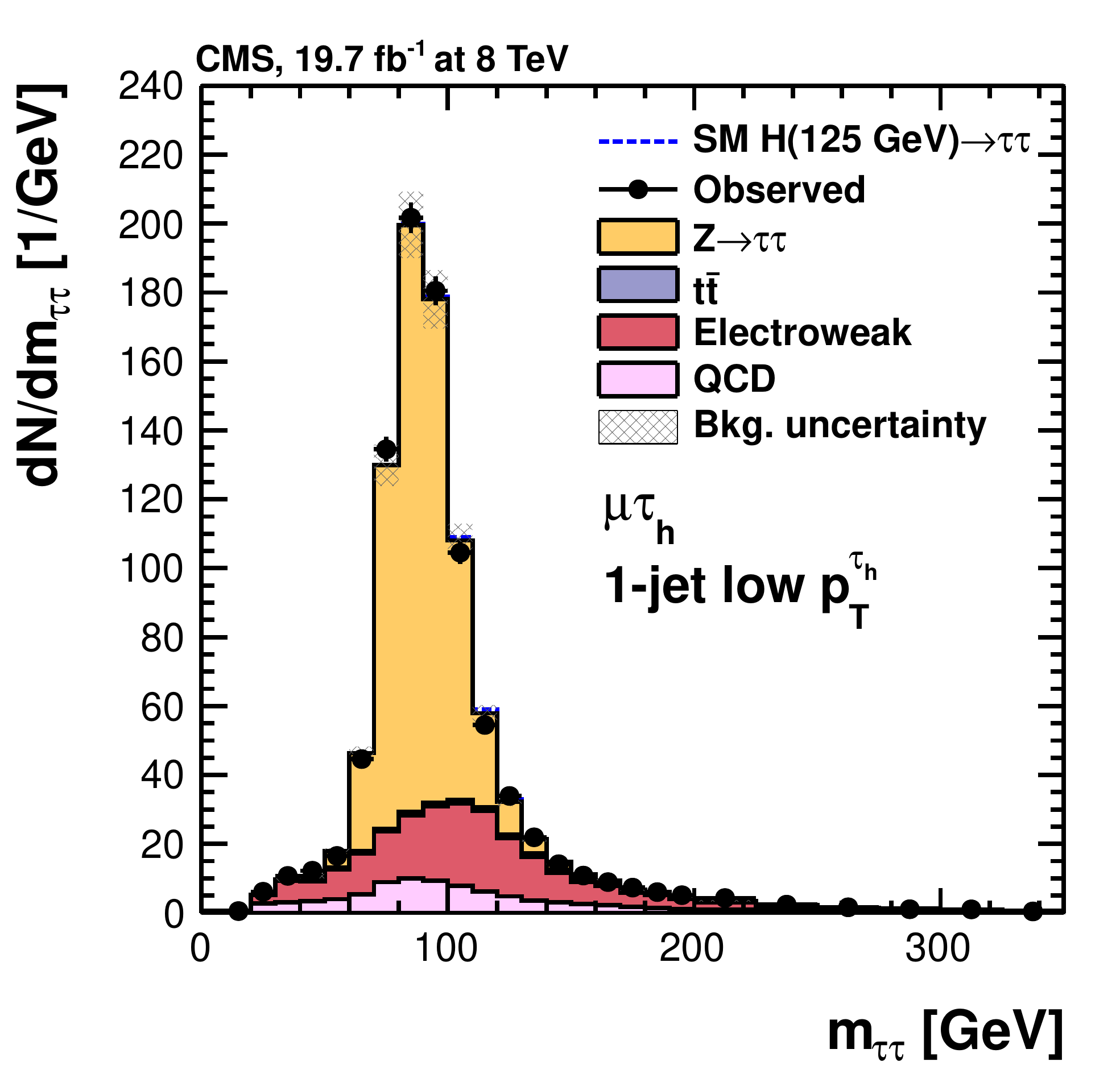}
\includegraphics[width=0.32\textwidth]{figures/plots_131206/mt/muTau_vbf_loose_postfit_8TeV_LIN.pdf}
\centering
     \caption{Observed and predicted $\mtt$ distributions in the $\mutau$ channel, for all categories used in the 8\TeV data analysis. The normalization of the predicted background distributions corresponds to the result of the global fit. The signal distribution, on the other hand, is normalized to the SM prediction. The signal and background histograms are stacked.}
     \label{fig:app_mt_8}
\end{figure}

\begin{figure}[b!htp]
\makebox[0.32\textwidth]{}
\includegraphics[width=0.32\textwidth]{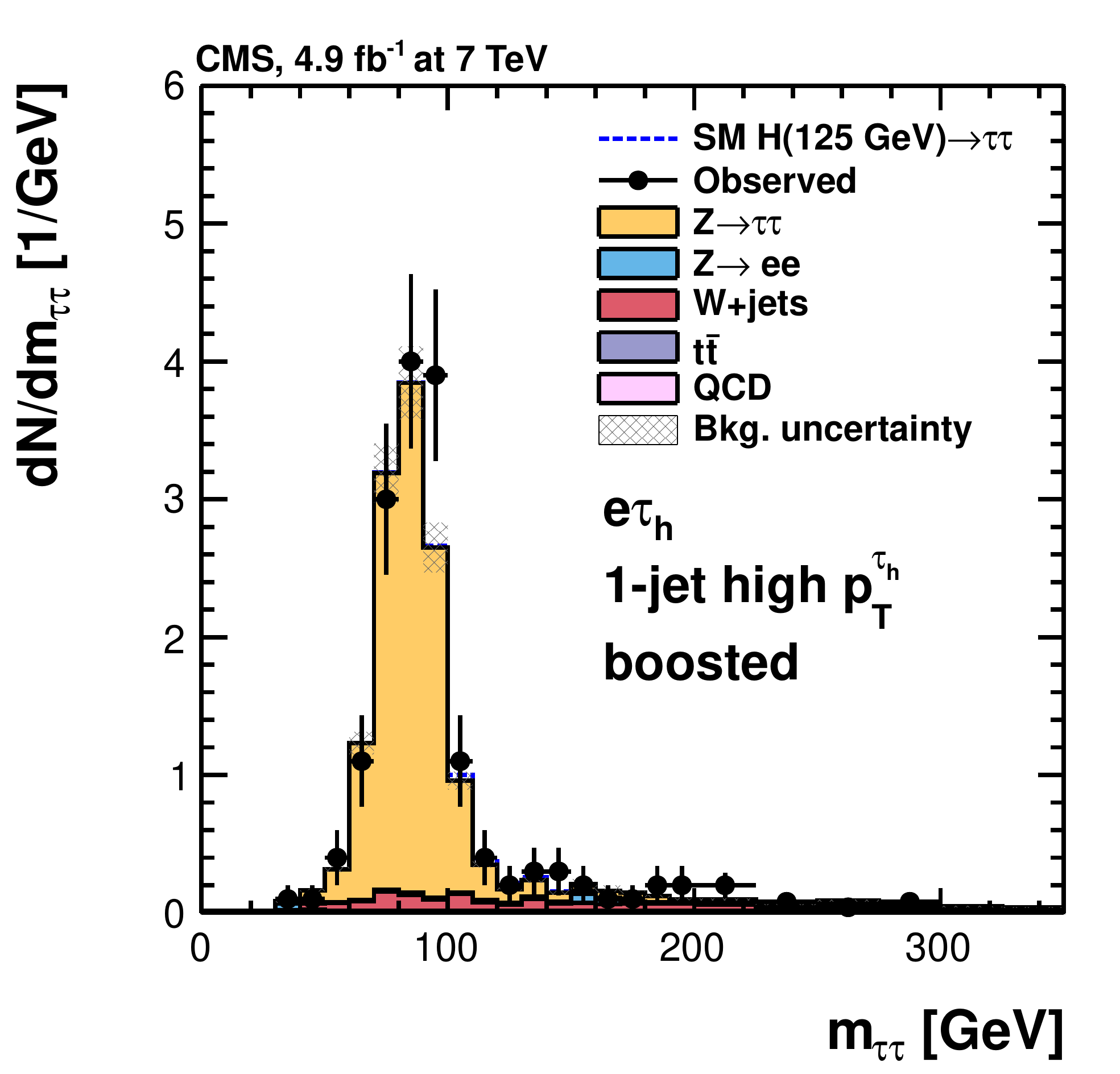}\\
\includegraphics[width=0.32\textwidth]{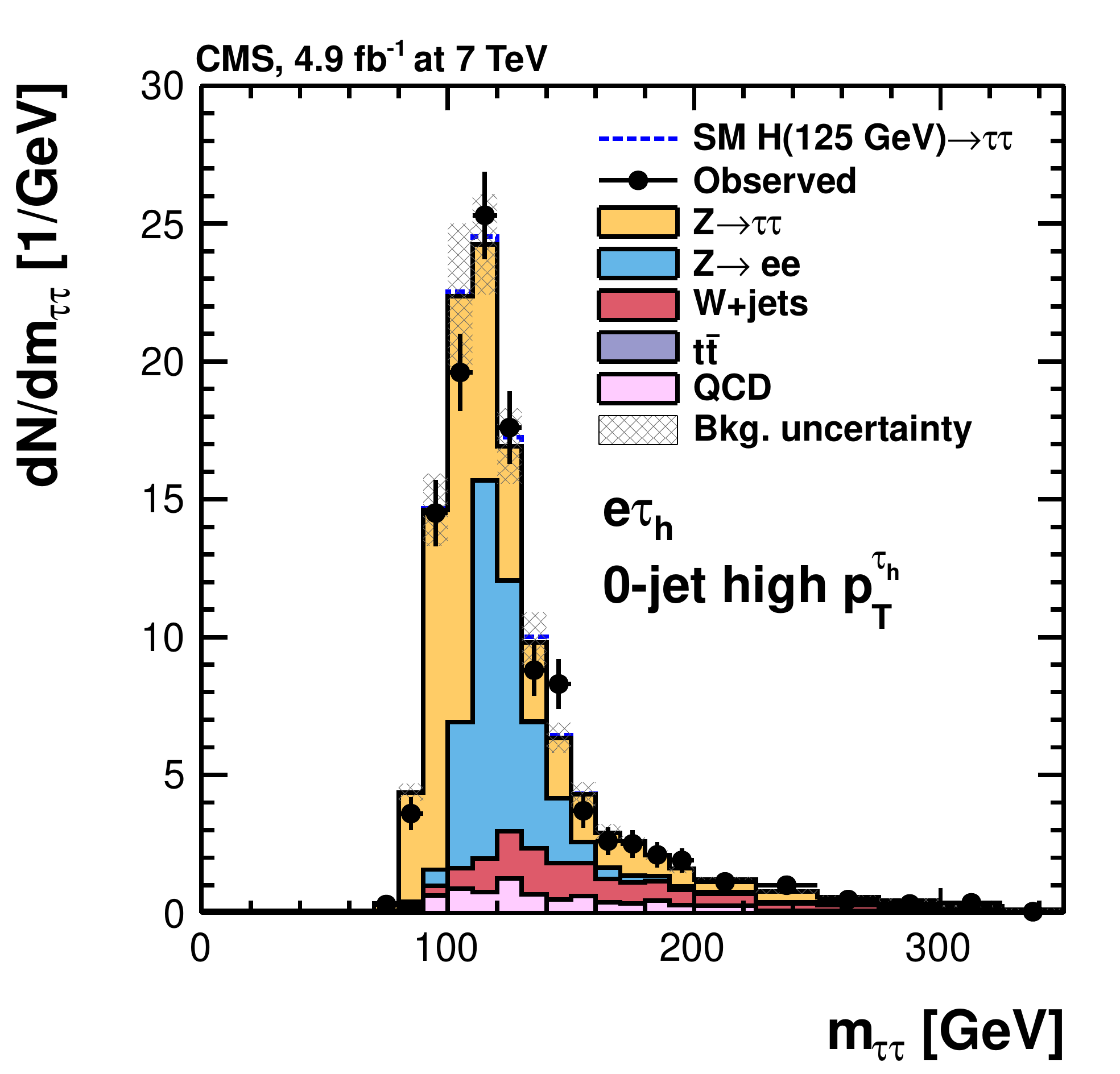}\\
\includegraphics[width=0.32\textwidth]{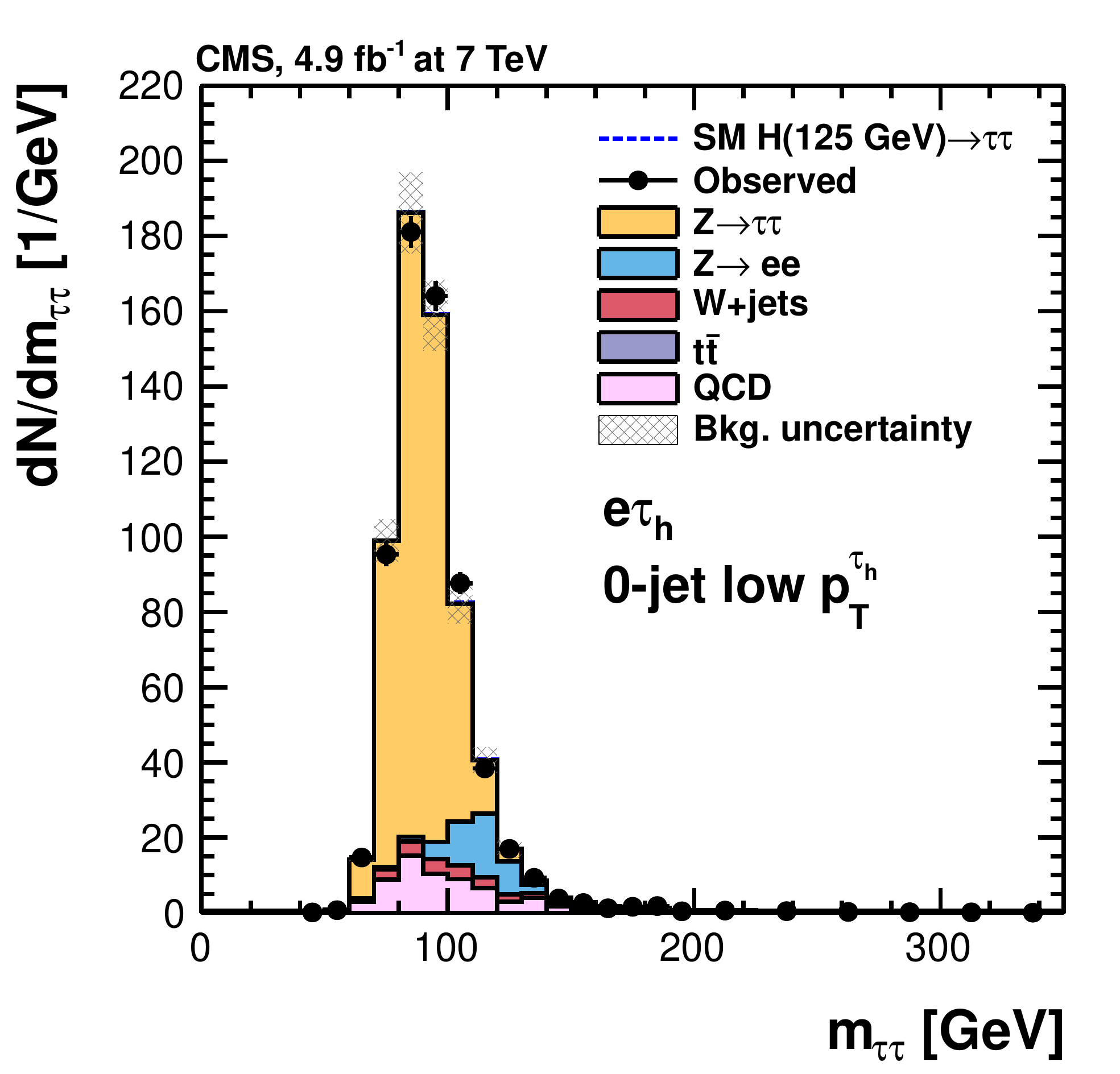}
\includegraphics[width=0.32\textwidth]{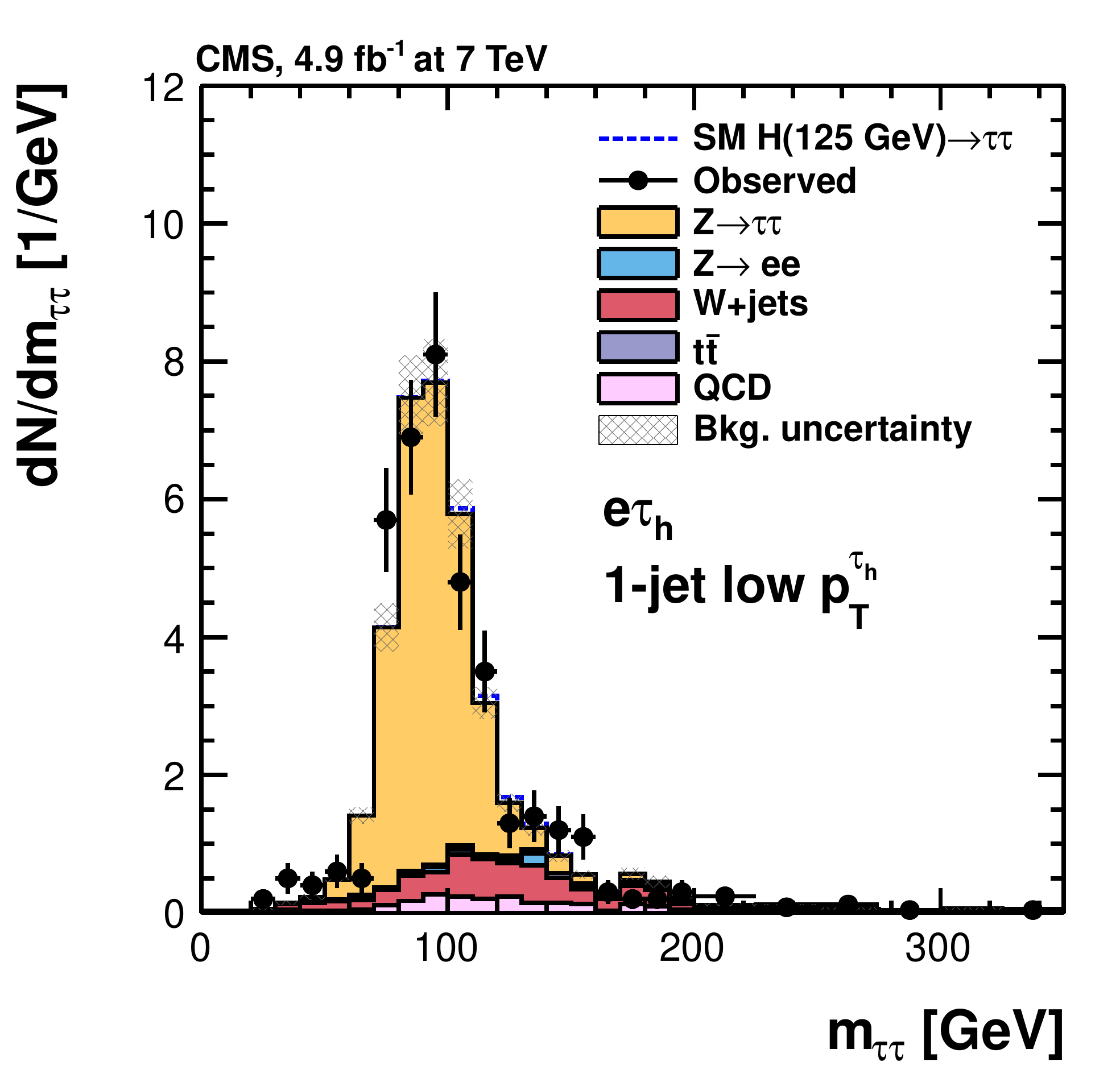}
\includegraphics[width=0.32\textwidth]{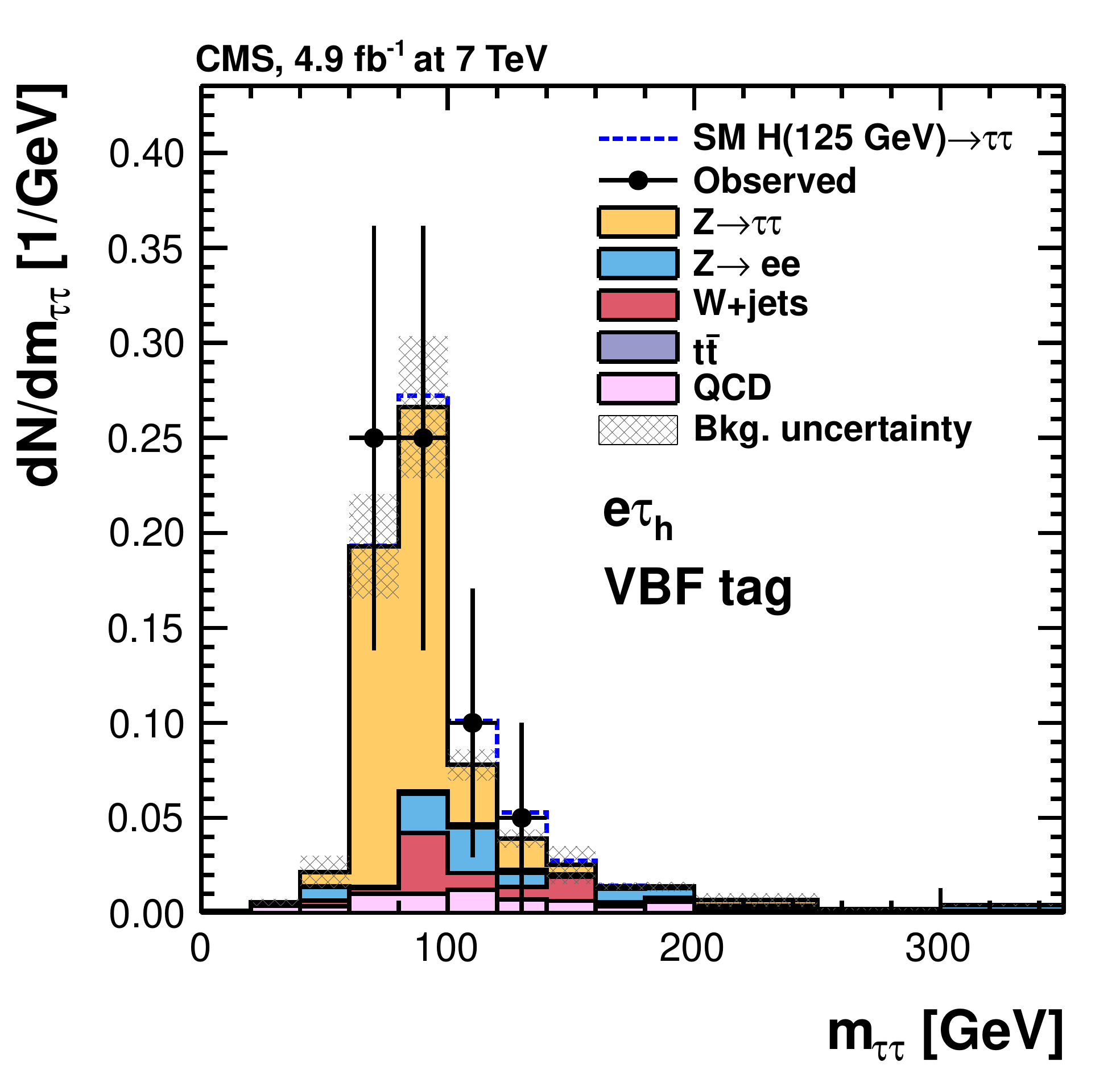}\\
     \caption{Observed and predicted $\mtt$ distributions in the $\etau$ channel, for all categories used in the 7\TeV data analysis. The normalization of the predicted background distributions corresponds to the result of the global fit. The signal distribution, on the other hand, is normalized to the SM prediction. The signal and background histograms are stacked.}
     \label{fig:app_et_7}
\end{figure}

\begin{figure}[b!htp]
\makebox[0.32\textwidth]{}
\includegraphics[width=0.32\textwidth]{figures/plots_131206/et/eleTau_1jet_high_mediumhiggs_postfit_8TeV_LIN.pdf}\\
\includegraphics[width=0.32\textwidth]{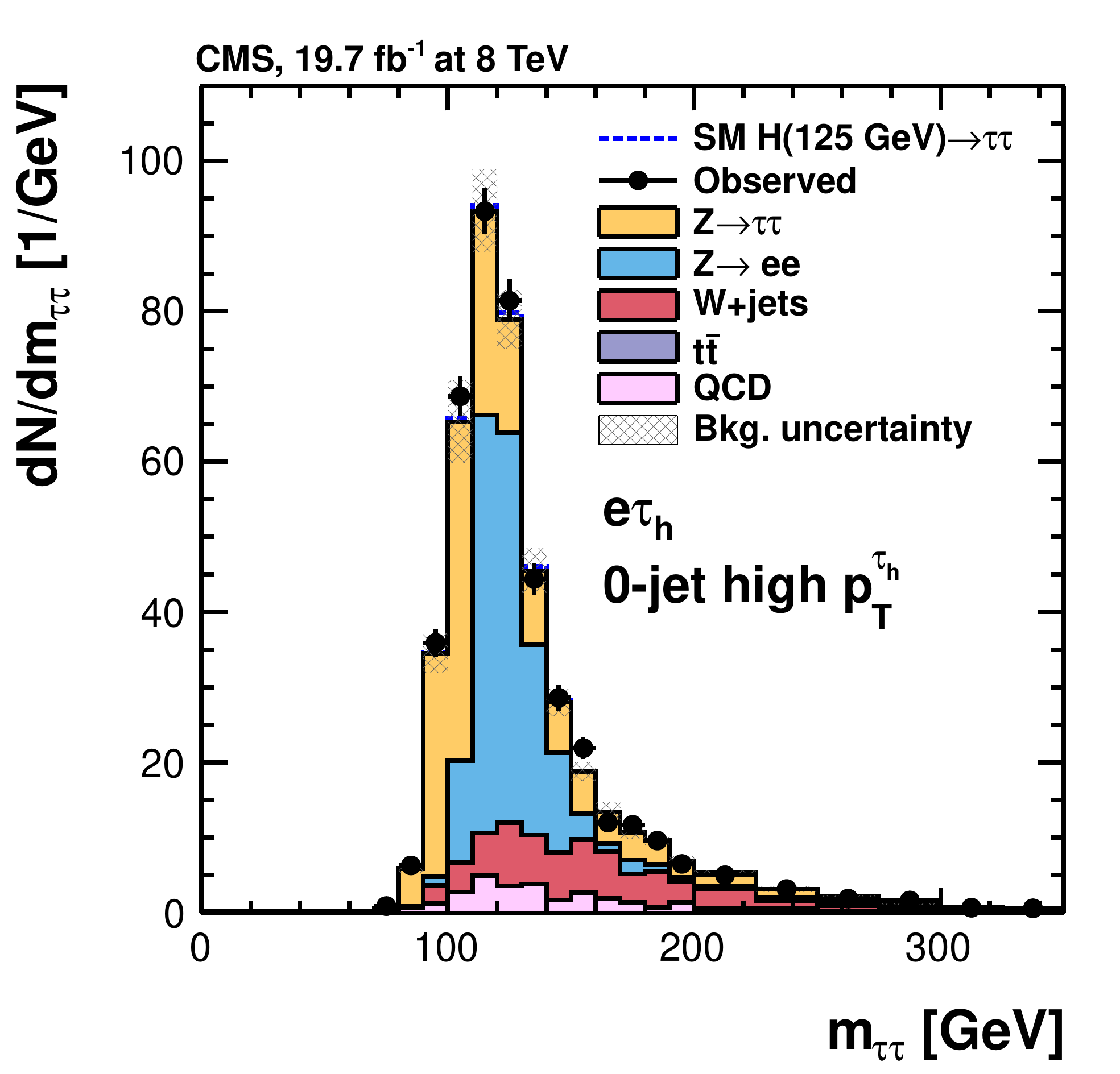}
\makebox[0.32\textwidth]{}
\includegraphics[width=0.32\textwidth]{figures/plots_131206/et/eleTau_vbf_tight_postfit_8TeV_LIN.pdf}\\
\includegraphics[width=0.32\textwidth]{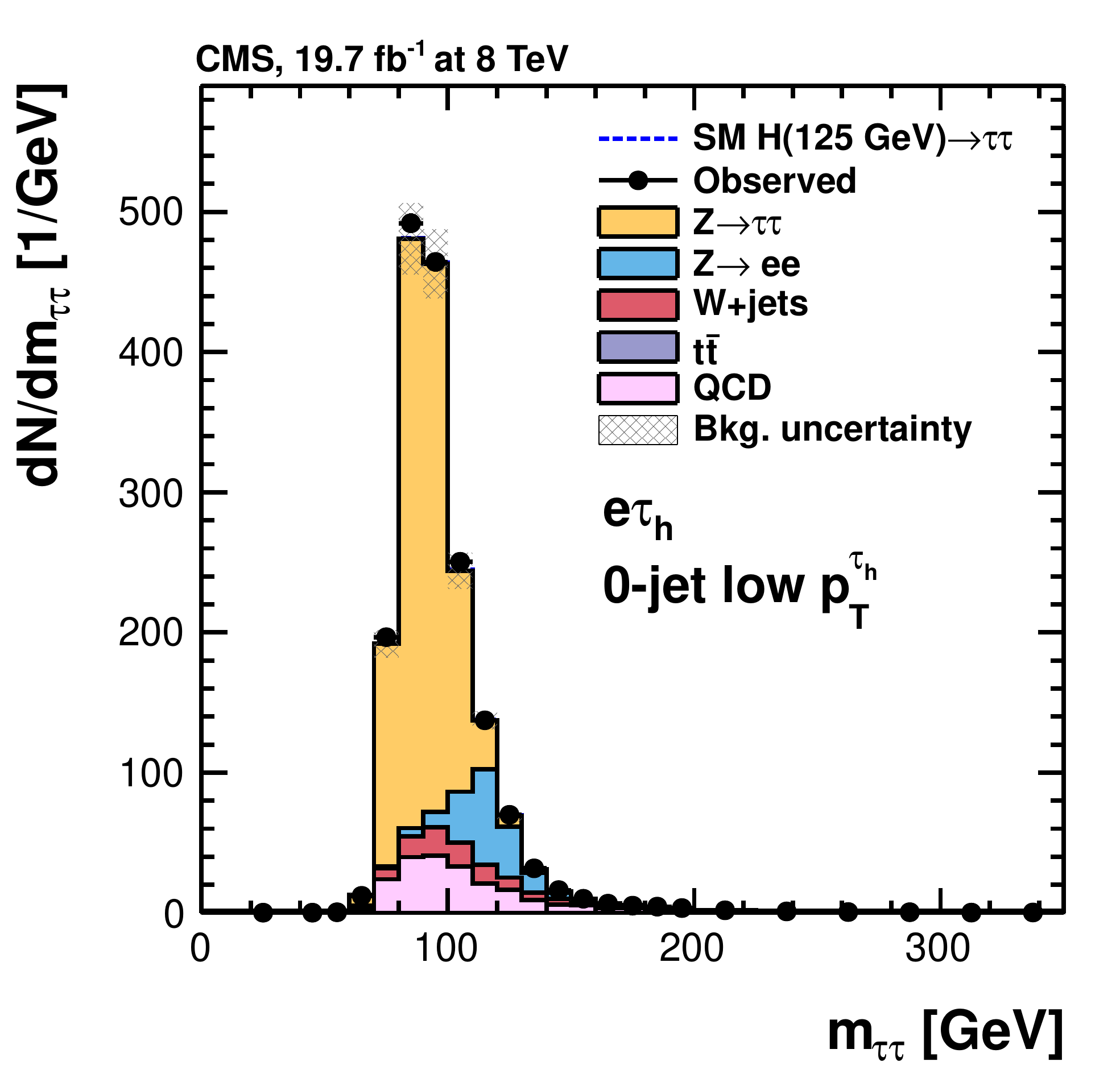}
\includegraphics[width=0.32\textwidth]{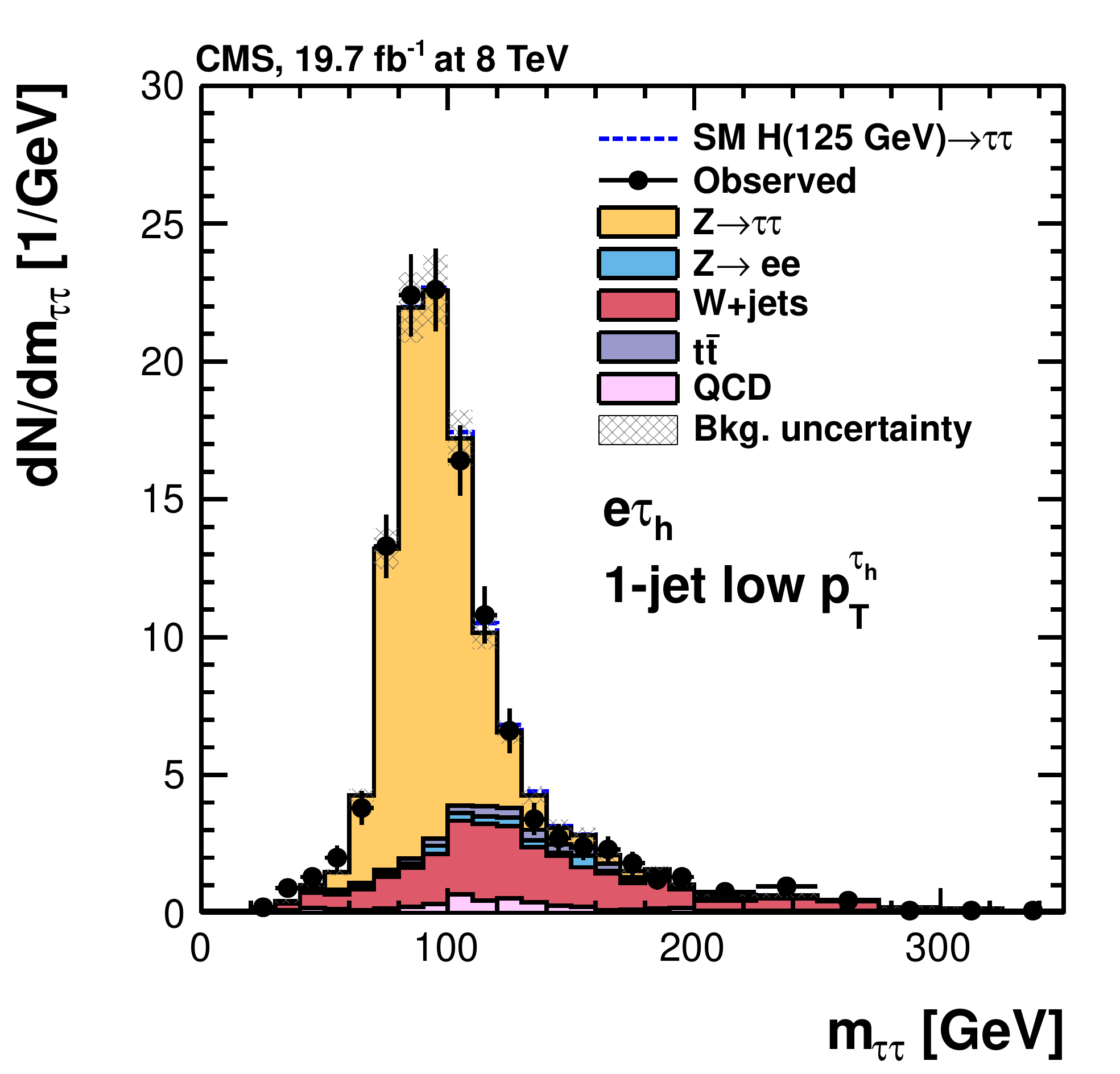}
\includegraphics[width=0.32\textwidth]{figures/plots_131206/et/eleTau_vbf_loose_postfit_8TeV_LIN.pdf}\\
     \caption{Observed and predicted $\mtt$ distributions in the $\etau$ channel, for all categories used in the 8\TeV data analysis. The normalization of the predicted background distributions corresponds to the result of the global fit. The signal distribution, on the other hand, is normalized to the SM prediction. The signal and background histograms are stacked.}
     \label{fig:app_et_8}
\end{figure}

\begin{figure}[bhtp]
     \includegraphics[width=0.32\textwidth]{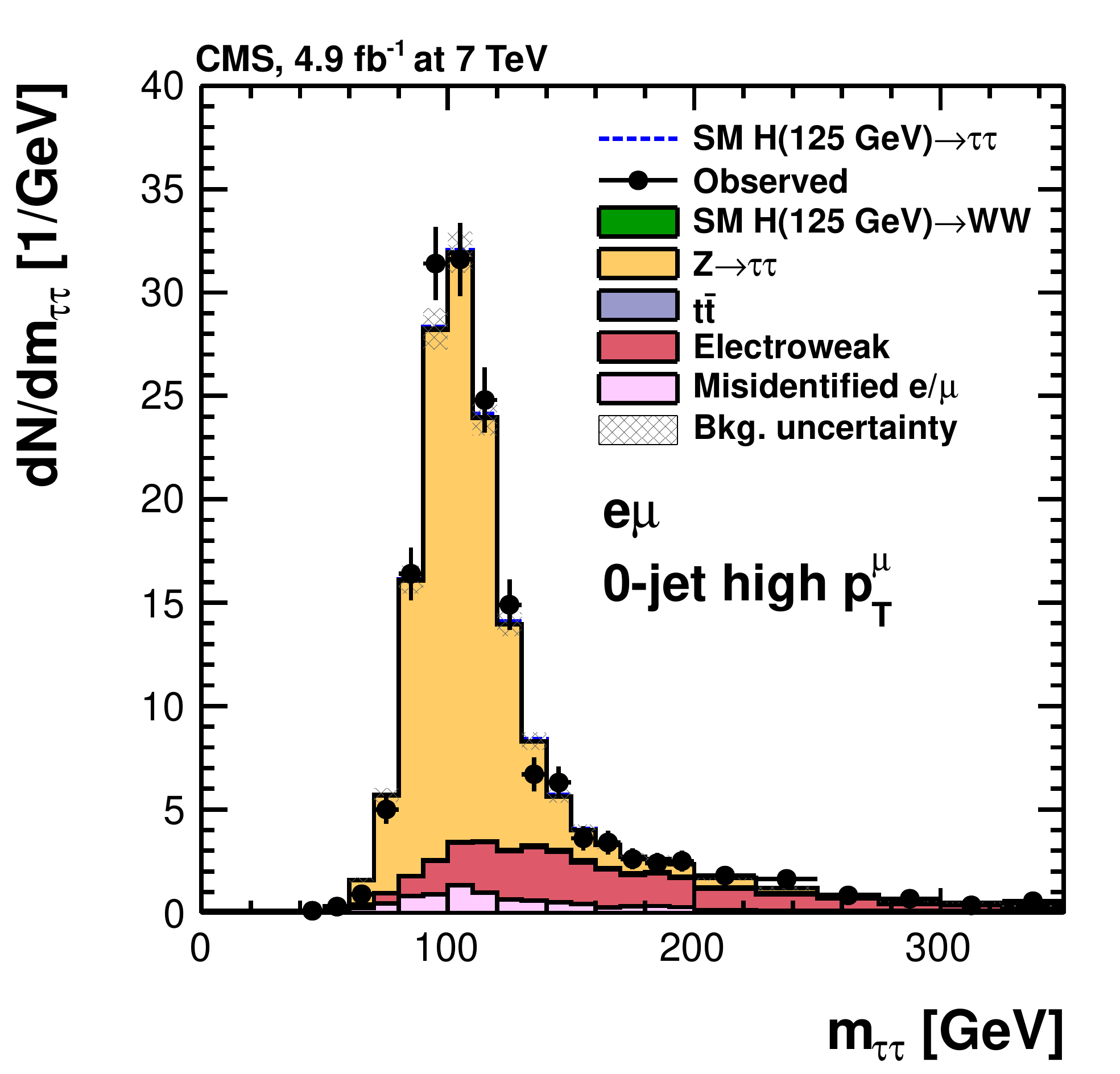}
     \includegraphics[width=0.32\textwidth]{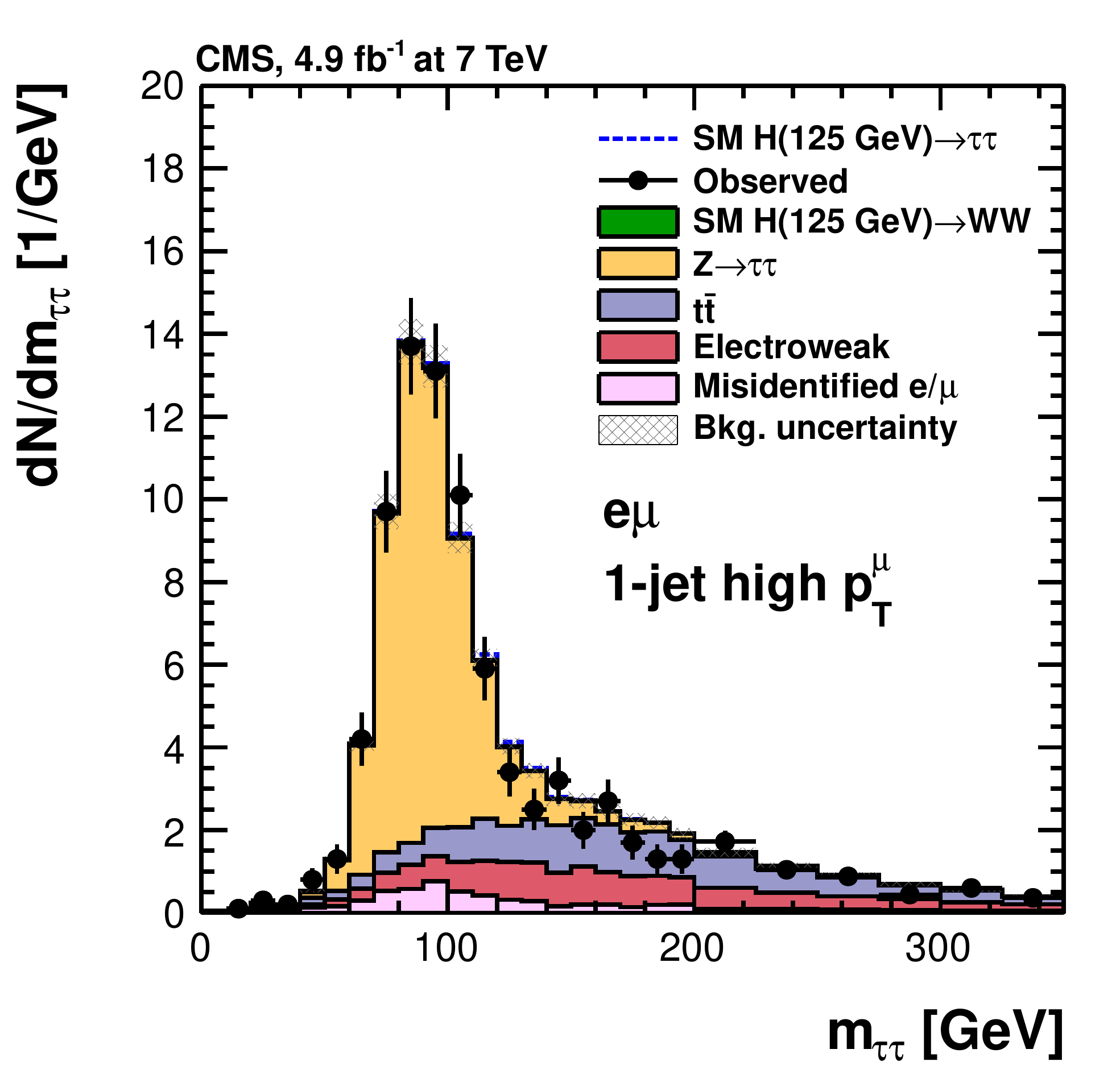}\\
     \includegraphics[width=0.32\textwidth]{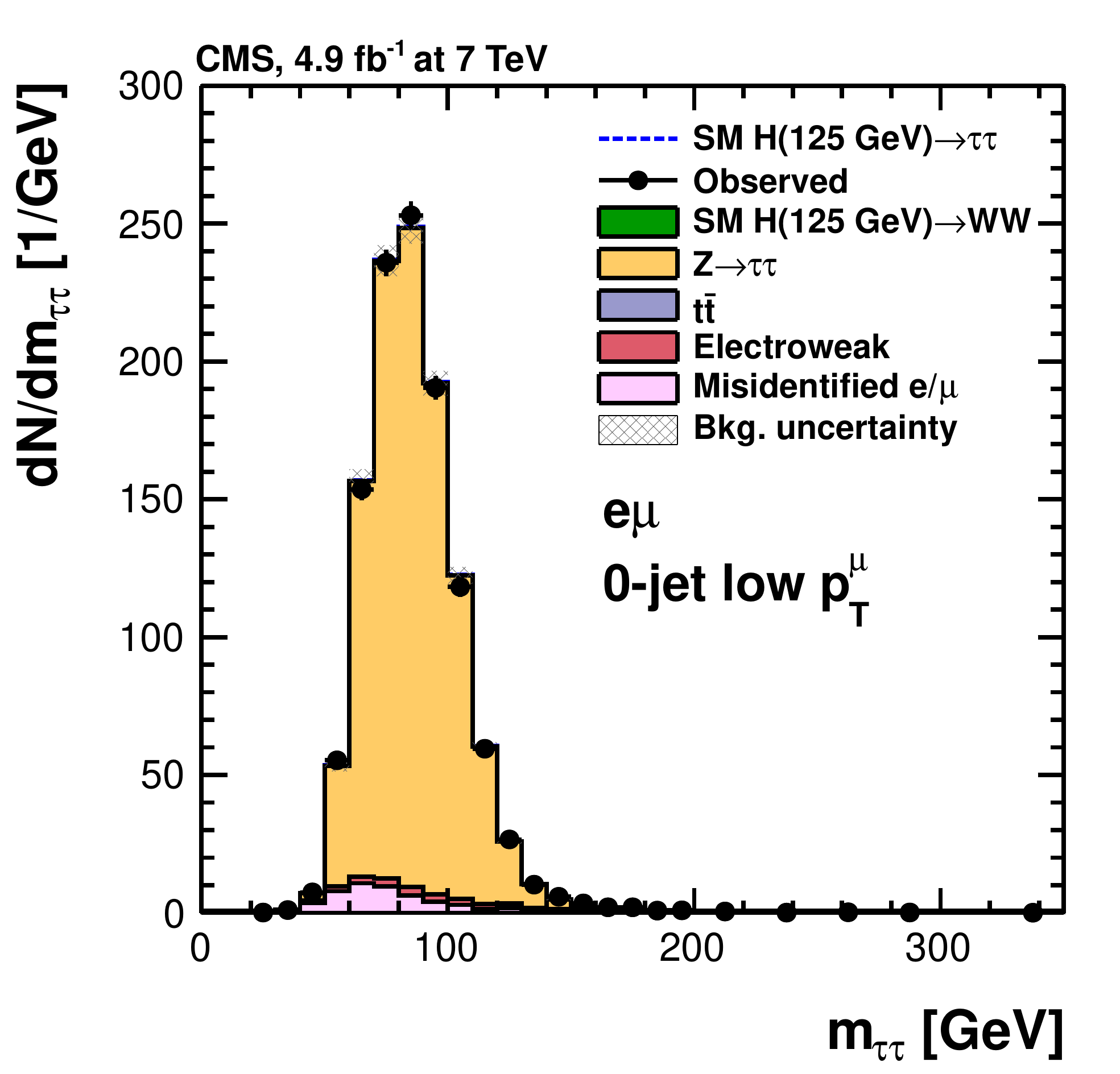}
     \includegraphics[width=0.32\textwidth]{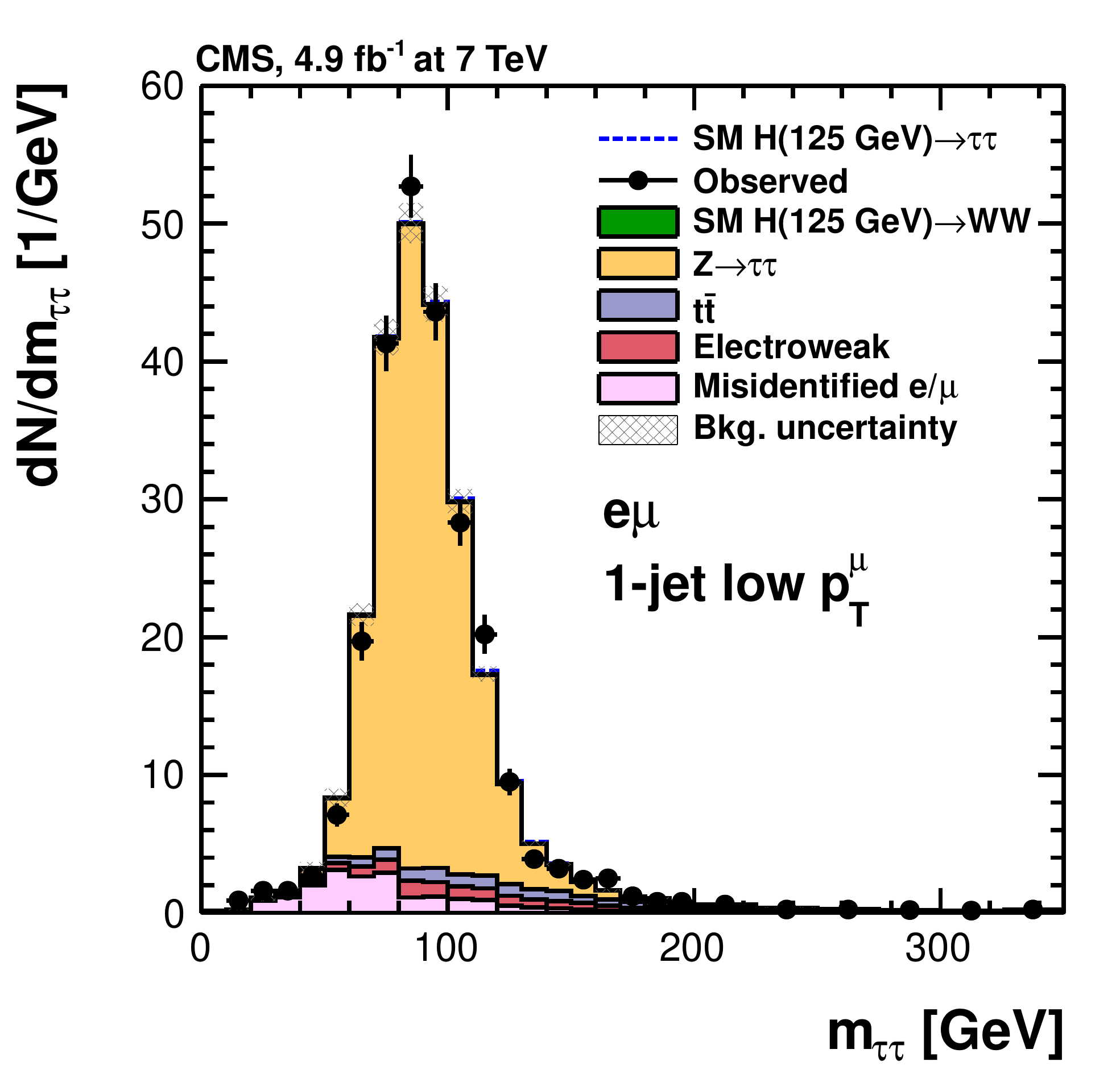}
     \includegraphics[width=0.32\textwidth]{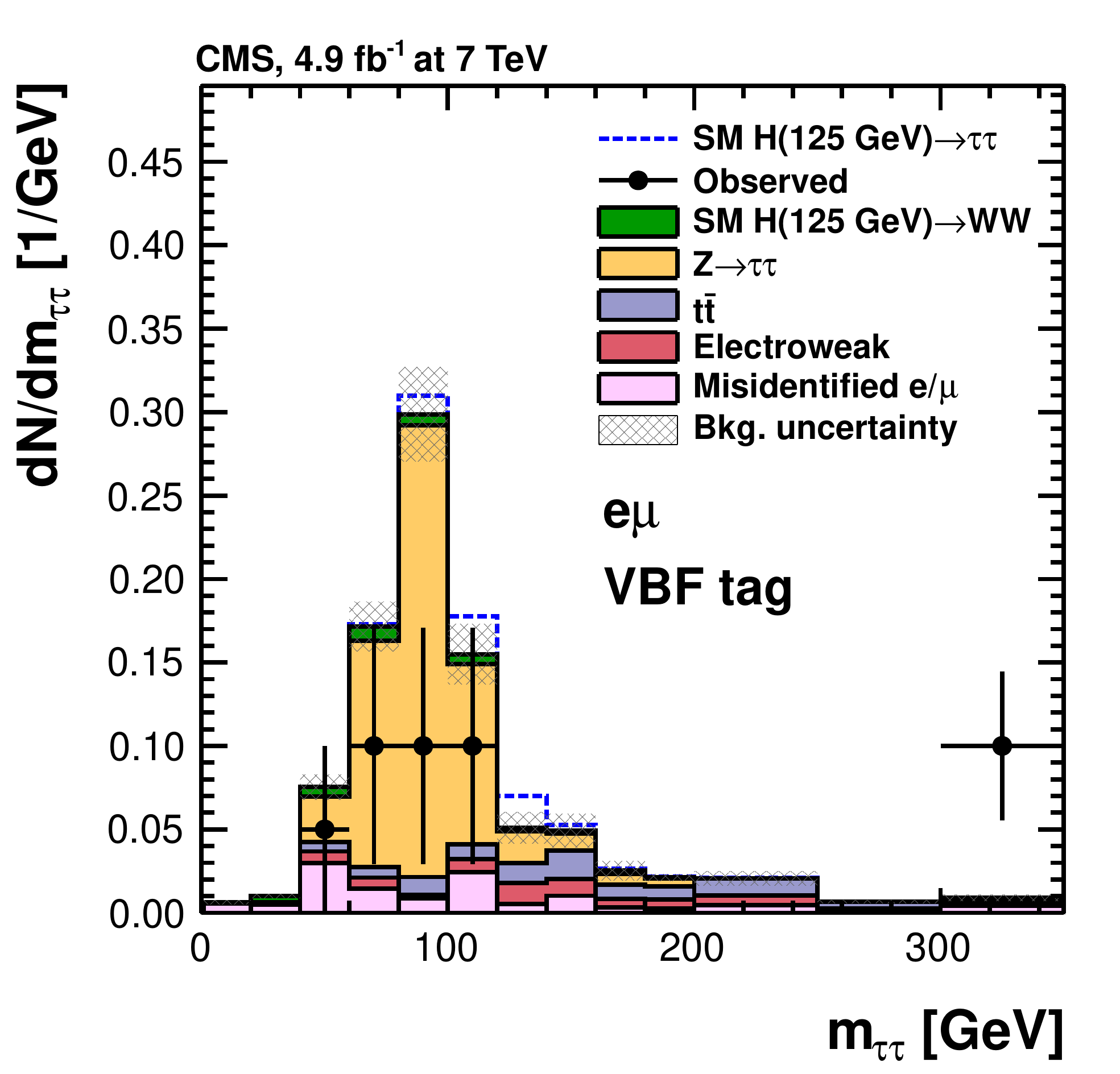}\\
     \caption{Observed and predicted $\mtt$ distributions in the $\emu$ channel, for all categories used in the 7\TeV data analysis. The normalization of the predicted background distributions corresponds to the result of the global fit. The signal distribution, on the other hand, is normalized to the SM prediction. The signal and background histograms are stacked.}
     \label{fig:app_em_7}
\end{figure}

\begin{figure}[bhtp]
     \includegraphics[width=0.32\textwidth]{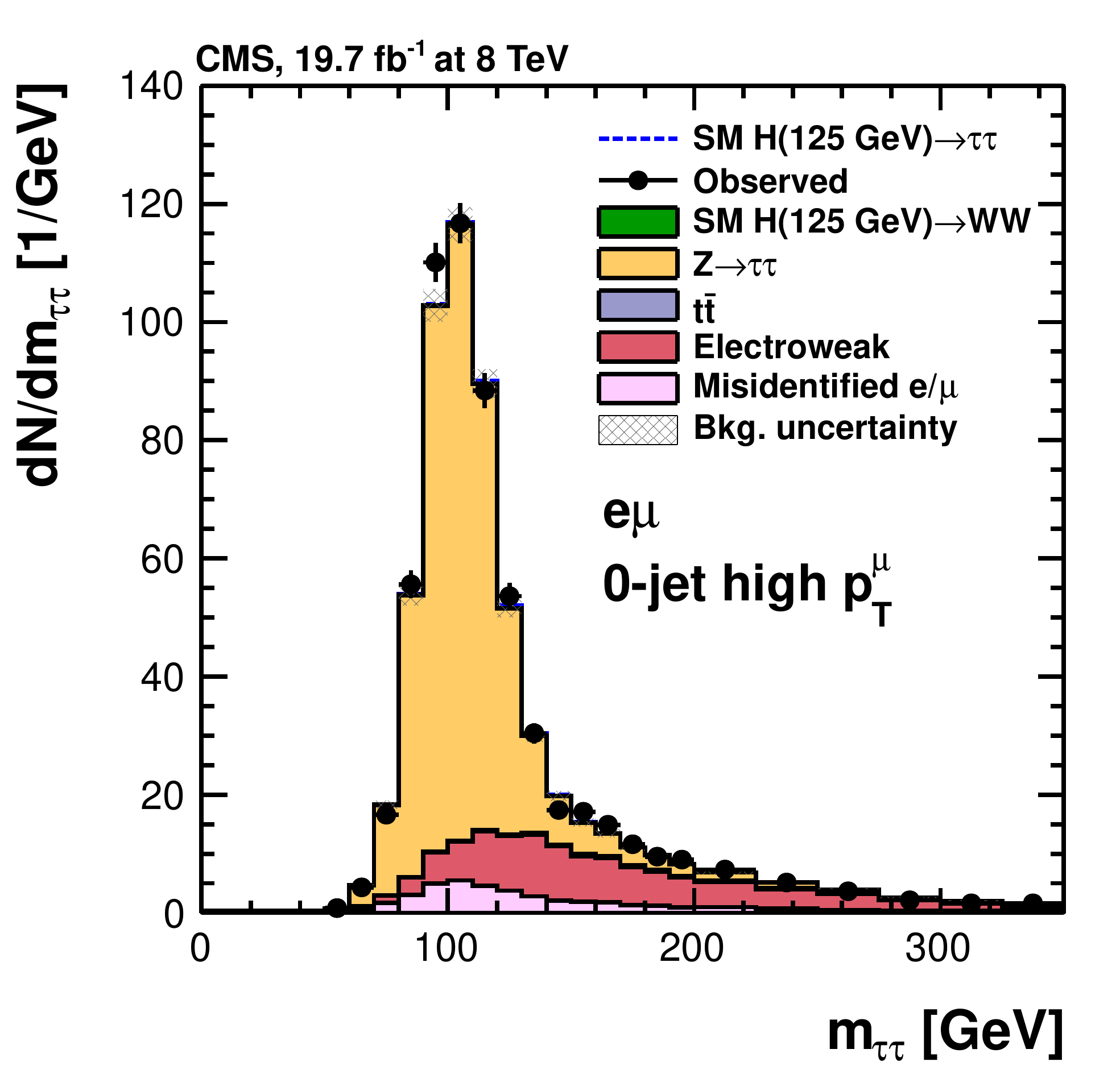}
     \includegraphics[width=0.32\textwidth]{figures/plots_131206/em/emu_1jet_high_postfit_8TeV_LIN.pdf}
     \includegraphics[width=0.32\textwidth]{figures/plots_131206/em/emu_vbf_tight_postfit_8TeV_LIN.pdf}\\
     \includegraphics[width=0.32\textwidth]{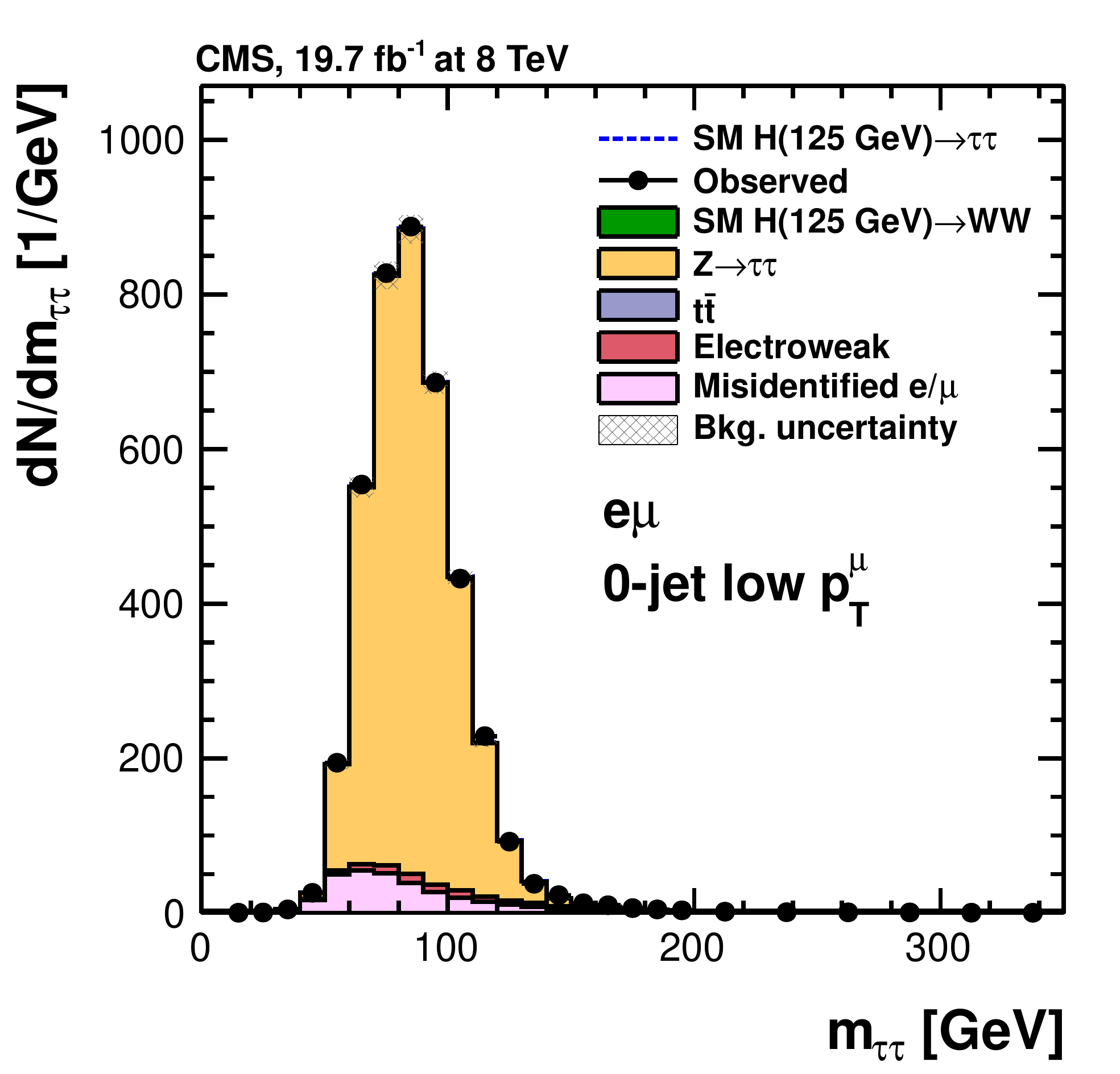}
     \includegraphics[width=0.32\textwidth]{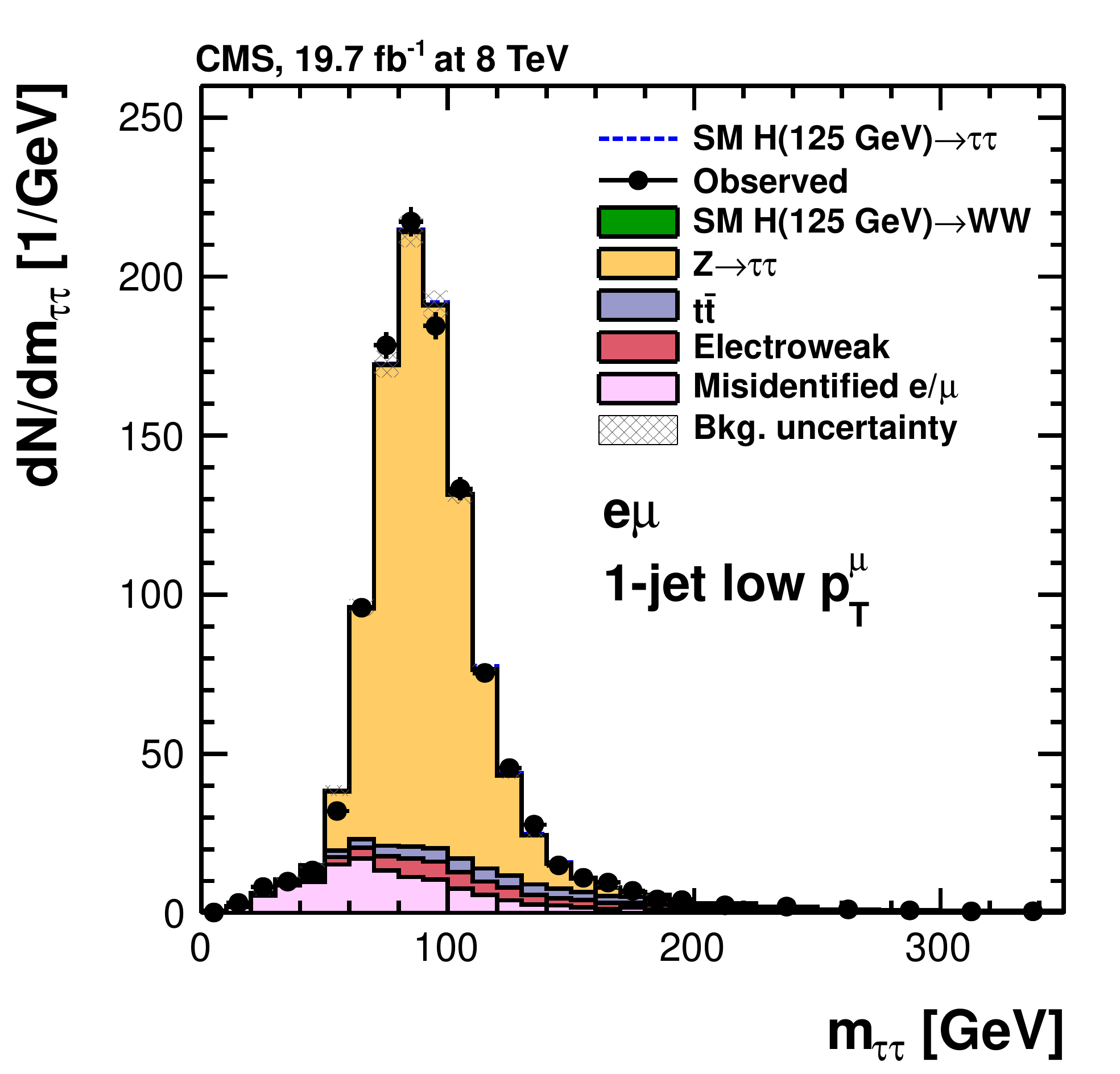}
     \includegraphics[width=0.32\textwidth]{figures/plots_131206/em/emu_vbf_loose_postfit_8TeV_LIN.pdf}\\
     \caption{Observed and predicted $\mtt$ distributions in the $\emu$ channel, for all categories used in the 8\TeV data analysis. The normalization of the predicted background distributions corresponds to the result of the global fit. The signal distribution, on the other hand, is normalized to the SM prediction. The signal and background histograms are stacked.}
     \label{fig:app_em_8}
\end{figure}

\begin{figure}[bhtp]
     \includegraphics[width=0.32\textwidth]{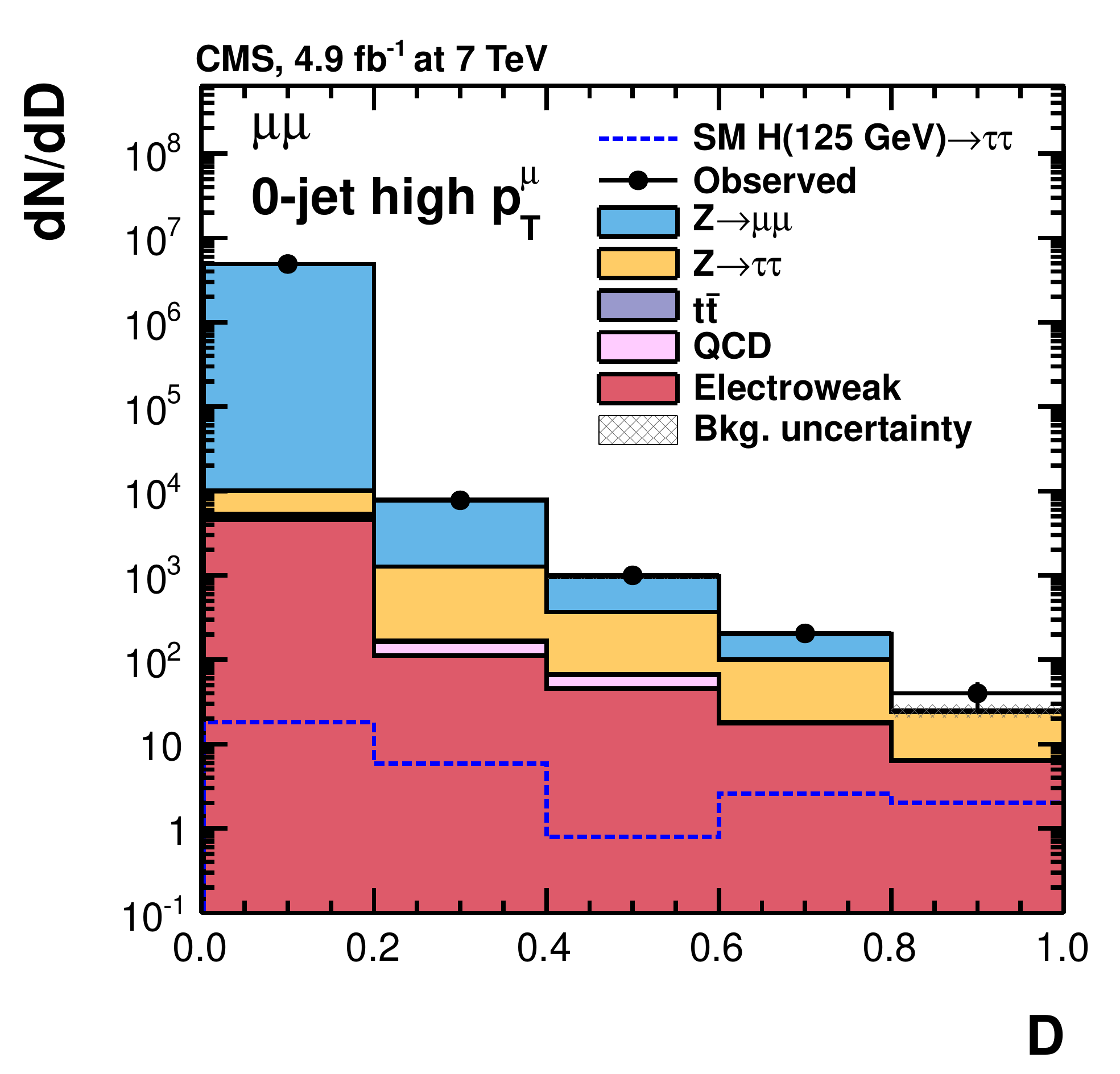}
     \includegraphics[width=0.32\textwidth]{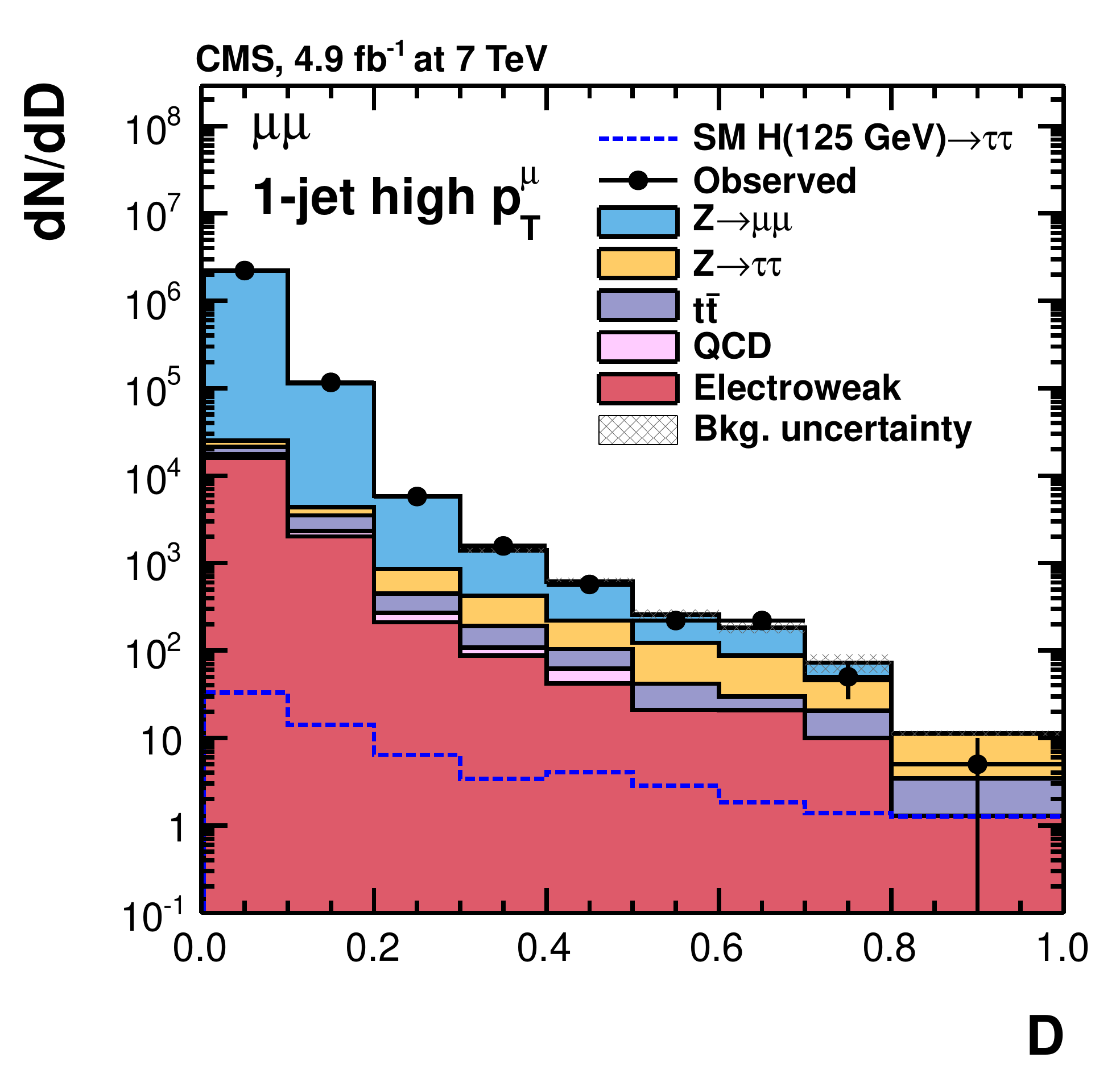}  \\
     \includegraphics[width=0.32\textwidth]{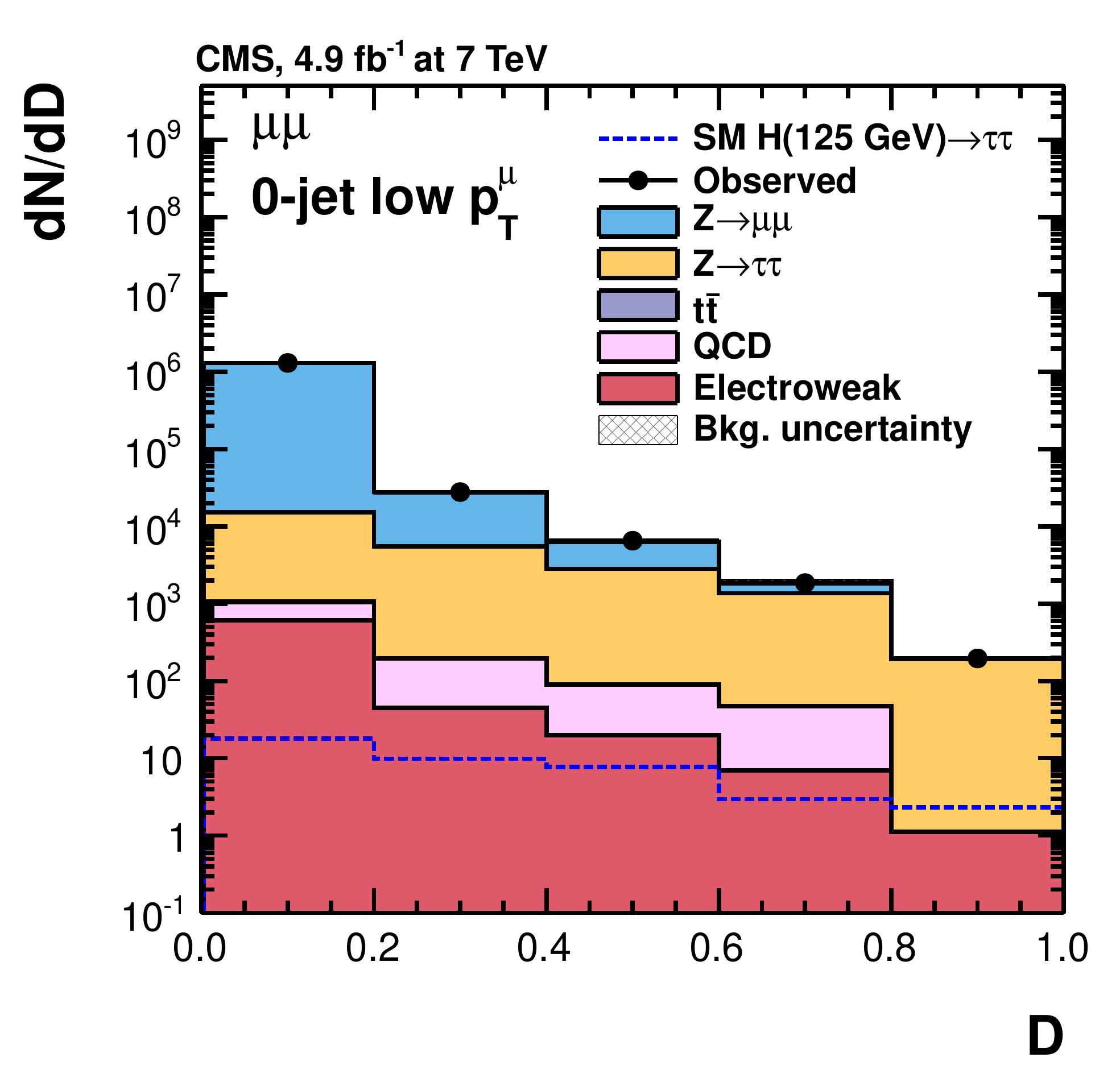}
     \includegraphics[width=0.32\textwidth]{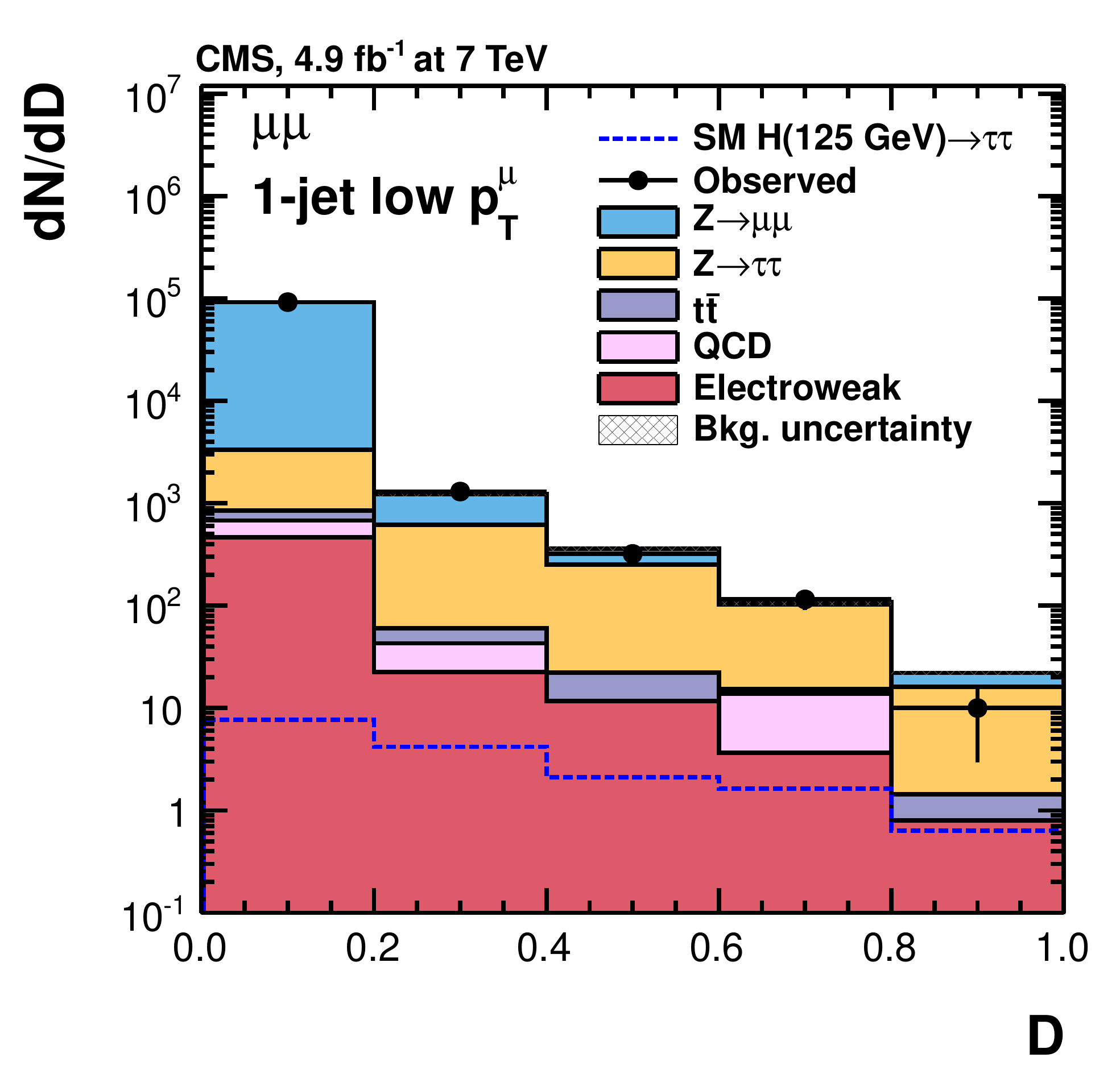}
     \includegraphics[width=0.32\textwidth]{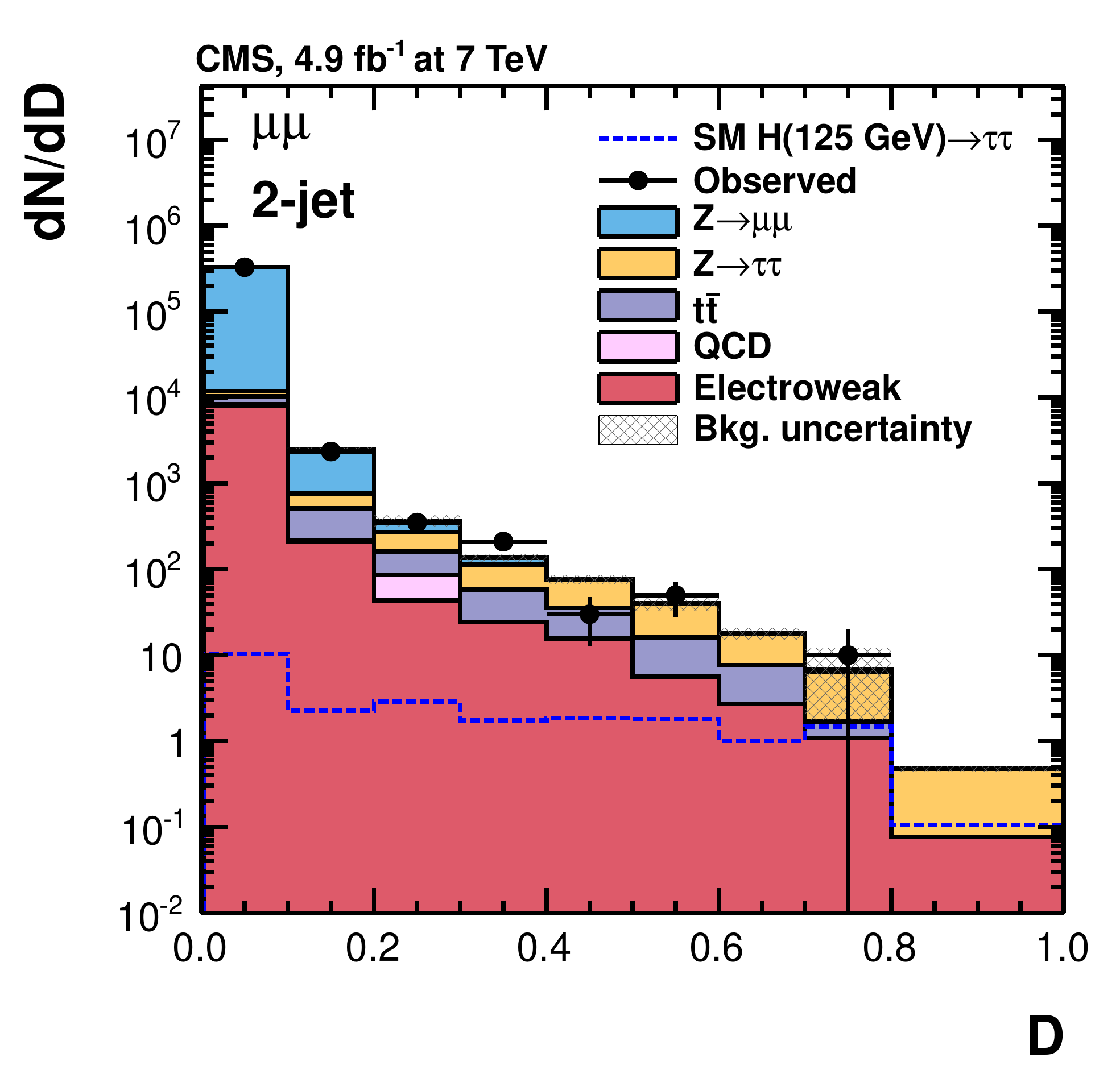}
     \caption{Observed and predicted $D$ distributions in the $\mumu$ channel, for all categories used in the 7\TeV data analysis. The normalization of the predicted background distributions corresponds to the result of the global fit. The signal distribution, on the other hand, is normalized to the SM prediction. The open signal histogram is shown superimposed to the background histograms, which are stacked.}
     \label{fig:app_mm_7}
\end{figure}

\begin{figure}[bhtp]
     \includegraphics[width=0.32\textwidth]{figures/plots_140111/mm/mumu_0jet_high_postfit_8TeV_LOG.pdf}
     \includegraphics[width=0.32\textwidth]{figures/plots_140111/mm/mumu_1jet_high_postfit_8TeV_LOG.pdf}  \\
     \includegraphics[width=0.32\textwidth]{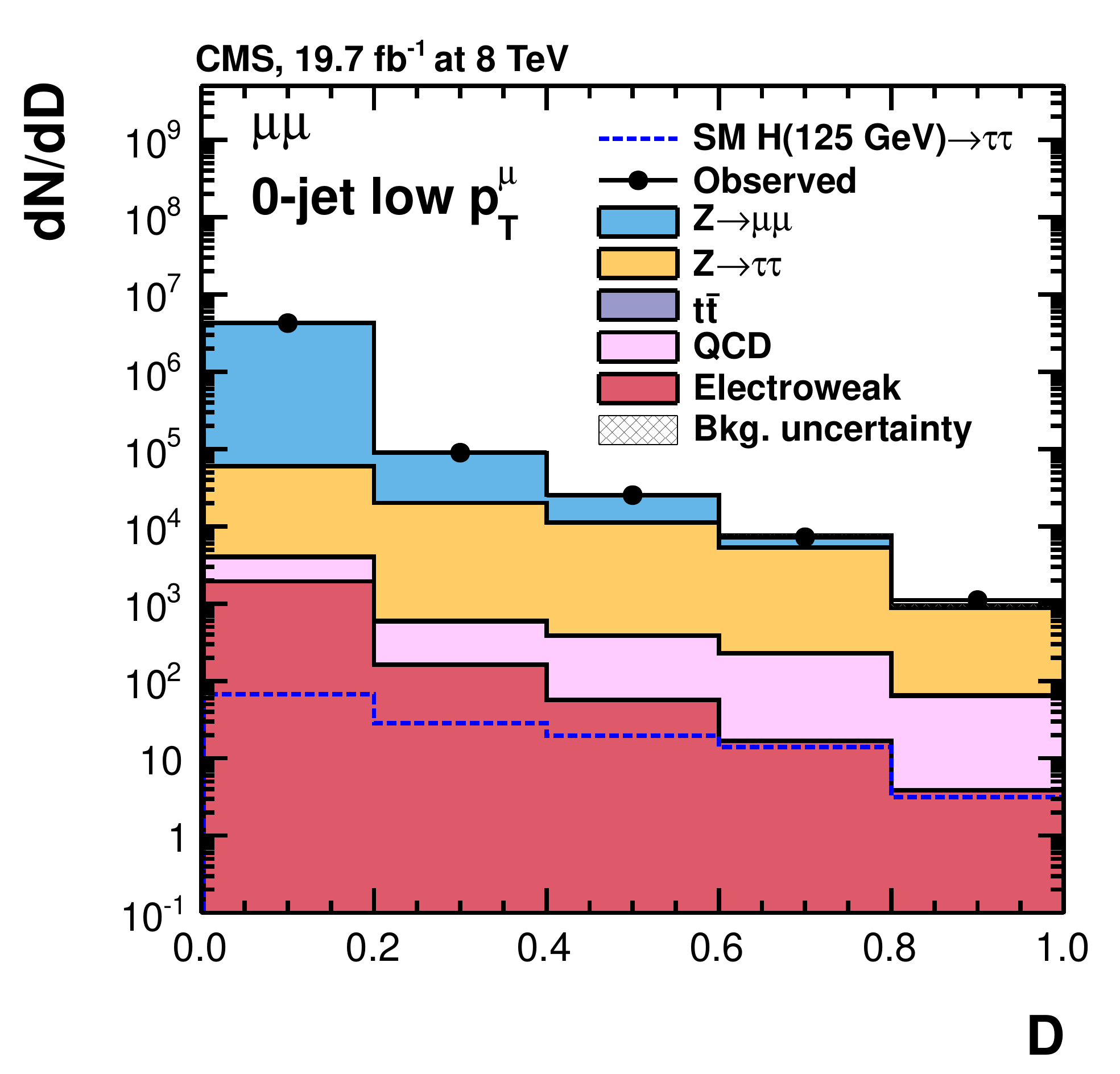}
     \includegraphics[width=0.32\textwidth]{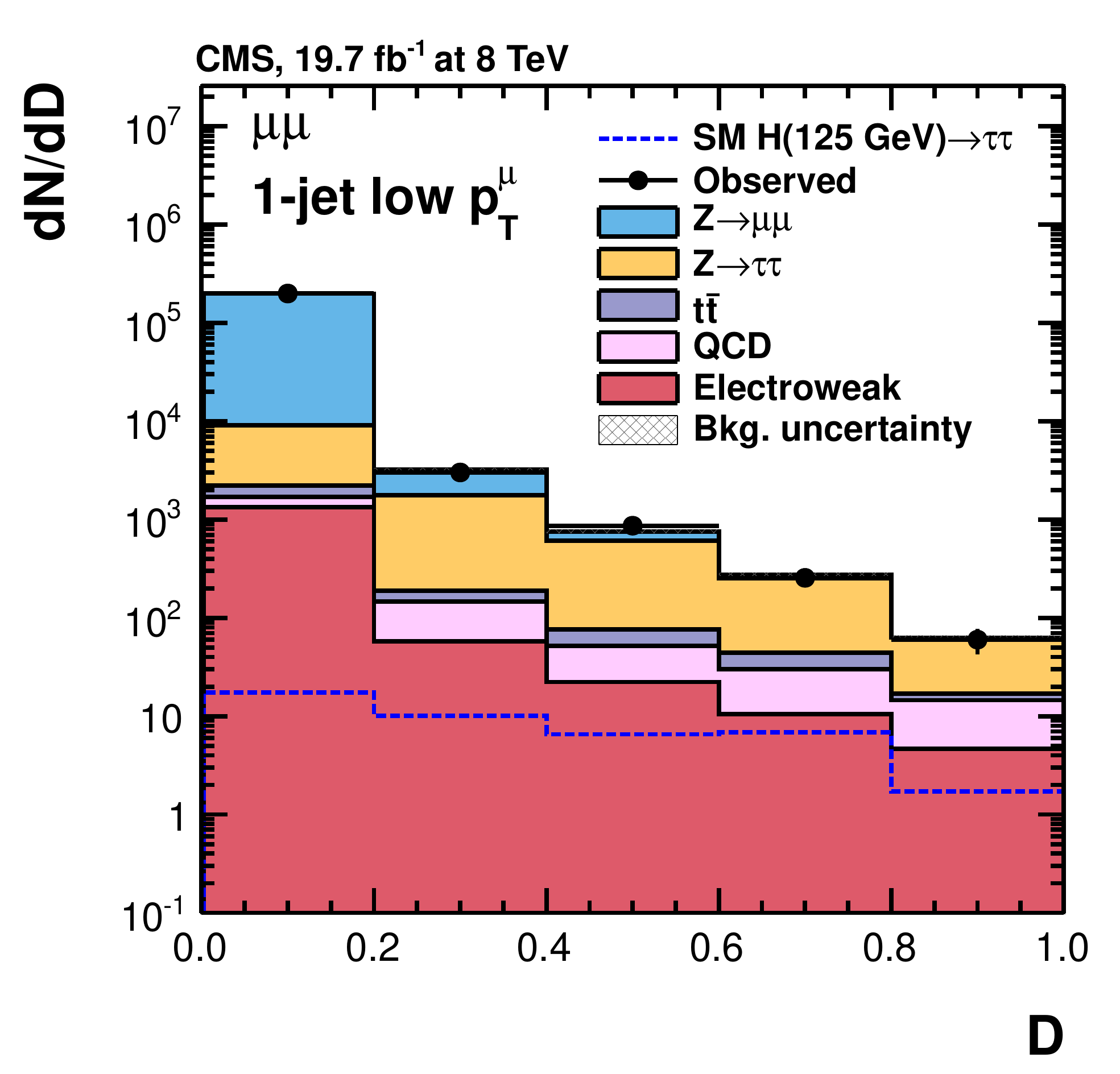}
     \includegraphics[width=0.32\textwidth]{figures/plots_140111/mm/mumu_vbf_postfit_8TeV_LOG.pdf}
     \caption{Observed and predicted $D$ distributions in the $\mumu$ channel, for all categories used in the 8\TeV data analysis. The normalization of the predicted background distributions corresponds to the result of the global fit. The signal distribution, on the other hand, is normalized to the SM prediction. The open signal histogram is shown superimposed to the background histograms, which are stacked.}
     \label{fig:app_mm_8}
\end{figure}

\begin{figure}[bhtp]
\includegraphics[width=0.32\textwidth]{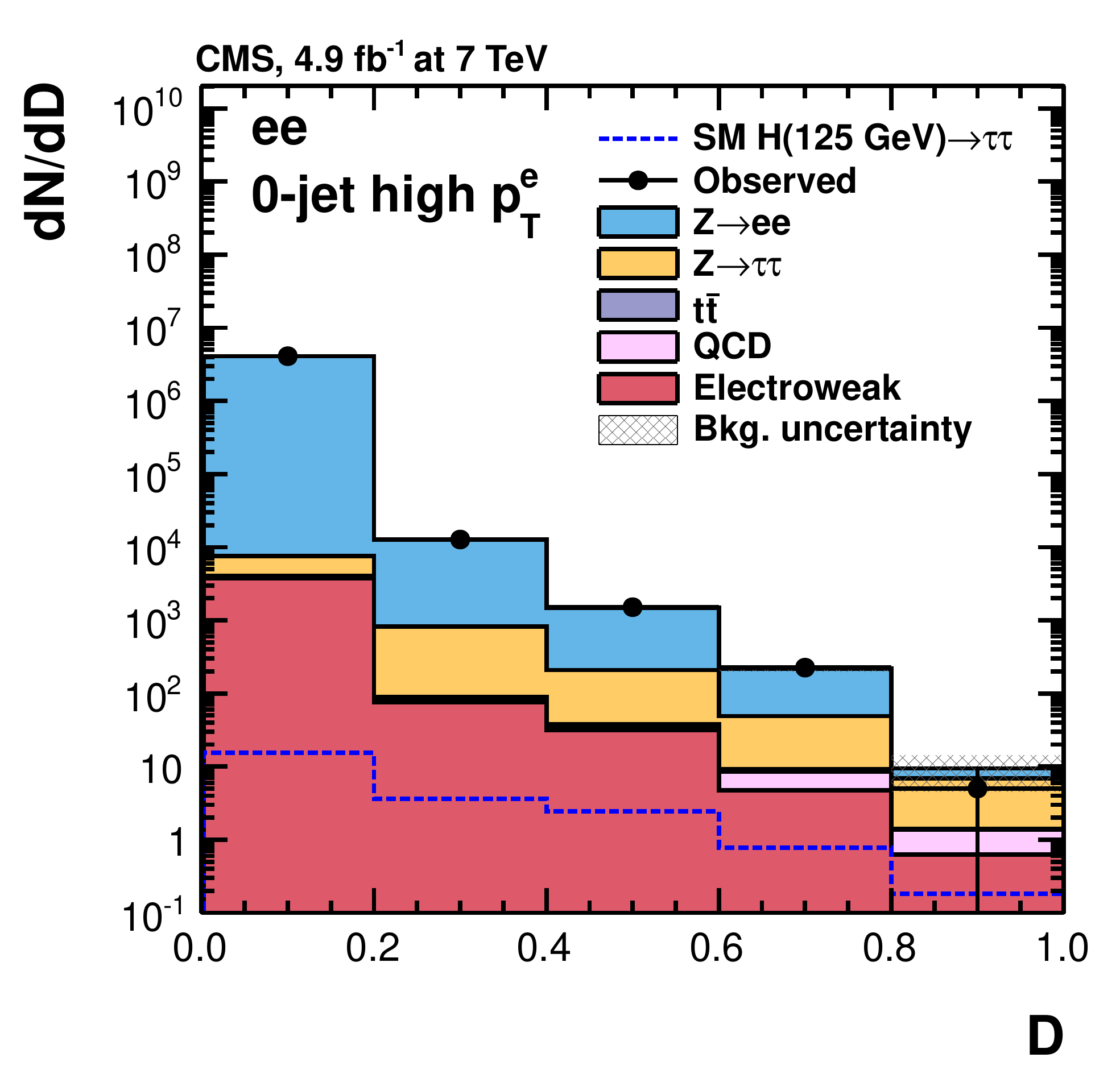}
\includegraphics[width=0.32\textwidth]{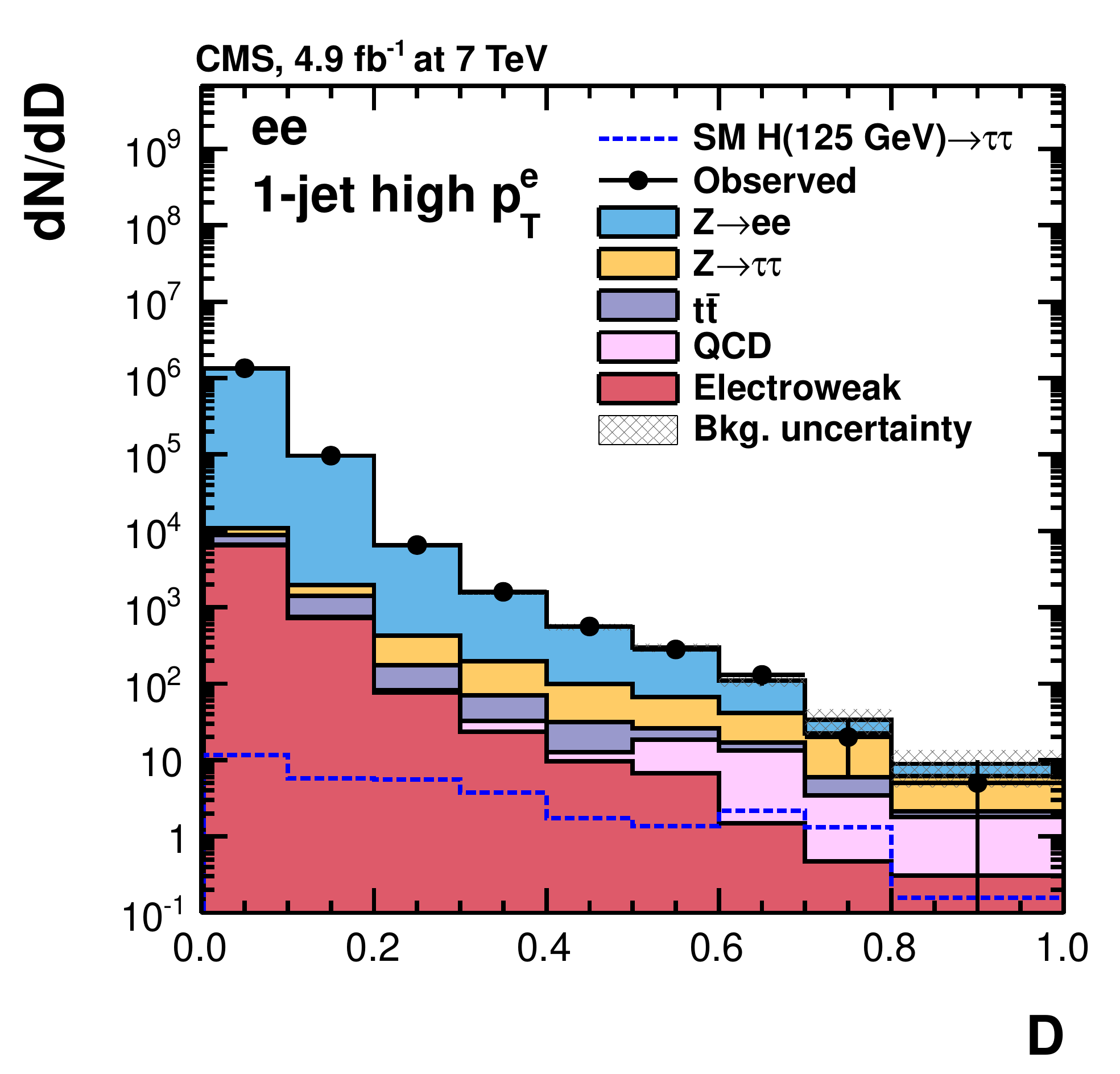}  \\
\includegraphics[width=0.32\textwidth]{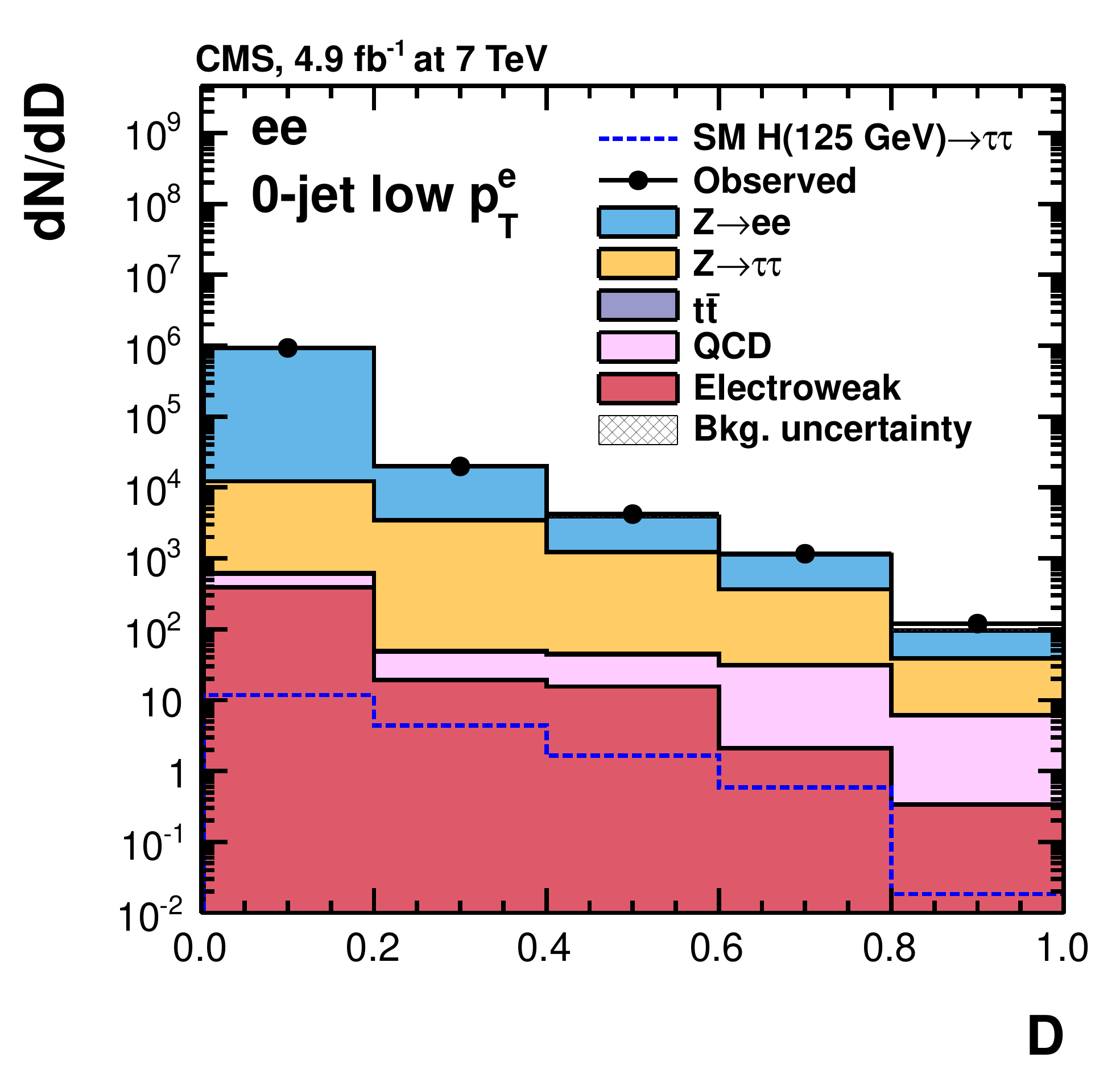}
\includegraphics[width=0.32\textwidth]{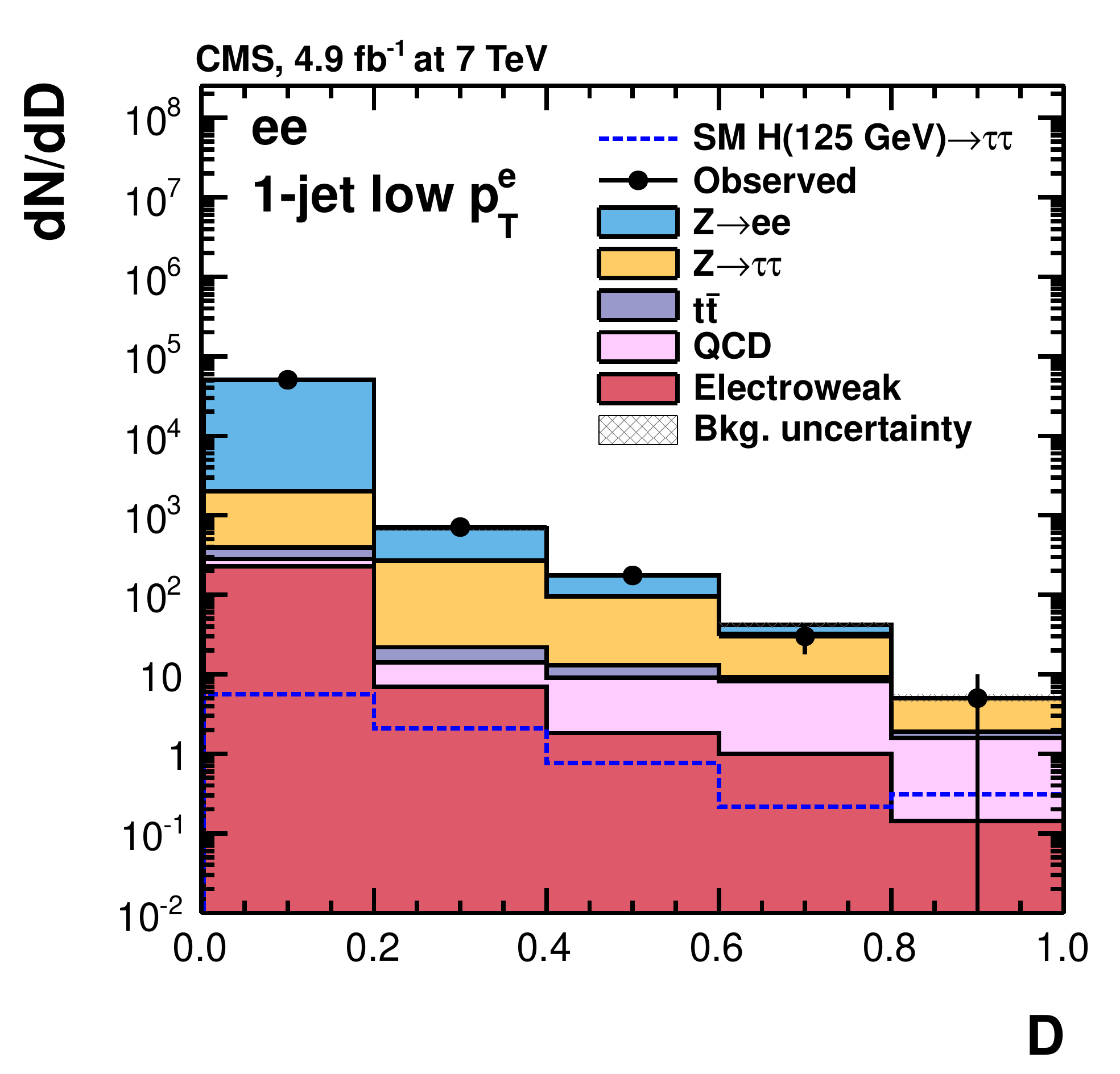}
\includegraphics[width=0.32\textwidth]{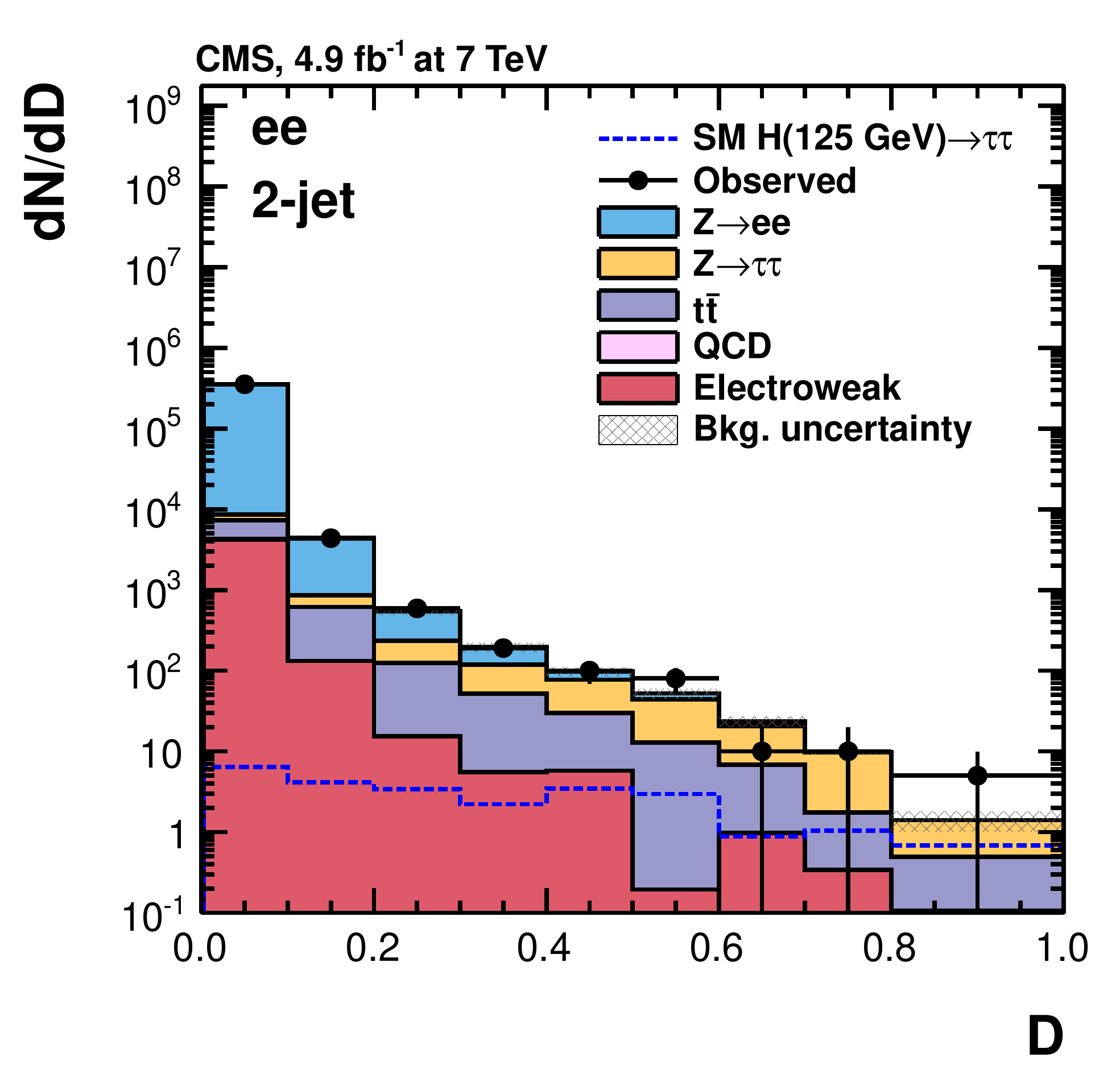}
\caption{Observed and predicted $D$ distributions in the $\ee$ channel, for all categories used in the 7\TeV data analysis. The normalization of the predicted background distributions corresponds to the result of the global fit. The signal distribution, on the other hand, is normalized to the SM prediction. The open signal histogram is shown superimposed to the background histograms, which are stacked.}
\label{fig:app_ee_7}
\end{figure}

\begin{figure}[bhtp]
 \includegraphics[width=0.32\textwidth]{figures/plots_140111/ee/ee_0jet_high_postfit_8TeV_LOG.pdf}
 \includegraphics[width=0.32\textwidth]{figures/plots_140111/ee/ee_1jet_high_postfit_8TeV_LOG.pdf}  \\
 \includegraphics[width=0.32\textwidth]{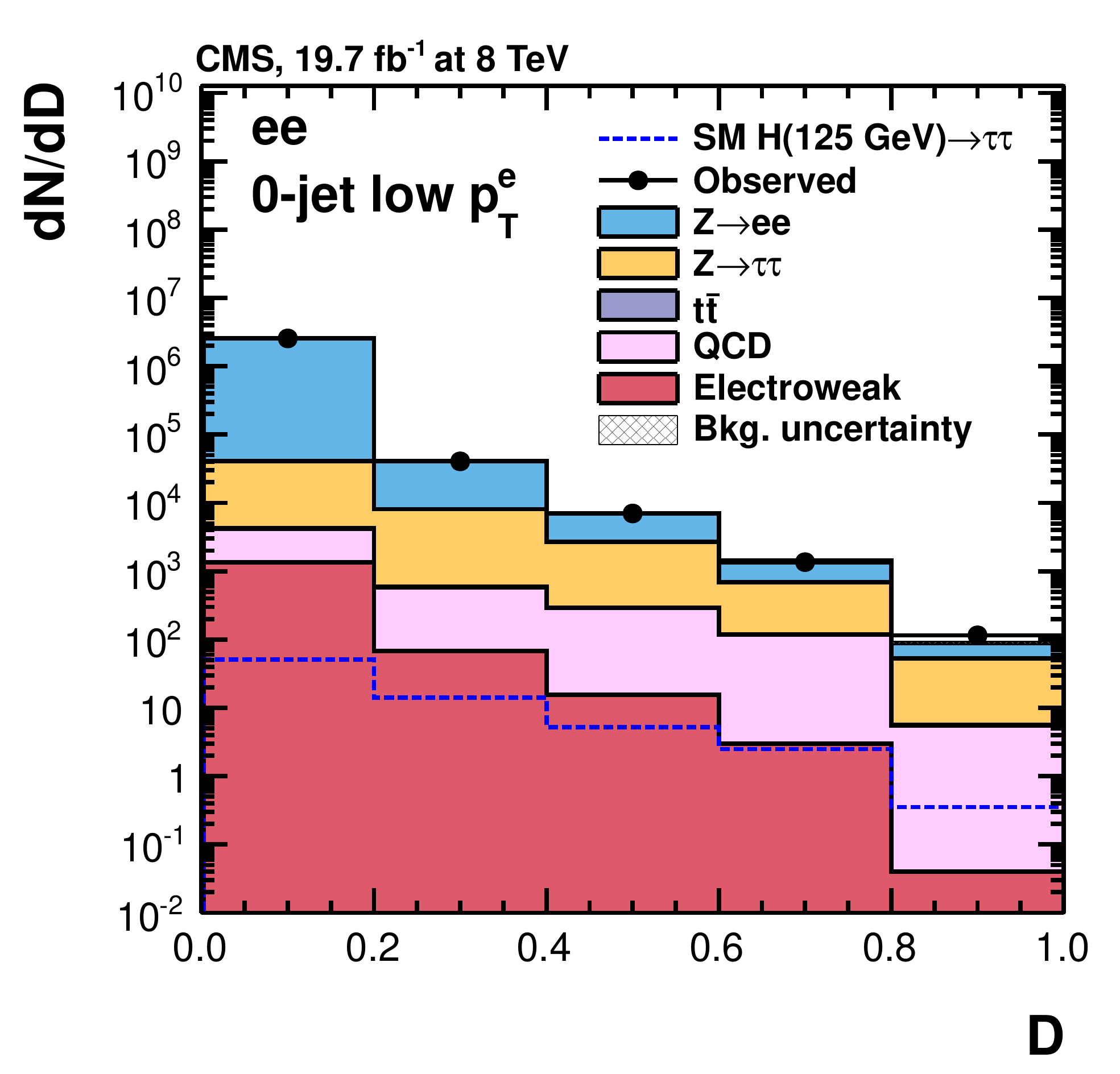}
 \includegraphics[width=0.32\textwidth]{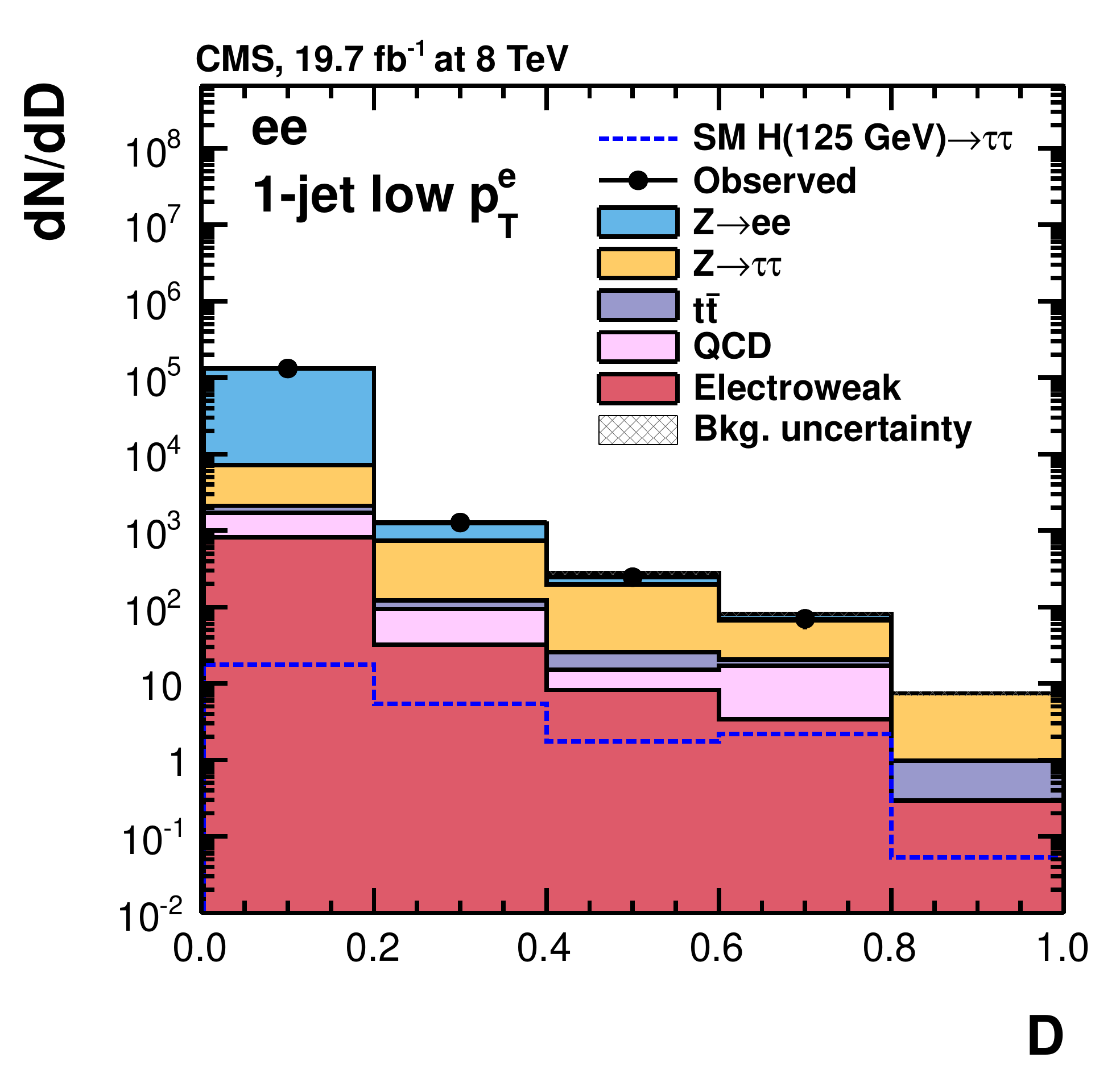}
 \includegraphics[width=0.32\textwidth]{figures/plots_140111/ee/ee_vbf_postfit_8TeV_LOG.pdf}
 \caption{Observed and predicted $D$ distributions in the $\ee$ channel, for all categories used in the 8\TeV data analysis. The normalization of the predicted background distributions corresponds to the result of the global fit. The signal distribution, on the other hand, is normalized to the SM prediction. The open signal histogram is shown superimposed to the background histograms, which are stacked.}
 \label{fig:app_ee_8}
\end{figure}

The $\mtt$ distributions obtained with the 7 and 8\TeV datasets are shown respectively
in figures~\ref{fig:app_mt_7} and~\ref{fig:app_mt_8} for the $\mutau$ channel,
in figures~\ref{fig:app_et_7} and~\ref{fig:app_et_8} for the $\etau$ channel,
and in figures~\ref{fig:app_em_7} and~\ref{fig:app_em_8} for the $\emu$ channel.
The distributions of the final discriminator $D$ obtained with the 7 and 8\TeV datasets are shown respectively
in figures~\ref{fig:app_mm_7} and~\ref{fig:app_mm_8} for the $\mumu$ channel,
and in figures~\ref{fig:app_ee_7} and~\ref{fig:app_ee_8} for the $\ee$ channel.
The figures are arranged according to the category layout of figure~\ref{fig:LL_categories}.

\begin{figure}[bhtp]
\centering
\includegraphics[width=0.4\textwidth]{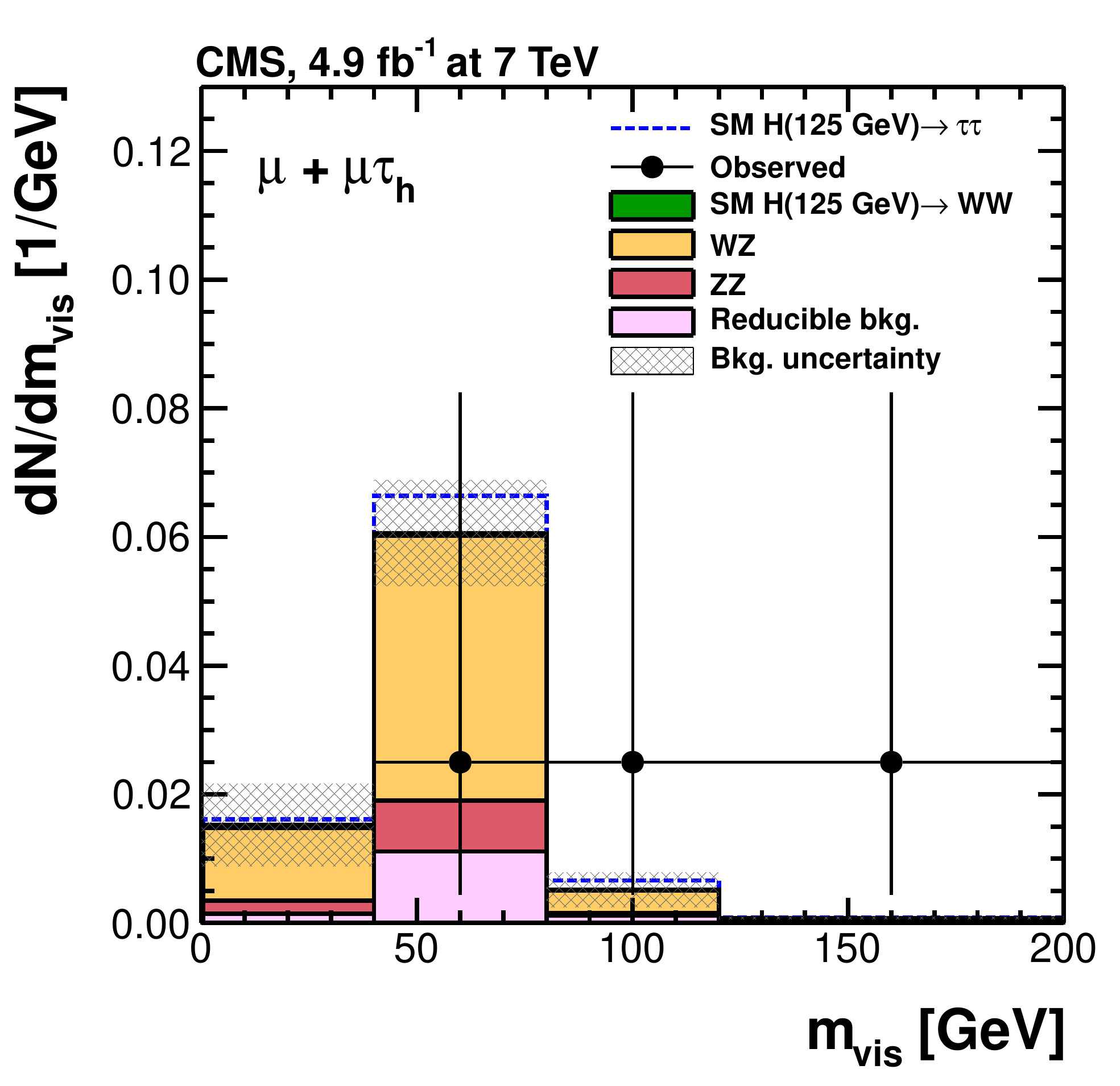}
\includegraphics[width=0.4\textwidth]{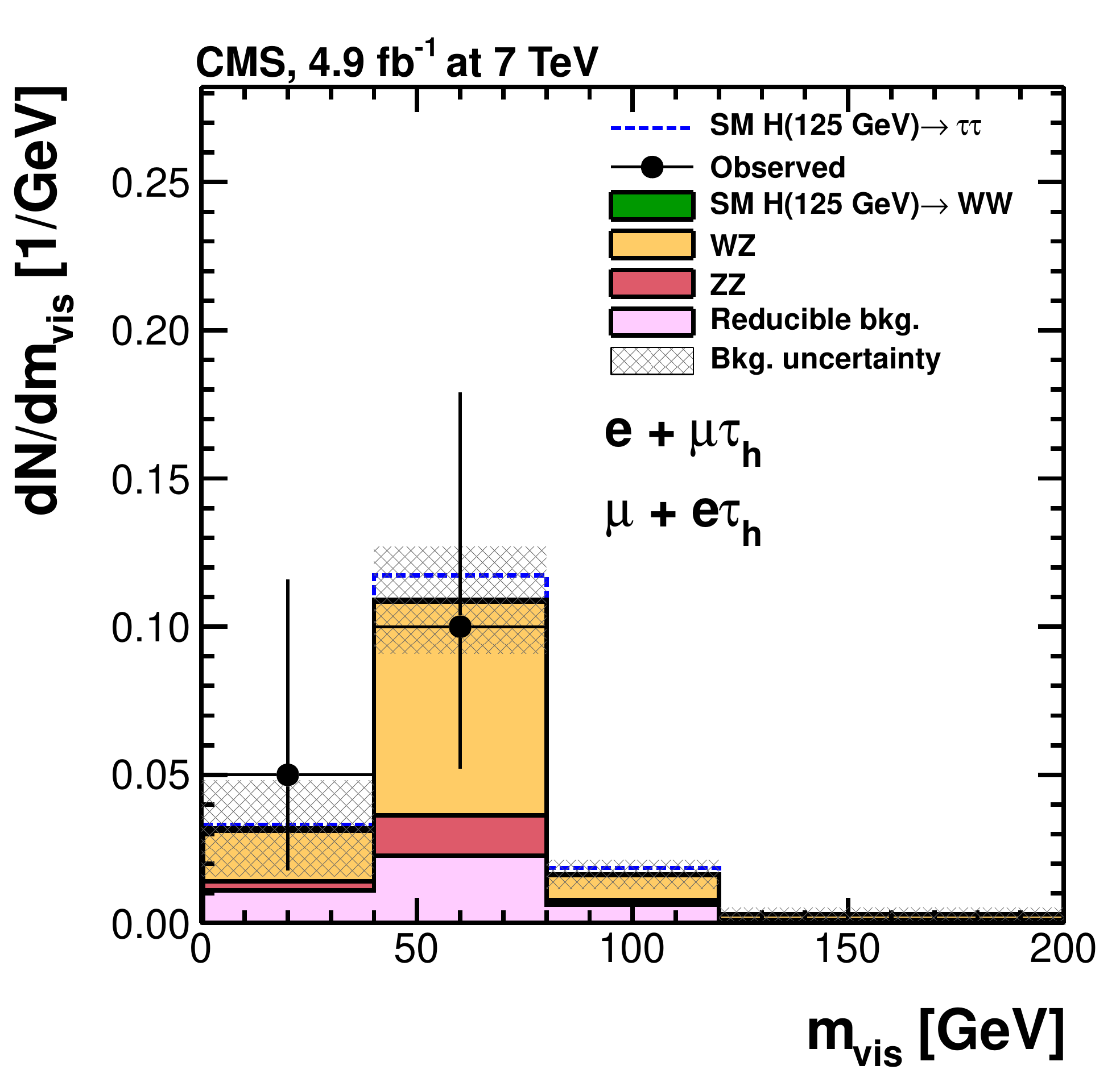} \\ \includegraphics[width=0.4\textwidth]{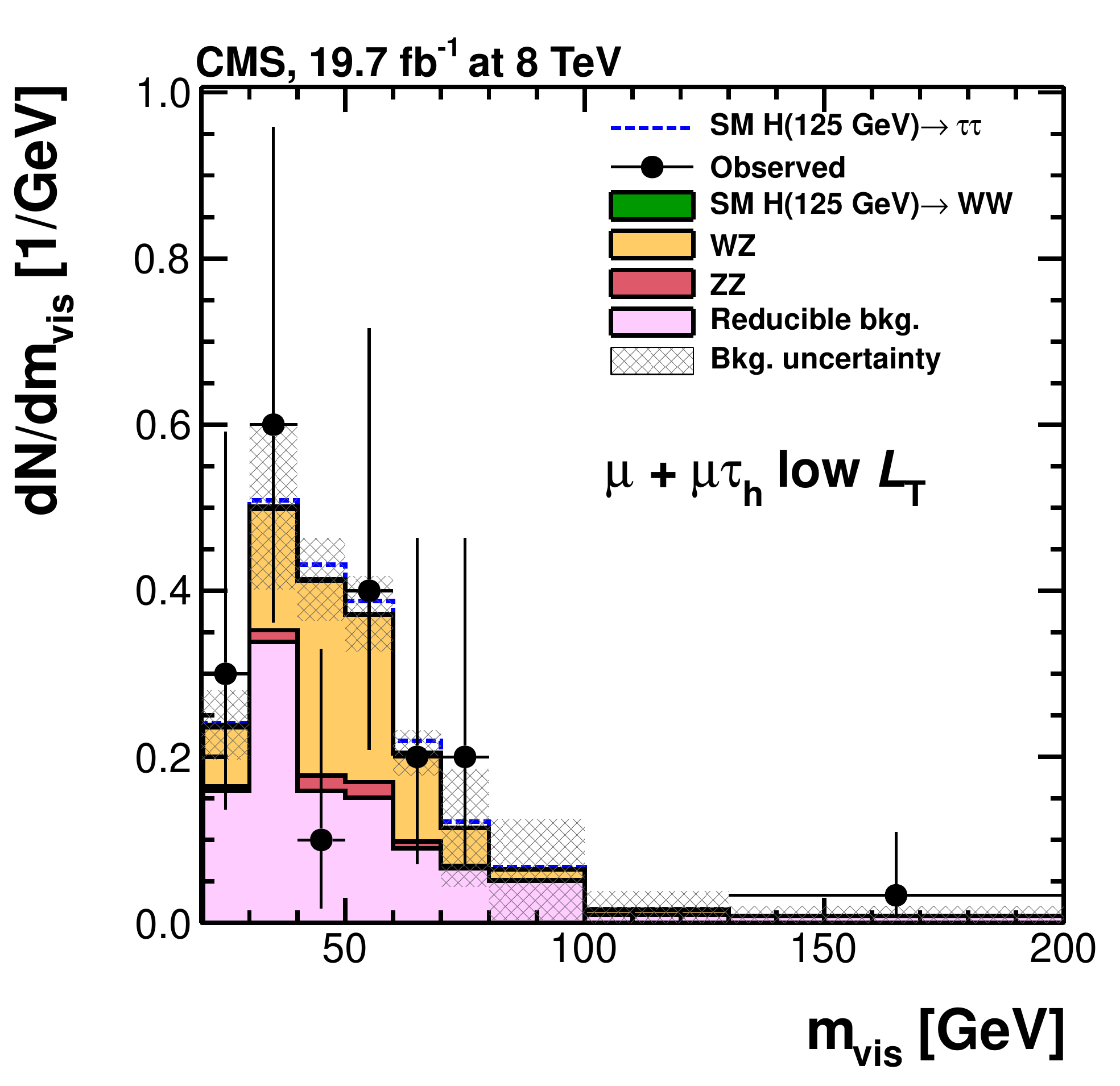} \includegraphics[width=0.4\textwidth]{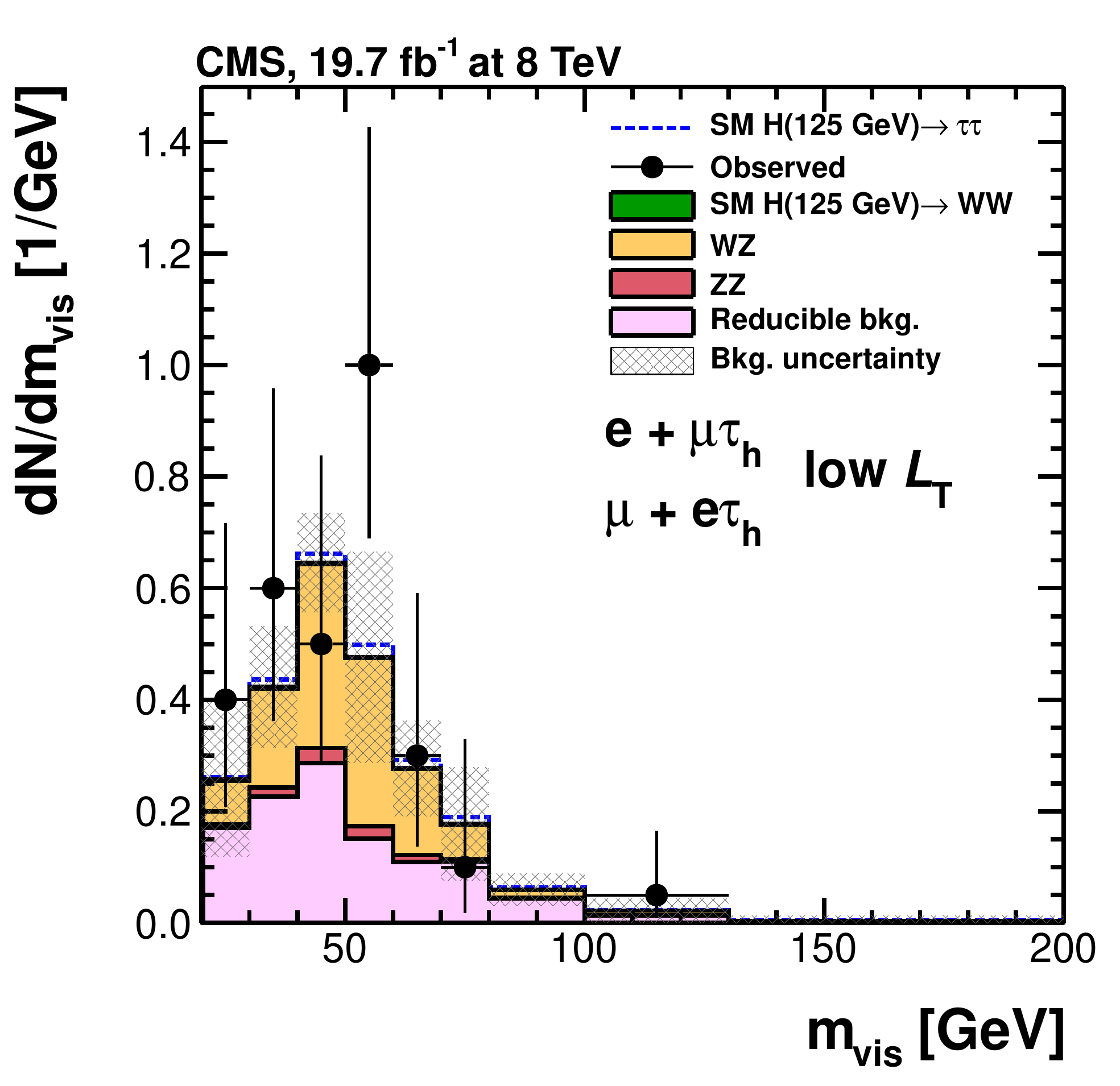} \\ \includegraphics[width=0.4\textwidth]{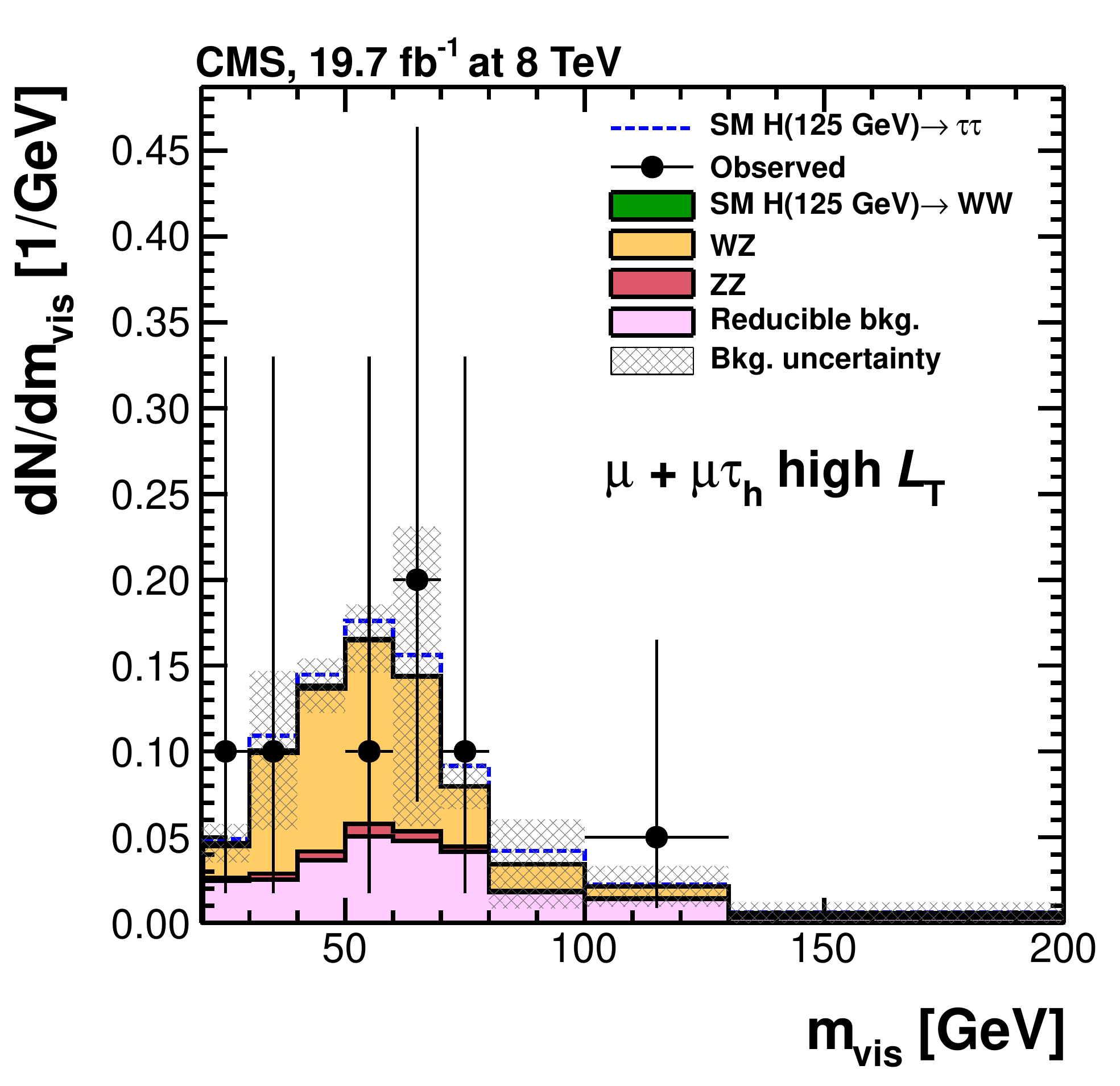} \includegraphics[width=0.4\textwidth]{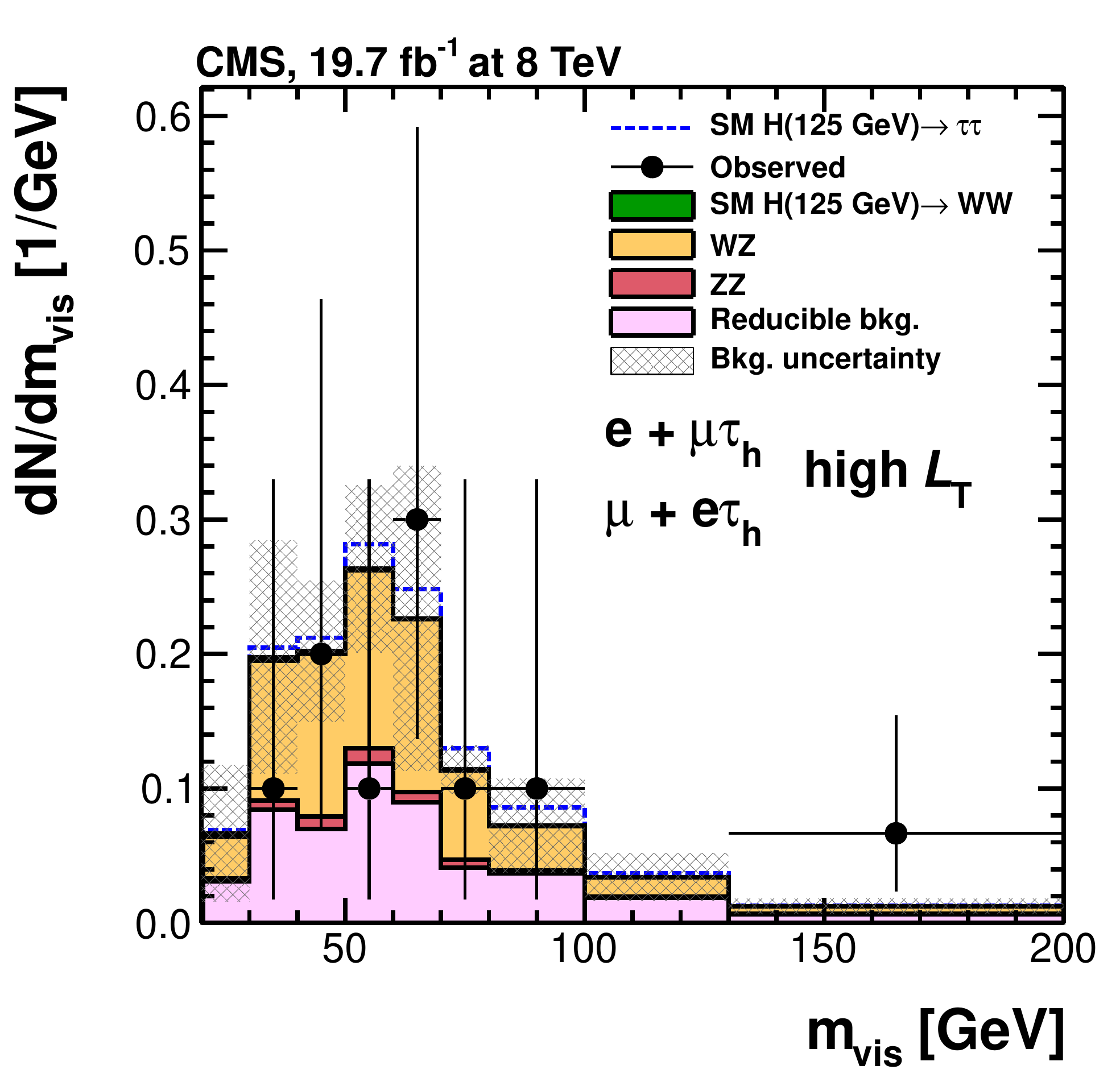}
\caption{Observed and predicted $\mvis$ distributions in the $\Pgm+\Pgm\Pgth$ (left) and $\Pe+\Pgm\Pgth/\Pgm+\Pe\Pgth$ (right) channels for the 7\TeV data analysis (top) and
in the low-$\LT$ (middle) and high-$\LT$ (bottom) categories used in the 8\TeV data analysis. The normalization of the predicted background distributions corresponds to the result of the global fit. The signal distribution, on the other hand, is normalized to the SM prediction. The signal and background histograms are stacked.}
\label{fig:app_wh_llt}
\end{figure}

\begin{figure}[bhtp]
\centering
\includegraphics[width=0.4\textwidth]{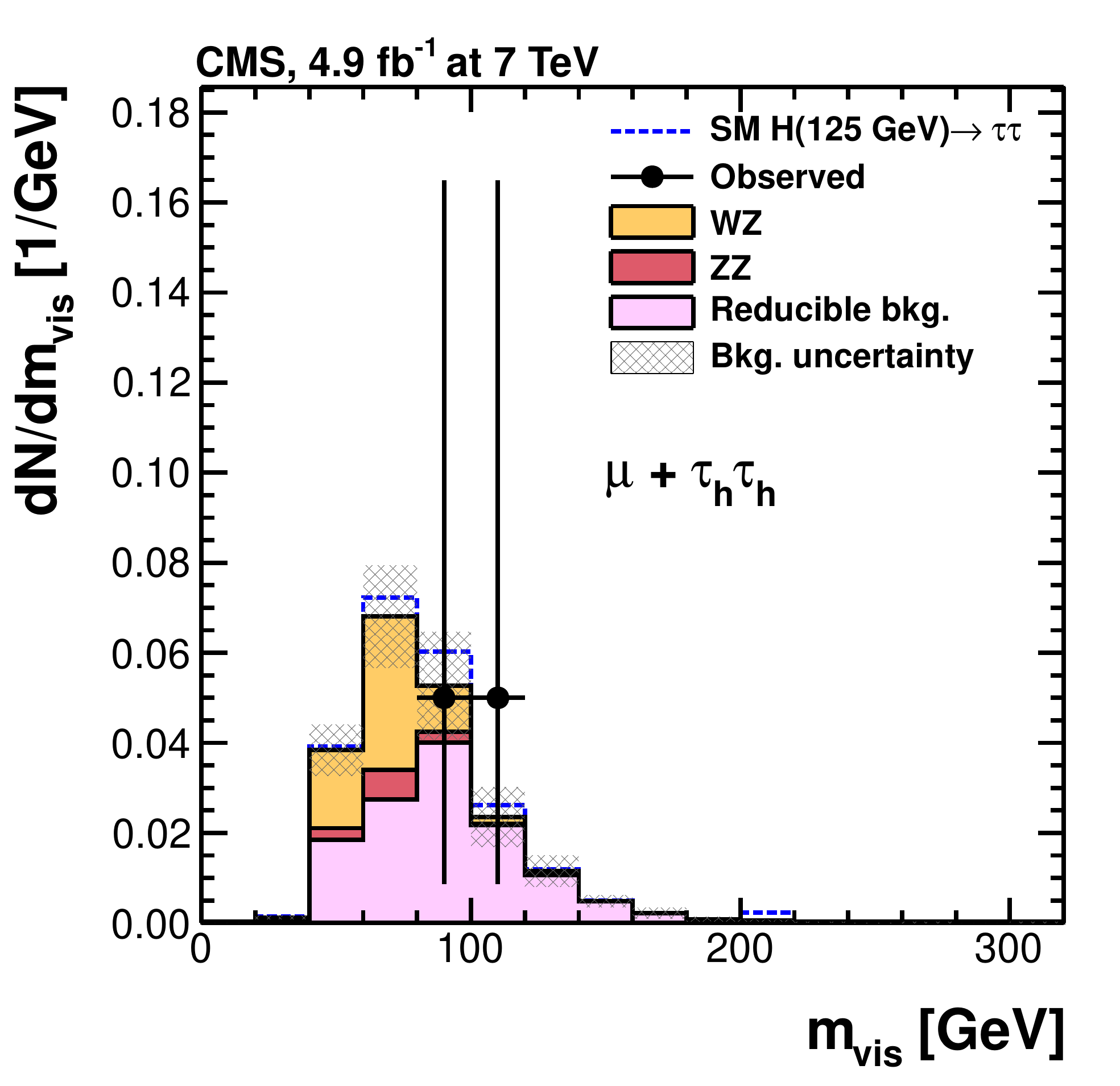}
\includegraphics[width=0.4\textwidth]{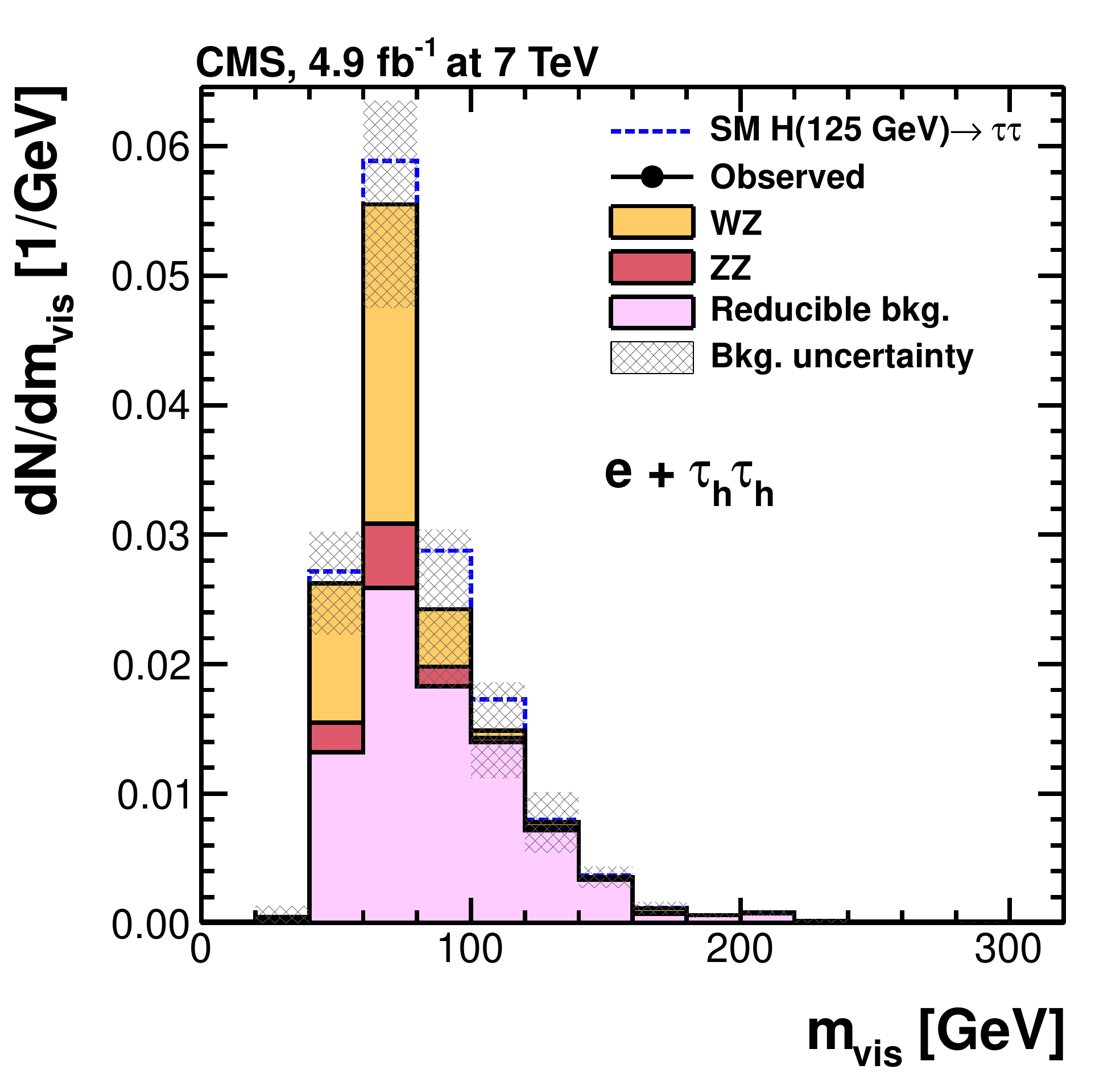} \\
\includegraphics[width=0.4\textwidth]{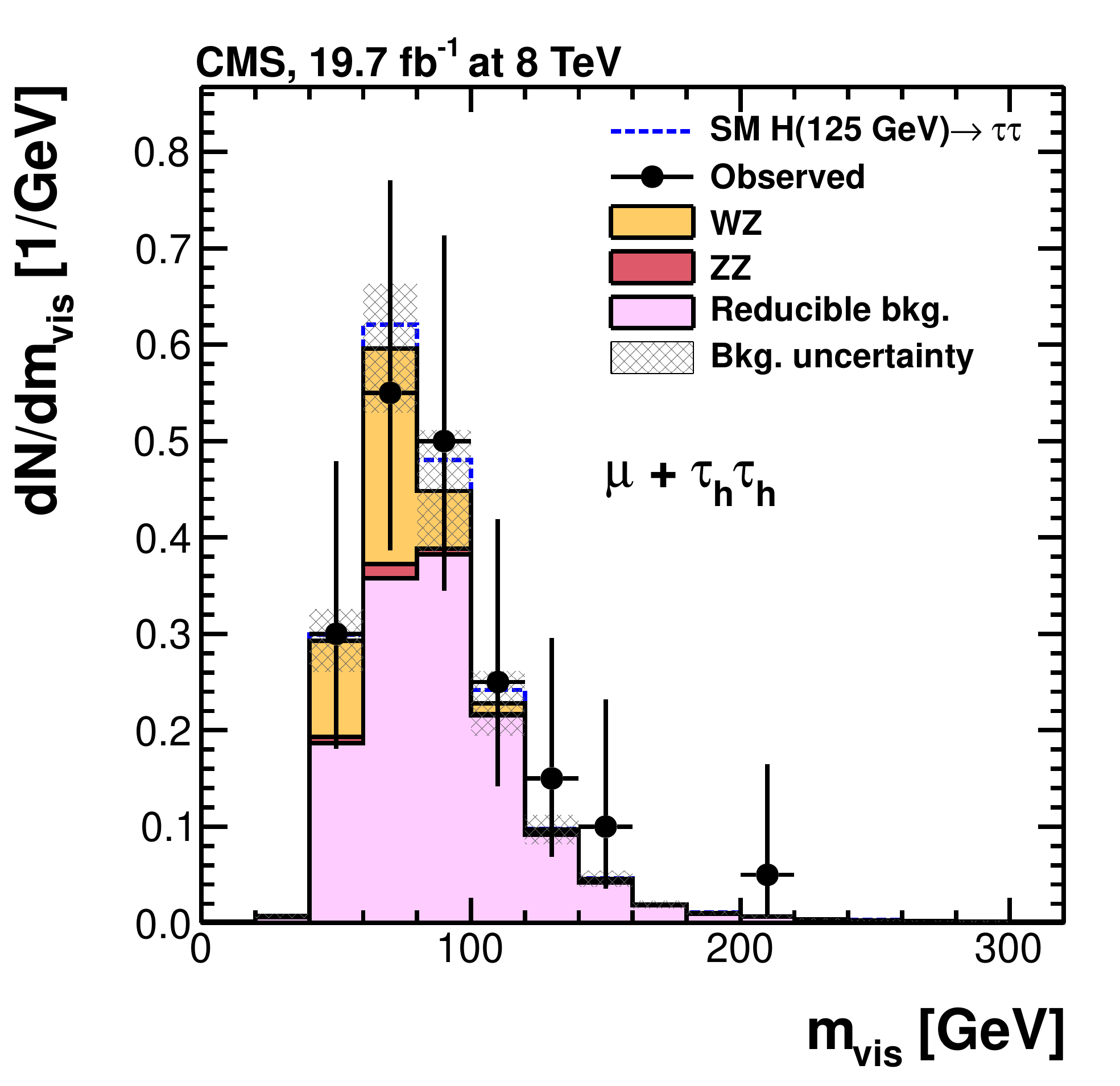}
\includegraphics[width=0.4\textwidth]{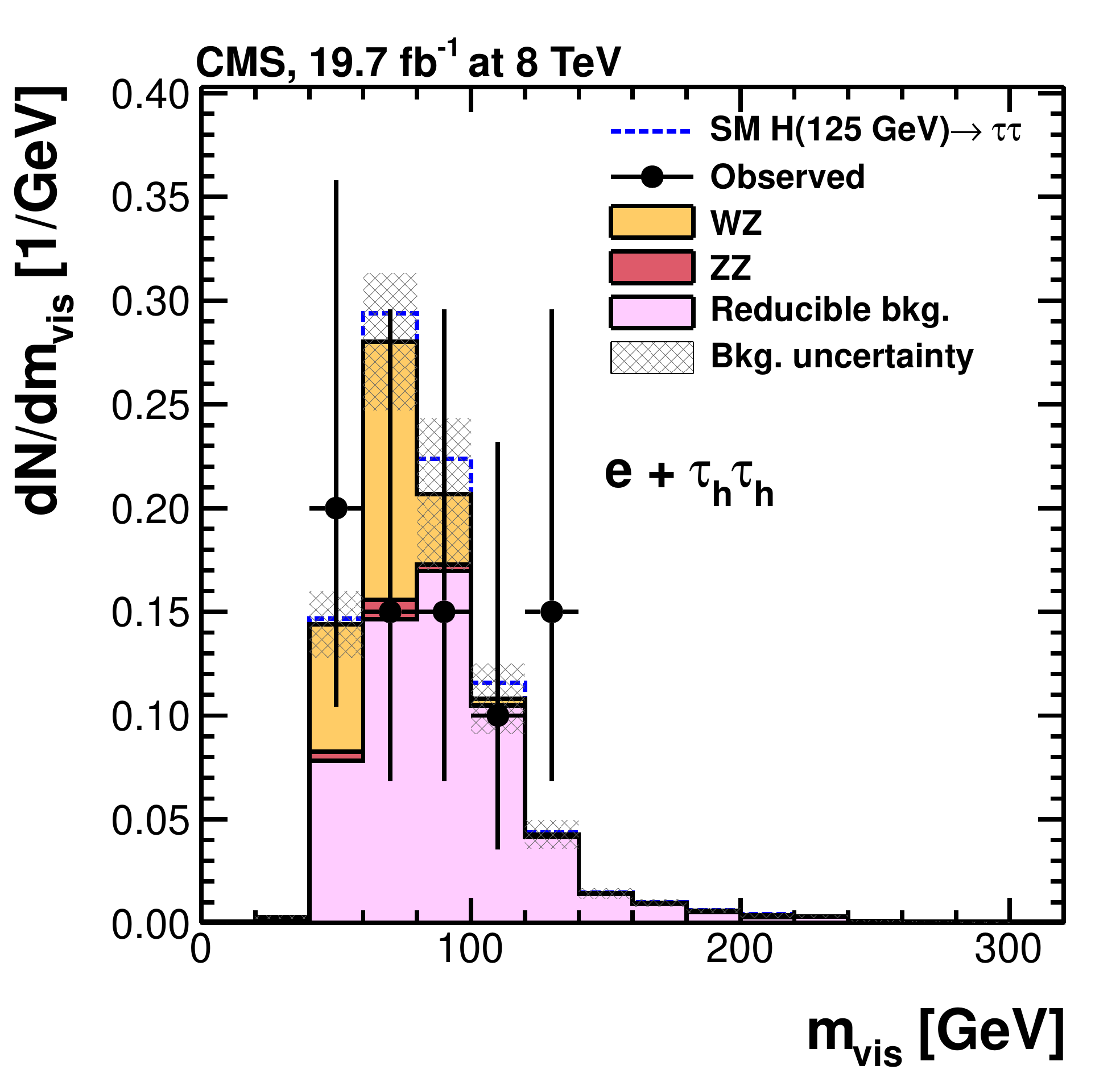}
\caption{Observed and predicted $\mvis$ distributions in the $\Pgm+\Pgth\Pgth$ (left) and $\Pe+\Pgth\Pgth$ (right) channels for the 7\TeV data analysis (top) and
the 8\TeV data analysis (bottom). The normalization of the predicted background distributions corresponds to the result of the global fit. The signal distribution, on the other hand, is normalized to the SM prediction. The signal and background histograms are stacked.}
\label{fig:app_wh_ltt}
\end{figure}

\begin{figure}[bhtp]
\centering
\includegraphics[width=0.34\textwidth]{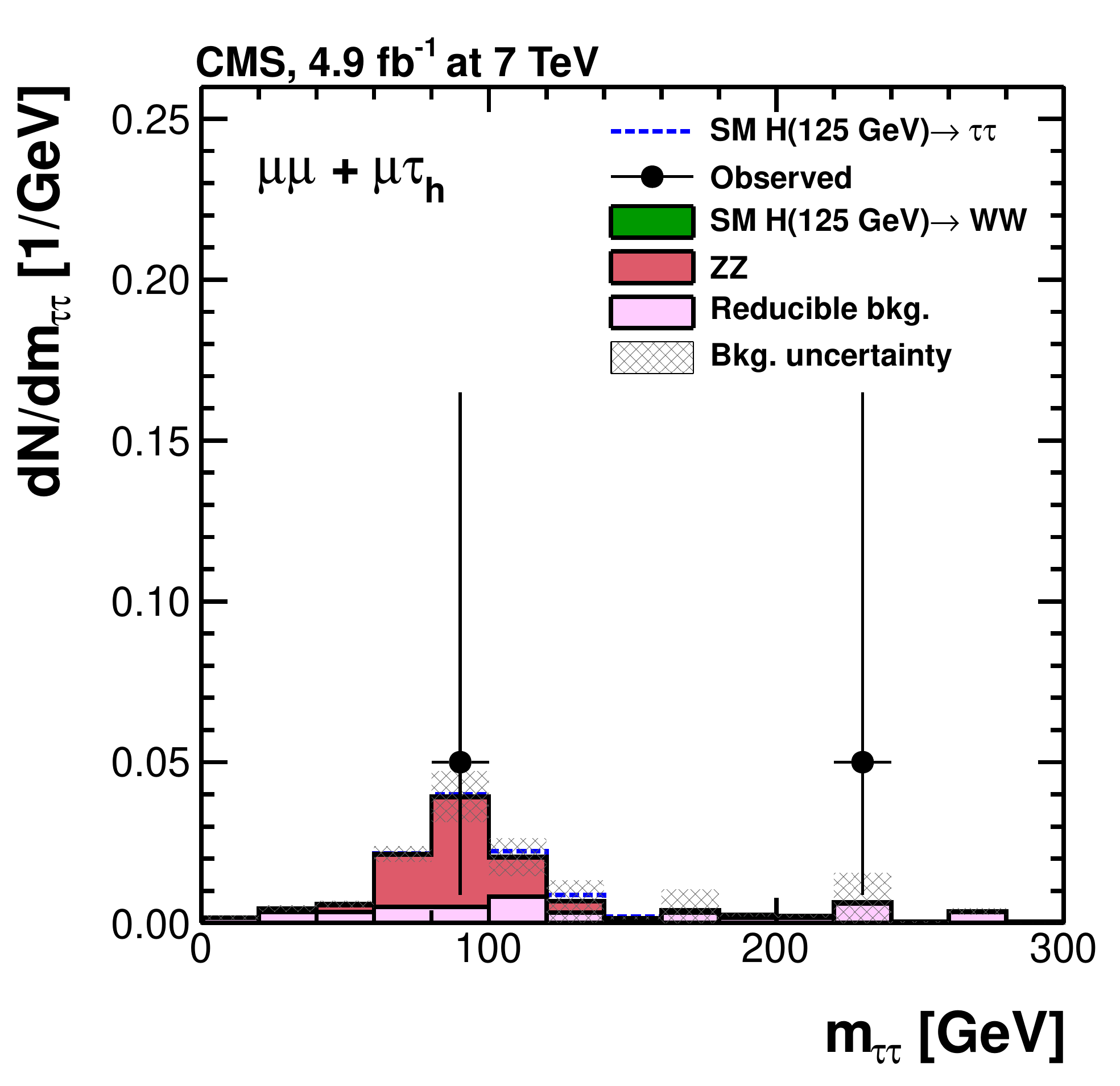}
\includegraphics[width=0.34\textwidth]{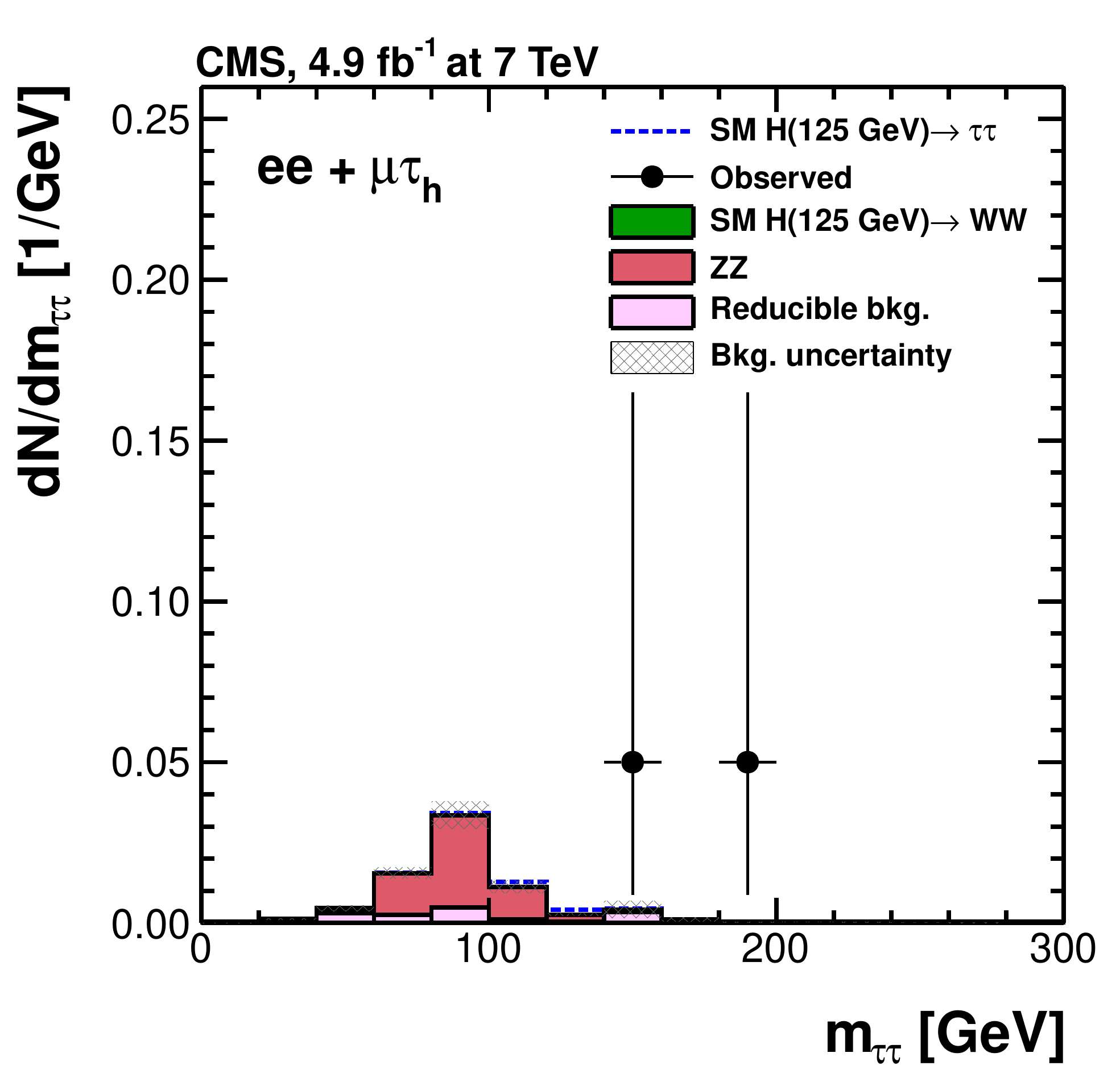} \\
\includegraphics[width=0.34\textwidth]{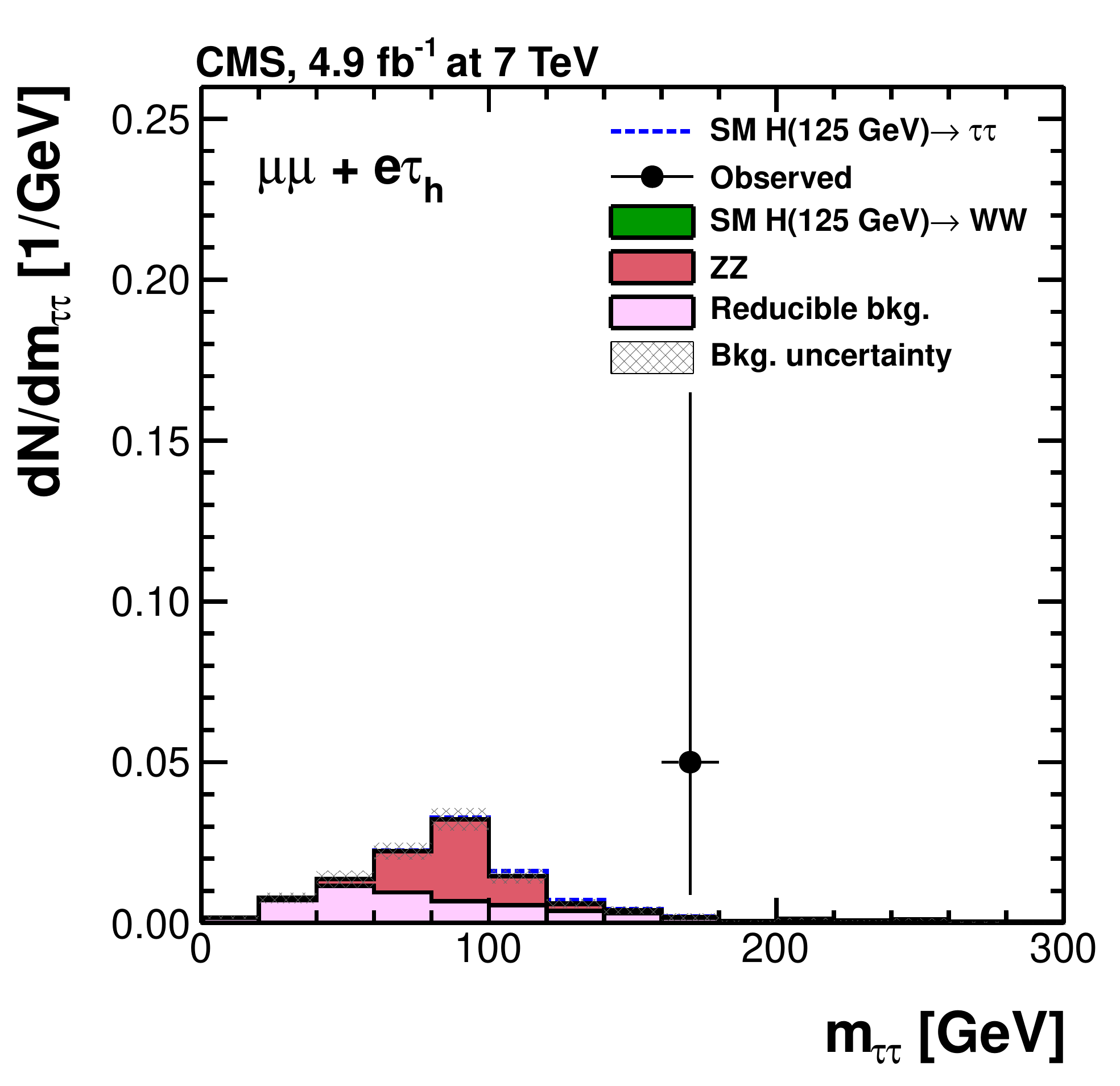}
\includegraphics[width=0.34\textwidth]{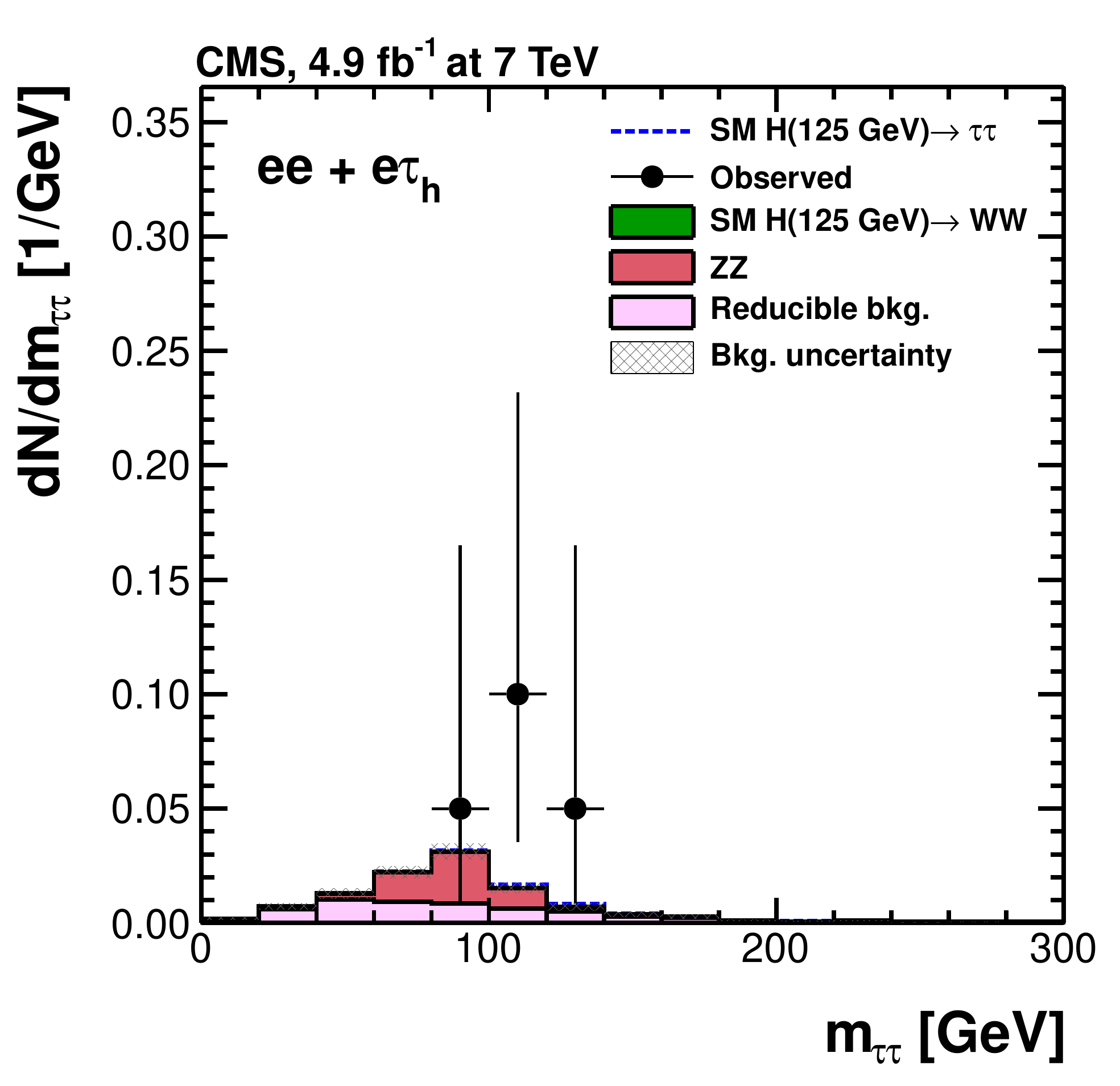} \\
\includegraphics[width=0.34\textwidth]{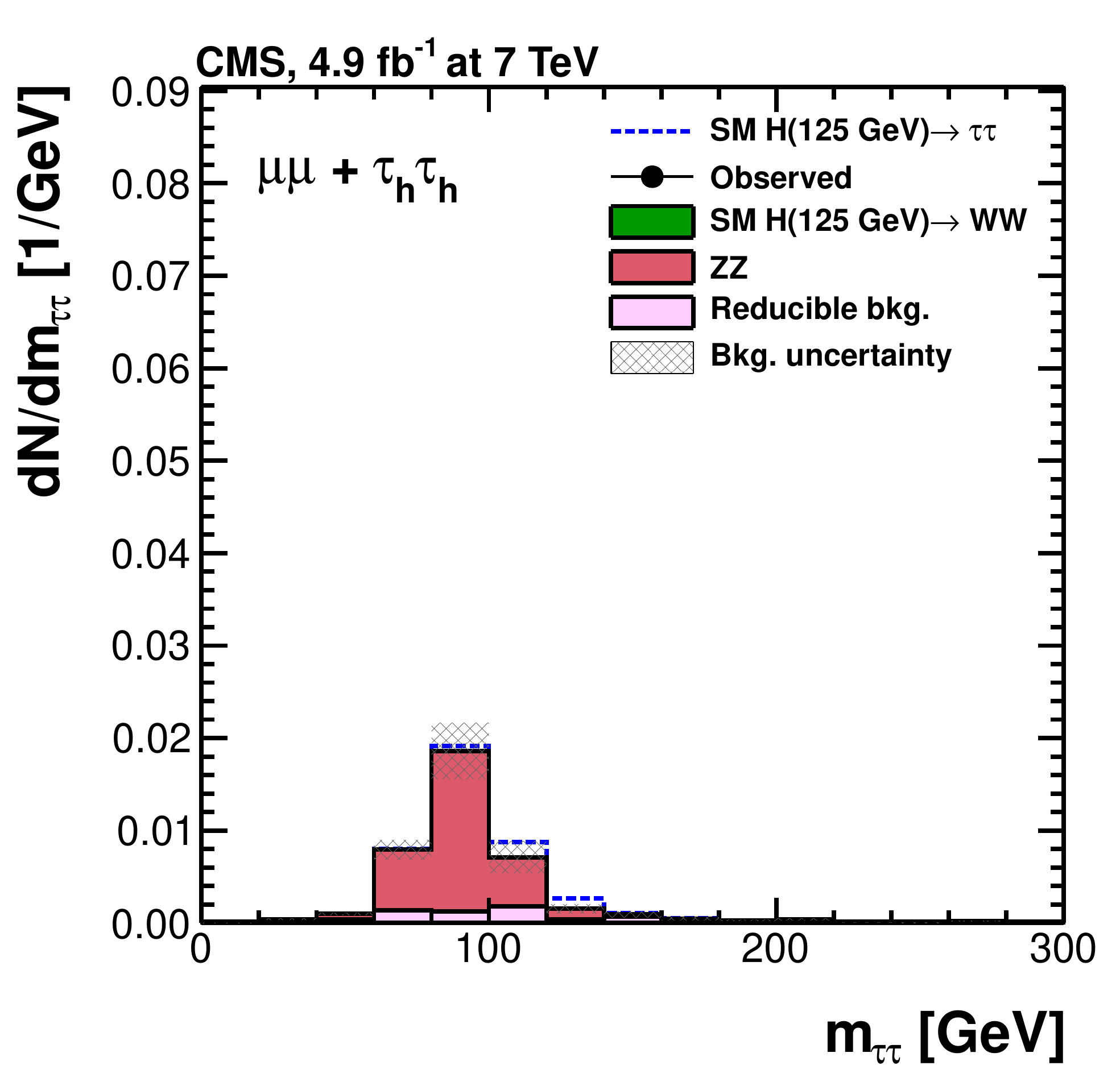}
\includegraphics[width=0.34\textwidth]{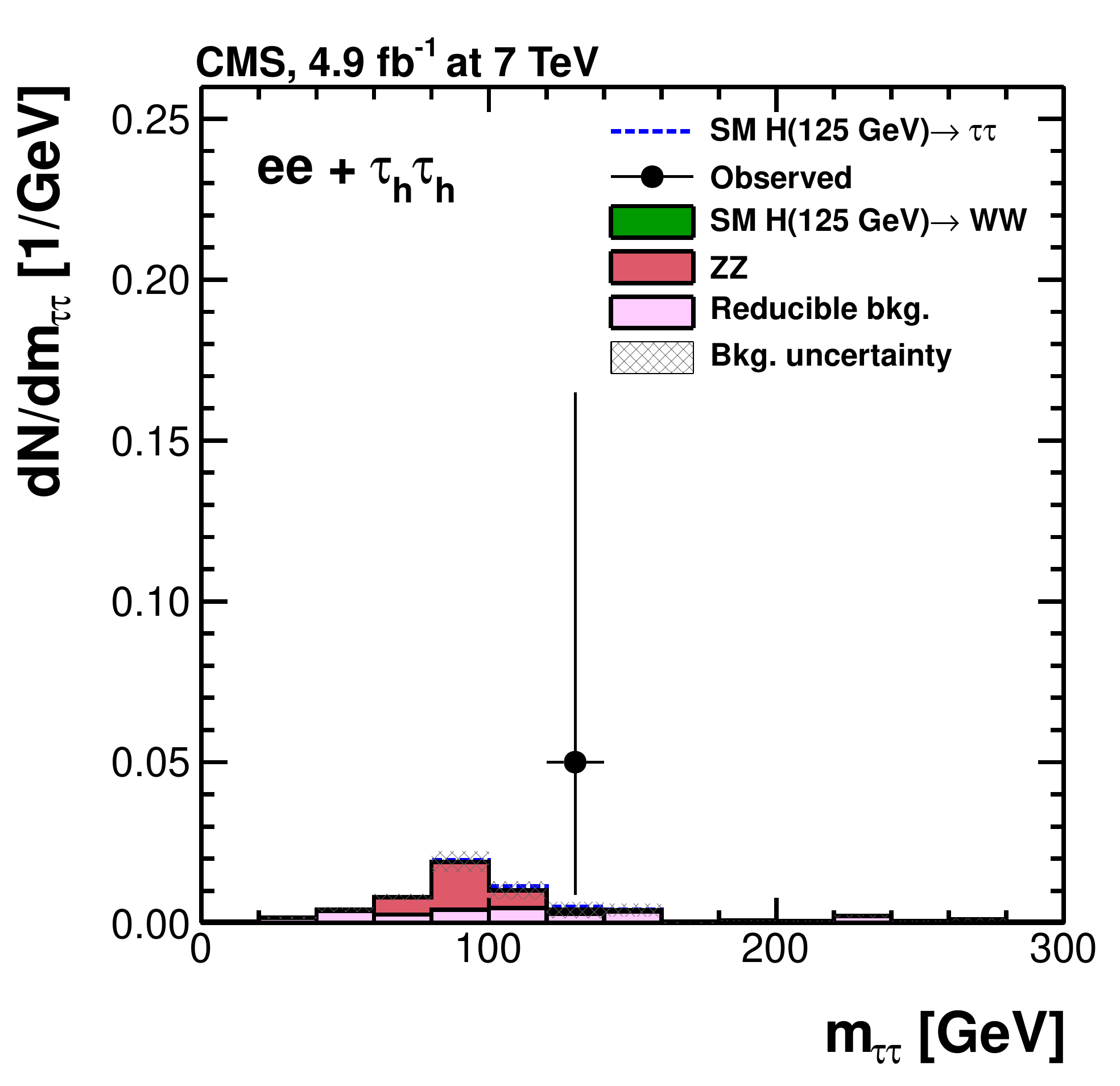} \\
\includegraphics[width=0.34\textwidth]{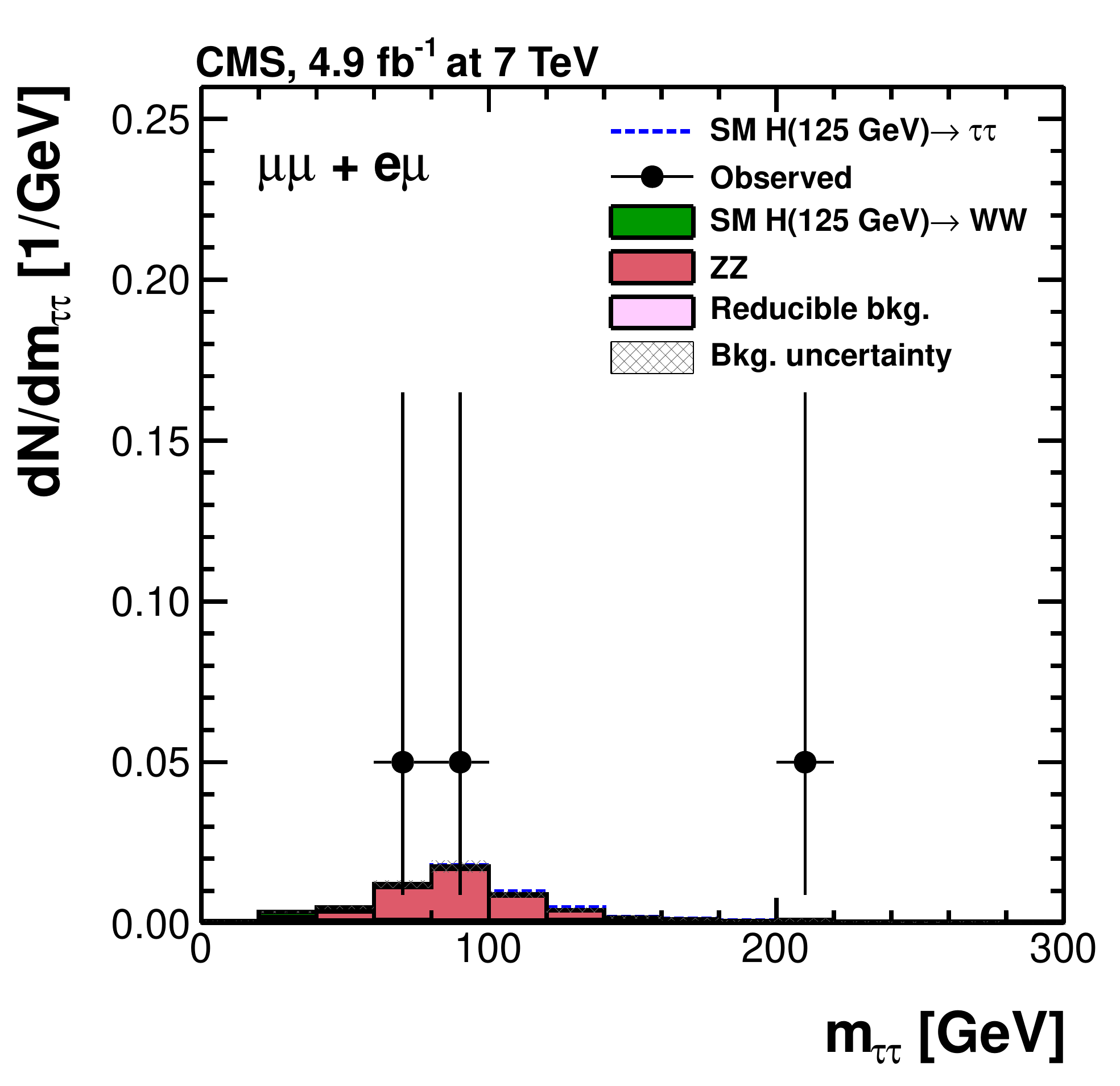}
\includegraphics[width=0.34\textwidth]{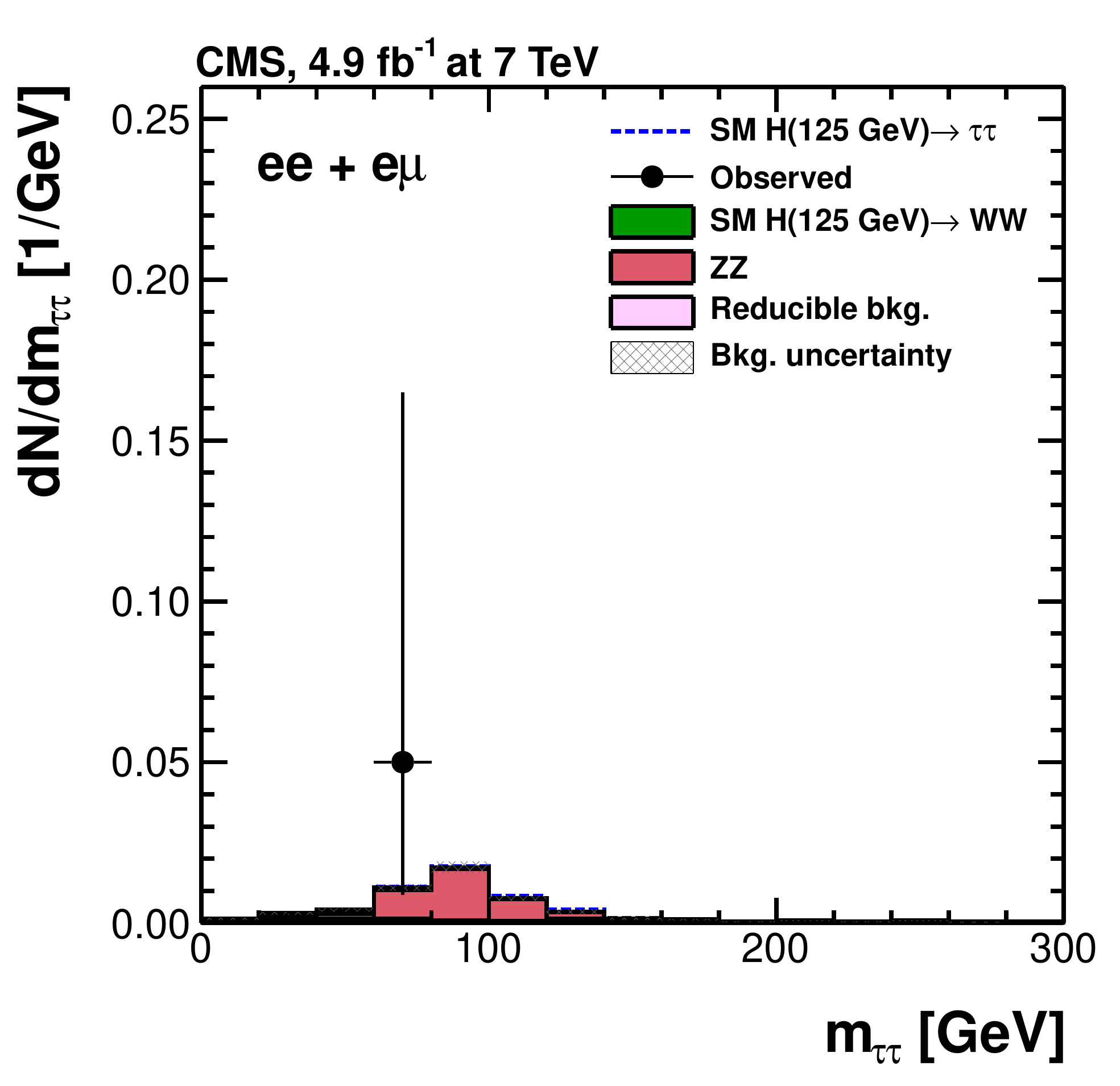}
\\
\caption{Observed and predicted $\mtt$ distributions for the four different $LL'$ final states of the $\Pe\Pe + LL'$ (left) and $\Pgm\Pgm + LL'$ (right) channels for the 7\TeV data analysis. The normalization of the predicted background distributions corresponds to the result of the global fit. The signal distribution, on the other hand, is normalized to the SM prediction. The signal and background histograms are stacked.}
\label{fig:app_zh_7}
\end{figure}

\begin{figure}[bhtp]
\centering
\includegraphics[width=0.34\textwidth]{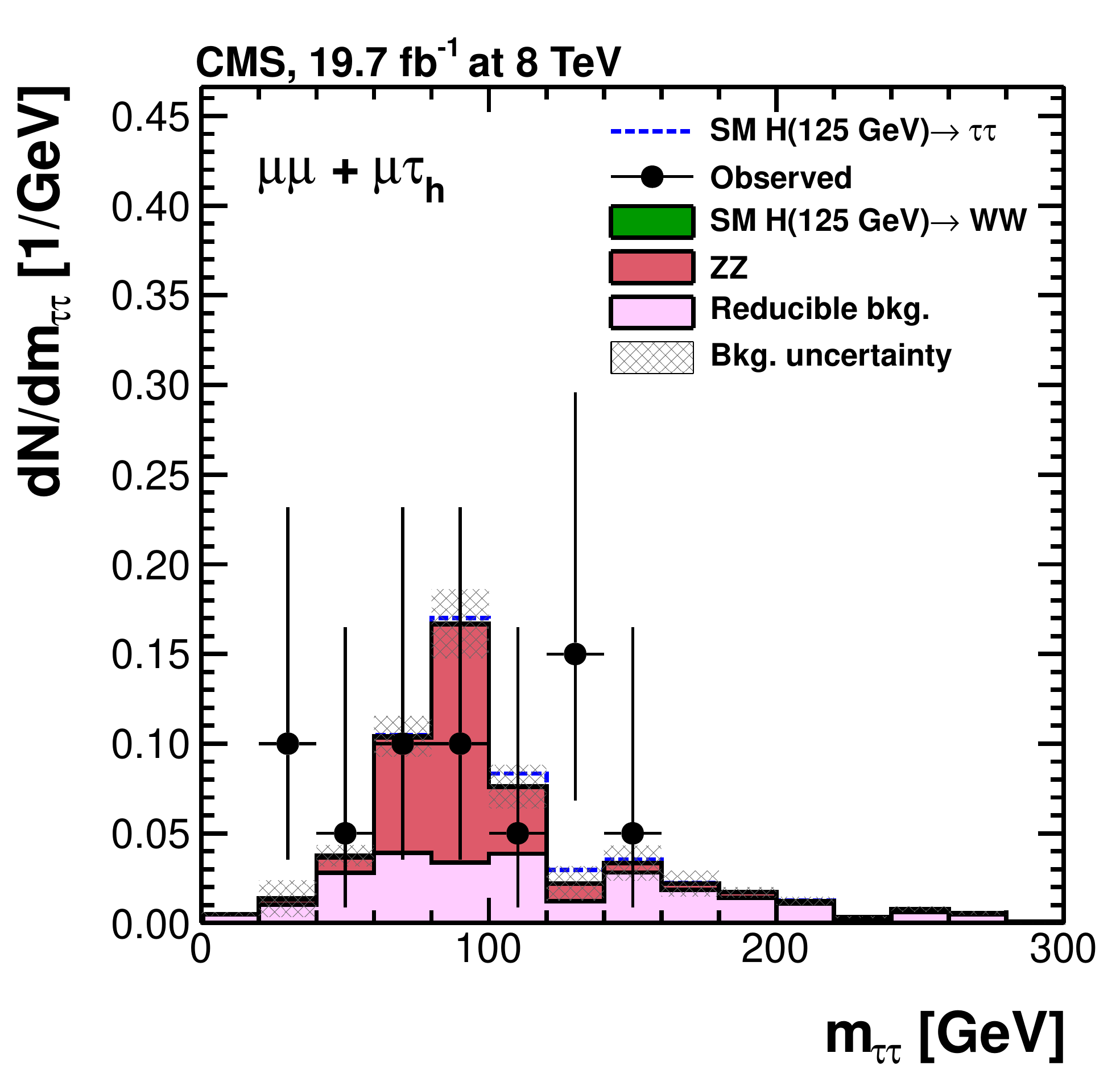}
\includegraphics[width=0.34\textwidth]{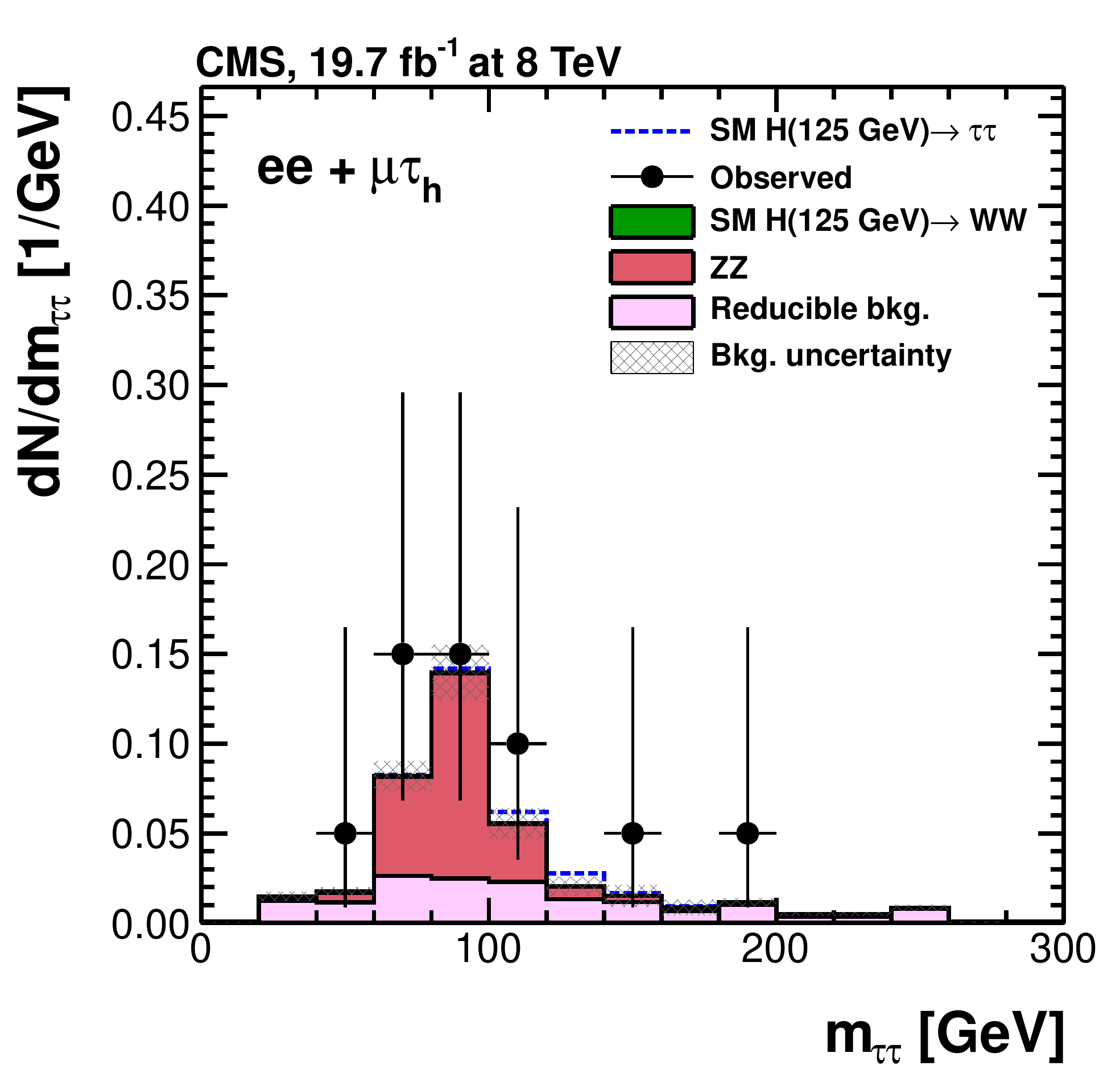} \\
\includegraphics[width=0.34\textwidth]{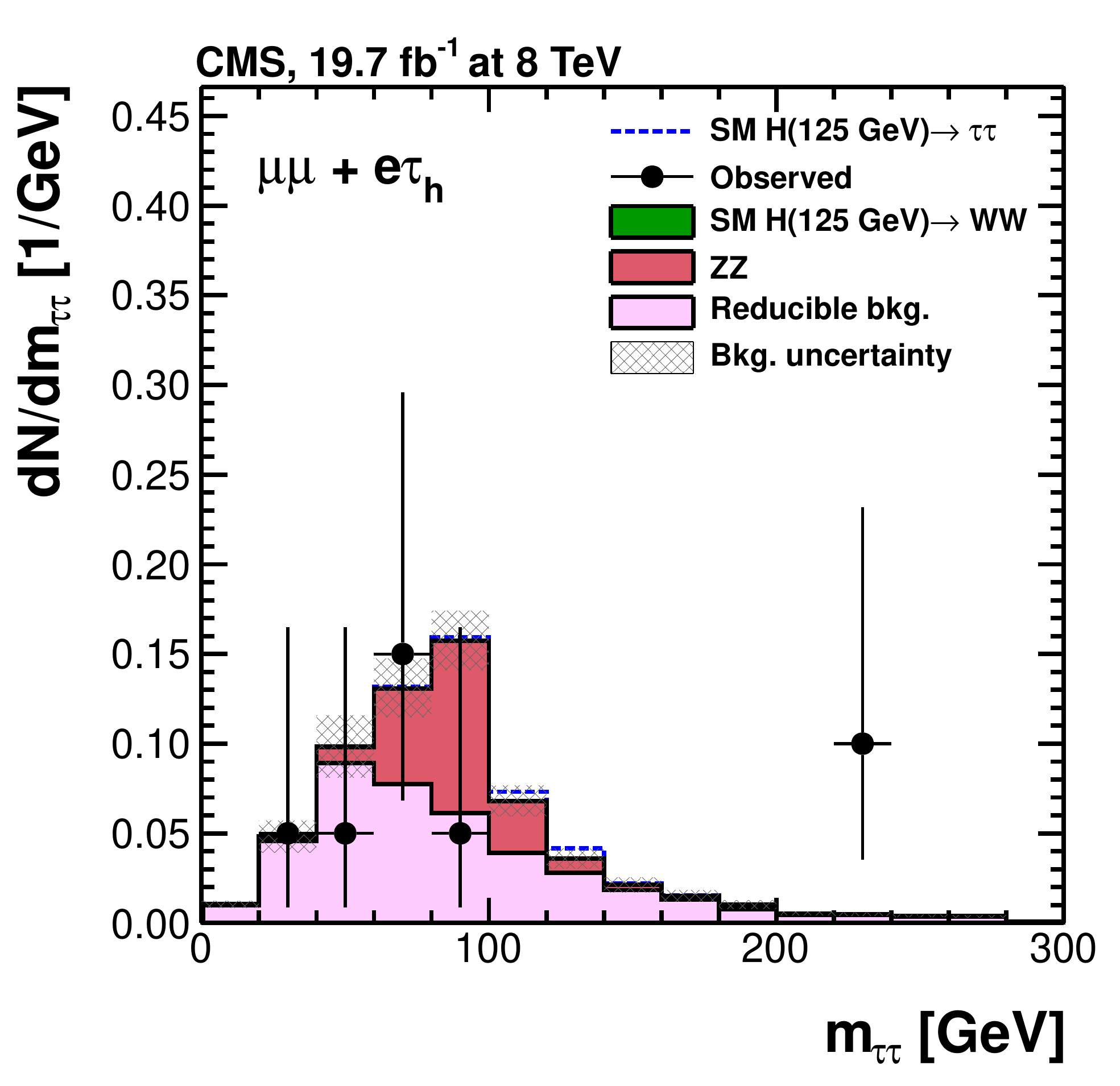}
\includegraphics[width=0.34\textwidth]{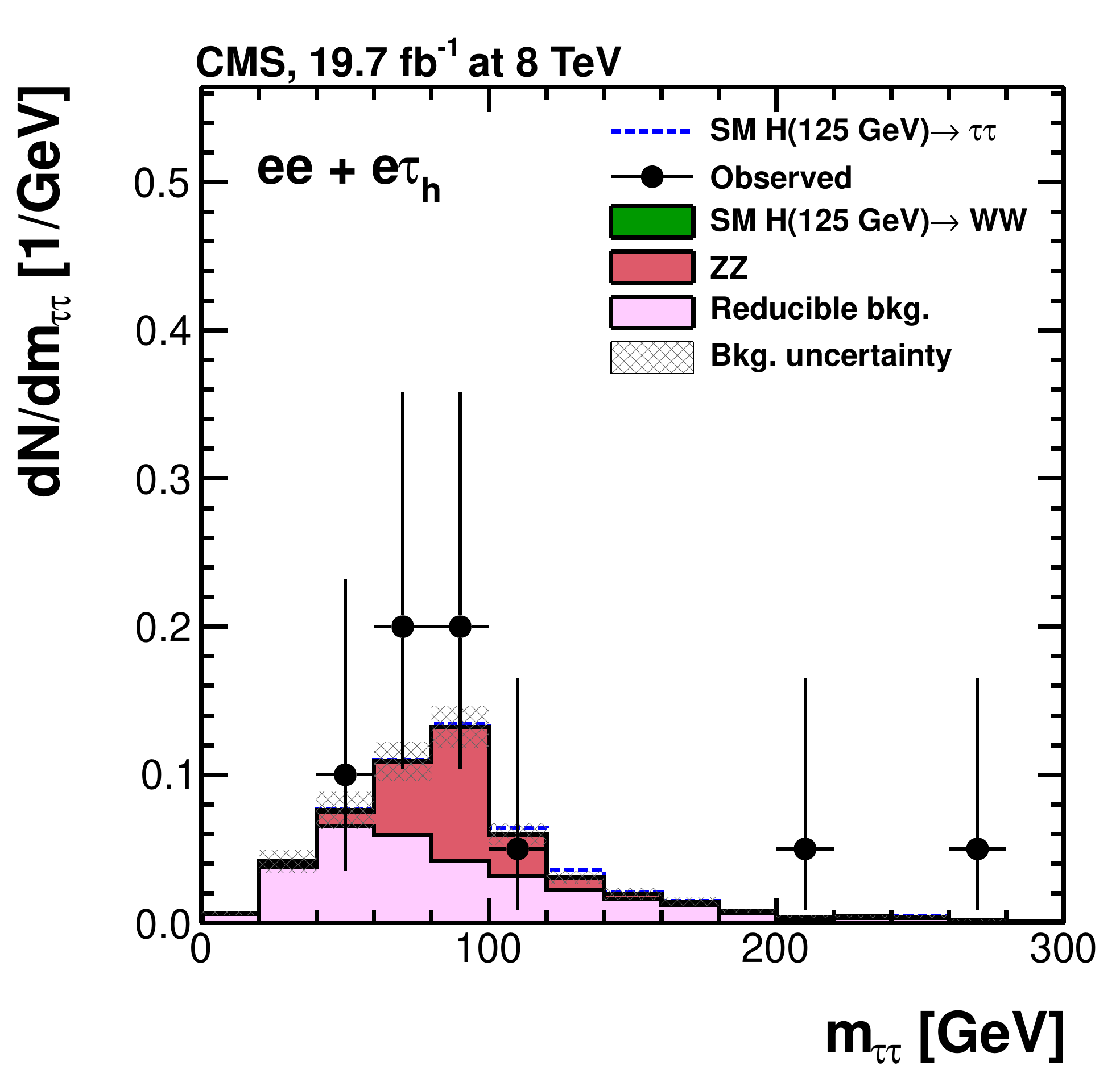} \\
\includegraphics[width=0.34\textwidth]{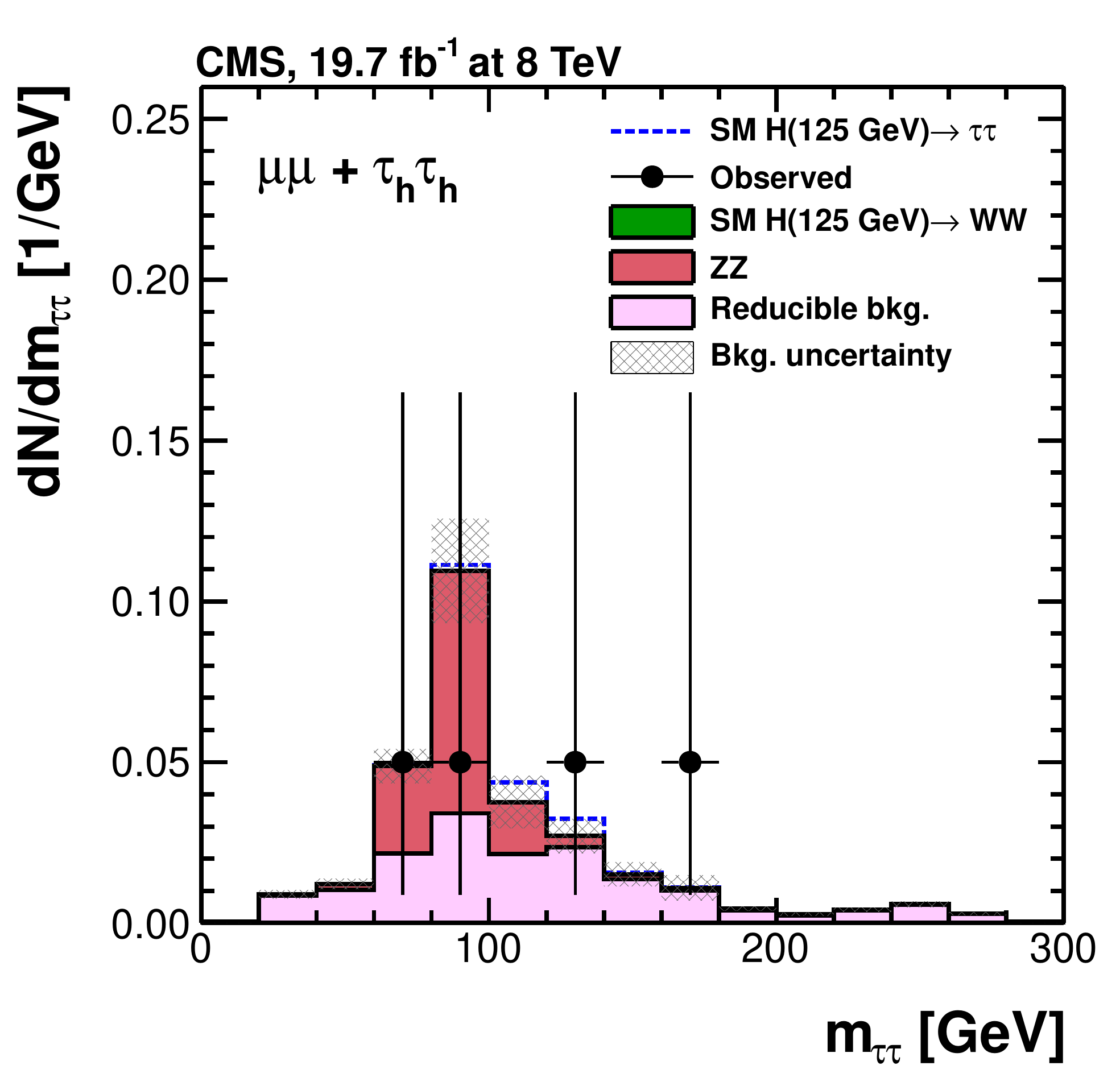}
\includegraphics[width=0.34\textwidth]{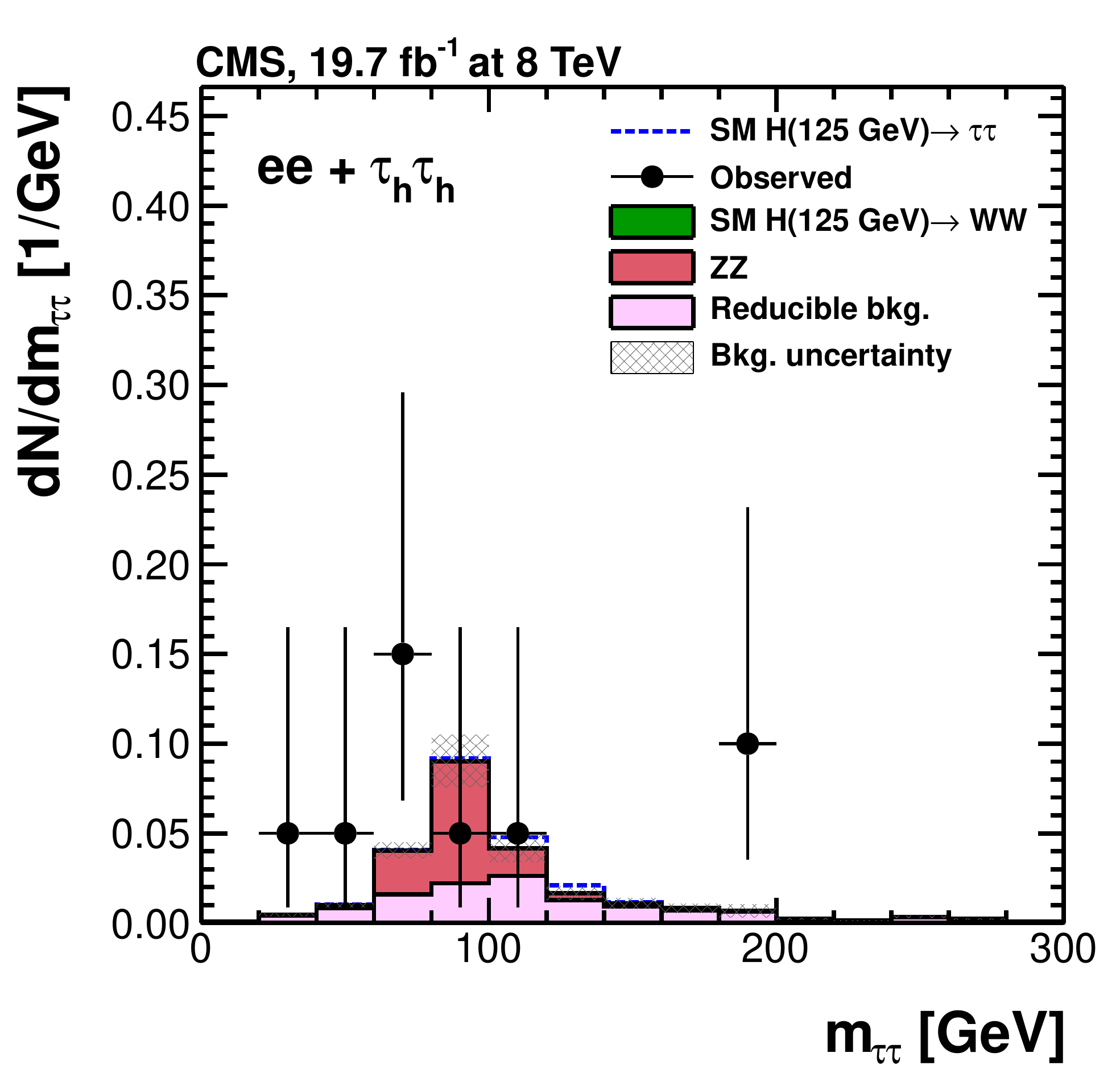} \\
\includegraphics[width=0.34\textwidth]{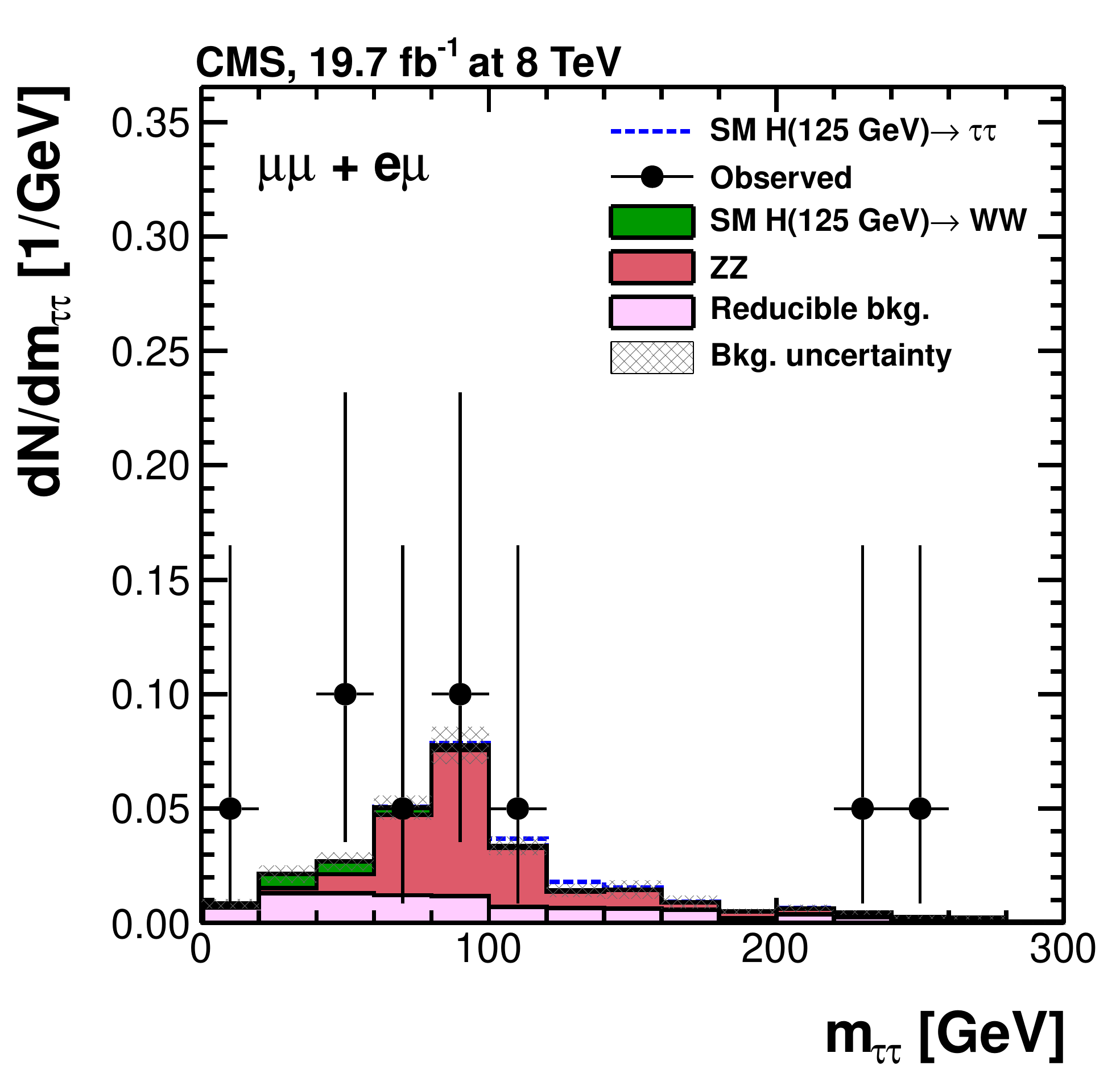}
\includegraphics[width=0.34\textwidth]{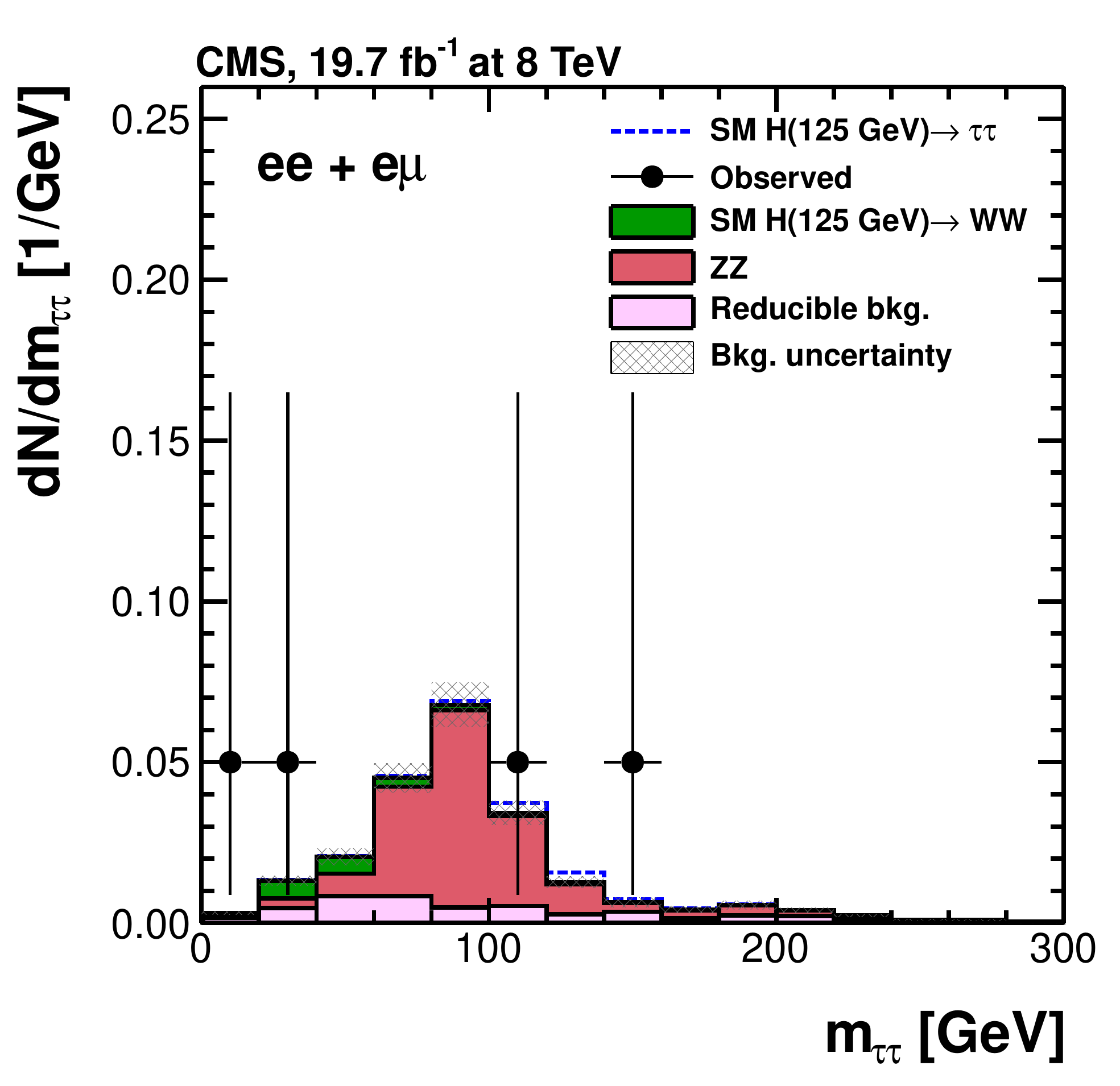}
\caption{Observed and predicted $\mtt$ distributions for the four different $LL'$ final states of the $\Pe\Pe + LL'$ (left) and $\Pgm\Pgm + LL'$ (right) channels for the 8\TeV data analysis. The normalization of the predicted background distributions corresponds to the result of the global fit. The signal distribution, on the other hand, is normalized to the SM prediction. The signal and background histograms are stacked.}
\label{fig:app_zh_8}
\end{figure}
\clearpage
\section{Event yields}
\label{sec:event_yields}
\begin{table}[htbp]
\topcaption{Observed and predicted event yields in all event categories of the $\mu\Pgth$, $\Pe\Pgth$, $\Pgth\Pgth$, and $\Pe\mu$ channels in the full \mtt mass range. The event yields of the predicted background distributions correspond to the result of the global fit. The signal yields, on the other hand, are normalized to the standard model prediction. The different signal processes are labelled as ggH (gluon-gluon fusion), VH (production in association with a $\PW$ or $\PZ$~boson), and VBF (vector-boson fusion). The $\frac{S}{S+B}$ variable denotes the ratio of the signal and the signal-plus-background yields in the central $\mtt$ range containing $68\%$ of the signal events for $\mH = 125$\GeV. The RMS variable denotes the standard deviation of the $\mtt$ distribution for corresponding signal events.
\label{tab:event_yields}
}
\small
\centering
\begin{tabular}{l|rrrr|r|r|lc}
 & \multicolumn{4}{c|}{ SM Higgs ($\mH = 125$\GeV) } & & & & \multicolumn{1}{c}{RMS} \\
Event category & ggH & VBF & VH & $\Sigma$ signal & \multicolumn{1}{c|}{Background} & \multicolumn{1}{c|}{Data} & $\frac{S}{S+B}$ & (\GeV)  \\
\hline
$\mu\Pgth$ & & & & & & & & \\

0-jet low-$\pt^{\Pgth}$ 7\TeV & $ 23.1$ & $  0.2$ & $  0.1$ & $    23.5 \pm     3.4 \phantom{0}$  & $   11950 \pm     590 \phantom{0} $ & $  11959$ & $0.002$ &  17.4  \\

0-jet low-$\pt^{\Pgth}$ 8\TeV & $ 83.0$ & $  0.8$ & $  0.4$ & $    85.0 \pm    11.0 \phantom{}$  & $   40800 \pm    1900 \phantom{} $ & $  40353$ & $0.003$ &  16.3  \\

0-jet high-$\pt^{\Pgth}$ 7\TeV & $ 17.5$ & $  0.2$ & $  0.2$ & $    17.9 \pm     2.6 \phantom{0}$  & $    1595 \pm      83 \phantom{00} $ & $   1594$ & $0.022$ &  15.1  \\

0-jet high-$\pt^{\Pgth}$ 8\TeV & $ 66.2$ & $  0.7$ & $  0.6$ & $    67.5 \pm     9.3 \phantom{0}$  & $    5990 \pm     250 \phantom{0} $ & $   5789$ & $0.020$ &  15.2  \\

1-jet low-$\pt^{\Pgth}$ 7\TeV & $  9.1$ & $  1.6$ & $  0.8$ & $    11.5 \pm     1.7 \phantom{0}$  & $    2020 \pm     120 \phantom{0} $ & $   2047$ & $0.012$ &  18.8  \\

1-jet low-$\pt^{\Pgth}$ 8\TeV & $ 36.0$ & $  6.0$ & $  3.0$ & $    45.0 \pm     6.0 \phantom{0}$  & $    9030 \pm     360 \phantom{0} $ & $   9010$ & $0.010$ &  18.6  \\

1-jet high-$\pt^{\Pgth}$ 7\TeV & $  7.7$ & $  1.1$ & $  0.6$ & $     9.4 \pm     1.3 \phantom{0}$  & $     796 \pm      39 \phantom{00} $ & $    817$ & $0.033$ &  19.1  \\

1-jet high-$\pt^{\Pgth}$ 8\TeV & $ 29.6$ & $  4.3$ & $  2.4$ & $    36.3 \pm     4.6 \phantom{0}$  & $    3180 \pm     130 \phantom{0} $ & $   3160$ & $0.029$ &  19.7  \\

1-jet high-$\pt^{\Pgth}$ boosted 7\TeV & $  2.6$ & $  0.8$ & $  0.5$ & $     3.9 \pm     0.6 \phantom{0}$  & $     282 \pm      16 \phantom{00} $ & $    269$ & $0.054$ &  17.7  \\

1-jet high-$\pt^{\Pgth}$ boosted 8\TeV & $ 11.5$ & $  2.9$ & $  2.0$ & $    16.5 \pm     2.6 \phantom{0}$  & $    1265 \pm      62 \phantom{00} $ & $   1253$ & $0.072$ &  17.2  \\

VBF tag 7\TeV                & $  0.2$ & $  1.3$ & $  -$ & $     1.6 \pm     0.1 \phantom{0}$  & $      22 \pm       2 \phantom{000} $ & $     23$ & $0.14$ &  19.6  \\

Loose VBF tag 8\TeV          & $  1.1$ & $  3.4$ & $  -$ & $     4.5 \pm     0.4 \phantom{0}$  & $      81 \pm       7 \phantom{000} $ & $     76$ & $0.17$ &  17.0  \\

Tight VBF tag 8\TeV          & $  0.3$ & $  2.0$ & $  -$ & $     2.4 \pm     0.2 \phantom{0}$  & $      15 \pm       2 \phantom{000} $ & $     20$ & $0.49$ &  18.1  \\
\hline

$\Pe\Pgth$ & & & & & & & & \\

0-jet low-$\pt^{\Pgth}$ 7\TeV & $ 11.8$ & $  0.1$ & $  0.1$ & $    12.0 \pm     1.8 \phantom{0}$  & $    6140 \pm     320 \phantom{0} $ & $   6238$ & $0.002$ &  16.4  \\

0-jet low-$\pt^{\Pgth}$ 8\TeV & $ 33.4$ & $  0.3$ & $  0.2$ & $    34.0 \pm     4.6 \phantom{0}$  & $   16750 \pm     750 \phantom{0} $ & $  17109$ & $0.002$ &  15.8  \\

0-jet high-$\pt^{\Pgth}$ 7\TeV & $ 11.1$ & $  0.1$ & $  0.1$ & $    11.3 \pm     1.7 \phantom{0}$  & $    1159 \pm      62 \phantom{00} $ & $   1191$ & $0.015$ &  14.3  \\

0-jet high-$\pt^{\Pgth}$ 8\TeV & $ 31.4$ & $  0.3$ & $  0.3$ & $    32.1 \pm     4.4 \phantom{0}$  & $    4380 \pm     170 \phantom{0} $ & $   4536$ & $0.010$ &  15.4  \\

1-jet low-$\pt^{\Pgth}$ 7\TeV & $  3.1$ & $  0.6$ & $  0.3$ & $     4.0 \pm     0.6 \phantom{0}$  & $     366 \pm      25 \phantom{00} $ & $    385$ & $0.029$ &  19.6  \\

1-jet low-$\pt^{\Pgth}$ 8\TeV & $  9.1$ & $  1.8$ & $  1.0$ & $    11.9 \pm     1.6 \phantom{0}$  & $    1200 \pm      56 \phantom{00} $ & $   1214$ & $0.025$ &  16.5  \\

1-jet high-$\pt^{\Pgth}$ boosted 7\TeV & $  1.2$ & $  0.3$ & $  0.2$ & $     1.8 \pm     0.3 \phantom{0}$  & $     150 \pm       9 \phantom{000} $ & $    167$ & $0.089$ &  15.5  \\

1-jet high-$\pt^{\Pgth}$ boosted 8\TeV & $  5.1$ & $  1.4$ & $  0.9$ & $     7.5 \pm     1.1 \phantom{0}$  & $     497 \pm      27 \phantom{00} $ & $    476$ & $0.11$ &  15.5  \\

VBF tag 7\TeV                & $  0.2$ & $  0.7$ & $  -$ & $     0.9 \pm     0.1 \phantom{0}$  & $      14 \pm       2 \phantom{000} $ & $     13$ & $0.24$ &  15.9  \\

Loose VBF tag 8\TeV          & $  0.6$ & $  1.8$ & $  -$ & $     2.4 \pm     0.2 \phantom{0}$  & $      45 \pm       4 \phantom{000} $ & $     40$ & $0.14$ &  16.7  \\

Tight VBF tag 8\TeV          & $  0.3$ & $  1.3$ & $  -$ & $     1.6 \pm     0.1 \phantom{0}$  & $       9 \pm       2 \phantom{000} $ & $      7$ & $0.51$ &  16.2  \\
\hline

$\Pgth\Pgth$ & & & & & & & & \\

1-jet boosted 8\TeV          & $  7.2$ & $  2.1$ & $  1.0$ & $    10.3 \pm     1.7 \phantom{0}$  & $    1133 \pm      49 \phantom{00} $ & $   1120$ & $0.054$ &  15.2  \\

1-jet highly-boosted 8\TeV   & $  5.6$ & $  1.6$ & $  1.2$ & $     8.4 \pm     1.2 \phantom{0}$  & $     380 \pm      23 \phantom{00} $ & $    366$ & $0.14$ &  13.1  \\

VBF tag 8\TeV                & $  0.5$ & $  2.4$ & $  -$ & $     3.0 \pm     0.3 \phantom{0}$  & $      29 \pm       4 \phantom{000} $ & $     34$ & $0.32$ &  14.3  \\
\hline

$\Pe\mu$ & & & & & & & & \\

0-jet low-$\pt^\mu$ 7\TeV    & $ 20.8$ & $  0.2$ & $  0.2$ & $    21.1 \pm     3.0 \phantom{0}$  & $   11320 \pm     260 \phantom{0} $ & $  11283$ & $0.002$ &  24.4  \\

0-jet low-$\pt^\mu$ 8\TeV    & $ 70.3$ & $  0.7$ & $  0.7$ & $    71.7 \pm     9.6 \phantom{0}$  & $   40410 \pm     830 \phantom{0} $ & $  40381$ & $0.002$ &  23.6  \\

0-jet high-$\pt^\mu$ 7\TeV   & $  7.5$ & $  0.1$ & $  0.1$ & $     7.8 \pm     1.1 \phantom{0}$  & $    1636 \pm      55 \phantom{00} $ & $   1676$ & $0.007$ &  22.7  \\

0-jet high-$\pt^\mu$ 8\TeV   & $ 24.0$ & $  0.2$ & $  0.5$ & $    24.7 \pm     3.3 \phantom{0}$  & $    6000 \pm     150 \phantom{0} $ & $   6095$ & $0.006$ &  20.7  \\

1-jet low-$\pt^\mu$ 7\TeV    & $  9.0$ & $  1.6$ & $  1.0$ & $    11.7 \pm     1.5 \phantom{0}$  & $    2475 \pm      74 \phantom{00} $ & $   2482$ & $0.009$ &  23.7  \\

1-jet low-$\pt^\mu$ 8\TeV    & $ 40.6$ & $  6.5$ & $  3.7$ & $    50.8 \pm     6.1 \phantom{0}$  & $   10910 \pm     250 \phantom{0} $ & $  10926$ & $0.007$ &  23.8  \\

1-jet high-$\pt^\mu$ 7\TeV   & $  4.7$ & $  1.0$ & $  0.6$ & $     6.2 \pm     0.8 \phantom{0}$  & $     928 \pm      37 \phantom{00} $ & $    901$ & $0.015$ &  23.3  \\

1-jet high-$\pt^\mu$ 8\TeV   & $ 18.0$ & $  3.4$ & $  2.6$ & $    23.9 \pm     2.9 \phantom{0}$  & $    4040 \pm     110 \phantom{0} $ & $   4050$ & $0.014$ &  23.1  \\

Loose VBF tag 7\TeV          & $  0.2$ & $  1.0$ & $  -$ & $     1.2 \pm     0.1 \phantom{0}$  & $      19 \pm       1 \phantom{000} $ & $     12$ & $0.13$ &  23.0  \\

Loose VBF tag 8\TeV          & $  0.6$ & $  2.6$ & $  -$ & $     3.3 \pm     0.3 \phantom{0}$  & $      99 \pm       6 \phantom{000} $ & $    112$ & $0.054$ &  23.5  \\

Tight VBF tag 8\TeV          & $  0.2$ & $  1.5$ & $  -$ & $     1.6 \pm     0.1 \phantom{0}$  & $      14 \pm       1 \phantom{000} $ & $     17$ & $0.31$ &  17.8  \\
\hline
\end{tabular}
\end{table}

\begin{table}[htbp]
\centering
\topcaption{Observed and predicted event yields in all event categories of the $\Pe\Pe$ and $\mu\mu$ channels for the full discriminator value $D$ region. The event yields of the predicted background distributions correspond to the result of the global fit. The signal yields, on the other hand, are normalized to the standard model prediction.  The different signal processes are labelled as ggH (gluon-gluon fusion), VH (production in association with a $\PW$ or $\PZ$~boson), and VBF (vector-boson fusion).
\label{tab:event_yields_eemumu}
}
\small
\begin{tabular}{l|rrrr|r|r}
\hline
 & \multicolumn{4}{c|}{ SM Higgs ($\mH = 125$\GeV) } & & \\
Event category & ggH & VBF & VH & $\Sigma$ signal & \multicolumn{1}{c|}{Background} & \multicolumn{1}{c}{Data}  \\
\hline

$\mu\mu$ & & & & & & \\
0-jet low-$\pt^\mu$ 7\TeV    & $  8.0$ & $  0.1$ & $  0.1$  & $     8.2 \pm     1.2 \phantom{0}$  & $  266200 \pm    1400 \phantom{} $ & $ 266365$  \\

0-jet low-$\pt^\mu$ 8\TeV    & $ 25.4$ & $  0.3$ & $  0.6$  & $    26.4 \pm     3.8 \phantom{0}$  & $  873200 \pm    2600 \phantom{} $ & $ 873709$  \\

0-jet high-$\pt^\mu$ 7\TeV   & $  5.5$ & $  0.1$ & $  0.3$  & $     5.9 \pm     0.8 \phantom{0}$  & $  982900 \pm    2100 \phantom{} $ & $ 982442$  \\

0-jet high-$\pt^\mu$ 8\TeV   & $ 30.6$ & $  0.4$ & $  3.5$  & $    34.6 \pm     4.6 \phantom{0}$  & $ 3775700 \pm    3100 \phantom{} $ & $3776365$  \\

1-jet low-$\pt^\mu$ 7\TeV    & $  2.5$ & $  0.4$ & $  0.3$  & $     3.2 \pm     0.4 \phantom{0}$  & $   18680 \pm     180 \phantom{0} $ & $  18757$  \\

1-jet low-$\pt^\mu$ 8\TeV    & $  7.0$ & $  1.0$ & $  0.6$  & $     8.6 \pm     1.1 \phantom{0}$  & $   40900 \pm     360 \phantom{0} $ & $  40606$  \\

1-jet high-$\pt^\mu$ 7\TeV   & $  3.7$ & $  1.4$ & $  1.9$  & $     7.0 \pm     0.6 \phantom{0}$  & $  233600 \pm    1200 \phantom{} $ & $ 234390$  \\

1-jet high-$\pt^\mu$ 8\TeV   & $ 15.1$ & $  2.2$ & $  4.4$  & $    21.7 \pm     2.3 \phantom{0}$  & $  646000 \pm    2500 \phantom{} $ & $ 646549$  \\

2-jet 7\TeV                  & $  1.4$ & $  0.2$ & $  0.7$  & $     2.4 \pm     0.3 \phantom{0}$  & $   33260 \pm     350 \phantom{0} $ & $  33186$  \\

2-jet 8\TeV                  & $  6.3$ & $  3.9$ & $  2.6$  & $    12.8 \pm     1.4 \phantom{0}$  & $  164100 \pm    1400 \phantom{} $ & $ 164469$  \\

\hline

$\Pe\Pe$ & & & & & & \\

0-jet low-$\pt^\Pe$ 7\TeV    & $  3.6$ & $  -$ & $  0.1$  & $     3.7 \pm     0.5 \phantom{0}$  & $  190900 \pm    1000 \phantom{} $ & $ 190890$  \\

0-jet low-$\pt^\Pe$ 8\TeV    & $ 14.3$ & $  0.2$ & $  0.3$  & $    14.7 \pm     2.2 \phantom{0}$  & $  519440 \pm     700 \phantom{0} $ & $ 519376$  \\

0-jet high-$\pt^\Pe$ 7\TeV   & $  4.0$ & $  -$ & $  0.5$  & $     4.5 \pm     0.6 \phantom{0}$  & $  819900 \pm    1700 \phantom{} $ & $ 820035$  \\

0-jet high-$\pt^\Pe$ 8\TeV   & $ 22.3$ & $  0.3$ & $  2.5$  & $    25.1 \pm     3.4 \phantom{0}$  & $ 3225000 \pm    2000 \phantom{} $ & $3225144$  \\

1-jet low-$\pt^\Pe$ 7\TeV    & $  1.5$ & $  0.2$ & $  0.1$  & $     1.8 \pm     0.2 \phantom{0}$  & $   10268 \pm      97 \phantom{00} $ & $  10300$  \\

1-jet low-$\pt^\Pe$ 8\TeV    & $  4.6$ & $  0.6$ & $  0.3$  & $     5.5 \pm     0.7 \phantom{0}$  & $   26570 \pm     180 \phantom{0} $ & $  26604$  \\

1-jet high-$\pt^\Pe$ 7\TeV   & $  2.4$ & $  0.4$ & $  0.6$  & $     3.4 \pm     0.4 \phantom{0}$  & $  144900 \pm     740 \phantom{0} $ & $ 144945$  \\

1-jet high-$\pt^\Pe$ 8\TeV   & $ 11.7$ & $  1.9$ & $  3.2$  & $    16.9 \pm     1.8 \phantom{0}$  & $  560000 \pm    1900 \phantom{} $ & $ 560104$  \\

2-jet 7\TeV                  & $  1.6$ & $  0.6$ & $  0.4$  & $     2.6 \pm     0.4 \phantom{0}$  & $   35800 \pm     280 \phantom{0} $ & $  35796$  \\

2-jet 8\TeV                  & $  5.0$ & $  2.8$ & $  1.6$  & $     9.4 \pm     1.1 \phantom{0}$  & $  140000 \pm    1200 \phantom{} $ & $ 140070$  \\
\hline
\end{tabular}
\end{table}

\begin{table}[htbp]
\centering
\topcaption{ Observed and predicted event yields in all event categories of the $\ell\ell + LL'$and $\ell + L\Pgth$ channels for the full $\mtt$ and $\mvis$ regions, respectively. The event yields of the predicted background distributions correspond to the result of the global fit. The signal yields, on the other hand, are normalized to the standard model prediction. Only SM Higgs boson production ($\mH = 125$\GeV) in association with a $\PW$ or $\PZ$~boson is considered as a signal process. The $\frac{S}{S+B}$ variable denotes the ratio of the signal and the signal-plus-background yields in the central $\mtt$ range containing $68\%$ of the signal events for $\mH = 125$\GeV.
\label{tab:event_yields_vh}
}
\small
\begin{tabular}{l|rr|r|r}
\hline
Event category & \multicolumn{1}{c}{Signal} & Background & Data &  \multicolumn{1}{c}{$\frac{S}{S+B}$} \\
\hline
$\ell\ell +LL'$ & & & & \\
$\Pgm\Pgm + \Pgm\Pgth$ 7\TeV & 0.111 $\pm$ 0.005 & 2.4 $\pm$ 0.3 & 2 & 0.103 \\
$\Pgm\Pgm + \Pgm\Pgth$ 8\TeV & 0.427 $\pm$ 0.021 & 10.5 $\pm$ 0.6 & 12 & 0.092 \\
$\Pe\Pe + \Pgm\Pgth$ 7\TeV & 0.087 $\pm$ 0.004 & 1.5 $\pm$ 0.1 & 2 & 0.135 \\
$\Pe\Pe + \Pgm\Pgth$ 8\TeV & 0.385 $\pm$ 0.018 & 7.6 $\pm$ 0.4 & 11 & 0.149 \\
$\Pgm\Pgm + \Pe\Pgth$ 7\TeV & 0.078 $\pm$ 0.004 & 2.2 $\pm$ 0.1 & 1 & 0.092\\
$\Pgm\Pgm + \Pe\Pgth$ 8\TeV & 0.293 $\pm$ 0.014 & 12.2 $\pm$ 0.6 & 8 & 0.081\\
$\Pe\Pe + \Pe\Pgth$ 7\TeV & 0.075 $\pm$ 0.004 & 2.2 $\pm$ 0.1 & 4 & 0.077\\
$\Pe\Pe + \Pe\Pgth$ 8\TeV & 0.279 $\pm$ 0.013 & 10.2 $\pm$ 0.5 & 13 & 0.063\\
$\Pgm\Pgm + \Pgth\Pgth$ 7\TeV & 0.073 $\pm$ 0.006 & 0.8 $\pm$ 0.1 & 0 & 0.195\\
$\Pgm\Pgm + \Pgth\Pgth$ 8\TeV & 0.285 $\pm$ 0.022 & 5.8 $\pm$ 0.4 & 4 & 0.150\\
$\Pe\Pe + \Pgth\Pgth$ 7\TeV & 0.061 $\pm$ 0.004 & 1.1 $\pm$ 0.1 & 1 & 0.127\\
$\Pe\Pe + \Pgth\Pgth$ 8\TeV & 0.260 $\pm$ 0.020 & 4.8 $\pm$ 0.4 & 9 & 0.148\\
$\Pgm\Pgm + \Pe\Pgm$ 7\TeV & 0.051 $\pm$ 0.002 & 1.0 $\pm$ 0.1 & 3 & 0.100\\
$\Pgm\Pgm + \Pe\Pgm$ 8\TeV & 0.202 $\pm$ 0.008 & 5.1 $\pm$ 0.3 & 9 & 0.105\\
$\Pe\Pe + \Pe\Pgm$ 7\TeV & 0.045 $\pm$ 0.002 & 1.0 $\pm$ 0.0 & 1 & 0.077 \\
$\Pe\Pe + \Pe\Pgm$ 8\TeV & 0.185 $\pm$ 0.007 & 4.0 $\pm$ 0.2 & 4 & 0.082 \\
\hline
$\ell + \Pgth\Pgth$ & & & \\
$\Pgm + \Pgth\Pgth$ 7\TeV & 0.35 $\pm$ 0.03 & 4.1 $\pm$ 0.4 & 2 & 0.098 \\
$\Pgm + \Pgth\Pgth$ 8\TeV & 1.57 $\pm$ 0.12 & 35.2 $\pm$ 2.1 & 38 & 0.054 \\
$\Pe + \Pgth\Pgth$ 7\TeV & 0.23 $\pm$ 0.02 & 2.7 $\pm$ 0.2 & 0 & 0.101 \\
$\Pe + \Pgth\Pgth$ 8\TeV & 0.87 $\pm$ 0.08 & 16.5 $\pm$ 1.1 & 15 & 0.062 \\
\hline
$\ell + \ell'\Pgth$ & & & & \\
$\Pgm + \Pgm\Pgth$ 7\TeV & 0.33 $\pm$ 0.02 & 3.2 $\pm$ 0.4 & 2 & 0.090 \\
$\Pgm + \Pgm\Pgth$ low \LT 8\TeV & 0.72 $\pm$ 0.03 & 20.7 $\pm$ 2.2 & 19 & 0.046 \\
$\Pgm + \Pgm\Pgth$ high \LT 8\TeV & 0.72 $\pm$ 0.02 & 8.4 $\pm$ 1.3 & 7 & 0.102\\
$\Pe + \Pgm\Pgth$/$\Pgm + \Pe\Pgth$ 7\TeV & 0.47 $\pm$ 0.03 & 6.2 $\pm$ 1.0 & 6 & 0.074 \\
$\Pe + \Pgm\Pgth$/$\Pgm + \Pe\Pgth$ low \LT 8\TeV & 0.92 $\pm$ 0.03 & 24.6 $\pm$ 3.2 & 30 & 0.041\\
$\Pe + \Pgm\Pgth$/$\Pgm + \Pe\Pgth$ high \LT 8\TeV & 1.15 $\pm$ 0.04 & 13.9 $\pm$ 2.0 & 11 & 0.109\\
\hline
\end{tabular}
\end{table}
\cleardoublepage \section{The CMS Collaboration \label{app:collab}}\begin{sloppypar}\hyphenpenalty=5000\widowpenalty=500\clubpenalty=5000\textbf{Yerevan Physics Institute,  Yerevan,  Armenia}\\*[0pt]
S.~Chatrchyan, V.~Khachatryan, A.M.~Sirunyan, A.~Tumasyan
\vskip\cmsinstskip
\textbf{Institut f\"{u}r Hochenergiephysik der OeAW,  Wien,  Austria}\\*[0pt]
W.~Adam, T.~Bergauer, M.~Dragicevic, J.~Er\"{o}, C.~Fabjan\cmsAuthorMark{1}, M.~Friedl, R.~Fr\"{u}hwirth\cmsAuthorMark{1}, V.M.~Ghete, C.~Hartl, N.~H\"{o}rmann, J.~Hrubec, M.~Jeitler\cmsAuthorMark{1}, W.~Kiesenhofer, V.~Kn\"{u}nz, M.~Krammer\cmsAuthorMark{1}, I.~Kr\"{a}tschmer, D.~Liko, I.~Mikulec, D.~Rabady\cmsAuthorMark{2}, B.~Rahbaran, H.~Rohringer, R.~Sch\"{o}fbeck, J.~Strauss, A.~Taurok, W.~Treberer-Treberspurg, W.~Waltenberger, C.-E.~Wulz\cmsAuthorMark{1}
% ...
\vskip\cmsinstskip
% ...
1:~~Also at Vienna University of Technology, Vienna, Austria\\
2:~~Also at CERN, European Organization for Nuclear Research, Geneva, Switzerland\\
% ...

\end{sloppypar}
\end{document}